\documentclass[12pt]{article}
\usepackage{publication}
\usepackage{calc}
\usepackage{graphicx}
\usepackage{amssymb}
\usepackage{nomencl}
\makenomenclature
\usepackage{tikz}
\usepackage{nameref}
\usetikzlibrary{shapes}
\usepackage{tabularx}
\usetikzlibrary{calc}
\usetikzlibrary{arrows}
\usetikzlibrary {arrows.meta}
\usepackage{tikz,pgfplots}
\usepackage{pgfplotstable}
\usetikzlibrary{plotmarks}
\usepackage{multicol}
\usepackage{placeins}
\usepackage{pifont}
\usetikzlibrary{decorations.pathreplacing}
\usepgfplotslibrary{groupplots}
\usepgfplotslibrary{fillbetween}
\usepackage{pgfplots}
\usepackage{xcolor}
\tikzstyle{process} = [rectangle, minimum width=9cm, minimum height=0.8cm, text width=9cm, text centered, draw=black, fill=blue!10, font=\scriptsize]
\tikzstyle{arrow} = [thick,->,>=stealth, font=\scriptsize]
\usepgfplotslibrary{colorbrewer}
    \pgfplotsset{
        colormap/Set1,
        Test1 cycle list/.style={
            cycle multiindex* list={
                mark list*\nextlist
                very thick \nextlist
                PuRd\nextlist%
            },
        },
    }
\pgfplotsset{width=11cm,compat=1.9}
\pgfplotsset{select coords between index/.style 2 args={
		x filter/.code={
			\ifnum\coordindex<#1\fi
			\ifnum\coordindex>#2\fi}}}
\usepackage{amsthm}
\usepackage{caption}
\usepackage{subcaption}
\usepackage{siunitx}
\usepackage[hidelinks,colorlinks,allcolors=blue,breaklinks]{hyperref}
\usepackage{xurl}
\usepackage{natbib}

\definecolor{mts1}{RGB}{253,141,60}
\definecolor{mts2}{RGB}{65,182,196}
\definecolor{mts3}{RGB}{37,52,148}
\definecolor{mts4}{RGB}{116,196,118}
\definecolor{mts5}{RGB}{0,109,44}
\definecolor{mts6}{RGB}{189,0,38}
\definecolor{mts7}{RGB}{240,59,32}

\definecolor{parula-1}{rgb}{0.2081,0.1663,0.5292}
\definecolor{parula-2}{rgb}{0.0146,0.3845,0.8813}
\definecolor{parula-3}{rgb}{0.0795,0.5159,0.8328}
\definecolor{parula-4}{rgb}{0.0228,0.6492,0.7823}
\definecolor{parula-5}{rgb}{0.1986,0.7214,0.6310}
\definecolor{parula-6}{rgb}{0.5456,0.7490,0.4597}
\definecolor{parula-7}{rgb}{0.8266,0.7320,0.3464}
\definecolor{parula-8}{rgb}{0.9948,0.7886,0.1943}
\definecolor{parula-9}{rgb}{0.9763,0.9831,0.0538}
\definecolor{YlGn-3-1}{RGB}{247,252,185}
\definecolor{YlGn-3-C}{RGB}{247,252,185}
\definecolor{YlGn-3-2}{RGB}{173,221,142}
\definecolor{YlGn-3-F}{RGB}{173,221,142}
\definecolor{YlGn-3-3}{RGB}{49,163,84}
\definecolor{YlGn-3-I}{RGB}{49,163,84}
\definecolor{YlGn-4-1}{RGB}{255,255,204}
\definecolor{YlGn-4-B}{RGB}{255,255,204}
\definecolor{YlGn-4-2}{RGB}{194,230,153}
\definecolor{YlGn-4-E}{RGB}{194,230,153}
\definecolor{YlGn-4-3}{RGB}{120,198,121}
\definecolor{YlGn-4-G}{RGB}{120,198,121}
\definecolor{YlGn-4-4}{RGB}{35,132,67}
\definecolor{YlGn-4-J}{RGB}{35,132,67}
\definecolor{YlGn-5-1}{RGB}{255,255,204}
\definecolor{YlGn-5-B}{RGB}{255,255,204}
\definecolor{YlGn-5-2}{RGB}{194,230,153}
\definecolor{YlGn-5-E}{RGB}{194,230,153}
\definecolor{YlGn-5-3}{RGB}{120,198,121}
\definecolor{YlGn-5-G}{RGB}{120,198,121}
\definecolor{YlGn-5-4}{RGB}{49,163,84}
\definecolor{YlGn-5-I}{RGB}{49,163,84}
\definecolor{YlGn-5-5}{RGB}{0,104,55}
\definecolor{YlGn-5-K}{RGB}{0,104,55}
\definecolor{YlGn-6-1}{RGB}{255,255,204}
\definecolor{YlGn-6-B}{RGB}{255,255,204}
\definecolor{YlGn-6-2}{RGB}{217,240,163}
\definecolor{YlGn-6-D}{RGB}{217,240,163}
\definecolor{YlGn-6-3}{RGB}{173,221,142}
\definecolor{YlGn-6-F}{RGB}{173,221,142}
\definecolor{YlGn-6-4}{RGB}{120,198,121}
\definecolor{YlGn-6-G}{RGB}{120,198,121}
\definecolor{YlGn-6-5}{RGB}{49,163,84}
\definecolor{YlGn-6-I}{RGB}{49,163,84}
\definecolor{YlGn-6-6}{RGB}{0,104,55}
\definecolor{YlGn-6-K}{RGB}{0,104,55}
\definecolor{YlGn-7-1}{RGB}{255,255,204}
\definecolor{YlGn-7-B}{RGB}{255,255,204}
\definecolor{YlGn-7-2}{RGB}{217,240,163}
\definecolor{YlGn-7-D}{RGB}{217,240,163}
\definecolor{YlGn-7-3}{RGB}{173,221,142}
\definecolor{YlGn-7-F}{RGB}{173,221,142}
\definecolor{YlGn-7-4}{RGB}{120,198,121}
\definecolor{YlGn-7-G}{RGB}{120,198,121}
\definecolor{YlGn-7-5}{RGB}{65,171,93}
\definecolor{YlGn-7-H}{RGB}{65,171,93}
\definecolor{YlGn-7-6}{RGB}{35,132,67}
\definecolor{YlGn-7-J}{RGB}{35,132,67}
\definecolor{YlGn-7-7}{RGB}{0,90,50}
\definecolor{YlGn-7-L}{RGB}{0,90,50}
\definecolor{YlGn-8-1}{RGB}{255,255,229}
\definecolor{YlGn-8-A}{RGB}{255,255,229}
\definecolor{YlGn-8-2}{RGB}{247,252,185}
\definecolor{YlGn-8-C}{RGB}{247,252,185}
\definecolor{YlGn-8-3}{RGB}{217,240,163}
\definecolor{YlGn-8-D}{RGB}{217,240,163}
\definecolor{YlGn-8-4}{RGB}{173,221,142}
\definecolor{YlGn-8-F}{RGB}{173,221,142}
\definecolor{YlGn-8-5}{RGB}{120,198,121}
\definecolor{YlGn-8-G}{RGB}{120,198,121}
\definecolor{YlGn-8-6}{RGB}{65,171,93}
\definecolor{YlGn-8-H}{RGB}{65,171,93}
\definecolor{YlGn-8-7}{RGB}{35,132,67}
\definecolor{YlGn-8-J}{RGB}{35,132,67}
\definecolor{YlGn-8-8}{RGB}{0,90,50}
\definecolor{YlGn-8-L}{RGB}{0,90,50}
\definecolor{YlGn-9-1}{RGB}{255,255,229}
\definecolor{YlGn-9-A}{RGB}{255,255,229}
\definecolor{YlGn-9-2}{RGB}{247,252,185}
\definecolor{YlGn-9-C}{RGB}{247,252,185}
\definecolor{YlGn-9-3}{RGB}{217,240,163}
\definecolor{YlGn-9-D}{RGB}{217,240,163}
\definecolor{YlGn-9-4}{RGB}{173,221,142}
\definecolor{YlGn-9-F}{RGB}{173,221,142}
\definecolor{YlGn-9-5}{RGB}{120,198,121}
\definecolor{YlGn-9-G}{RGB}{120,198,121}
\definecolor{YlGn-9-6}{RGB}{65,171,93}
\definecolor{YlGn-9-H}{RGB}{65,171,93}
\definecolor{YlGn-9-7}{RGB}{35,132,67}
\definecolor{YlGn-9-J}{RGB}{35,132,67}
\definecolor{YlGn-9-8}{RGB}{0,104,55}
\definecolor{YlGn-9-K}{RGB}{0,104,55}
\definecolor{YlGn-9-9}{RGB}{0,69,41}
\definecolor{YlGn-9-M}{RGB}{0,69,41}
\definecolor{YlGnBu-3-1}{RGB}{237,248,177}
\definecolor{YlGnBu-3-C}{RGB}{237,248,177}
\definecolor{YlGnBu-3-2}{RGB}{127,205,187}
\definecolor{YlGnBu-3-F}{RGB}{127,205,187}
\definecolor{YlGnBu-3-3}{RGB}{44,127,184}
\definecolor{YlGnBu-3-I}{RGB}{44,127,184}
\definecolor{YlGnBu-4-1}{RGB}{255,255,204}
\definecolor{YlGnBu-4-B}{RGB}{255,255,204}
\definecolor{YlGnBu-4-2}{RGB}{161,218,180}
\definecolor{YlGnBu-4-E}{RGB}{161,218,180}
\definecolor{YlGnBu-4-3}{RGB}{65,182,196}
\definecolor{YlGnBu-4-G}{RGB}{65,182,196}
\definecolor{YlGnBu-4-4}{RGB}{34,94,168}
\definecolor{YlGnBu-4-J}{RGB}{34,94,168}
\definecolor{YlGnBu-5-1}{RGB}{255,255,204}
\definecolor{YlGnBu-5-B}{RGB}{255,255,204}
\definecolor{YlGnBu-5-2}{RGB}{161,218,180}
\definecolor{YlGnBu-5-E}{RGB}{161,218,180}
\definecolor{YlGnBu-5-3}{RGB}{65,182,196}
\definecolor{YlGnBu-5-G}{RGB}{65,182,196}
\definecolor{YlGnBu-5-4}{RGB}{44,127,184}
\definecolor{YlGnBu-5-I}{RGB}{44,127,184}
\definecolor{YlGnBu-5-5}{RGB}{37,52,148}
\definecolor{YlGnBu-5-K}{RGB}{37,52,148}
\definecolor{YlGnBu-6-1}{RGB}{255,255,204}
\definecolor{YlGnBu-6-B}{RGB}{255,255,204}
\definecolor{YlGnBu-6-2}{RGB}{199,233,180}
\definecolor{YlGnBu-6-D}{RGB}{199,233,180}
\definecolor{YlGnBu-6-3}{RGB}{127,205,187}
\definecolor{YlGnBu-6-F}{RGB}{127,205,187}
\definecolor{YlGnBu-6-4}{RGB}{65,182,196}
\definecolor{YlGnBu-6-G}{RGB}{65,182,196}
\definecolor{YlGnBu-6-5}{RGB}{44,127,184}
\definecolor{YlGnBu-6-I}{RGB}{44,127,184}
\definecolor{YlGnBu-6-6}{RGB}{37,52,148}
\definecolor{YlGnBu-6-K}{RGB}{37,52,148}
\definecolor{YlGnBu-7-1}{RGB}{255,255,204}
\definecolor{YlGnBu-7-B}{RGB}{255,255,204}
\definecolor{YlGnBu-7-2}{RGB}{199,233,180}
\definecolor{YlGnBu-7-D}{RGB}{199,233,180}
\definecolor{YlGnBu-7-3}{RGB}{127,205,187}
\definecolor{YlGnBu-7-F}{RGB}{127,205,187}
\definecolor{YlGnBu-7-4}{RGB}{65,182,196}
\definecolor{YlGnBu-7-G}{RGB}{65,182,196}
\definecolor{YlGnBu-7-5}{RGB}{29,145,192}
\definecolor{YlGnBu-7-H}{RGB}{29,145,192}
\definecolor{YlGnBu-7-6}{RGB}{34,94,168}
\definecolor{YlGnBu-7-J}{RGB}{34,94,168}
\definecolor{YlGnBu-7-7}{RGB}{12,44,132}
\definecolor{YlGnBu-7-L}{RGB}{12,44,132}
\definecolor{YlGnBu-8-1}{RGB}{255,255,217}
\definecolor{YlGnBu-8-A}{RGB}{255,255,217}
\definecolor{YlGnBu-8-2}{RGB}{237,248,177}
\definecolor{YlGnBu-8-C}{RGB}{237,248,177}
\definecolor{YlGnBu-8-3}{RGB}{199,233,180}
\definecolor{YlGnBu-8-D}{RGB}{199,233,180}
\definecolor{YlGnBu-8-4}{RGB}{127,205,187}
\definecolor{YlGnBu-8-F}{RGB}{127,205,187}
\definecolor{YlGnBu-8-5}{RGB}{65,182,196}
\definecolor{YlGnBu-8-G}{RGB}{65,182,196}
\definecolor{YlGnBu-8-6}{RGB}{29,145,192}
\definecolor{YlGnBu-8-H}{RGB}{29,145,192}
\definecolor{YlGnBu-8-7}{RGB}{34,94,168}
\definecolor{YlGnBu-8-J}{RGB}{34,94,168}
\definecolor{YlGnBu-8-8}{RGB}{12,44,132}
\definecolor{YlGnBu-8-L}{RGB}{12,44,132}
\definecolor{YlGnBu-9-1}{RGB}{255,255,217}
\definecolor{YlGnBu-9-A}{RGB}{255,255,217}
\definecolor{YlGnBu-9-2}{RGB}{237,248,177}
\definecolor{YlGnBu-9-C}{RGB}{237,248,177}
\definecolor{YlGnBu-9-3}{RGB}{199,233,180}
\definecolor{YlGnBu-9-D}{RGB}{199,233,180}
\definecolor{YlGnBu-9-4}{RGB}{127,205,187}
\definecolor{YlGnBu-9-F}{RGB}{127,205,187}
\definecolor{YlGnBu-9-5}{RGB}{65,182,196}
\definecolor{YlGnBu-9-G}{RGB}{65,182,196}
\definecolor{YlGnBu-9-6}{RGB}{29,145,192}
\definecolor{YlGnBu-9-H}{RGB}{29,145,192}
\definecolor{YlGnBu-9-7}{RGB}{34,94,168}
\definecolor{YlGnBu-9-J}{RGB}{34,94,168}
\definecolor{YlGnBu-9-8}{RGB}{37,52,148}
\definecolor{YlGnBu-9-K}{RGB}{37,52,148}
\definecolor{YlGnBu-9-9}{RGB}{8,29,88}
\definecolor{YlGnBu-9-M}{RGB}{8,29,88}
\definecolor{GnBu-3-1}{RGB}{224,243,219}
\definecolor{GnBu-3-C}{RGB}{224,243,219}
\definecolor{GnBu-3-2}{RGB}{168,221,181}
\definecolor{GnBu-3-F}{RGB}{168,221,181}
\definecolor{GnBu-3-3}{RGB}{67,162,202}
\definecolor{GnBu-3-I}{RGB}{67,162,202}
\definecolor{GnBu-4-1}{RGB}{240,249,232}
\definecolor{GnBu-4-B}{RGB}{240,249,232}
\definecolor{GnBu-4-2}{RGB}{186,228,188}
\definecolor{GnBu-4-E}{RGB}{186,228,188}
\definecolor{GnBu-4-3}{RGB}{123,204,196}
\definecolor{GnBu-4-G}{RGB}{123,204,196}
\definecolor{GnBu-4-4}{RGB}{43,140,190}
\definecolor{GnBu-4-J}{RGB}{43,140,190}
\definecolor{GnBu-5-1}{RGB}{240,249,232}
\definecolor{GnBu-5-B}{RGB}{240,249,232}
\definecolor{GnBu-5-2}{RGB}{186,228,188}
\definecolor{GnBu-5-E}{RGB}{186,228,188}
\definecolor{GnBu-5-3}{RGB}{123,204,196}
\definecolor{GnBu-5-G}{RGB}{123,204,196}
\definecolor{GnBu-5-4}{RGB}{67,162,202}
\definecolor{GnBu-5-I}{RGB}{67,162,202}
\definecolor{GnBu-5-5}{RGB}{8,104,172}
\definecolor{GnBu-5-K}{RGB}{8,104,172}
\definecolor{GnBu-6-1}{RGB}{240,249,232}
\definecolor{GnBu-6-B}{RGB}{240,249,232}
\definecolor{GnBu-6-2}{RGB}{204,235,197}
\definecolor{GnBu-6-D}{RGB}{204,235,197}
\definecolor{GnBu-6-3}{RGB}{168,221,181}
\definecolor{GnBu-6-F}{RGB}{168,221,181}
\definecolor{GnBu-6-4}{RGB}{123,204,196}
\definecolor{GnBu-6-G}{RGB}{123,204,196}
\definecolor{GnBu-6-5}{RGB}{67,162,202}
\definecolor{GnBu-6-I}{RGB}{67,162,202}
\definecolor{GnBu-6-6}{RGB}{8,104,172}
\definecolor{GnBu-6-K}{RGB}{8,104,172}
\definecolor{GnBu-7-1}{RGB}{240,249,232}
\definecolor{GnBu-7-B}{RGB}{240,249,232}
\definecolor{GnBu-7-2}{RGB}{204,235,197}
\definecolor{GnBu-7-D}{RGB}{204,235,197}
\definecolor{GnBu-7-3}{RGB}{168,221,181}
\definecolor{GnBu-7-F}{RGB}{168,221,181}
\definecolor{GnBu-7-4}{RGB}{123,204,196}
\definecolor{GnBu-7-G}{RGB}{123,204,196}
\definecolor{GnBu-7-5}{RGB}{78,179,211}
\definecolor{GnBu-7-H}{RGB}{78,179,211}
\definecolor{GnBu-7-6}{RGB}{43,140,190}
\definecolor{GnBu-7-J}{RGB}{43,140,190}
\definecolor{GnBu-7-7}{RGB}{8,88,158}
\definecolor{GnBu-7-L}{RGB}{8,88,158}
\definecolor{GnBu-8-1}{RGB}{247,252,240}
\definecolor{GnBu-8-A}{RGB}{247,252,240}
\definecolor{GnBu-8-2}{RGB}{224,243,219}
\definecolor{GnBu-8-C}{RGB}{224,243,219}
\definecolor{GnBu-8-3}{RGB}{204,235,197}
\definecolor{GnBu-8-D}{RGB}{204,235,197}
\definecolor{GnBu-8-4}{RGB}{168,221,181}
\definecolor{GnBu-8-F}{RGB}{168,221,181}
\definecolor{GnBu-8-5}{RGB}{123,204,196}
\definecolor{GnBu-8-G}{RGB}{123,204,196}
\definecolor{GnBu-8-6}{RGB}{78,179,211}
\definecolor{GnBu-8-H}{RGB}{78,179,211}
\definecolor{GnBu-8-7}{RGB}{43,140,190}
\definecolor{GnBu-8-J}{RGB}{43,140,190}
\definecolor{GnBu-8-8}{RGB}{8,88,158}
\definecolor{GnBu-8-L}{RGB}{8,88,158}
\definecolor{GnBu-9-1}{RGB}{247,252,240}
\definecolor{GnBu-9-A}{RGB}{247,252,240}
\definecolor{GnBu-9-2}{RGB}{224,243,219}
\definecolor{GnBu-9-C}{RGB}{224,243,219}
\definecolor{GnBu-9-3}{RGB}{204,235,197}
\definecolor{GnBu-9-D}{RGB}{204,235,197}
\definecolor{GnBu-9-4}{RGB}{168,221,181}
\definecolor{GnBu-9-F}{RGB}{168,221,181}
\definecolor{GnBu-9-5}{RGB}{123,204,196}
\definecolor{GnBu-9-G}{RGB}{123,204,196}
\definecolor{GnBu-9-6}{RGB}{78,179,211}
\definecolor{GnBu-9-H}{RGB}{78,179,211}
\definecolor{GnBu-9-7}{RGB}{43,140,190}
\definecolor{GnBu-9-J}{RGB}{43,140,190}
\definecolor{GnBu-9-8}{RGB}{8,104,172}
\definecolor{GnBu-9-K}{RGB}{8,104,172}
\definecolor{GnBu-9-9}{RGB}{8,64,129}
\definecolor{GnBu-9-M}{RGB}{8,64,129}
\definecolor{BuGn-3-1}{RGB}{229,245,249}
\definecolor{BuGn-3-C}{RGB}{229,245,249}
\definecolor{BuGn-3-2}{RGB}{153,216,201}
\definecolor{BuGn-3-F}{RGB}{153,216,201}
\definecolor{BuGn-3-3}{RGB}{44,162,95}
\definecolor{BuGn-3-I}{RGB}{44,162,95}
\definecolor{BuGn-4-1}{RGB}{237,248,251}
\definecolor{BuGn-4-B}{RGB}{237,248,251}
\definecolor{BuGn-4-2}{RGB}{178,226,226}
\definecolor{BuGn-4-E}{RGB}{178,226,226}
\definecolor{BuGn-4-3}{RGB}{102,194,164}
\definecolor{BuGn-4-G}{RGB}{102,194,164}
\definecolor{BuGn-4-4}{RGB}{35,139,69}
\definecolor{BuGn-4-J}{RGB}{35,139,69}
\definecolor{BuGn-5-1}{RGB}{237,248,251}
\definecolor{BuGn-5-B}{RGB}{237,248,251}
\definecolor{BuGn-5-2}{RGB}{178,226,226}
\definecolor{BuGn-5-E}{RGB}{178,226,226}
\definecolor{BuGn-5-3}{RGB}{102,194,164}
\definecolor{BuGn-5-G}{RGB}{102,194,164}
\definecolor{BuGn-5-4}{RGB}{44,162,95}
\definecolor{BuGn-5-I}{RGB}{44,162,95}
\definecolor{BuGn-5-5}{RGB}{0,109,44}
\definecolor{BuGn-5-K}{RGB}{0,109,44}
\definecolor{BuGn-6-1}{RGB}{237,248,251}
\definecolor{BuGn-6-B}{RGB}{237,248,251}
\definecolor{BuGn-6-2}{RGB}{204,236,230}
\definecolor{BuGn-6-D}{RGB}{204,236,230}
\definecolor{BuGn-6-3}{RGB}{153,216,201}
\definecolor{BuGn-6-F}{RGB}{153,216,201}
\definecolor{BuGn-6-4}{RGB}{102,194,164}
\definecolor{BuGn-6-G}{RGB}{102,194,164}
\definecolor{BuGn-6-5}{RGB}{44,162,95}
\definecolor{BuGn-6-I}{RGB}{44,162,95}
\definecolor{BuGn-6-6}{RGB}{0,109,44}
\definecolor{BuGn-6-K}{RGB}{0,109,44}
\definecolor{BuGn-7-1}{RGB}{237,248,251}
\definecolor{BuGn-7-B}{RGB}{237,248,251}
\definecolor{BuGn-7-2}{RGB}{204,236,230}
\definecolor{BuGn-7-D}{RGB}{204,236,230}
\definecolor{BuGn-7-3}{RGB}{153,216,201}
\definecolor{BuGn-7-F}{RGB}{153,216,201}
\definecolor{BuGn-7-4}{RGB}{102,194,164}
\definecolor{BuGn-7-G}{RGB}{102,194,164}
\definecolor{BuGn-7-5}{RGB}{65,174,118}
\definecolor{BuGn-7-H}{RGB}{65,174,118}
\definecolor{BuGn-7-6}{RGB}{35,139,69}
\definecolor{BuGn-7-J}{RGB}{35,139,69}
\definecolor{BuGn-7-7}{RGB}{0,88,36}
\definecolor{BuGn-7-L}{RGB}{0,88,36}
\definecolor{BuGn-8-1}{RGB}{247,252,253}
\definecolor{BuGn-8-A}{RGB}{247,252,253}
\definecolor{BuGn-8-2}{RGB}{229,245,249}
\definecolor{BuGn-8-C}{RGB}{229,245,249}
\definecolor{BuGn-8-3}{RGB}{204,236,230}
\definecolor{BuGn-8-D}{RGB}{204,236,230}
\definecolor{BuGn-8-4}{RGB}{153,216,201}
\definecolor{BuGn-8-F}{RGB}{153,216,201}
\definecolor{BuGn-8-5}{RGB}{102,194,164}
\definecolor{BuGn-8-G}{RGB}{102,194,164}
\definecolor{BuGn-8-6}{RGB}{65,174,118}
\definecolor{BuGn-8-H}{RGB}{65,174,118}
\definecolor{BuGn-8-7}{RGB}{35,139,69}
\definecolor{BuGn-8-J}{RGB}{35,139,69}
\definecolor{BuGn-8-8}{RGB}{0,88,36}
\definecolor{BuGn-8-L}{RGB}{0,88,36}
\definecolor{BuGn-9-1}{RGB}{247,252,253}
\definecolor{BuGn-9-A}{RGB}{247,252,253}
\definecolor{BuGn-9-2}{RGB}{229,245,249}
\definecolor{BuGn-9-C}{RGB}{229,245,249}
\definecolor{BuGn-9-3}{RGB}{204,236,230}
\definecolor{BuGn-9-D}{RGB}{204,236,230}
\definecolor{BuGn-9-4}{RGB}{153,216,201}
\definecolor{BuGn-9-F}{RGB}{153,216,201}
\definecolor{BuGn-9-5}{RGB}{102,194,164}
\definecolor{BuGn-9-G}{RGB}{102,194,164}
\definecolor{BuGn-9-6}{RGB}{65,174,118}
\definecolor{BuGn-9-H}{RGB}{65,174,118}
\definecolor{BuGn-9-7}{RGB}{35,139,69}
\definecolor{BuGn-9-J}{RGB}{35,139,69}
\definecolor{BuGn-9-8}{RGB}{0,109,44}
\definecolor{BuGn-9-K}{RGB}{0,109,44}
\definecolor{BuGn-9-9}{RGB}{0,68,27}
\definecolor{BuGn-9-M}{RGB}{0,68,27}
\definecolor{PuBuGn-3-1}{RGB}{236,226,240}
\definecolor{PuBuGn-3-C}{RGB}{236,226,240}
\definecolor{PuBuGn-3-2}{RGB}{166,189,219}
\definecolor{PuBuGn-3-F}{RGB}{166,189,219}
\definecolor{PuBuGn-3-3}{RGB}{28,144,153}
\definecolor{PuBuGn-3-I}{RGB}{28,144,153}
\definecolor{PuBuGn-4-1}{RGB}{246,239,247}
\definecolor{PuBuGn-4-B}{RGB}{246,239,247}
\definecolor{PuBuGn-4-2}{RGB}{189,201,225}
\definecolor{PuBuGn-4-E}{RGB}{189,201,225}
\definecolor{PuBuGn-4-3}{RGB}{103,169,207}
\definecolor{PuBuGn-4-G}{RGB}{103,169,207}
\definecolor{PuBuGn-4-4}{RGB}{2,129,138}
\definecolor{PuBuGn-4-J}{RGB}{2,129,138}
\definecolor{PuBuGn-5-1}{RGB}{246,239,247}
\definecolor{PuBuGn-5-B}{RGB}{246,239,247}
\definecolor{PuBuGn-5-2}{RGB}{189,201,225}
\definecolor{PuBuGn-5-E}{RGB}{189,201,225}
\definecolor{PuBuGn-5-3}{RGB}{103,169,207}
\definecolor{PuBuGn-5-G}{RGB}{103,169,207}
\definecolor{PuBuGn-5-4}{RGB}{28,144,153}
\definecolor{PuBuGn-5-I}{RGB}{28,144,153}
\definecolor{PuBuGn-5-5}{RGB}{1,108,89}
\definecolor{PuBuGn-5-K}{RGB}{1,108,89}
\definecolor{PuBuGn-6-1}{RGB}{246,239,247}
\definecolor{PuBuGn-6-B}{RGB}{246,239,247}
\definecolor{PuBuGn-6-2}{RGB}{208,209,230}
\definecolor{PuBuGn-6-D}{RGB}{208,209,230}
\definecolor{PuBuGn-6-3}{RGB}{166,189,219}
\definecolor{PuBuGn-6-F}{RGB}{166,189,219}
\definecolor{PuBuGn-6-4}{RGB}{103,169,207}
\definecolor{PuBuGn-6-G}{RGB}{103,169,207}
\definecolor{PuBuGn-6-5}{RGB}{28,144,153}
\definecolor{PuBuGn-6-I}{RGB}{28,144,153}
\definecolor{PuBuGn-6-6}{RGB}{1,108,89}
\definecolor{PuBuGn-6-K}{RGB}{1,108,89}
\definecolor{PuBuGn-7-1}{RGB}{246,239,247}
\definecolor{PuBuGn-7-B}{RGB}{246,239,247}
\definecolor{PuBuGn-7-2}{RGB}{208,209,230}
\definecolor{PuBuGn-7-D}{RGB}{208,209,230}
\definecolor{PuBuGn-7-3}{RGB}{166,189,219}
\definecolor{PuBuGn-7-F}{RGB}{166,189,219}
\definecolor{PuBuGn-7-4}{RGB}{103,169,207}
\definecolor{PuBuGn-7-G}{RGB}{103,169,207}
\definecolor{PuBuGn-7-5}{RGB}{54,144,192}
\definecolor{PuBuGn-7-H}{RGB}{54,144,192}
\definecolor{PuBuGn-7-6}{RGB}{2,129,138}
\definecolor{PuBuGn-7-J}{RGB}{2,129,138}
\definecolor{PuBuGn-7-7}{RGB}{1,100,80}
\definecolor{PuBuGn-7-L}{RGB}{1,100,80}
\definecolor{PuBuGn-8-1}{RGB}{255,247,251}
\definecolor{PuBuGn-8-A}{RGB}{255,247,251}
\definecolor{PuBuGn-8-2}{RGB}{236,226,240}
\definecolor{PuBuGn-8-C}{RGB}{236,226,240}
\definecolor{PuBuGn-8-3}{RGB}{208,209,230}
\definecolor{PuBuGn-8-D}{RGB}{208,209,230}
\definecolor{PuBuGn-8-4}{RGB}{166,189,219}
\definecolor{PuBuGn-8-F}{RGB}{166,189,219}
\definecolor{PuBuGn-8-5}{RGB}{103,169,207}
\definecolor{PuBuGn-8-G}{RGB}{103,169,207}
\definecolor{PuBuGn-8-6}{RGB}{54,144,192}
\definecolor{PuBuGn-8-H}{RGB}{54,144,192}
\definecolor{PuBuGn-8-7}{RGB}{2,129,138}
\definecolor{PuBuGn-8-J}{RGB}{2,129,138}
\definecolor{PuBuGn-8-8}{RGB}{1,100,80}
\definecolor{PuBuGn-8-L}{RGB}{1,100,80}
\definecolor{PuBuGn-9-1}{RGB}{255,247,251}
\definecolor{PuBuGn-9-A}{RGB}{255,247,251}
\definecolor{PuBuGn-9-2}{RGB}{236,226,240}
\definecolor{PuBuGn-9-C}{RGB}{236,226,240}
\definecolor{PuBuGn-9-3}{RGB}{208,209,230}
\definecolor{PuBuGn-9-D}{RGB}{208,209,230}
\definecolor{PuBuGn-9-4}{RGB}{166,189,219}
\definecolor{PuBuGn-9-F}{RGB}{166,189,219}
\definecolor{PuBuGn-9-5}{RGB}{103,169,207}
\definecolor{PuBuGn-9-G}{RGB}{103,169,207}
\definecolor{PuBuGn-9-6}{RGB}{54,144,192}
\definecolor{PuBuGn-9-H}{RGB}{54,144,192}
\definecolor{PuBuGn-9-7}{RGB}{2,129,138}
\definecolor{PuBuGn-9-J}{RGB}{2,129,138}
\definecolor{PuBuGn-9-8}{RGB}{1,108,89}
\definecolor{PuBuGn-9-K}{RGB}{1,108,89}
\definecolor{PuBuGn-9-9}{RGB}{1,70,54}
\definecolor{PuBuGn-9-M}{RGB}{1,70,54}
\definecolor{PuBu-3-1}{RGB}{236,231,242}
\definecolor{PuBu-3-C}{RGB}{236,231,242}
\definecolor{PuBu-3-2}{RGB}{166,189,219}
\definecolor{PuBu-3-F}{RGB}{166,189,219}
\definecolor{PuBu-3-3}{RGB}{43,140,190}
\definecolor{PuBu-3-I}{RGB}{43,140,190}
\definecolor{PuBu-4-1}{RGB}{241,238,246}
\definecolor{PuBu-4-B}{RGB}{241,238,246}
\definecolor{PuBu-4-2}{RGB}{189,201,225}
\definecolor{PuBu-4-E}{RGB}{189,201,225}
\definecolor{PuBu-4-3}{RGB}{116,169,207}
\definecolor{PuBu-4-G}{RGB}{116,169,207}
\definecolor{PuBu-4-4}{RGB}{5,112,176}
\definecolor{PuBu-4-J}{RGB}{5,112,176}
\definecolor{PuBu-5-1}{RGB}{241,238,246}
\definecolor{PuBu-5-B}{RGB}{241,238,246}
\definecolor{PuBu-5-2}{RGB}{189,201,225}
\definecolor{PuBu-5-E}{RGB}{189,201,225}
\definecolor{PuBu-5-3}{RGB}{116,169,207}
\definecolor{PuBu-5-G}{RGB}{116,169,207}
\definecolor{PuBu-5-4}{RGB}{43,140,190}
\definecolor{PuBu-5-I}{RGB}{43,140,190}
\definecolor{PuBu-5-5}{RGB}{4,90,141}
\definecolor{PuBu-5-K}{RGB}{4,90,141}
\definecolor{PuBu-6-1}{RGB}{241,238,246}
\definecolor{PuBu-6-B}{RGB}{241,238,246}
\definecolor{PuBu-6-2}{RGB}{208,209,230}
\definecolor{PuBu-6-D}{RGB}{208,209,230}
\definecolor{PuBu-6-3}{RGB}{166,189,219}
\definecolor{PuBu-6-F}{RGB}{166,189,219}
\definecolor{PuBu-6-4}{RGB}{116,169,207}
\definecolor{PuBu-6-G}{RGB}{116,169,207}
\definecolor{PuBu-6-5}{RGB}{43,140,190}
\definecolor{PuBu-6-I}{RGB}{43,140,190}
\definecolor{PuBu-6-6}{RGB}{4,90,141}
\definecolor{PuBu-6-K}{RGB}{4,90,141}
\definecolor{PuBu-7-1}{RGB}{241,238,246}
\definecolor{PuBu-7-B}{RGB}{241,238,246}
\definecolor{PuBu-7-2}{RGB}{208,209,230}
\definecolor{PuBu-7-D}{RGB}{208,209,230}
\definecolor{PuBu-7-3}{RGB}{166,189,219}
\definecolor{PuBu-7-F}{RGB}{166,189,219}
\definecolor{PuBu-7-4}{RGB}{116,169,207}
\definecolor{PuBu-7-G}{RGB}{116,169,207}
\definecolor{PuBu-7-5}{RGB}{54,144,192}
\definecolor{PuBu-7-H}{RGB}{54,144,192}
\definecolor{PuBu-7-6}{RGB}{5,112,176}
\definecolor{PuBu-7-J}{RGB}{5,112,176}
\definecolor{PuBu-7-7}{RGB}{3,78,123}
\definecolor{PuBu-7-L}{RGB}{3,78,123}
\definecolor{PuBu-8-1}{RGB}{255,247,251}
\definecolor{PuBu-8-A}{RGB}{255,247,251}
\definecolor{PuBu-8-2}{RGB}{236,231,242}
\definecolor{PuBu-8-C}{RGB}{236,231,242}
\definecolor{PuBu-8-3}{RGB}{208,209,230}
\definecolor{PuBu-8-D}{RGB}{208,209,230}
\definecolor{PuBu-8-4}{RGB}{166,189,219}
\definecolor{PuBu-8-F}{RGB}{166,189,219}
\definecolor{PuBu-8-5}{RGB}{116,169,207}
\definecolor{PuBu-8-G}{RGB}{116,169,207}
\definecolor{PuBu-8-6}{RGB}{54,144,192}
\definecolor{PuBu-8-H}{RGB}{54,144,192}
\definecolor{PuBu-8-7}{RGB}{5,112,176}
\definecolor{PuBu-8-J}{RGB}{5,112,176}
\definecolor{PuBu-8-8}{RGB}{3,78,123}
\definecolor{PuBu-8-L}{RGB}{3,78,123}
\definecolor{PuBu-9-1}{RGB}{255,247,251}
\definecolor{PuBu-9-A}{RGB}{255,247,251}
\definecolor{PuBu-9-2}{RGB}{236,231,242}
\definecolor{PuBu-9-C}{RGB}{236,231,242}
\definecolor{PuBu-9-3}{RGB}{208,209,230}
\definecolor{PuBu-9-D}{RGB}{208,209,230}
\definecolor{PuBu-9-4}{RGB}{166,189,219}
\definecolor{PuBu-9-F}{RGB}{166,189,219}
\definecolor{PuBu-9-5}{RGB}{116,169,207}
\definecolor{PuBu-9-G}{RGB}{116,169,207}
\definecolor{PuBu-9-6}{RGB}{54,144,192}
\definecolor{PuBu-9-H}{RGB}{54,144,192}
\definecolor{PuBu-9-7}{RGB}{5,112,176}
\definecolor{PuBu-9-J}{RGB}{5,112,176}
\definecolor{PuBu-9-8}{RGB}{4,90,141}
\definecolor{PuBu-9-K}{RGB}{4,90,141}
\definecolor{PuBu-9-9}{RGB}{2,56,88}
\definecolor{PuBu-9-M}{RGB}{2,56,88}
\definecolor{BuPu-3-1}{RGB}{224,236,244}
\definecolor{BuPu-3-C}{RGB}{224,236,244}
\definecolor{BuPu-3-2}{RGB}{158,188,218}
\definecolor{BuPu-3-F}{RGB}{158,188,218}
\definecolor{BuPu-3-3}{RGB}{136,86,167}
\definecolor{BuPu-3-I}{RGB}{136,86,167}
\definecolor{BuPu-4-1}{RGB}{237,248,251}
\definecolor{BuPu-4-B}{RGB}{237,248,251}
\definecolor{BuPu-4-2}{RGB}{179,205,227}
\definecolor{BuPu-4-E}{RGB}{179,205,227}
\definecolor{BuPu-4-3}{RGB}{140,150,198}
\definecolor{BuPu-4-G}{RGB}{140,150,198}
\definecolor{BuPu-4-4}{RGB}{136,65,157}
\definecolor{BuPu-4-J}{RGB}{136,65,157}
\definecolor{BuPu-5-1}{RGB}{237,248,251}
\definecolor{BuPu-5-B}{RGB}{237,248,251}
\definecolor{BuPu-5-2}{RGB}{179,205,227}
\definecolor{BuPu-5-E}{RGB}{179,205,227}
\definecolor{BuPu-5-3}{RGB}{140,150,198}
\definecolor{BuPu-5-G}{RGB}{140,150,198}
\definecolor{BuPu-5-4}{RGB}{136,86,167}
\definecolor{BuPu-5-I}{RGB}{136,86,167}
\definecolor{BuPu-5-5}{RGB}{129,15,124}
\definecolor{BuPu-5-K}{RGB}{129,15,124}
\definecolor{BuPu-6-1}{RGB}{237,248,251}
\definecolor{BuPu-6-B}{RGB}{237,248,251}
\definecolor{BuPu-6-2}{RGB}{191,211,230}
\definecolor{BuPu-6-D}{RGB}{191,211,230}
\definecolor{BuPu-6-3}{RGB}{158,188,218}
\definecolor{BuPu-6-F}{RGB}{158,188,218}
\definecolor{BuPu-6-4}{RGB}{140,150,198}
\definecolor{BuPu-6-G}{RGB}{140,150,198}
\definecolor{BuPu-6-5}{RGB}{136,86,167}
\definecolor{BuPu-6-I}{RGB}{136,86,167}
\definecolor{BuPu-6-6}{RGB}{129,15,124}
\definecolor{BuPu-6-K}{RGB}{129,15,124}
\definecolor{BuPu-7-1}{RGB}{237,248,251}
\definecolor{BuPu-7-B}{RGB}{237,248,251}
\definecolor{BuPu-7-2}{RGB}{191,211,230}
\definecolor{BuPu-7-D}{RGB}{191,211,230}
\definecolor{BuPu-7-3}{RGB}{158,188,218}
\definecolor{BuPu-7-F}{RGB}{158,188,218}
\definecolor{BuPu-7-4}{RGB}{140,150,198}
\definecolor{BuPu-7-G}{RGB}{140,150,198}
\definecolor{BuPu-7-5}{RGB}{140,107,177}
\definecolor{BuPu-7-H}{RGB}{140,107,177}
\definecolor{BuPu-7-6}{RGB}{136,65,157}
\definecolor{BuPu-7-J}{RGB}{136,65,157}
\definecolor{BuPu-7-7}{RGB}{110,1,107}
\definecolor{BuPu-7-L}{RGB}{110,1,107}
\definecolor{BuPu-8-1}{RGB}{247,252,253}
\definecolor{BuPu-8-A}{RGB}{247,252,253}
\definecolor{BuPu-8-2}{RGB}{224,236,244}
\definecolor{BuPu-8-C}{RGB}{224,236,244}
\definecolor{BuPu-8-3}{RGB}{191,211,230}
\definecolor{BuPu-8-D}{RGB}{191,211,230}
\definecolor{BuPu-8-4}{RGB}{158,188,218}
\definecolor{BuPu-8-F}{RGB}{158,188,218}
\definecolor{BuPu-8-5}{RGB}{140,150,198}
\definecolor{BuPu-8-G}{RGB}{140,150,198}
\definecolor{BuPu-8-6}{RGB}{140,107,177}
\definecolor{BuPu-8-H}{RGB}{140,107,177}
\definecolor{BuPu-8-7}{RGB}{136,65,157}
\definecolor{BuPu-8-J}{RGB}{136,65,157}
\definecolor{BuPu-8-8}{RGB}{110,1,107}
\definecolor{BuPu-8-L}{RGB}{110,1,107}
\definecolor{BuPu-9-1}{RGB}{247,252,253}
\definecolor{BuPu-9-A}{RGB}{247,252,253}
\definecolor{BuPu-9-2}{RGB}{224,236,244}
\definecolor{BuPu-9-C}{RGB}{224,236,244}
\definecolor{BuPu-9-3}{RGB}{191,211,230}
\definecolor{BuPu-9-D}{RGB}{191,211,230}
\definecolor{BuPu-9-4}{RGB}{158,188,218}
\definecolor{BuPu-9-F}{RGB}{158,188,218}
\definecolor{BuPu-9-5}{RGB}{140,150,198}
\definecolor{BuPu-9-G}{RGB}{140,150,198}
\definecolor{BuPu-9-6}{RGB}{140,107,177}
\definecolor{BuPu-9-H}{RGB}{140,107,177}
\definecolor{BuPu-9-7}{RGB}{136,65,157}
\definecolor{BuPu-9-J}{RGB}{136,65,157}
\definecolor{BuPu-9-8}{RGB}{129,15,124}
\definecolor{BuPu-9-K}{RGB}{129,15,124}
\definecolor{BuPu-9-9}{RGB}{77,0,75}
\definecolor{BuPu-9-M}{RGB}{77,0,75}
\definecolor{RdPu-3-1}{RGB}{253,224,221}
\definecolor{RdPu-3-C}{RGB}{253,224,221}
\definecolor{RdPu-3-2}{RGB}{250,159,181}
\definecolor{RdPu-3-F}{RGB}{250,159,181}
\definecolor{RdPu-3-3}{RGB}{197,27,138}
\definecolor{RdPu-3-I}{RGB}{197,27,138}
\definecolor{RdPu-4-1}{RGB}{254,235,226}
\definecolor{RdPu-4-B}{RGB}{254,235,226}
\definecolor{RdPu-4-2}{RGB}{251,180,185}
\definecolor{RdPu-4-E}{RGB}{251,180,185}
\definecolor{RdPu-4-3}{RGB}{247,104,161}
\definecolor{RdPu-4-G}{RGB}{247,104,161}
\definecolor{RdPu-4-4}{RGB}{174,1,126}
\definecolor{RdPu-4-J}{RGB}{174,1,126}
\definecolor{RdPu-5-1}{RGB}{254,235,226}
\definecolor{RdPu-5-B}{RGB}{254,235,226}
\definecolor{RdPu-5-2}{RGB}{251,180,185}
\definecolor{RdPu-5-E}{RGB}{251,180,185}
\definecolor{RdPu-5-3}{RGB}{247,104,161}
\definecolor{RdPu-5-G}{RGB}{247,104,161}
\definecolor{RdPu-5-4}{RGB}{197,27,138}
\definecolor{RdPu-5-I}{RGB}{197,27,138}
\definecolor{RdPu-5-5}{RGB}{122,1,119}
\definecolor{RdPu-5-K}{RGB}{122,1,119}
\definecolor{RdPu-6-1}{RGB}{254,235,226}
\definecolor{RdPu-6-B}{RGB}{254,235,226}
\definecolor{RdPu-6-2}{RGB}{252,197,192}
\definecolor{RdPu-6-D}{RGB}{252,197,192}
\definecolor{RdPu-6-3}{RGB}{250,159,181}
\definecolor{RdPu-6-F}{RGB}{250,159,181}
\definecolor{RdPu-6-4}{RGB}{247,104,161}
\definecolor{RdPu-6-G}{RGB}{247,104,161}
\definecolor{RdPu-6-5}{RGB}{197,27,138}
\definecolor{RdPu-6-I}{RGB}{197,27,138}
\definecolor{RdPu-6-6}{RGB}{122,1,119}
\definecolor{RdPu-6-K}{RGB}{122,1,119}
\definecolor{RdPu-7-1}{RGB}{254,235,226}
\definecolor{RdPu-7-B}{RGB}{254,235,226}
\definecolor{RdPu-7-2}{RGB}{252,197,192}
\definecolor{RdPu-7-D}{RGB}{252,197,192}
\definecolor{RdPu-7-3}{RGB}{250,159,181}
\definecolor{RdPu-7-F}{RGB}{250,159,181}
\definecolor{RdPu-7-4}{RGB}{247,104,161}
\definecolor{RdPu-7-G}{RGB}{247,104,161}
\definecolor{RdPu-7-5}{RGB}{221,52,151}
\definecolor{RdPu-7-H}{RGB}{221,52,151}
\definecolor{RdPu-7-6}{RGB}{174,1,126}
\definecolor{RdPu-7-J}{RGB}{174,1,126}
\definecolor{RdPu-7-7}{RGB}{122,1,119}
\definecolor{RdPu-7-L}{RGB}{122,1,119}
\definecolor{RdPu-8-1}{RGB}{255,247,243}
\definecolor{RdPu-8-A}{RGB}{255,247,243}
\definecolor{RdPu-8-2}{RGB}{253,224,221}
\definecolor{RdPu-8-C}{RGB}{253,224,221}
\definecolor{RdPu-8-3}{RGB}{252,197,192}
\definecolor{RdPu-8-D}{RGB}{252,197,192}
\definecolor{RdPu-8-4}{RGB}{250,159,181}
\definecolor{RdPu-8-F}{RGB}{250,159,181}
\definecolor{RdPu-8-5}{RGB}{247,104,161}
\definecolor{RdPu-8-G}{RGB}{247,104,161}
\definecolor{RdPu-8-6}{RGB}{221,52,151}
\definecolor{RdPu-8-H}{RGB}{221,52,151}
\definecolor{RdPu-8-7}{RGB}{174,1,126}
\definecolor{RdPu-8-J}{RGB}{174,1,126}
\definecolor{RdPu-8-8}{RGB}{122,1,119}
\definecolor{RdPu-8-L}{RGB}{122,1,119}
\definecolor{RdPu-9-1}{RGB}{255,247,243}
\definecolor{RdPu-9-A}{RGB}{255,247,243}
\definecolor{RdPu-9-2}{RGB}{253,224,221}
\definecolor{RdPu-9-C}{RGB}{253,224,221}
\definecolor{RdPu-9-3}{RGB}{252,197,192}
\definecolor{RdPu-9-D}{RGB}{252,197,192}
\definecolor{RdPu-9-4}{RGB}{250,159,181}
\definecolor{RdPu-9-F}{RGB}{250,159,181}
\definecolor{RdPu-9-5}{RGB}{247,104,161}
\definecolor{RdPu-9-G}{RGB}{247,104,161}
\definecolor{RdPu-9-6}{RGB}{221,52,151}
\definecolor{RdPu-9-H}{RGB}{221,52,151}
\definecolor{RdPu-9-7}{RGB}{174,1,126}
\definecolor{RdPu-9-J}{RGB}{174,1,126}
\definecolor{RdPu-9-8}{RGB}{122,1,119}
\definecolor{RdPu-9-K}{RGB}{122,1,119}
\definecolor{RdPu-9-9}{RGB}{73,0,106}
\definecolor{RdPu-9-M}{RGB}{73,0,106}
\definecolor{PuRd-3-1}{RGB}{231,225,239}
\definecolor{PuRd-3-C}{RGB}{231,225,239}
\definecolor{PuRd-3-2}{RGB}{201,148,199}
\definecolor{PuRd-3-F}{RGB}{201,148,199}
\definecolor{PuRd-3-3}{RGB}{221,28,119}
\definecolor{PuRd-3-I}{RGB}{221,28,119}
\definecolor{PuRd-4-1}{RGB}{241,238,246}
\definecolor{PuRd-4-B}{RGB}{241,238,246}
\definecolor{PuRd-4-2}{RGB}{215,181,216}
\definecolor{PuRd-4-E}{RGB}{215,181,216}
\definecolor{PuRd-4-3}{RGB}{223,101,176}
\definecolor{PuRd-4-G}{RGB}{223,101,176}
\definecolor{PuRd-4-4}{RGB}{206,18,86}
\definecolor{PuRd-4-J}{RGB}{206,18,86}
\definecolor{PuRd-5-1}{RGB}{241,238,246}
\definecolor{PuRd-5-B}{RGB}{241,238,246}
\definecolor{PuRd-5-2}{RGB}{215,181,216}
\definecolor{PuRd-5-E}{RGB}{215,181,216}
\definecolor{PuRd-5-3}{RGB}{223,101,176}
\definecolor{PuRd-5-G}{RGB}{223,101,176}
\definecolor{PuRd-5-4}{RGB}{221,28,119}
\definecolor{PuRd-5-I}{RGB}{221,28,119}
\definecolor{PuRd-5-5}{RGB}{152,0,67}
\definecolor{PuRd-5-K}{RGB}{152,0,67}
\definecolor{PuRd-6-1}{RGB}{241,238,246}
\definecolor{PuRd-6-B}{RGB}{241,238,246}
\definecolor{PuRd-6-2}{RGB}{212,185,218}
\definecolor{PuRd-6-D}{RGB}{212,185,218}
\definecolor{PuRd-6-3}{RGB}{201,148,199}
\definecolor{PuRd-6-F}{RGB}{201,148,199}
\definecolor{PuRd-6-4}{RGB}{223,101,176}
\definecolor{PuRd-6-G}{RGB}{223,101,176}
\definecolor{PuRd-6-5}{RGB}{221,28,119}
\definecolor{PuRd-6-I}{RGB}{221,28,119}
\definecolor{PuRd-6-6}{RGB}{152,0,67}
\definecolor{PuRd-6-K}{RGB}{152,0,67}
\definecolor{PuRd-7-1}{RGB}{241,238,246}
\definecolor{PuRd-7-B}{RGB}{241,238,246}
\definecolor{PuRd-7-2}{RGB}{212,185,218}
\definecolor{PuRd-7-D}{RGB}{212,185,218}
\definecolor{PuRd-7-3}{RGB}{201,148,199}
\definecolor{PuRd-7-F}{RGB}{201,148,199}
\definecolor{PuRd-7-4}{RGB}{223,101,176}
\definecolor{PuRd-7-G}{RGB}{223,101,176}
\definecolor{PuRd-7-5}{RGB}{231,41,138}
\definecolor{PuRd-7-H}{RGB}{231,41,138}
\definecolor{PuRd-7-6}{RGB}{206,18,86}
\definecolor{PuRd-7-J}{RGB}{206,18,86}
\definecolor{PuRd-7-7}{RGB}{145,0,63}
\definecolor{PuRd-7-L}{RGB}{145,0,63}
\definecolor{PuRd-8-1}{RGB}{247,244,249}
\definecolor{PuRd-8-A}{RGB}{247,244,249}
\definecolor{PuRd-8-2}{RGB}{231,225,239}
\definecolor{PuRd-8-C}{RGB}{231,225,239}
\definecolor{PuRd-8-3}{RGB}{212,185,218}
\definecolor{PuRd-8-D}{RGB}{212,185,218}
\definecolor{PuRd-8-4}{RGB}{201,148,199}
\definecolor{PuRd-8-F}{RGB}{201,148,199}
\definecolor{PuRd-8-5}{RGB}{223,101,176}
\definecolor{PuRd-8-G}{RGB}{223,101,176}
\definecolor{PuRd-8-6}{RGB}{231,41,138}
\definecolor{PuRd-8-H}{RGB}{231,41,138}
\definecolor{PuRd-8-7}{RGB}{206,18,86}
\definecolor{PuRd-8-J}{RGB}{206,18,86}
\definecolor{PuRd-8-8}{RGB}{145,0,63}
\definecolor{PuRd-8-L}{RGB}{145,0,63}
\definecolor{PuRd-9-1}{RGB}{247,244,249}
\definecolor{PuRd-9-A}{RGB}{247,244,249}
\definecolor{PuRd-9-2}{RGB}{231,225,239}
\definecolor{PuRd-9-C}{RGB}{231,225,239}
\definecolor{PuRd-9-3}{RGB}{212,185,218}
\definecolor{PuRd-9-D}{RGB}{212,185,218}
\definecolor{PuRd-9-4}{RGB}{201,148,199}
\definecolor{PuRd-9-F}{RGB}{201,148,199}
\definecolor{PuRd-9-5}{RGB}{223,101,176}
\definecolor{PuRd-9-G}{RGB}{223,101,176}
\definecolor{PuRd-9-6}{RGB}{231,41,138}
\definecolor{PuRd-9-H}{RGB}{231,41,138}
\definecolor{PuRd-9-7}{RGB}{206,18,86}
\definecolor{PuRd-9-J}{RGB}{206,18,86}
\definecolor{PuRd-9-8}{RGB}{152,0,67}
\definecolor{PuRd-9-K}{RGB}{152,0,67}
\definecolor{PuRd-9-9}{RGB}{103,0,31}
\definecolor{PuRd-9-M}{RGB}{103,0,31}
\definecolor{OrRd-3-1}{RGB}{254,232,200}
\definecolor{OrRd-3-C}{RGB}{254,232,200}
\definecolor{OrRd-3-2}{RGB}{253,187,132}
\definecolor{OrRd-3-F}{RGB}{253,187,132}
\definecolor{OrRd-3-3}{RGB}{227,74,51}
\definecolor{OrRd-3-I}{RGB}{227,74,51}
\definecolor{OrRd-4-1}{RGB}{254,240,217}
\definecolor{OrRd-4-B}{RGB}{254,240,217}
\definecolor{OrRd-4-2}{RGB}{253,204,138}
\definecolor{OrRd-4-E}{RGB}{253,204,138}
\definecolor{OrRd-4-3}{RGB}{252,141,89}
\definecolor{OrRd-4-G}{RGB}{252,141,89}
\definecolor{OrRd-4-4}{RGB}{215,48,31}
\definecolor{OrRd-4-J}{RGB}{215,48,31}
\definecolor{OrRd-5-1}{RGB}{254,240,217}
\definecolor{OrRd-5-B}{RGB}{254,240,217}
\definecolor{OrRd-5-2}{RGB}{253,204,138}
\definecolor{OrRd-5-E}{RGB}{253,204,138}
\definecolor{OrRd-5-3}{RGB}{252,141,89}
\definecolor{OrRd-5-G}{RGB}{252,141,89}
\definecolor{OrRd-5-4}{RGB}{227,74,51}
\definecolor{OrRd-5-I}{RGB}{227,74,51}
\definecolor{OrRd-5-5}{RGB}{179,0,0}
\definecolor{OrRd-5-K}{RGB}{179,0,0}
\definecolor{OrRd-6-1}{RGB}{254,240,217}
\definecolor{OrRd-6-B}{RGB}{254,240,217}
\definecolor{OrRd-6-2}{RGB}{253,212,158}
\definecolor{OrRd-6-D}{RGB}{253,212,158}
\definecolor{OrRd-6-3}{RGB}{253,187,132}
\definecolor{OrRd-6-F}{RGB}{253,187,132}
\definecolor{OrRd-6-4}{RGB}{252,141,89}
\definecolor{OrRd-6-G}{RGB}{252,141,89}
\definecolor{OrRd-6-5}{RGB}{227,74,51}
\definecolor{OrRd-6-I}{RGB}{227,74,51}
\definecolor{OrRd-6-6}{RGB}{179,0,0}
\definecolor{OrRd-6-K}{RGB}{179,0,0}
\definecolor{OrRd-7-1}{RGB}{254,240,217}
\definecolor{OrRd-7-B}{RGB}{254,240,217}
\definecolor{OrRd-7-2}{RGB}{253,212,158}
\definecolor{OrRd-7-D}{RGB}{253,212,158}
\definecolor{OrRd-7-3}{RGB}{253,187,132}
\definecolor{OrRd-7-F}{RGB}{253,187,132}
\definecolor{OrRd-7-4}{RGB}{252,141,89}
\definecolor{OrRd-7-G}{RGB}{252,141,89}
\definecolor{OrRd-7-5}{RGB}{239,101,72}
\definecolor{OrRd-7-H}{RGB}{239,101,72}
\definecolor{OrRd-7-6}{RGB}{215,48,31}
\definecolor{OrRd-7-J}{RGB}{215,48,31}
\definecolor{OrRd-7-7}{RGB}{153,0,0}
\definecolor{OrRd-7-L}{RGB}{153,0,0}
\definecolor{OrRd-8-1}{RGB}{255,247,236}
\definecolor{OrRd-8-A}{RGB}{255,247,236}
\definecolor{OrRd-8-2}{RGB}{254,232,200}
\definecolor{OrRd-8-C}{RGB}{254,232,200}
\definecolor{OrRd-8-3}{RGB}{253,212,158}
\definecolor{OrRd-8-D}{RGB}{253,212,158}
\definecolor{OrRd-8-4}{RGB}{253,187,132}
\definecolor{OrRd-8-F}{RGB}{253,187,132}
\definecolor{OrRd-8-5}{RGB}{252,141,89}
\definecolor{OrRd-8-G}{RGB}{252,141,89}
\definecolor{OrRd-8-6}{RGB}{239,101,72}
\definecolor{OrRd-8-H}{RGB}{239,101,72}
\definecolor{OrRd-8-7}{RGB}{215,48,31}
\definecolor{OrRd-8-J}{RGB}{215,48,31}
\definecolor{OrRd-8-8}{RGB}{153,0,0}
\definecolor{OrRd-8-L}{RGB}{153,0,0}
\definecolor{OrRd-9-1}{RGB}{255,247,236}
\definecolor{OrRd-9-A}{RGB}{255,247,236}
\definecolor{OrRd-9-2}{RGB}{254,232,200}
\definecolor{OrRd-9-C}{RGB}{254,232,200}
\definecolor{OrRd-9-3}{RGB}{253,212,158}
\definecolor{OrRd-9-D}{RGB}{253,212,158}
\definecolor{OrRd-9-4}{RGB}{253,187,132}
\definecolor{OrRd-9-F}{RGB}{253,187,132}
\definecolor{OrRd-9-5}{RGB}{252,141,89}
\definecolor{OrRd-9-G}{RGB}{252,141,89}
\definecolor{OrRd-9-6}{RGB}{239,101,72}
\definecolor{OrRd-9-H}{RGB}{239,101,72}
\definecolor{OrRd-9-7}{RGB}{215,48,31}
\definecolor{OrRd-9-J}{RGB}{215,48,31}
\definecolor{OrRd-9-8}{RGB}{179,0,0}
\definecolor{OrRd-9-K}{RGB}{179,0,0}
\definecolor{OrRd-9-9}{RGB}{127,0,0}
\definecolor{OrRd-9-M}{RGB}{127,0,0}
\definecolor{YlOrRd-3-1}{RGB}{255,237,160}
\definecolor{YlOrRd-3-C}{RGB}{255,237,160}
\definecolor{YlOrRd-3-2}{RGB}{254,178,76}
\definecolor{YlOrRd-3-F}{RGB}{254,178,76}
\definecolor{YlOrRd-3-3}{RGB}{240,59,32}
\definecolor{YlOrRd-3-I}{RGB}{240,59,32}
\definecolor{YlOrRd-4-1}{RGB}{255,255,178}
\definecolor{YlOrRd-4-B}{RGB}{255,255,178}
\definecolor{YlOrRd-4-2}{RGB}{254,204,92}
\definecolor{YlOrRd-4-E}{RGB}{254,204,92}
\definecolor{YlOrRd-4-3}{RGB}{253,141,60}
\definecolor{YlOrRd-4-G}{RGB}{253,141,60}
\definecolor{YlOrRd-4-4}{RGB}{227,26,28}
\definecolor{YlOrRd-4-J}{RGB}{227,26,28}
\definecolor{YlOrRd-5-1}{RGB}{255,255,178}
\definecolor{YlOrRd-5-B}{RGB}{255,255,178}
\definecolor{YlOrRd-5-2}{RGB}{254,204,92}
\definecolor{YlOrRd-5-E}{RGB}{254,204,92}
\definecolor{YlOrRd-5-3}{RGB}{253,141,60}
\definecolor{YlOrRd-5-G}{RGB}{253,141,60}
\definecolor{YlOrRd-5-4}{RGB}{240,59,32}
\definecolor{YlOrRd-5-I}{RGB}{240,59,32}
\definecolor{YlOrRd-5-5}{RGB}{189,0,38}
\definecolor{YlOrRd-5-K}{RGB}{189,0,38}
\definecolor{YlOrRd-6-1}{RGB}{255,255,178}
\definecolor{YlOrRd-6-B}{RGB}{255,255,178}
\definecolor{YlOrRd-6-2}{RGB}{254,217,118}
\definecolor{YlOrRd-6-D}{RGB}{254,217,118}
\definecolor{YlOrRd-6-3}{RGB}{254,178,76}
\definecolor{YlOrRd-6-F}{RGB}{254,178,76}
\definecolor{YlOrRd-6-4}{RGB}{253,141,60}
\definecolor{YlOrRd-6-G}{RGB}{253,141,60}
\definecolor{YlOrRd-6-5}{RGB}{240,59,32}
\definecolor{YlOrRd-6-I}{RGB}{240,59,32}
\definecolor{YlOrRd-6-6}{RGB}{189,0,38}
\definecolor{YlOrRd-6-K}{RGB}{189,0,38}
\definecolor{YlOrRd-7-1}{RGB}{255,255,178}
\definecolor{YlOrRd-7-B}{RGB}{255,255,178}
\definecolor{YlOrRd-7-2}{RGB}{254,217,118}
\definecolor{YlOrRd-7-D}{RGB}{254,217,118}
\definecolor{YlOrRd-7-3}{RGB}{254,178,76}
\definecolor{YlOrRd-7-F}{RGB}{254,178,76}
\definecolor{YlOrRd-7-4}{RGB}{253,141,60}
\definecolor{YlOrRd-7-G}{RGB}{253,141,60}
\definecolor{YlOrRd-7-5}{RGB}{252,78,42}
\definecolor{YlOrRd-7-H}{RGB}{252,78,42}
\definecolor{YlOrRd-7-6}{RGB}{227,26,28}
\definecolor{YlOrRd-7-J}{RGB}{227,26,28}
\definecolor{YlOrRd-7-7}{RGB}{177,0,38}
\definecolor{YlOrRd-7-L}{RGB}{177,0,38}
\definecolor{YlOrRd-8-1}{RGB}{255,255,204}
\definecolor{YlOrRd-8-A}{RGB}{255,255,204}
\definecolor{YlOrRd-8-2}{RGB}{255,237,160}
\definecolor{YlOrRd-8-C}{RGB}{255,237,160}
\definecolor{YlOrRd-8-3}{RGB}{254,217,118}
\definecolor{YlOrRd-8-D}{RGB}{254,217,118}
\definecolor{YlOrRd-8-4}{RGB}{254,178,76}
\definecolor{YlOrRd-8-F}{RGB}{254,178,76}
\definecolor{YlOrRd-8-5}{RGB}{253,141,60}
\definecolor{YlOrRd-8-G}{RGB}{253,141,60}
\definecolor{YlOrRd-8-6}{RGB}{252,78,42}
\definecolor{YlOrRd-8-H}{RGB}{252,78,42}
\definecolor{YlOrRd-8-7}{RGB}{227,26,28}
\definecolor{YlOrRd-8-J}{RGB}{227,26,28}
\definecolor{YlOrRd-8-8}{RGB}{177,0,38}
\definecolor{YlOrRd-8-L}{RGB}{177,0,38}
\definecolor{YlOrRd-9-1}{RGB}{255,255,204}
\definecolor{YlOrRd-9-A}{RGB}{255,255,204}
\definecolor{YlOrRd-9-2}{RGB}{255,237,160}
\definecolor{YlOrRd-9-C}{RGB}{255,237,160}
\definecolor{YlOrRd-9-3}{RGB}{254,217,118}
\definecolor{YlOrRd-9-D}{RGB}{254,217,118}
\definecolor{YlOrRd-9-4}{RGB}{254,178,76}
\definecolor{YlOrRd-9-F}{RGB}{254,178,76}
\definecolor{YlOrRd-9-5}{RGB}{253,141,60}
\definecolor{YlOrRd-9-G}{RGB}{253,141,60}
\definecolor{YlOrRd-9-6}{RGB}{252,78,42}
\definecolor{YlOrRd-9-H}{RGB}{252,78,42}
\definecolor{YlOrRd-9-7}{RGB}{227,26,28}
\definecolor{YlOrRd-9-J}{RGB}{227,26,28}
\definecolor{YlOrRd-9-8}{RGB}{189,0,38}
\definecolor{YlOrRd-9-K}{RGB}{189,0,38}
\definecolor{YlOrRd-9-9}{RGB}{128,0,38}
\definecolor{YlOrRd-9-M}{RGB}{128,0,38}
\definecolor{YlOrBr-3-1}{RGB}{255,247,188}
\definecolor{YlOrBr-3-C}{RGB}{255,247,188}
\definecolor{YlOrBr-3-2}{RGB}{254,196,79}
\definecolor{YlOrBr-3-F}{RGB}{254,196,79}
\definecolor{YlOrBr-3-3}{RGB}{217,95,14}
\definecolor{YlOrBr-3-I}{RGB}{217,95,14}
\definecolor{YlOrBr-4-1}{RGB}{255,255,212}
\definecolor{YlOrBr-4-B}{RGB}{255,255,212}
\definecolor{YlOrBr-4-2}{RGB}{254,217,142}
\definecolor{YlOrBr-4-E}{RGB}{254,217,142}
\definecolor{YlOrBr-4-3}{RGB}{254,153,41}
\definecolor{YlOrBr-4-G}{RGB}{254,153,41}
\definecolor{YlOrBr-4-4}{RGB}{204,76,2}
\definecolor{YlOrBr-4-J}{RGB}{204,76,2}
\definecolor{YlOrBr-5-1}{RGB}{255,255,212}
\definecolor{YlOrBr-5-B}{RGB}{255,255,212}
\definecolor{YlOrBr-5-2}{RGB}{254,217,142}
\definecolor{YlOrBr-5-E}{RGB}{254,217,142}
\definecolor{YlOrBr-5-3}{RGB}{254,153,41}
\definecolor{YlOrBr-5-G}{RGB}{254,153,41}
\definecolor{YlOrBr-5-4}{RGB}{217,95,14}
\definecolor{YlOrBr-5-I}{RGB}{217,95,14}
\definecolor{YlOrBr-5-5}{RGB}{153,52,4}
\definecolor{YlOrBr-5-K}{RGB}{153,52,4}
\definecolor{YlOrBr-6-1}{RGB}{255,255,212}
\definecolor{YlOrBr-6-B}{RGB}{255,255,212}
\definecolor{YlOrBr-6-2}{RGB}{254,227,145}
\definecolor{YlOrBr-6-D}{RGB}{254,227,145}
\definecolor{YlOrBr-6-3}{RGB}{254,196,79}
\definecolor{YlOrBr-6-F}{RGB}{254,196,79}
\definecolor{YlOrBr-6-4}{RGB}{254,153,41}
\definecolor{YlOrBr-6-G}{RGB}{254,153,41}
\definecolor{YlOrBr-6-5}{RGB}{217,95,14}
\definecolor{YlOrBr-6-I}{RGB}{217,95,14}
\definecolor{YlOrBr-6-6}{RGB}{153,52,4}
\definecolor{YlOrBr-6-K}{RGB}{153,52,4}
\definecolor{YlOrBr-7-1}{RGB}{255,255,212}
\definecolor{YlOrBr-7-B}{RGB}{255,255,212}
\definecolor{YlOrBr-7-2}{RGB}{254,227,145}
\definecolor{YlOrBr-7-D}{RGB}{254,227,145}
\definecolor{YlOrBr-7-3}{RGB}{254,196,79}
\definecolor{YlOrBr-7-F}{RGB}{254,196,79}
\definecolor{YlOrBr-7-4}{RGB}{254,153,41}
\definecolor{YlOrBr-7-G}{RGB}{254,153,41}
\definecolor{YlOrBr-7-5}{RGB}{236,112,20}
\definecolor{YlOrBr-7-H}{RGB}{236,112,20}
\definecolor{YlOrBr-7-6}{RGB}{204,76,2}
\definecolor{YlOrBr-7-J}{RGB}{204,76,2}
\definecolor{YlOrBr-7-7}{RGB}{140,45,4}
\definecolor{YlOrBr-7-L}{RGB}{140,45,4}
\definecolor{YlOrBr-8-1}{RGB}{255,255,229}
\definecolor{YlOrBr-8-A}{RGB}{255,255,229}
\definecolor{YlOrBr-8-2}{RGB}{255,247,188}
\definecolor{YlOrBr-8-C}{RGB}{255,247,188}
\definecolor{YlOrBr-8-3}{RGB}{254,227,145}
\definecolor{YlOrBr-8-D}{RGB}{254,227,145}
\definecolor{YlOrBr-8-4}{RGB}{254,196,79}
\definecolor{YlOrBr-8-F}{RGB}{254,196,79}
\definecolor{YlOrBr-8-5}{RGB}{254,153,41}
\definecolor{YlOrBr-8-G}{RGB}{254,153,41}
\definecolor{YlOrBr-8-6}{RGB}{236,112,20}
\definecolor{YlOrBr-8-H}{RGB}{236,112,20}
\definecolor{YlOrBr-8-7}{RGB}{204,76,2}
\definecolor{YlOrBr-8-J}{RGB}{204,76,2}
\definecolor{YlOrBr-8-8}{RGB}{140,45,4}
\definecolor{YlOrBr-8-L}{RGB}{140,45,4}
\definecolor{YlOrBr-9-1}{RGB}{255,255,229}
\definecolor{YlOrBr-9-A}{RGB}{255,255,229}
\definecolor{YlOrBr-9-2}{RGB}{255,247,188}
\definecolor{YlOrBr-9-C}{RGB}{255,247,188}
\definecolor{YlOrBr-9-3}{RGB}{254,227,145}
\definecolor{YlOrBr-9-D}{RGB}{254,227,145}
\definecolor{YlOrBr-9-4}{RGB}{254,196,79}
\definecolor{YlOrBr-9-F}{RGB}{254,196,79}
\definecolor{YlOrBr-9-5}{RGB}{254,153,41}
\definecolor{YlOrBr-9-G}{RGB}{254,153,41}
\definecolor{YlOrBr-9-6}{RGB}{236,112,20}
\definecolor{YlOrBr-9-H}{RGB}{236,112,20}
\definecolor{YlOrBr-9-7}{RGB}{204,76,2}
\definecolor{YlOrBr-9-J}{RGB}{204,76,2}
\definecolor{YlOrBr-9-8}{RGB}{153,52,4}
\definecolor{YlOrBr-9-K}{RGB}{153,52,4}
\definecolor{YlOrBr-9-9}{RGB}{102,37,6}
\definecolor{YlOrBr-9-M}{RGB}{102,37,6}
\definecolor{Purples-3-1}{RGB}{239,237,245}
\definecolor{Purples-3-C}{RGB}{239,237,245}
\definecolor{Purples-3-2}{RGB}{188,189,220}
\definecolor{Purples-3-F}{RGB}{188,189,220}
\definecolor{Purples-3-3}{RGB}{117,107,177}
\definecolor{Purples-3-I}{RGB}{117,107,177}
\definecolor{Purples-4-1}{RGB}{242,240,247}
\definecolor{Purples-4-B}{RGB}{242,240,247}
\definecolor{Purples-4-2}{RGB}{203,201,226}
\definecolor{Purples-4-E}{RGB}{203,201,226}
\definecolor{Purples-4-3}{RGB}{158,154,200}
\definecolor{Purples-4-G}{RGB}{158,154,200}
\definecolor{Purples-4-4}{RGB}{106,81,163}
\definecolor{Purples-4-J}{RGB}{106,81,163}
\definecolor{Purples-5-1}{RGB}{242,240,247}
\definecolor{Purples-5-B}{RGB}{242,240,247}
\definecolor{Purples-5-2}{RGB}{203,201,226}
\definecolor{Purples-5-E}{RGB}{203,201,226}
\definecolor{Purples-5-3}{RGB}{158,154,200}
\definecolor{Purples-5-G}{RGB}{158,154,200}
\definecolor{Purples-5-4}{RGB}{117,107,177}
\definecolor{Purples-5-I}{RGB}{117,107,177}
\definecolor{Purples-5-5}{RGB}{84,39,143}
\definecolor{Purples-5-K}{RGB}{84,39,143}
\definecolor{Purples-6-1}{RGB}{242,240,247}
\definecolor{Purples-6-B}{RGB}{242,240,247}
\definecolor{Purples-6-2}{RGB}{218,218,235}
\definecolor{Purples-6-D}{RGB}{218,218,235}
\definecolor{Purples-6-3}{RGB}{188,189,220}
\definecolor{Purples-6-F}{RGB}{188,189,220}
\definecolor{Purples-6-4}{RGB}{158,154,200}
\definecolor{Purples-6-G}{RGB}{158,154,200}
\definecolor{Purples-6-5}{RGB}{117,107,177}
\definecolor{Purples-6-I}{RGB}{117,107,177}
\definecolor{Purples-6-6}{RGB}{84,39,143}
\definecolor{Purples-6-K}{RGB}{84,39,143}
\definecolor{Purples-7-1}{RGB}{242,240,247}
\definecolor{Purples-7-B}{RGB}{242,240,247}
\definecolor{Purples-7-2}{RGB}{218,218,235}
\definecolor{Purples-7-D}{RGB}{218,218,235}
\definecolor{Purples-7-3}{RGB}{188,189,220}
\definecolor{Purples-7-F}{RGB}{188,189,220}
\definecolor{Purples-7-4}{RGB}{158,154,200}
\definecolor{Purples-7-G}{RGB}{158,154,200}
\definecolor{Purples-7-5}{RGB}{128,125,186}
\definecolor{Purples-7-H}{RGB}{128,125,186}
\definecolor{Purples-7-6}{RGB}{106,81,163}
\definecolor{Purples-7-J}{RGB}{106,81,163}
\definecolor{Purples-7-7}{RGB}{74,20,134}
\definecolor{Purples-7-L}{RGB}{74,20,134}
\definecolor{Purples-8-1}{RGB}{252,251,253}
\definecolor{Purples-8-A}{RGB}{252,251,253}
\definecolor{Purples-8-2}{RGB}{239,237,245}
\definecolor{Purples-8-C}{RGB}{239,237,245}
\definecolor{Purples-8-3}{RGB}{218,218,235}
\definecolor{Purples-8-D}{RGB}{218,218,235}
\definecolor{Purples-8-4}{RGB}{188,189,220}
\definecolor{Purples-8-F}{RGB}{188,189,220}
\definecolor{Purples-8-5}{RGB}{158,154,200}
\definecolor{Purples-8-G}{RGB}{158,154,200}
\definecolor{Purples-8-6}{RGB}{128,125,186}
\definecolor{Purples-8-H}{RGB}{128,125,186}
\definecolor{Purples-8-7}{RGB}{106,81,163}
\definecolor{Purples-8-J}{RGB}{106,81,163}
\definecolor{Purples-8-8}{RGB}{74,20,134}
\definecolor{Purples-8-L}{RGB}{74,20,134}
\definecolor{Purples-9-1}{RGB}{252,251,253}
\definecolor{Purples-9-A}{RGB}{252,251,253}
\definecolor{Purples-9-2}{RGB}{239,237,245}
\definecolor{Purples-9-C}{RGB}{239,237,245}
\definecolor{Purples-9-3}{RGB}{218,218,235}
\definecolor{Purples-9-D}{RGB}{218,218,235}
\definecolor{Purples-9-4}{RGB}{188,189,220}
\definecolor{Purples-9-F}{RGB}{188,189,220}
\definecolor{Purples-9-5}{RGB}{158,154,200}
\definecolor{Purples-9-G}{RGB}{158,154,200}
\definecolor{Purples-9-6}{RGB}{128,125,186}
\definecolor{Purples-9-H}{RGB}{128,125,186}
\definecolor{Purples-9-7}{RGB}{106,81,163}
\definecolor{Purples-9-J}{RGB}{106,81,163}
\definecolor{Purples-9-8}{RGB}{84,39,143}
\definecolor{Purples-9-K}{RGB}{84,39,143}
\definecolor{Purples-9-9}{RGB}{63,0,125}
\definecolor{Purples-9-M}{RGB}{63,0,125}
\definecolor{Blues-3-1}{RGB}{222,235,247}
\definecolor{Blues-3-C}{RGB}{222,235,247}
\definecolor{Blues-3-2}{RGB}{158,202,225}
\definecolor{Blues-3-F}{RGB}{158,202,225}
\definecolor{Blues-3-3}{RGB}{49,130,189}
\definecolor{Blues-3-I}{RGB}{49,130,189}
\definecolor{Blues-4-1}{RGB}{239,243,255}
\definecolor{Blues-4-B}{RGB}{239,243,255}
\definecolor{Blues-4-2}{RGB}{189,215,231}
\definecolor{Blues-4-E}{RGB}{189,215,231}
\definecolor{Blues-4-3}{RGB}{107,174,214}
\definecolor{Blues-4-G}{RGB}{107,174,214}
\definecolor{Blues-4-4}{RGB}{33,113,181}
\definecolor{Blues-4-J}{RGB}{33,113,181}
\definecolor{Blues-5-1}{RGB}{239,243,255}
\definecolor{Blues-5-B}{RGB}{239,243,255}
\definecolor{Blues-5-2}{RGB}{189,215,231}
\definecolor{Blues-5-E}{RGB}{189,215,231}
\definecolor{Blues-5-3}{RGB}{107,174,214}
\definecolor{Blues-5-G}{RGB}{107,174,214}
\definecolor{Blues-5-4}{RGB}{49,130,189}
\definecolor{Blues-5-I}{RGB}{49,130,189}
\definecolor{Blues-5-5}{RGB}{8,81,156}
\definecolor{Blues-5-K}{RGB}{8,81,156}
\definecolor{Blues-6-1}{RGB}{239,243,255}
\definecolor{Blues-6-B}{RGB}{239,243,255}
\definecolor{Blues-6-2}{RGB}{198,219,239}
\definecolor{Blues-6-D}{RGB}{198,219,239}
\definecolor{Blues-6-3}{RGB}{158,202,225}
\definecolor{Blues-6-F}{RGB}{158,202,225}
\definecolor{Blues-6-4}{RGB}{107,174,214}
\definecolor{Blues-6-G}{RGB}{107,174,214}
\definecolor{Blues-6-5}{RGB}{49,130,189}
\definecolor{Blues-6-I}{RGB}{49,130,189}
\definecolor{Blues-6-6}{RGB}{8,81,156}
\definecolor{Blues-6-K}{RGB}{8,81,156}
\definecolor{Blues-7-1}{RGB}{239,243,255}
\definecolor{Blues-7-B}{RGB}{239,243,255}
\definecolor{Blues-7-2}{RGB}{198,219,239}
\definecolor{Blues-7-D}{RGB}{198,219,239}
\definecolor{Blues-7-3}{RGB}{158,202,225}
\definecolor{Blues-7-F}{RGB}{158,202,225}
\definecolor{Blues-7-4}{RGB}{107,174,214}
\definecolor{Blues-7-G}{RGB}{107,174,214}
\definecolor{Blues-7-5}{RGB}{66,146,198}
\definecolor{Blues-7-H}{RGB}{66,146,198}
\definecolor{Blues-7-6}{RGB}{33,113,181}
\definecolor{Blues-7-J}{RGB}{33,113,181}
\definecolor{Blues-7-7}{RGB}{8,69,148}
\definecolor{Blues-7-L}{RGB}{8,69,148}
\definecolor{Blues-8-1}{RGB}{247,251,255}
\definecolor{Blues-8-A}{RGB}{247,251,255}
\definecolor{Blues-8-2}{RGB}{222,235,247}
\definecolor{Blues-8-C}{RGB}{222,235,247}
\definecolor{Blues-8-3}{RGB}{198,219,239}
\definecolor{Blues-8-D}{RGB}{198,219,239}
\definecolor{Blues-8-4}{RGB}{158,202,225}
\definecolor{Blues-8-F}{RGB}{158,202,225}
\definecolor{Blues-8-5}{RGB}{107,174,214}
\definecolor{Blues-8-G}{RGB}{107,174,214}
\definecolor{Blues-8-6}{RGB}{66,146,198}
\definecolor{Blues-8-H}{RGB}{66,146,198}
\definecolor{Blues-8-7}{RGB}{33,113,181}
\definecolor{Blues-8-J}{RGB}{33,113,181}
\definecolor{Blues-8-8}{RGB}{8,69,148}
\definecolor{Blues-8-L}{RGB}{8,69,148}
\definecolor{Blues-9-1}{RGB}{247,251,255}
\definecolor{Blues-9-A}{RGB}{247,251,255}
\definecolor{Blues-9-2}{RGB}{222,235,247}
\definecolor{Blues-9-C}{RGB}{222,235,247}
\definecolor{Blues-9-3}{RGB}{198,219,239}
\definecolor{Blues-9-D}{RGB}{198,219,239}
\definecolor{Blues-9-4}{RGB}{158,202,225}
\definecolor{Blues-9-F}{RGB}{158,202,225}
\definecolor{Blues-9-5}{RGB}{107,174,214}
\definecolor{Blues-9-G}{RGB}{107,174,214}
\definecolor{Blues-9-6}{RGB}{66,146,198}
\definecolor{Blues-9-H}{RGB}{66,146,198}
\definecolor{Blues-9-7}{RGB}{33,113,181}
\definecolor{Blues-9-J}{RGB}{33,113,181}
\definecolor{Blues-9-8}{RGB}{8,81,156}
\definecolor{Blues-9-K}{RGB}{8,81,156}
\definecolor{Blues-9-9}{RGB}{8,48,107}
\definecolor{Blues-9-M}{RGB}{8,48,107}
\definecolor{Greens-3-1}{RGB}{229,245,224}
\definecolor{Greens-3-C}{RGB}{229,245,224}
\definecolor{Greens-3-2}{RGB}{161,217,155}
\definecolor{Greens-3-F}{RGB}{161,217,155}
\definecolor{Greens-3-3}{RGB}{49,163,84}
\definecolor{Greens-3-I}{RGB}{49,163,84}
\definecolor{Greens-4-1}{RGB}{237,248,233}
\definecolor{Greens-4-B}{RGB}{237,248,233}
\definecolor{Greens-4-2}{RGB}{186,228,179}
\definecolor{Greens-4-E}{RGB}{186,228,179}
\definecolor{Greens-4-3}{RGB}{116,196,118}
\definecolor{Greens-4-G}{RGB}{116,196,118}
\definecolor{Greens-4-4}{RGB}{35,139,69}
\definecolor{Greens-4-J}{RGB}{35,139,69}
\definecolor{Greens-5-1}{RGB}{237,248,233}
\definecolor{Greens-5-B}{RGB}{237,248,233}
\definecolor{Greens-5-2}{RGB}{186,228,179}
\definecolor{Greens-5-E}{RGB}{186,228,179}
\definecolor{Greens-5-3}{RGB}{116,196,118}
\definecolor{Greens-5-G}{RGB}{116,196,118}
\definecolor{Greens-5-4}{RGB}{49,163,84}
\definecolor{Greens-5-I}{RGB}{49,163,84}
\definecolor{Greens-5-5}{RGB}{0,109,44}
\definecolor{Greens-5-K}{RGB}{0,109,44}
\definecolor{Greens-6-1}{RGB}{237,248,233}
\definecolor{Greens-6-B}{RGB}{237,248,233}
\definecolor{Greens-6-2}{RGB}{199,233,192}
\definecolor{Greens-6-D}{RGB}{199,233,192}
\definecolor{Greens-6-3}{RGB}{161,217,155}
\definecolor{Greens-6-F}{RGB}{161,217,155}
\definecolor{Greens-6-4}{RGB}{116,196,118}
\definecolor{Greens-6-G}{RGB}{116,196,118}
\definecolor{Greens-6-5}{RGB}{49,163,84}
\definecolor{Greens-6-I}{RGB}{49,163,84}
\definecolor{Greens-6-6}{RGB}{0,109,44}
\definecolor{Greens-6-K}{RGB}{0,109,44}
\definecolor{Greens-7-1}{RGB}{237,248,233}
\definecolor{Greens-7-B}{RGB}{237,248,233}
\definecolor{Greens-7-2}{RGB}{199,233,192}
\definecolor{Greens-7-D}{RGB}{199,233,192}
\definecolor{Greens-7-3}{RGB}{161,217,155}
\definecolor{Greens-7-F}{RGB}{161,217,155}
\definecolor{Greens-7-4}{RGB}{116,196,118}
\definecolor{Greens-7-G}{RGB}{116,196,118}
\definecolor{Greens-7-5}{RGB}{65,171,93}
\definecolor{Greens-7-H}{RGB}{65,171,93}
\definecolor{Greens-7-6}{RGB}{35,139,69}
\definecolor{Greens-7-J}{RGB}{35,139,69}
\definecolor{Greens-7-7}{RGB}{0,90,50}
\definecolor{Greens-7-L}{RGB}{0,90,50}
\definecolor{Greens-8-1}{RGB}{247,252,245}
\definecolor{Greens-8-A}{RGB}{247,252,245}
\definecolor{Greens-8-2}{RGB}{229,245,224}
\definecolor{Greens-8-C}{RGB}{229,245,224}
\definecolor{Greens-8-3}{RGB}{199,233,192}
\definecolor{Greens-8-D}{RGB}{199,233,192}
\definecolor{Greens-8-4}{RGB}{161,217,155}
\definecolor{Greens-8-F}{RGB}{161,217,155}
\definecolor{Greens-8-5}{RGB}{116,196,118}
\definecolor{Greens-8-G}{RGB}{116,196,118}
\definecolor{Greens-8-6}{RGB}{65,171,93}
\definecolor{Greens-8-H}{RGB}{65,171,93}
\definecolor{Greens-8-7}{RGB}{35,139,69}
\definecolor{Greens-8-J}{RGB}{35,139,69}
\definecolor{Greens-8-8}{RGB}{0,90,50}
\definecolor{Greens-8-L}{RGB}{0,90,50}
\definecolor{Greens-9-1}{RGB}{247,252,245}
\definecolor{Greens-9-A}{RGB}{247,252,245}
\definecolor{Greens-9-2}{RGB}{229,245,224}
\definecolor{Greens-9-C}{RGB}{229,245,224}
\definecolor{Greens-9-3}{RGB}{199,233,192}
\definecolor{Greens-9-D}{RGB}{199,233,192}
\definecolor{Greens-9-4}{RGB}{161,217,155}
\definecolor{Greens-9-F}{RGB}{161,217,155}
\definecolor{Greens-9-5}{RGB}{116,196,118}
\definecolor{Greens-9-G}{RGB}{116,196,118}
\definecolor{Greens-9-6}{RGB}{65,171,93}
\definecolor{Greens-9-H}{RGB}{65,171,93}
\definecolor{Greens-9-7}{RGB}{35,139,69}
\definecolor{Greens-9-J}{RGB}{35,139,69}
\definecolor{Greens-9-8}{RGB}{0,109,44}
\definecolor{Greens-9-K}{RGB}{0,109,44}
\definecolor{Greens-9-9}{RGB}{0,68,27}
\definecolor{Greens-9-M}{RGB}{0,68,27}
\definecolor{Oranges-3-1}{RGB}{254,230,206}
\definecolor{Oranges-3-C}{RGB}{254,230,206}
\definecolor{Oranges-3-2}{RGB}{253,174,107}
\definecolor{Oranges-3-F}{RGB}{253,174,107}
\definecolor{Oranges-3-3}{RGB}{230,85,13}
\definecolor{Oranges-3-I}{RGB}{230,85,13}
\definecolor{Oranges-4-1}{RGB}{254,237,222}
\definecolor{Oranges-4-B}{RGB}{254,237,222}
\definecolor{Oranges-4-2}{RGB}{253,190,133}
\definecolor{Oranges-4-E}{RGB}{253,190,133}
\definecolor{Oranges-4-3}{RGB}{253,141,60}
\definecolor{Oranges-4-G}{RGB}{253,141,60}
\definecolor{Oranges-4-4}{RGB}{217,71,1}
\definecolor{Oranges-4-J}{RGB}{217,71,1}
\definecolor{Oranges-5-1}{RGB}{254,237,222}
\definecolor{Oranges-5-B}{RGB}{254,237,222}
\definecolor{Oranges-5-2}{RGB}{253,190,133}
\definecolor{Oranges-5-E}{RGB}{253,190,133}
\definecolor{Oranges-5-3}{RGB}{253,141,60}
\definecolor{Oranges-5-G}{RGB}{253,141,60}
\definecolor{Oranges-5-4}{RGB}{230,85,13}
\definecolor{Oranges-5-I}{RGB}{230,85,13}
\definecolor{Oranges-5-5}{RGB}{166,54,3}
\definecolor{Oranges-5-K}{RGB}{166,54,3}
\definecolor{Oranges-6-1}{RGB}{254,237,222}
\definecolor{Oranges-6-B}{RGB}{254,237,222}
\definecolor{Oranges-6-2}{RGB}{253,208,162}
\definecolor{Oranges-6-D}{RGB}{253,208,162}
\definecolor{Oranges-6-3}{RGB}{253,174,107}
\definecolor{Oranges-6-F}{RGB}{253,174,107}
\definecolor{Oranges-6-4}{RGB}{253,141,60}
\definecolor{Oranges-6-G}{RGB}{253,141,60}
\definecolor{Oranges-6-5}{RGB}{230,85,13}
\definecolor{Oranges-6-I}{RGB}{230,85,13}
\definecolor{Oranges-6-6}{RGB}{166,54,3}
\definecolor{Oranges-6-K}{RGB}{166,54,3}
\definecolor{Oranges-7-1}{RGB}{254,237,222}
\definecolor{Oranges-7-B}{RGB}{254,237,222}
\definecolor{Oranges-7-2}{RGB}{253,208,162}
\definecolor{Oranges-7-D}{RGB}{253,208,162}
\definecolor{Oranges-7-3}{RGB}{253,174,107}
\definecolor{Oranges-7-F}{RGB}{253,174,107}
\definecolor{Oranges-7-4}{RGB}{253,141,60}
\definecolor{Oranges-7-G}{RGB}{253,141,60}
\definecolor{Oranges-7-5}{RGB}{241,105,19}
\definecolor{Oranges-7-H}{RGB}{241,105,19}
\definecolor{Oranges-7-6}{RGB}{217,72,1}
\definecolor{Oranges-7-J}{RGB}{217,72,1}
\definecolor{Oranges-7-7}{RGB}{140,45,4}
\definecolor{Oranges-7-L}{RGB}{140,45,4}
\definecolor{Oranges-8-1}{RGB}{255,245,235}
\definecolor{Oranges-8-A}{RGB}{255,245,235}
\definecolor{Oranges-8-2}{RGB}{254,230,206}
\definecolor{Oranges-8-C}{RGB}{254,230,206}
\definecolor{Oranges-8-3}{RGB}{253,208,162}
\definecolor{Oranges-8-D}{RGB}{253,208,162}
\definecolor{Oranges-8-4}{RGB}{253,174,107}
\definecolor{Oranges-8-F}{RGB}{253,174,107}
\definecolor{Oranges-8-5}{RGB}{253,141,60}
\definecolor{Oranges-8-G}{RGB}{253,141,60}
\definecolor{Oranges-8-6}{RGB}{241,105,19}
\definecolor{Oranges-8-H}{RGB}{241,105,19}
\definecolor{Oranges-8-7}{RGB}{217,72,1}
\definecolor{Oranges-8-J}{RGB}{217,72,1}
\definecolor{Oranges-8-8}{RGB}{140,45,4}
\definecolor{Oranges-8-L}{RGB}{140,45,4}
\definecolor{Oranges-9-1}{RGB}{255,245,235}
\definecolor{Oranges-9-A}{RGB}{255,245,235}
\definecolor{Oranges-9-2}{RGB}{254,230,206}
\definecolor{Oranges-9-C}{RGB}{254,230,206}
\definecolor{Oranges-9-3}{RGB}{253,208,162}
\definecolor{Oranges-9-D}{RGB}{253,208,162}
\definecolor{Oranges-9-4}{RGB}{253,174,107}
\definecolor{Oranges-9-F}{RGB}{253,174,107}
\definecolor{Oranges-9-5}{RGB}{253,141,60}
\definecolor{Oranges-9-G}{RGB}{253,141,60}
\definecolor{Oranges-9-6}{RGB}{241,105,19}
\definecolor{Oranges-9-H}{RGB}{241,105,19}
\definecolor{Oranges-9-7}{RGB}{217,72,1}
\definecolor{Oranges-9-J}{RGB}{217,72,1}
\definecolor{Oranges-9-8}{RGB}{166,54,3}
\definecolor{Oranges-9-K}{RGB}{166,54,3}
\definecolor{Oranges-9-9}{RGB}{127,39,4}
\definecolor{Oranges-9-M}{RGB}{127,39,4}
\definecolor{Reds-3-1}{RGB}{254,224,210}
\definecolor{Reds-3-C}{RGB}{254,224,210}
\definecolor{Reds-3-2}{RGB}{252,146,114}
\definecolor{Reds-3-F}{RGB}{252,146,114}
\definecolor{Reds-3-3}{RGB}{222,45,38}
\definecolor{Reds-3-I}{RGB}{222,45,38}
\definecolor{Reds-4-1}{RGB}{254,229,217}
\definecolor{Reds-4-B}{RGB}{254,229,217}
\definecolor{Reds-4-2}{RGB}{252,174,145}
\definecolor{Reds-4-E}{RGB}{252,174,145}
\definecolor{Reds-4-3}{RGB}{251,106,74}
\definecolor{Reds-4-G}{RGB}{251,106,74}
\definecolor{Reds-4-4}{RGB}{203,24,29}
\definecolor{Reds-4-J}{RGB}{203,24,29}
\definecolor{Reds-5-1}{RGB}{254,229,217}
\definecolor{Reds-5-B}{RGB}{254,229,217}
\definecolor{Reds-5-2}{RGB}{252,174,145}
\definecolor{Reds-5-E}{RGB}{252,174,145}
\definecolor{Reds-5-3}{RGB}{251,106,74}
\definecolor{Reds-5-G}{RGB}{251,106,74}
\definecolor{Reds-5-4}{RGB}{222,45,38}
\definecolor{Reds-5-I}{RGB}{222,45,38}
\definecolor{Reds-5-5}{RGB}{165,15,21}
\definecolor{Reds-5-K}{RGB}{165,15,21}
\definecolor{Reds-6-1}{RGB}{254,229,217}
\definecolor{Reds-6-B}{RGB}{254,229,217}
\definecolor{Reds-6-2}{RGB}{252,187,161}
\definecolor{Reds-6-D}{RGB}{252,187,161}
\definecolor{Reds-6-3}{RGB}{252,146,114}
\definecolor{Reds-6-F}{RGB}{252,146,114}
\definecolor{Reds-6-4}{RGB}{251,106,74}
\definecolor{Reds-6-G}{RGB}{251,106,74}
\definecolor{Reds-6-5}{RGB}{222,45,38}
\definecolor{Reds-6-I}{RGB}{222,45,38}
\definecolor{Reds-6-6}{RGB}{165,15,21}
\definecolor{Reds-6-K}{RGB}{165,15,21}
\definecolor{Reds-7-1}{RGB}{254,229,217}
\definecolor{Reds-7-B}{RGB}{254,229,217}
\definecolor{Reds-7-2}{RGB}{252,187,161}
\definecolor{Reds-7-D}{RGB}{252,187,161}
\definecolor{Reds-7-3}{RGB}{252,146,114}
\definecolor{Reds-7-F}{RGB}{252,146,114}
\definecolor{Reds-7-4}{RGB}{251,106,74}
\definecolor{Reds-7-G}{RGB}{251,106,74}
\definecolor{Reds-7-5}{RGB}{239,59,44}
\definecolor{Reds-7-H}{RGB}{239,59,44}
\definecolor{Reds-7-6}{RGB}{203,24,29}
\definecolor{Reds-7-J}{RGB}{203,24,29}
\definecolor{Reds-7-7}{RGB}{153,0,13}
\definecolor{Reds-7-L}{RGB}{153,0,13}
\definecolor{Reds-8-1}{RGB}{255,245,240}
\definecolor{Reds-8-A}{RGB}{255,245,240}
\definecolor{Reds-8-2}{RGB}{254,224,210}
\definecolor{Reds-8-C}{RGB}{254,224,210}
\definecolor{Reds-8-3}{RGB}{252,187,161}
\definecolor{Reds-8-D}{RGB}{252,187,161}
\definecolor{Reds-8-4}{RGB}{252,146,114}
\definecolor{Reds-8-F}{RGB}{252,146,114}
\definecolor{Reds-8-5}{RGB}{251,106,74}
\definecolor{Reds-8-G}{RGB}{251,106,74}
\definecolor{Reds-8-6}{RGB}{239,59,44}
\definecolor{Reds-8-H}{RGB}{239,59,44}
\definecolor{Reds-8-7}{RGB}{203,24,29}
\definecolor{Reds-8-J}{RGB}{203,24,29}
\definecolor{Reds-8-8}{RGB}{153,0,13}
\definecolor{Reds-8-L}{RGB}{153,0,13}
\definecolor{Reds-9-1}{RGB}{255,245,240}
\definecolor{Reds-9-A}{RGB}{255,245,240}
\definecolor{Reds-9-2}{RGB}{254,224,210}
\definecolor{Reds-9-C}{RGB}{254,224,210}
\definecolor{Reds-9-3}{RGB}{252,187,161}
\definecolor{Reds-9-D}{RGB}{252,187,161}
\definecolor{Reds-9-4}{RGB}{252,146,114}
\definecolor{Reds-9-F}{RGB}{252,146,114}
\definecolor{Reds-9-5}{RGB}{251,106,74}
\definecolor{Reds-9-G}{RGB}{251,106,74}
\definecolor{Reds-9-6}{RGB}{239,59,44}
\definecolor{Reds-9-H}{RGB}{239,59,44}
\definecolor{Reds-9-7}{RGB}{203,24,29}
\definecolor{Reds-9-J}{RGB}{203,24,29}
\definecolor{Reds-9-8}{RGB}{165,15,21}
\definecolor{Reds-9-K}{RGB}{165,15,21}
\definecolor{Reds-9-9}{RGB}{103,0,13}
\definecolor{Reds-9-M}{RGB}{103,0,13}
\definecolor{Greys-3-1}{RGB}{240,240,240}
\definecolor{Greys-3-C}{RGB}{240,240,240}
\definecolor{Greys-3-2}{RGB}{189,189,189}
\definecolor{Greys-3-F}{RGB}{189,189,189}
\definecolor{Greys-3-3}{RGB}{99,99,99}
\definecolor{Greys-3-I}{RGB}{99,99,99}
\definecolor{Greys-4-1}{RGB}{247,247,247}
\definecolor{Greys-4-B}{RGB}{247,247,247}
\definecolor{Greys-4-2}{RGB}{204,204,204}
\definecolor{Greys-4-E}{RGB}{204,204,204}
\definecolor{Greys-4-3}{RGB}{150,150,150}
\definecolor{Greys-4-G}{RGB}{150,150,150}
\definecolor{Greys-4-4}{RGB}{82,82,82}
\definecolor{Greys-4-J}{RGB}{82,82,82}
\definecolor{Greys-5-1}{RGB}{247,247,247}
\definecolor{Greys-5-B}{RGB}{247,247,247}
\definecolor{Greys-5-2}{RGB}{204,204,204}
\definecolor{Greys-5-E}{RGB}{204,204,204}
\definecolor{Greys-5-3}{RGB}{150,150,150}
\definecolor{Greys-5-G}{RGB}{150,150,150}
\definecolor{Greys-5-4}{RGB}{99,99,99}
\definecolor{Greys-5-I}{RGB}{99,99,99}
\definecolor{Greys-5-5}{RGB}{37,37,37}
\definecolor{Greys-5-K}{RGB}{37,37,37}
\definecolor{Greys-6-1}{RGB}{247,247,247}
\definecolor{Greys-6-B}{RGB}{247,247,247}
\definecolor{Greys-6-2}{RGB}{217,217,217}
\definecolor{Greys-6-D}{RGB}{217,217,217}
\definecolor{Greys-6-3}{RGB}{189,189,189}
\definecolor{Greys-6-F}{RGB}{189,189,189}
\definecolor{Greys-6-4}{RGB}{150,150,150}
\definecolor{Greys-6-G}{RGB}{150,150,150}
\definecolor{Greys-6-5}{RGB}{99,99,99}
\definecolor{Greys-6-I}{RGB}{99,99,99}
\definecolor{Greys-6-6}{RGB}{37,37,37}
\definecolor{Greys-6-K}{RGB}{37,37,37}
\definecolor{Greys-7-1}{RGB}{247,247,247}
\definecolor{Greys-7-B}{RGB}{247,247,247}
\definecolor{Greys-7-2}{RGB}{217,217,217}
\definecolor{Greys-7-D}{RGB}{217,217,217}
\definecolor{Greys-7-3}{RGB}{189,189,189}
\definecolor{Greys-7-F}{RGB}{189,189,189}
\definecolor{Greys-7-4}{RGB}{150,150,150}
\definecolor{Greys-7-G}{RGB}{150,150,150}
\definecolor{Greys-7-5}{RGB}{115,115,115}
\definecolor{Greys-7-H}{RGB}{115,115,115}
\definecolor{Greys-7-6}{RGB}{82,82,82}
\definecolor{Greys-7-J}{RGB}{82,82,82}
\definecolor{Greys-7-7}{RGB}{37,37,37}
\definecolor{Greys-7-L}{RGB}{37,37,37}
\definecolor{Greys-8-1}{RGB}{255,255,255}
\definecolor{Greys-8-A}{RGB}{255,255,255}
\definecolor{Greys-8-2}{RGB}{240,240,240}
\definecolor{Greys-8-C}{RGB}{240,240,240}
\definecolor{Greys-8-3}{RGB}{217,217,217}
\definecolor{Greys-8-D}{RGB}{217,217,217}
\definecolor{Greys-8-4}{RGB}{189,189,189}
\definecolor{Greys-8-F}{RGB}{189,189,189}
\definecolor{Greys-8-5}{RGB}{150,150,150}
\definecolor{Greys-8-G}{RGB}{150,150,150}
\definecolor{Greys-8-6}{RGB}{115,115,115}
\definecolor{Greys-8-H}{RGB}{115,115,115}
\definecolor{Greys-8-7}{RGB}{82,82,82}
\definecolor{Greys-8-J}{RGB}{82,82,82}
\definecolor{Greys-8-8}{RGB}{37,37,37}
\definecolor{Greys-8-L}{RGB}{37,37,37}
\definecolor{Greys-9-1}{RGB}{255,255,255}
\definecolor{Greys-9-A}{RGB}{255,255,255}
\definecolor{Greys-9-2}{RGB}{240,240,240}
\definecolor{Greys-9-C}{RGB}{240,240,240}
\definecolor{Greys-9-3}{RGB}{217,217,217}
\definecolor{Greys-9-D}{RGB}{217,217,217}
\definecolor{Greys-9-4}{RGB}{189,189,189}
\definecolor{Greys-9-F}{RGB}{189,189,189}
\definecolor{Greys-9-5}{RGB}{150,150,150}
\definecolor{Greys-9-G}{RGB}{150,150,150}
\definecolor{Greys-9-6}{RGB}{115,115,115}
\definecolor{Greys-9-H}{RGB}{115,115,115}
\definecolor{Greys-9-7}{RGB}{82,82,82}
\definecolor{Greys-9-J}{RGB}{82,82,82}
\definecolor{Greys-9-8}{RGB}{37,37,37}
\definecolor{Greys-9-K}{RGB}{37,37,37}
\definecolor{Greys-9-9}{RGB}{0,0,0}
\definecolor{Greys-9-M}{RGB}{0,0,0}
\definecolor{PuOr-3-1}{RGB}{241,163,64}
\definecolor{PuOr-3-E}{RGB}{241,163,64}
\definecolor{PuOr-3-2}{RGB}{247,247,247}
\definecolor{PuOr-3-H}{RGB}{247,247,247}
\definecolor{PuOr-3-3}{RGB}{153,142,195}
\definecolor{PuOr-3-K}{RGB}{153,142,195}
\definecolor{PuOr-4-1}{RGB}{230,97,1}
\definecolor{PuOr-4-C}{RGB}{230,97,1}
\definecolor{PuOr-4-2}{RGB}{253,184,99}
\definecolor{PuOr-4-F}{RGB}{253,184,99}
\definecolor{PuOr-4-3}{RGB}{178,171,210}
\definecolor{PuOr-4-J}{RGB}{178,171,210}
\definecolor{PuOr-4-4}{RGB}{94,60,153}
\definecolor{PuOr-4-M}{RGB}{94,60,153}
\definecolor{PuOr-5-1}{RGB}{230,97,1}
\definecolor{PuOr-5-C}{RGB}{230,97,1}
\definecolor{PuOr-5-2}{RGB}{253,184,99}
\definecolor{PuOr-5-F}{RGB}{253,184,99}
\definecolor{PuOr-5-3}{RGB}{247,247,247}
\definecolor{PuOr-5-H}{RGB}{247,247,247}
\definecolor{PuOr-5-4}{RGB}{178,171,210}
\definecolor{PuOr-5-J}{RGB}{178,171,210}
\definecolor{PuOr-5-5}{RGB}{94,60,153}
\definecolor{PuOr-5-M}{RGB}{94,60,153}
\definecolor{PuOr-6-1}{RGB}{179,88,6}
\definecolor{PuOr-6-B}{RGB}{179,88,6}
\definecolor{PuOr-6-2}{RGB}{241,163,64}
\definecolor{PuOr-6-E}{RGB}{241,163,64}
\definecolor{PuOr-6-3}{RGB}{254,224,182}
\definecolor{PuOr-6-G}{RGB}{254,224,182}
\definecolor{PuOr-6-4}{RGB}{216,218,235}
\definecolor{PuOr-6-I}{RGB}{216,218,235}
\definecolor{PuOr-6-5}{RGB}{153,142,195}
\definecolor{PuOr-6-K}{RGB}{153,142,195}
\definecolor{PuOr-6-6}{RGB}{84,39,136}
\definecolor{PuOr-6-N}{RGB}{84,39,136}
\definecolor{PuOr-7-1}{RGB}{179,88,6}
\definecolor{PuOr-7-B}{RGB}{179,88,6}
\definecolor{PuOr-7-2}{RGB}{241,163,64}
\definecolor{PuOr-7-E}{RGB}{241,163,64}
\definecolor{PuOr-7-3}{RGB}{254,224,182}
\definecolor{PuOr-7-G}{RGB}{254,224,182}
\definecolor{PuOr-7-4}{RGB}{247,247,247}
\definecolor{PuOr-7-H}{RGB}{247,247,247}
\definecolor{PuOr-7-5}{RGB}{216,218,235}
\definecolor{PuOr-7-I}{RGB}{216,218,235}
\definecolor{PuOr-7-6}{RGB}{153,142,195}
\definecolor{PuOr-7-K}{RGB}{153,142,195}
\definecolor{PuOr-7-7}{RGB}{84,39,136}
\definecolor{PuOr-7-N}{RGB}{84,39,136}
\definecolor{PuOr-8-1}{RGB}{179,88,6}
\definecolor{PuOr-8-B}{RGB}{179,88,6}
\definecolor{PuOr-8-2}{RGB}{224,130,20}
\definecolor{PuOr-8-D}{RGB}{224,130,20}
\definecolor{PuOr-8-3}{RGB}{253,184,99}
\definecolor{PuOr-8-F}{RGB}{253,184,99}
\definecolor{PuOr-8-4}{RGB}{254,224,182}
\definecolor{PuOr-8-G}{RGB}{254,224,182}
\definecolor{PuOr-8-5}{RGB}{216,218,235}
\definecolor{PuOr-8-I}{RGB}{216,218,235}
\definecolor{PuOr-8-6}{RGB}{178,171,210}
\definecolor{PuOr-8-J}{RGB}{178,171,210}
\definecolor{PuOr-8-7}{RGB}{128,115,172}
\definecolor{PuOr-8-L}{RGB}{128,115,172}
\definecolor{PuOr-8-8}{RGB}{84,39,136}
\definecolor{PuOr-8-N}{RGB}{84,39,136}
\definecolor{PuOr-9-1}{RGB}{179,88,6}
\definecolor{PuOr-9-B}{RGB}{179,88,6}
\definecolor{PuOr-9-2}{RGB}{224,130,20}
\definecolor{PuOr-9-D}{RGB}{224,130,20}
\definecolor{PuOr-9-3}{RGB}{253,184,99}
\definecolor{PuOr-9-F}{RGB}{253,184,99}
\definecolor{PuOr-9-4}{RGB}{254,224,182}
\definecolor{PuOr-9-G}{RGB}{254,224,182}
\definecolor{PuOr-9-5}{RGB}{247,247,247}
\definecolor{PuOr-9-H}{RGB}{247,247,247}
\definecolor{PuOr-9-6}{RGB}{216,218,235}
\definecolor{PuOr-9-I}{RGB}{216,218,235}
\definecolor{PuOr-9-7}{RGB}{178,171,210}
\definecolor{PuOr-9-J}{RGB}{178,171,210}
\definecolor{PuOr-9-8}{RGB}{128,115,172}
\definecolor{PuOr-9-L}{RGB}{128,115,172}
\definecolor{PuOr-9-9}{RGB}{84,39,136}
\definecolor{PuOr-9-N}{RGB}{84,39,136}
\definecolor{PuOr-10-1}{RGB}{127,59,8}
\definecolor{PuOr-10-A}{RGB}{127,59,8}
\definecolor{PuOr-10-2}{RGB}{179,88,6}
\definecolor{PuOr-10-B}{RGB}{179,88,6}
\definecolor{PuOr-10-3}{RGB}{224,130,20}
\definecolor{PuOr-10-D}{RGB}{224,130,20}
\definecolor{PuOr-10-4}{RGB}{253,184,99}
\definecolor{PuOr-10-F}{RGB}{253,184,99}
\definecolor{PuOr-10-5}{RGB}{254,224,182}
\definecolor{PuOr-10-G}{RGB}{254,224,182}
\definecolor{PuOr-10-6}{RGB}{216,218,235}
\definecolor{PuOr-10-I}{RGB}{216,218,235}
\definecolor{PuOr-10-7}{RGB}{178,171,210}
\definecolor{PuOr-10-J}{RGB}{178,171,210}
\definecolor{PuOr-10-8}{RGB}{128,115,172}
\definecolor{PuOr-10-L}{RGB}{128,115,172}
\definecolor{PuOr-10-9}{RGB}{84,39,136}
\definecolor{PuOr-10-N}{RGB}{84,39,136}
\definecolor{PuOr-10-10}{RGB}{45,0,75}
\definecolor{PuOr-10-O}{RGB}{45,0,75}
\definecolor{PuOr-11-1}{RGB}{127,59,8}
\definecolor{PuOr-11-A}{RGB}{127,59,8}
\definecolor{PuOr-11-2}{RGB}{179,88,6}
\definecolor{PuOr-11-B}{RGB}{179,88,6}
\definecolor{PuOr-11-3}{RGB}{224,130,20}
\definecolor{PuOr-11-D}{RGB}{224,130,20}
\definecolor{PuOr-11-4}{RGB}{253,184,99}
\definecolor{PuOr-11-F}{RGB}{253,184,99}
\definecolor{PuOr-11-5}{RGB}{254,224,182}
\definecolor{PuOr-11-G}{RGB}{254,224,182}
\definecolor{PuOr-11-6}{RGB}{247,247,247}
\definecolor{PuOr-11-H}{RGB}{247,247,247}
\definecolor{PuOr-11-7}{RGB}{216,218,235}
\definecolor{PuOr-11-I}{RGB}{216,218,235}
\definecolor{PuOr-11-8}{RGB}{178,171,210}
\definecolor{PuOr-11-J}{RGB}{178,171,210}
\definecolor{PuOr-11-9}{RGB}{128,115,172}
\definecolor{PuOr-11-L}{RGB}{128,115,172}
\definecolor{PuOr-11-10}{RGB}{84,39,136}
\definecolor{PuOr-11-N}{RGB}{84,39,136}
\definecolor{PuOr-11-11}{RGB}{45,0,75}
\definecolor{PuOr-11-O}{RGB}{45,0,75}
\definecolor{BrBG-3-1}{RGB}{216,179,101}
\definecolor{BrBG-3-E}{RGB}{216,179,101}
\definecolor{BrBG-3-2}{RGB}{245,245,245}
\definecolor{BrBG-3-H}{RGB}{245,245,245}
\definecolor{BrBG-3-3}{RGB}{90,180,172}
\definecolor{BrBG-3-K}{RGB}{90,180,172}
\definecolor{BrBG-4-1}{RGB}{166,97,26}
\definecolor{BrBG-4-C}{RGB}{166,97,26}
\definecolor{BrBG-4-2}{RGB}{223,194,125}
\definecolor{BrBG-4-F}{RGB}{223,194,125}
\definecolor{BrBG-4-3}{RGB}{128,205,193}
\definecolor{BrBG-4-J}{RGB}{128,205,193}
\definecolor{BrBG-4-4}{RGB}{1,133,113}
\definecolor{BrBG-4-M}{RGB}{1,133,113}
\definecolor{BrBG-5-1}{RGB}{166,97,26}
\definecolor{BrBG-5-C}{RGB}{166,97,26}
\definecolor{BrBG-5-2}{RGB}{223,194,125}
\definecolor{BrBG-5-F}{RGB}{223,194,125}
\definecolor{BrBG-5-3}{RGB}{245,245,245}
\definecolor{BrBG-5-H}{RGB}{245,245,245}
\definecolor{BrBG-5-4}{RGB}{128,205,193}
\definecolor{BrBG-5-J}{RGB}{128,205,193}
\definecolor{BrBG-5-5}{RGB}{1,133,113}
\definecolor{BrBG-5-M}{RGB}{1,133,113}
\definecolor{BrBG-6-1}{RGB}{140,81,10}
\definecolor{BrBG-6-B}{RGB}{140,81,10}
\definecolor{BrBG-6-2}{RGB}{216,179,101}
\definecolor{BrBG-6-E}{RGB}{216,179,101}
\definecolor{BrBG-6-3}{RGB}{246,232,195}
\definecolor{BrBG-6-G}{RGB}{246,232,195}
\definecolor{BrBG-6-4}{RGB}{199,234,229}
\definecolor{BrBG-6-I}{RGB}{199,234,229}
\definecolor{BrBG-6-5}{RGB}{90,180,172}
\definecolor{BrBG-6-K}{RGB}{90,180,172}
\definecolor{BrBG-6-6}{RGB}{1,102,94}
\definecolor{BrBG-6-N}{RGB}{1,102,94}
\definecolor{BrBG-7-1}{RGB}{140,81,10}
\definecolor{BrBG-7-B}{RGB}{140,81,10}
\definecolor{BrBG-7-2}{RGB}{216,179,101}
\definecolor{BrBG-7-E}{RGB}{216,179,101}
\definecolor{BrBG-7-3}{RGB}{246,232,195}
\definecolor{BrBG-7-G}{RGB}{246,232,195}
\definecolor{BrBG-7-4}{RGB}{245,245,245}
\definecolor{BrBG-7-H}{RGB}{245,245,245}
\definecolor{BrBG-7-5}{RGB}{199,234,229}
\definecolor{BrBG-7-I}{RGB}{199,234,229}
\definecolor{BrBG-7-6}{RGB}{90,180,172}
\definecolor{BrBG-7-K}{RGB}{90,180,172}
\definecolor{BrBG-7-7}{RGB}{1,102,94}
\definecolor{BrBG-7-N}{RGB}{1,102,94}
\definecolor{BrBG-8-1}{RGB}{140,81,10}
\definecolor{BrBG-8-B}{RGB}{140,81,10}
\definecolor{BrBG-8-2}{RGB}{191,129,45}
\definecolor{BrBG-8-D}{RGB}{191,129,45}
\definecolor{BrBG-8-3}{RGB}{223,194,125}
\definecolor{BrBG-8-F}{RGB}{223,194,125}
\definecolor{BrBG-8-4}{RGB}{246,232,195}
\definecolor{BrBG-8-G}{RGB}{246,232,195}
\definecolor{BrBG-8-5}{RGB}{199,234,229}
\definecolor{BrBG-8-I}{RGB}{199,234,229}
\definecolor{BrBG-8-6}{RGB}{128,205,193}
\definecolor{BrBG-8-J}{RGB}{128,205,193}
\definecolor{BrBG-8-7}{RGB}{53,151,143}
\definecolor{BrBG-8-L}{RGB}{53,151,143}
\definecolor{BrBG-8-8}{RGB}{1,102,94}
\definecolor{BrBG-8-N}{RGB}{1,102,94}
\definecolor{BrBG-9-1}{RGB}{140,81,10}
\definecolor{BrBG-9-B}{RGB}{140,81,10}
\definecolor{BrBG-9-2}{RGB}{191,129,45}
\definecolor{BrBG-9-D}{RGB}{191,129,45}
\definecolor{BrBG-9-3}{RGB}{223,194,125}
\definecolor{BrBG-9-F}{RGB}{223,194,125}
\definecolor{BrBG-9-4}{RGB}{246,232,195}
\definecolor{BrBG-9-G}{RGB}{246,232,195}
\definecolor{BrBG-9-5}{RGB}{245,245,245}
\definecolor{BrBG-9-H}{RGB}{245,245,245}
\definecolor{BrBG-9-6}{RGB}{199,234,229}
\definecolor{BrBG-9-I}{RGB}{199,234,229}
\definecolor{BrBG-9-7}{RGB}{128,205,193}
\definecolor{BrBG-9-J}{RGB}{128,205,193}
\definecolor{BrBG-9-8}{RGB}{53,151,143}
\definecolor{BrBG-9-L}{RGB}{53,151,143}
\definecolor{BrBG-9-9}{RGB}{1,102,94}
\definecolor{BrBG-9-N}{RGB}{1,102,94}
\definecolor{BrBG-10-1}{RGB}{84,48,5}
\definecolor{BrBG-10-A}{RGB}{84,48,5}
\definecolor{BrBG-10-2}{RGB}{140,81,10}
\definecolor{BrBG-10-B}{RGB}{140,81,10}
\definecolor{BrBG-10-3}{RGB}{191,129,45}
\definecolor{BrBG-10-D}{RGB}{191,129,45}
\definecolor{BrBG-10-4}{RGB}{223,194,125}
\definecolor{BrBG-10-F}{RGB}{223,194,125}
\definecolor{BrBG-10-5}{RGB}{246,232,195}
\definecolor{BrBG-10-G}{RGB}{246,232,195}
\definecolor{BrBG-10-6}{RGB}{199,234,229}
\definecolor{BrBG-10-I}{RGB}{199,234,229}
\definecolor{BrBG-10-7}{RGB}{128,205,193}
\definecolor{BrBG-10-J}{RGB}{128,205,193}
\definecolor{BrBG-10-8}{RGB}{53,151,143}
\definecolor{BrBG-10-L}{RGB}{53,151,143}
\definecolor{BrBG-10-9}{RGB}{1,102,94}
\definecolor{BrBG-10-N}{RGB}{1,102,94}
\definecolor{BrBG-10-10}{RGB}{0,60,48}
\definecolor{BrBG-10-O}{RGB}{0,60,48}
\definecolor{BrBG-11-1}{RGB}{84,48,5}
\definecolor{BrBG-11-A}{RGB}{84,48,5}
\definecolor{BrBG-11-2}{RGB}{140,81,10}
\definecolor{BrBG-11-B}{RGB}{140,81,10}
\definecolor{BrBG-11-3}{RGB}{191,129,45}
\definecolor{BrBG-11-D}{RGB}{191,129,45}
\definecolor{BrBG-11-4}{RGB}{223,194,125}
\definecolor{BrBG-11-F}{RGB}{223,194,125}
\definecolor{BrBG-11-5}{RGB}{246,232,195}
\definecolor{BrBG-11-G}{RGB}{246,232,195}
\definecolor{BrBG-11-6}{RGB}{245,245,245}
\definecolor{BrBG-11-H}{RGB}{245,245,245}
\definecolor{BrBG-11-7}{RGB}{199,234,229}
\definecolor{BrBG-11-I}{RGB}{199,234,229}
\definecolor{BrBG-11-8}{RGB}{128,205,193}
\definecolor{BrBG-11-J}{RGB}{128,205,193}
\definecolor{BrBG-11-9}{RGB}{53,151,143}
\definecolor{BrBG-11-L}{RGB}{53,151,143}
\definecolor{BrBG-11-10}{RGB}{1,102,94}
\definecolor{BrBG-11-N}{RGB}{1,102,94}
\definecolor{BrBG-11-11}{RGB}{0,60,48}
\definecolor{BrBG-11-O}{RGB}{0,60,48}
\definecolor{PRGn-3-1}{RGB}{175,141,195}
\definecolor{PRGn-3-E}{RGB}{175,141,195}
\definecolor{PRGn-3-2}{RGB}{247,247,247}
\definecolor{PRGn-3-H}{RGB}{247,247,247}
\definecolor{PRGn-3-3}{RGB}{127,191,123}
\definecolor{PRGn-3-K}{RGB}{127,191,123}
\definecolor{PRGn-4-1}{RGB}{123,50,148}
\definecolor{PRGn-4-C}{RGB}{123,50,148}
\definecolor{PRGn-4-2}{RGB}{194,165,207}
\definecolor{PRGn-4-F}{RGB}{194,165,207}
\definecolor{PRGn-4-3}{RGB}{166,219,160}
\definecolor{PRGn-4-J}{RGB}{166,219,160}
\definecolor{PRGn-4-4}{RGB}{0,136,55}
\definecolor{PRGn-4-M}{RGB}{0,136,55}
\definecolor{PRGn-5-1}{RGB}{123,50,148}
\definecolor{PRGn-5-C}{RGB}{123,50,148}
\definecolor{PRGn-5-2}{RGB}{194,165,207}
\definecolor{PRGn-5-F}{RGB}{194,165,207}
\definecolor{PRGn-5-3}{RGB}{247,247,247}
\definecolor{PRGn-5-H}{RGB}{247,247,247}
\definecolor{PRGn-5-4}{RGB}{166,219,160}
\definecolor{PRGn-5-J}{RGB}{166,219,160}
\definecolor{PRGn-5-5}{RGB}{0,136,55}
\definecolor{PRGn-5-M}{RGB}{0,136,55}
\definecolor{PRGn-6-1}{RGB}{118,42,131}
\definecolor{PRGn-6-B}{RGB}{118,42,131}
\definecolor{PRGn-6-2}{RGB}{175,141,195}
\definecolor{PRGn-6-E}{RGB}{175,141,195}
\definecolor{PRGn-6-3}{RGB}{231,212,232}
\definecolor{PRGn-6-G}{RGB}{231,212,232}
\definecolor{PRGn-6-4}{RGB}{217,240,211}
\definecolor{PRGn-6-I}{RGB}{217,240,211}
\definecolor{PRGn-6-5}{RGB}{127,191,123}
\definecolor{PRGn-6-K}{RGB}{127,191,123}
\definecolor{PRGn-6-6}{RGB}{27,120,55}
\definecolor{PRGn-6-N}{RGB}{27,120,55}
\definecolor{PRGn-7-1}{RGB}{118,42,131}
\definecolor{PRGn-7-B}{RGB}{118,42,131}
\definecolor{PRGn-7-2}{RGB}{175,141,195}
\definecolor{PRGn-7-E}{RGB}{175,141,195}
\definecolor{PRGn-7-3}{RGB}{231,212,232}
\definecolor{PRGn-7-G}{RGB}{231,212,232}
\definecolor{PRGn-7-4}{RGB}{247,247,247}
\definecolor{PRGn-7-H}{RGB}{247,247,247}
\definecolor{PRGn-7-5}{RGB}{217,240,211}
\definecolor{PRGn-7-I}{RGB}{217,240,211}
\definecolor{PRGn-7-6}{RGB}{127,191,123}
\definecolor{PRGn-7-K}{RGB}{127,191,123}
\definecolor{PRGn-7-7}{RGB}{27,120,55}
\definecolor{PRGn-7-N}{RGB}{27,120,55}
\definecolor{PRGn-8-1}{RGB}{118,42,131}
\definecolor{PRGn-8-B}{RGB}{118,42,131}
\definecolor{PRGn-8-2}{RGB}{153,112,171}
\definecolor{PRGn-8-D}{RGB}{153,112,171}
\definecolor{PRGn-8-3}{RGB}{194,165,207}
\definecolor{PRGn-8-F}{RGB}{194,165,207}
\definecolor{PRGn-8-4}{RGB}{231,212,232}
\definecolor{PRGn-8-G}{RGB}{231,212,232}
\definecolor{PRGn-8-5}{RGB}{217,240,211}
\definecolor{PRGn-8-I}{RGB}{217,240,211}
\definecolor{PRGn-8-6}{RGB}{166,219,160}
\definecolor{PRGn-8-J}{RGB}{166,219,160}
\definecolor{PRGn-8-7}{RGB}{90,174,97}
\definecolor{PRGn-8-L}{RGB}{90,174,97}
\definecolor{PRGn-8-8}{RGB}{27,120,55}
\definecolor{PRGn-8-N}{RGB}{27,120,55}
\definecolor{PRGn-9-1}{RGB}{118,42,131}
\definecolor{PRGn-9-B}{RGB}{118,42,131}
\definecolor{PRGn-9-2}{RGB}{153,112,171}
\definecolor{PRGn-9-D}{RGB}{153,112,171}
\definecolor{PRGn-9-3}{RGB}{194,165,207}
\definecolor{PRGn-9-F}{RGB}{194,165,207}
\definecolor{PRGn-9-4}{RGB}{231,212,232}
\definecolor{PRGn-9-G}{RGB}{231,212,232}
\definecolor{PRGn-9-5}{RGB}{247,247,247}
\definecolor{PRGn-9-H}{RGB}{247,247,247}
\definecolor{PRGn-9-6}{RGB}{217,240,211}
\definecolor{PRGn-9-I}{RGB}{217,240,211}
\definecolor{PRGn-9-7}{RGB}{166,219,160}
\definecolor{PRGn-9-J}{RGB}{166,219,160}
\definecolor{PRGn-9-8}{RGB}{90,174,97}
\definecolor{PRGn-9-L}{RGB}{90,174,97}
\definecolor{PRGn-9-9}{RGB}{27,120,55}
\definecolor{PRGn-9-N}{RGB}{27,120,55}
\definecolor{PRGn-10-1}{RGB}{64,0,75}
\definecolor{PRGn-10-A}{RGB}{64,0,75}
\definecolor{PRGn-10-2}{RGB}{118,42,131}
\definecolor{PRGn-10-B}{RGB}{118,42,131}
\definecolor{PRGn-10-3}{RGB}{153,112,171}
\definecolor{PRGn-10-D}{RGB}{153,112,171}
\definecolor{PRGn-10-4}{RGB}{194,165,207}
\definecolor{PRGn-10-F}{RGB}{194,165,207}
\definecolor{PRGn-10-5}{RGB}{231,212,232}
\definecolor{PRGn-10-G}{RGB}{231,212,232}
\definecolor{PRGn-10-6}{RGB}{217,240,211}
\definecolor{PRGn-10-I}{RGB}{217,240,211}
\definecolor{PRGn-10-7}{RGB}{166,219,160}
\definecolor{PRGn-10-J}{RGB}{166,219,160}
\definecolor{PRGn-10-8}{RGB}{90,174,97}
\definecolor{PRGn-10-L}{RGB}{90,174,97}
\definecolor{PRGn-10-9}{RGB}{27,120,55}
\definecolor{PRGn-10-N}{RGB}{27,120,55}
\definecolor{PRGn-10-10}{RGB}{0,68,27}
\definecolor{PRGn-10-O}{RGB}{0,68,27}
\definecolor{PRGn-11-1}{RGB}{64,0,75}
\definecolor{PRGn-11-A}{RGB}{64,0,75}
\definecolor{PRGn-11-2}{RGB}{118,42,131}
\definecolor{PRGn-11-B}{RGB}{118,42,131}
\definecolor{PRGn-11-3}{RGB}{153,112,171}
\definecolor{PRGn-11-D}{RGB}{153,112,171}
\definecolor{PRGn-11-4}{RGB}{194,165,207}
\definecolor{PRGn-11-F}{RGB}{194,165,207}
\definecolor{PRGn-11-5}{RGB}{231,212,232}
\definecolor{PRGn-11-G}{RGB}{231,212,232}
\definecolor{PRGn-11-6}{RGB}{247,247,247}
\definecolor{PRGn-11-H}{RGB}{247,247,247}
\definecolor{PRGn-11-7}{RGB}{217,240,211}
\definecolor{PRGn-11-I}{RGB}{217,240,211}
\definecolor{PRGn-11-8}{RGB}{166,219,160}
\definecolor{PRGn-11-J}{RGB}{166,219,160}
\definecolor{PRGn-11-9}{RGB}{90,174,97}
\definecolor{PRGn-11-L}{RGB}{90,174,97}
\definecolor{PRGn-11-10}{RGB}{27,120,55}
\definecolor{PRGn-11-N}{RGB}{27,120,55}
\definecolor{PRGn-11-11}{RGB}{0,68,27}
\definecolor{PRGn-11-O}{RGB}{0,68,27}
\definecolor{PiYG-3-1}{RGB}{233,163,201}
\definecolor{PiYG-3-E}{RGB}{233,163,201}
\definecolor{PiYG-3-2}{RGB}{247,247,247}
\definecolor{PiYG-3-H}{RGB}{247,247,247}
\definecolor{PiYG-3-3}{RGB}{161,215,106}
\definecolor{PiYG-3-K}{RGB}{161,215,106}
\definecolor{PiYG-4-1}{RGB}{208,28,139}
\definecolor{PiYG-4-C}{RGB}{208,28,139}
\definecolor{PiYG-4-2}{RGB}{241,182,218}
\definecolor{PiYG-4-F}{RGB}{241,182,218}
\definecolor{PiYG-4-3}{RGB}{184,225,134}
\definecolor{PiYG-4-J}{RGB}{184,225,134}
\definecolor{PiYG-4-4}{RGB}{77,172,38}
\definecolor{PiYG-4-M}{RGB}{77,172,38}
\definecolor{PiYG-5-1}{RGB}{208,28,139}
\definecolor{PiYG-5-C}{RGB}{208,28,139}
\definecolor{PiYG-5-2}{RGB}{241,182,218}
\definecolor{PiYG-5-F}{RGB}{241,182,218}
\definecolor{PiYG-5-3}{RGB}{247,247,247}
\definecolor{PiYG-5-H}{RGB}{247,247,247}
\definecolor{PiYG-5-4}{RGB}{184,225,134}
\definecolor{PiYG-5-J}{RGB}{184,225,134}
\definecolor{PiYG-5-5}{RGB}{77,172,38}
\definecolor{PiYG-5-M}{RGB}{77,172,38}
\definecolor{PiYG-6-1}{RGB}{197,27,125}
\definecolor{PiYG-6-B}{RGB}{197,27,125}
\definecolor{PiYG-6-2}{RGB}{233,163,201}
\definecolor{PiYG-6-E}{RGB}{233,163,201}
\definecolor{PiYG-6-3}{RGB}{253,224,239}
\definecolor{PiYG-6-G}{RGB}{253,224,239}
\definecolor{PiYG-6-4}{RGB}{230,245,208}
\definecolor{PiYG-6-I}{RGB}{230,245,208}
\definecolor{PiYG-6-5}{RGB}{161,215,106}
\definecolor{PiYG-6-K}{RGB}{161,215,106}
\definecolor{PiYG-6-6}{RGB}{77,146,33}
\definecolor{PiYG-6-N}{RGB}{77,146,33}
\definecolor{PiYG-7-1}{RGB}{197,27,125}
\definecolor{PiYG-7-B}{RGB}{197,27,125}
\definecolor{PiYG-7-2}{RGB}{233,163,201}
\definecolor{PiYG-7-E}{RGB}{233,163,201}
\definecolor{PiYG-7-3}{RGB}{253,224,239}
\definecolor{PiYG-7-G}{RGB}{253,224,239}
\definecolor{PiYG-7-4}{RGB}{247,247,247}
\definecolor{PiYG-7-H}{RGB}{247,247,247}
\definecolor{PiYG-7-5}{RGB}{230,245,208}
\definecolor{PiYG-7-I}{RGB}{230,245,208}
\definecolor{PiYG-7-6}{RGB}{161,215,106}
\definecolor{PiYG-7-K}{RGB}{161,215,106}
\definecolor{PiYG-7-7}{RGB}{77,146,33}
\definecolor{PiYG-7-N}{RGB}{77,146,33}
\definecolor{PiYG-8-1}{RGB}{197,27,125}
\definecolor{PiYG-8-B}{RGB}{197,27,125}
\definecolor{PiYG-8-2}{RGB}{222,119,174}
\definecolor{PiYG-8-D}{RGB}{222,119,174}
\definecolor{PiYG-8-3}{RGB}{241,182,218}
\definecolor{PiYG-8-F}{RGB}{241,182,218}
\definecolor{PiYG-8-4}{RGB}{253,224,239}
\definecolor{PiYG-8-G}{RGB}{253,224,239}
\definecolor{PiYG-8-5}{RGB}{230,245,208}
\definecolor{PiYG-8-I}{RGB}{230,245,208}
\definecolor{PiYG-8-6}{RGB}{184,225,134}
\definecolor{PiYG-8-J}{RGB}{184,225,134}
\definecolor{PiYG-8-7}{RGB}{127,188,65}
\definecolor{PiYG-8-L}{RGB}{127,188,65}
\definecolor{PiYG-8-8}{RGB}{77,146,33}
\definecolor{PiYG-8-N}{RGB}{77,146,33}
\definecolor{PiYG-9-1}{RGB}{197,27,125}
\definecolor{PiYG-9-B}{RGB}{197,27,125}
\definecolor{PiYG-9-2}{RGB}{222,119,174}
\definecolor{PiYG-9-D}{RGB}{222,119,174}
\definecolor{PiYG-9-3}{RGB}{241,182,218}
\definecolor{PiYG-9-F}{RGB}{241,182,218}
\definecolor{PiYG-9-4}{RGB}{253,224,239}
\definecolor{PiYG-9-G}{RGB}{253,224,239}
\definecolor{PiYG-9-5}{RGB}{247,247,247}
\definecolor{PiYG-9-H}{RGB}{247,247,247}
\definecolor{PiYG-9-6}{RGB}{230,245,208}
\definecolor{PiYG-9-I}{RGB}{230,245,208}
\definecolor{PiYG-9-7}{RGB}{184,225,134}
\definecolor{PiYG-9-J}{RGB}{184,225,134}
\definecolor{PiYG-9-8}{RGB}{127,188,65}
\definecolor{PiYG-9-L}{RGB}{127,188,65}
\definecolor{PiYG-9-9}{RGB}{77,146,33}
\definecolor{PiYG-9-N}{RGB}{77,146,33}
\definecolor{PiYG-10-1}{RGB}{142,1,82}
\definecolor{PiYG-10-A}{RGB}{142,1,82}
\definecolor{PiYG-10-2}{RGB}{197,27,125}
\definecolor{PiYG-10-B}{RGB}{197,27,125}
\definecolor{PiYG-10-3}{RGB}{222,119,174}
\definecolor{PiYG-10-D}{RGB}{222,119,174}
\definecolor{PiYG-10-4}{RGB}{241,182,218}
\definecolor{PiYG-10-F}{RGB}{241,182,218}
\definecolor{PiYG-10-5}{RGB}{253,224,239}
\definecolor{PiYG-10-G}{RGB}{253,224,239}
\definecolor{PiYG-10-6}{RGB}{230,245,208}
\definecolor{PiYG-10-I}{RGB}{230,245,208}
\definecolor{PiYG-10-7}{RGB}{184,225,134}
\definecolor{PiYG-10-J}{RGB}{184,225,134}
\definecolor{PiYG-10-8}{RGB}{127,188,65}
\definecolor{PiYG-10-L}{RGB}{127,188,65}
\definecolor{PiYG-10-9}{RGB}{77,146,33}
\definecolor{PiYG-10-N}{RGB}{77,146,33}
\definecolor{PiYG-10-10}{RGB}{39,100,25}
\definecolor{PiYG-10-O}{RGB}{39,100,25}
\definecolor{PiYG-11-1}{RGB}{142,1,82}
\definecolor{PiYG-11-A}{RGB}{142,1,82}
\definecolor{PiYG-11-2}{RGB}{197,27,125}
\definecolor{PiYG-11-B}{RGB}{197,27,125}
\definecolor{PiYG-11-3}{RGB}{222,119,174}
\definecolor{PiYG-11-D}{RGB}{222,119,174}
\definecolor{PiYG-11-4}{RGB}{241,182,218}
\definecolor{PiYG-11-F}{RGB}{241,182,218}
\definecolor{PiYG-11-5}{RGB}{253,224,239}
\definecolor{PiYG-11-G}{RGB}{253,224,239}
\definecolor{PiYG-11-6}{RGB}{247,247,247}
\definecolor{PiYG-11-H}{RGB}{247,247,247}
\definecolor{PiYG-11-7}{RGB}{230,245,208}
\definecolor{PiYG-11-I}{RGB}{230,245,208}
\definecolor{PiYG-11-8}{RGB}{184,225,134}
\definecolor{PiYG-11-J}{RGB}{184,225,134}
\definecolor{PiYG-11-9}{RGB}{127,188,65}
\definecolor{PiYG-11-L}{RGB}{127,188,65}
\definecolor{PiYG-11-10}{RGB}{77,146,33}
\definecolor{PiYG-11-N}{RGB}{77,146,33}
\definecolor{PiYG-11-11}{RGB}{39,100,25}
\definecolor{PiYG-11-O}{RGB}{39,100,25}
\definecolor{RdBu-3-1}{RGB}{239,138,98}
\definecolor{RdBu-3-E}{RGB}{239,138,98}
\definecolor{RdBu-3-2}{RGB}{247,247,247}
\definecolor{RdBu-3-H}{RGB}{247,247,247}
\definecolor{RdBu-3-3}{RGB}{103,169,207}
\definecolor{RdBu-3-K}{RGB}{103,169,207}
\definecolor{RdBu-4-1}{RGB}{202,0,32}
\definecolor{RdBu-4-C}{RGB}{202,0,32}
\definecolor{RdBu-4-2}{RGB}{244,165,130}
\definecolor{RdBu-4-F}{RGB}{244,165,130}
\definecolor{RdBu-4-3}{RGB}{146,197,222}
\definecolor{RdBu-4-J}{RGB}{146,197,222}
\definecolor{RdBu-4-4}{RGB}{5,113,176}
\definecolor{RdBu-4-M}{RGB}{5,113,176}
\definecolor{RdBu-5-1}{RGB}{202,0,32}
\definecolor{RdBu-5-C}{RGB}{202,0,32}
\definecolor{RdBu-5-2}{RGB}{244,165,130}
\definecolor{RdBu-5-F}{RGB}{244,165,130}
\definecolor{RdBu-5-3}{RGB}{247,247,247}
\definecolor{RdBu-5-H}{RGB}{247,247,247}
\definecolor{RdBu-5-4}{RGB}{146,197,222}
\definecolor{RdBu-5-J}{RGB}{146,197,222}
\definecolor{RdBu-5-5}{RGB}{5,113,176}
\definecolor{RdBu-5-M}{RGB}{5,113,176}
\definecolor{RdBu-6-1}{RGB}{178,24,43}
\definecolor{RdBu-6-B}{RGB}{178,24,43}
\definecolor{RdBu-6-2}{RGB}{239,138,98}
\definecolor{RdBu-6-E}{RGB}{239,138,98}
\definecolor{RdBu-6-3}{RGB}{253,219,199}
\definecolor{RdBu-6-G}{RGB}{253,219,199}
\definecolor{RdBu-6-4}{RGB}{209,229,240}
\definecolor{RdBu-6-I}{RGB}{209,229,240}
\definecolor{RdBu-6-5}{RGB}{103,169,207}
\definecolor{RdBu-6-K}{RGB}{103,169,207}
\definecolor{RdBu-6-6}{RGB}{33,102,172}
\definecolor{RdBu-6-N}{RGB}{33,102,172}
\definecolor{RdBu-7-1}{RGB}{178,24,43}
\definecolor{RdBu-7-B}{RGB}{178,24,43}
\definecolor{RdBu-7-2}{RGB}{239,138,98}
\definecolor{RdBu-7-E}{RGB}{239,138,98}
\definecolor{RdBu-7-3}{RGB}{253,219,199}
\definecolor{RdBu-7-G}{RGB}{253,219,199}
\definecolor{RdBu-7-4}{RGB}{247,247,247}
\definecolor{RdBu-7-H}{RGB}{247,247,247}
\definecolor{RdBu-7-5}{RGB}{209,229,240}
\definecolor{RdBu-7-I}{RGB}{209,229,240}
\definecolor{RdBu-7-6}{RGB}{103,169,207}
\definecolor{RdBu-7-K}{RGB}{103,169,207}
\definecolor{RdBu-7-7}{RGB}{33,102,172}
\definecolor{RdBu-7-N}{RGB}{33,102,172}
\definecolor{RdBu-8-1}{RGB}{178,24,43}
\definecolor{RdBu-8-B}{RGB}{178,24,43}
\definecolor{RdBu-8-2}{RGB}{214,96,77}
\definecolor{RdBu-8-D}{RGB}{214,96,77}
\definecolor{RdBu-8-3}{RGB}{244,165,130}
\definecolor{RdBu-8-F}{RGB}{244,165,130}
\definecolor{RdBu-8-4}{RGB}{253,219,199}
\definecolor{RdBu-8-G}{RGB}{253,219,199}
\definecolor{RdBu-8-5}{RGB}{209,229,240}
\definecolor{RdBu-8-I}{RGB}{209,229,240}
\definecolor{RdBu-8-6}{RGB}{146,197,222}
\definecolor{RdBu-8-J}{RGB}{146,197,222}
\definecolor{RdBu-8-7}{RGB}{67,147,195}
\definecolor{RdBu-8-L}{RGB}{67,147,195}
\definecolor{RdBu-8-8}{RGB}{33,102,172}
\definecolor{RdBu-8-N}{RGB}{33,102,172}
\definecolor{RdBu-9-1}{RGB}{178,24,43}
\definecolor{RdBu-9-B}{RGB}{178,24,43}
\definecolor{RdBu-9-2}{RGB}{214,96,77}
\definecolor{RdBu-9-D}{RGB}{214,96,77}
\definecolor{RdBu-9-3}{RGB}{244,165,130}
\definecolor{RdBu-9-F}{RGB}{244,165,130}
\definecolor{RdBu-9-4}{RGB}{253,219,199}
\definecolor{RdBu-9-G}{RGB}{253,219,199}
\definecolor{RdBu-9-5}{RGB}{247,247,247}
\definecolor{RdBu-9-H}{RGB}{247,247,247}
\definecolor{RdBu-9-6}{RGB}{209,229,240}
\definecolor{RdBu-9-I}{RGB}{209,229,240}
\definecolor{RdBu-9-7}{RGB}{146,197,222}
\definecolor{RdBu-9-J}{RGB}{146,197,222}
\definecolor{RdBu-9-8}{RGB}{67,147,195}
\definecolor{RdBu-9-L}{RGB}{67,147,195}
\definecolor{RdBu-9-9}{RGB}{33,102,172}
\definecolor{RdBu-9-N}{RGB}{33,102,172}
\definecolor{RdBu-10-1}{RGB}{103,0,31}
\definecolor{RdBu-10-A}{RGB}{103,0,31}
\definecolor{RdBu-10-2}{RGB}{178,24,43}
\definecolor{RdBu-10-B}{RGB}{178,24,43}
\definecolor{RdBu-10-3}{RGB}{214,96,77}
\definecolor{RdBu-10-D}{RGB}{214,96,77}
\definecolor{RdBu-10-4}{RGB}{244,165,130}
\definecolor{RdBu-10-F}{RGB}{244,165,130}
\definecolor{RdBu-10-5}{RGB}{253,219,199}
\definecolor{RdBu-10-G}{RGB}{253,219,199}
\definecolor{RdBu-10-6}{RGB}{209,229,240}
\definecolor{RdBu-10-I}{RGB}{209,229,240}
\definecolor{RdBu-10-7}{RGB}{146,197,222}
\definecolor{RdBu-10-J}{RGB}{146,197,222}
\definecolor{RdBu-10-8}{RGB}{67,147,195}
\definecolor{RdBu-10-L}{RGB}{67,147,195}
\definecolor{RdBu-10-9}{RGB}{33,102,172}
\definecolor{RdBu-10-N}{RGB}{33,102,172}
\definecolor{RdBu-10-10}{RGB}{5,48,97}
\definecolor{RdBu-10-O}{RGB}{5,48,97}
\definecolor{RdBu-11-1}{RGB}{103,0,31}
\definecolor{RdBu-11-A}{RGB}{103,0,31}
\definecolor{RdBu-11-2}{RGB}{178,24,43}
\definecolor{RdBu-11-B}{RGB}{178,24,43}
\definecolor{RdBu-11-3}{RGB}{214,96,77}
\definecolor{RdBu-11-D}{RGB}{214,96,77}
\definecolor{RdBu-11-4}{RGB}{244,165,130}
\definecolor{RdBu-11-F}{RGB}{244,165,130}
\definecolor{RdBu-11-5}{RGB}{253,219,199}
\definecolor{RdBu-11-G}{RGB}{253,219,199}
\definecolor{RdBu-11-6}{RGB}{247,247,247}
\definecolor{RdBu-11-H}{RGB}{247,247,247}
\definecolor{RdBu-11-7}{RGB}{209,229,240}
\definecolor{RdBu-11-I}{RGB}{209,229,240}
\definecolor{RdBu-11-8}{RGB}{146,197,222}
\definecolor{RdBu-11-J}{RGB}{146,197,222}
\definecolor{RdBu-11-9}{RGB}{67,147,195}
\definecolor{RdBu-11-L}{RGB}{67,147,195}
\definecolor{RdBu-11-10}{RGB}{33,102,172}
\definecolor{RdBu-11-N}{RGB}{33,102,172}
\definecolor{RdBu-11-11}{RGB}{5,48,97}
\definecolor{RdBu-11-O}{RGB}{5,48,97}
\definecolor{RdGy-3-1}{RGB}{239,138,98}
\definecolor{RdGy-3-E}{RGB}{239,138,98}
\definecolor{RdGy-3-2}{RGB}{255,255,255}
\definecolor{RdGy-3-H}{RGB}{255,255,255}
\definecolor{RdGy-3-3}{RGB}{153,153,153}
\definecolor{RdGy-3-K}{RGB}{153,153,153}
\definecolor{RdGy-4-1}{RGB}{202,0,32}
\definecolor{RdGy-4-C}{RGB}{202,0,32}
\definecolor{RdGy-4-2}{RGB}{244,165,130}
\definecolor{RdGy-4-F}{RGB}{244,165,130}
\definecolor{RdGy-4-3}{RGB}{186,186,186}
\definecolor{RdGy-4-J}{RGB}{186,186,186}
\definecolor{RdGy-4-4}{RGB}{64,64,64}
\definecolor{RdGy-4-M}{RGB}{64,64,64}
\definecolor{RdGy-5-1}{RGB}{202,0,32}
\definecolor{RdGy-5-C}{RGB}{202,0,32}
\definecolor{RdGy-5-2}{RGB}{244,165,130}
\definecolor{RdGy-5-F}{RGB}{244,165,130}
\definecolor{RdGy-5-3}{RGB}{255,255,255}
\definecolor{RdGy-5-H}{RGB}{255,255,255}
\definecolor{RdGy-5-4}{RGB}{186,186,186}
\definecolor{RdGy-5-J}{RGB}{186,186,186}
\definecolor{RdGy-5-5}{RGB}{64,64,64}
\definecolor{RdGy-5-M}{RGB}{64,64,64}
\definecolor{RdGy-6-1}{RGB}{178,24,43}
\definecolor{RdGy-6-B}{RGB}{178,24,43}
\definecolor{RdGy-6-2}{RGB}{239,138,98}
\definecolor{RdGy-6-E}{RGB}{239,138,98}
\definecolor{RdGy-6-3}{RGB}{253,219,199}
\definecolor{RdGy-6-G}{RGB}{253,219,199}
\definecolor{RdGy-6-4}{RGB}{224,224,224}
\definecolor{RdGy-6-I}{RGB}{224,224,224}
\definecolor{RdGy-6-5}{RGB}{153,153,153}
\definecolor{RdGy-6-K}{RGB}{153,153,153}
\definecolor{RdGy-6-6}{RGB}{77,77,77}
\definecolor{RdGy-6-N}{RGB}{77,77,77}
\definecolor{RdGy-7-1}{RGB}{178,24,43}
\definecolor{RdGy-7-B}{RGB}{178,24,43}
\definecolor{RdGy-7-2}{RGB}{239,138,98}
\definecolor{RdGy-7-E}{RGB}{239,138,98}
\definecolor{RdGy-7-3}{RGB}{253,219,199}
\definecolor{RdGy-7-G}{RGB}{253,219,199}
\definecolor{RdGy-7-4}{RGB}{255,255,255}
\definecolor{RdGy-7-H}{RGB}{255,255,255}
\definecolor{RdGy-7-5}{RGB}{224,224,224}
\definecolor{RdGy-7-I}{RGB}{224,224,224}
\definecolor{RdGy-7-6}{RGB}{153,153,153}
\definecolor{RdGy-7-K}{RGB}{153,153,153}
\definecolor{RdGy-7-7}{RGB}{77,77,77}
\definecolor{RdGy-7-N}{RGB}{77,77,77}
\definecolor{RdGy-8-1}{RGB}{178,24,43}
\definecolor{RdGy-8-B}{RGB}{178,24,43}
\definecolor{RdGy-8-2}{RGB}{214,96,77}
\definecolor{RdGy-8-D}{RGB}{214,96,77}
\definecolor{RdGy-8-3}{RGB}{244,165,130}
\definecolor{RdGy-8-F}{RGB}{244,165,130}
\definecolor{RdGy-8-4}{RGB}{253,219,199}
\definecolor{RdGy-8-G}{RGB}{253,219,199}
\definecolor{RdGy-8-5}{RGB}{224,224,224}
\definecolor{RdGy-8-I}{RGB}{224,224,224}
\definecolor{RdGy-8-6}{RGB}{186,186,186}
\definecolor{RdGy-8-J}{RGB}{186,186,186}
\definecolor{RdGy-8-7}{RGB}{135,135,135}
\definecolor{RdGy-8-L}{RGB}{135,135,135}
\definecolor{RdGy-8-8}{RGB}{77,77,77}
\definecolor{RdGy-8-N}{RGB}{77,77,77}
\definecolor{RdGy-9-1}{RGB}{178,24,43}
\definecolor{RdGy-9-B}{RGB}{178,24,43}
\definecolor{RdGy-9-2}{RGB}{214,96,77}
\definecolor{RdGy-9-D}{RGB}{214,96,77}
\definecolor{RdGy-9-3}{RGB}{244,165,130}
\definecolor{RdGy-9-F}{RGB}{244,165,130}
\definecolor{RdGy-9-4}{RGB}{253,219,199}
\definecolor{RdGy-9-G}{RGB}{253,219,199}
\definecolor{RdGy-9-5}{RGB}{255,255,255}
\definecolor{RdGy-9-H}{RGB}{255,255,255}
\definecolor{RdGy-9-6}{RGB}{224,224,224}
\definecolor{RdGy-9-I}{RGB}{224,224,224}
\definecolor{RdGy-9-7}{RGB}{186,186,186}
\definecolor{RdGy-9-J}{RGB}{186,186,186}
\definecolor{RdGy-9-8}{RGB}{135,135,135}
\definecolor{RdGy-9-L}{RGB}{135,135,135}
\definecolor{RdGy-9-9}{RGB}{77,77,77}
\definecolor{RdGy-9-N}{RGB}{77,77,77}
\definecolor{RdGy-10-1}{RGB}{103,0,31}
\definecolor{RdGy-10-A}{RGB}{103,0,31}
\definecolor{RdGy-10-2}{RGB}{178,24,43}
\definecolor{RdGy-10-B}{RGB}{178,24,43}
\definecolor{RdGy-10-3}{RGB}{214,96,77}
\definecolor{RdGy-10-D}{RGB}{214,96,77}
\definecolor{RdGy-10-4}{RGB}{244,165,130}
\definecolor{RdGy-10-F}{RGB}{244,165,130}
\definecolor{RdGy-10-5}{RGB}{253,219,199}
\definecolor{RdGy-10-G}{RGB}{253,219,199}
\definecolor{RdGy-10-6}{RGB}{224,224,224}
\definecolor{RdGy-10-I}{RGB}{224,224,224}
\definecolor{RdGy-10-7}{RGB}{186,186,186}
\definecolor{RdGy-10-J}{RGB}{186,186,186}
\definecolor{RdGy-10-8}{RGB}{135,135,135}
\definecolor{RdGy-10-L}{RGB}{135,135,135}
\definecolor{RdGy-10-9}{RGB}{77,77,77}
\definecolor{RdGy-10-N}{RGB}{77,77,77}
\definecolor{RdGy-10-10}{RGB}{26,26,26}
\definecolor{RdGy-10-O}{RGB}{26,26,26}
\definecolor{RdGy-11-1}{RGB}{103,0,31}
\definecolor{RdGy-11-A}{RGB}{103,0,31}
\definecolor{RdGy-11-2}{RGB}{178,24,43}
\definecolor{RdGy-11-B}{RGB}{178,24,43}
\definecolor{RdGy-11-3}{RGB}{214,96,77}
\definecolor{RdGy-11-D}{RGB}{214,96,77}
\definecolor{RdGy-11-4}{RGB}{244,165,130}
\definecolor{RdGy-11-F}{RGB}{244,165,130}
\definecolor{RdGy-11-5}{RGB}{253,219,199}
\definecolor{RdGy-11-G}{RGB}{253,219,199}
\definecolor{RdGy-11-6}{RGB}{255,255,255}
\definecolor{RdGy-11-H}{RGB}{255,255,255}
\definecolor{RdGy-11-7}{RGB}{224,224,224}
\definecolor{RdGy-11-I}{RGB}{224,224,224}
\definecolor{RdGy-11-8}{RGB}{186,186,186}
\definecolor{RdGy-11-J}{RGB}{186,186,186}
\definecolor{RdGy-11-9}{RGB}{135,135,135}
\definecolor{RdGy-11-L}{RGB}{135,135,135}
\definecolor{RdGy-11-10}{RGB}{77,77,77}
\definecolor{RdGy-11-N}{RGB}{77,77,77}
\definecolor{RdGy-11-11}{RGB}{26,26,26}
\definecolor{RdGy-11-O}{RGB}{26,26,26}
\definecolor{RdYlBu-3-1}{RGB}{252,141,89}
\definecolor{RdYlBu-3-E}{RGB}{252,141,89}
\definecolor{RdYlBu-3-2}{RGB}{255,255,191}
\definecolor{RdYlBu-3-H}{RGB}{255,255,191}
\definecolor{RdYlBu-3-3}{RGB}{145,191,219}
\definecolor{RdYlBu-3-K}{RGB}{145,191,219}
\definecolor{RdYlBu-4-1}{RGB}{215,25,28}
\definecolor{RdYlBu-4-C}{RGB}{215,25,28}
\definecolor{RdYlBu-4-2}{RGB}{253,174,97}
\definecolor{RdYlBu-4-F}{RGB}{253,174,97}
\definecolor{RdYlBu-4-3}{RGB}{171,217,233}
\definecolor{RdYlBu-4-J}{RGB}{171,217,233}
\definecolor{RdYlBu-4-4}{RGB}{44,123,182}
\definecolor{RdYlBu-4-M}{RGB}{44,123,182}
\definecolor{RdYlBu-5-1}{RGB}{215,25,28}
\definecolor{RdYlBu-5-C}{RGB}{215,25,28}
\definecolor{RdYlBu-5-2}{RGB}{253,174,97}
\definecolor{RdYlBu-5-F}{RGB}{253,174,97}
\definecolor{RdYlBu-5-3}{RGB}{255,255,191}
\definecolor{RdYlBu-5-H}{RGB}{255,255,191}
\definecolor{RdYlBu-5-4}{RGB}{171,217,233}
\definecolor{RdYlBu-5-J}{RGB}{171,217,233}
\definecolor{RdYlBu-5-5}{RGB}{44,123,182}
\definecolor{RdYlBu-5-M}{RGB}{44,123,182}
\definecolor{RdYlBu-6-1}{RGB}{215,48,39}
\definecolor{RdYlBu-6-B}{RGB}{215,48,39}
\definecolor{RdYlBu-6-2}{RGB}{252,141,89}
\definecolor{RdYlBu-6-E}{RGB}{252,141,89}
\definecolor{RdYlBu-6-3}{RGB}{254,224,144}
\definecolor{RdYlBu-6-G}{RGB}{254,224,144}
\definecolor{RdYlBu-6-4}{RGB}{224,243,248}
\definecolor{RdYlBu-6-I}{RGB}{224,243,248}
\definecolor{RdYlBu-6-5}{RGB}{145,191,219}
\definecolor{RdYlBu-6-K}{RGB}{145,191,219}
\definecolor{RdYlBu-6-6}{RGB}{69,117,180}
\definecolor{RdYlBu-6-N}{RGB}{69,117,180}
\definecolor{RdYlBu-7-1}{RGB}{215,48,39}
\definecolor{RdYlBu-7-B}{RGB}{215,48,39}
\definecolor{RdYlBu-7-2}{RGB}{252,141,89}
\definecolor{RdYlBu-7-E}{RGB}{252,141,89}
\definecolor{RdYlBu-7-3}{RGB}{254,224,144}
\definecolor{RdYlBu-7-G}{RGB}{254,224,144}
\definecolor{RdYlBu-7-4}{RGB}{255,255,191}
\definecolor{RdYlBu-7-H}{RGB}{255,255,191}
\definecolor{RdYlBu-7-5}{RGB}{224,243,248}
\definecolor{RdYlBu-7-I}{RGB}{224,243,248}
\definecolor{RdYlBu-7-6}{RGB}{145,191,219}
\definecolor{RdYlBu-7-K}{RGB}{145,191,219}
\definecolor{RdYlBu-7-7}{RGB}{69,117,180}
\definecolor{RdYlBu-7-N}{RGB}{69,117,180}
\definecolor{RdYlBu-8-1}{RGB}{215,48,39}
\definecolor{RdYlBu-8-B}{RGB}{215,48,39}
\definecolor{RdYlBu-8-2}{RGB}{244,109,67}
\definecolor{RdYlBu-8-D}{RGB}{244,109,67}
\definecolor{RdYlBu-8-3}{RGB}{253,174,97}
\definecolor{RdYlBu-8-F}{RGB}{253,174,97}
\definecolor{RdYlBu-8-4}{RGB}{254,224,144}
\definecolor{RdYlBu-8-G}{RGB}{254,224,144}
\definecolor{RdYlBu-8-5}{RGB}{224,243,248}
\definecolor{RdYlBu-8-I}{RGB}{224,243,248}
\definecolor{RdYlBu-8-6}{RGB}{171,217,233}
\definecolor{RdYlBu-8-J}{RGB}{171,217,233}
\definecolor{RdYlBu-8-7}{RGB}{116,173,209}
\definecolor{RdYlBu-8-L}{RGB}{116,173,209}
\definecolor{RdYlBu-8-8}{RGB}{69,117,180}
\definecolor{RdYlBu-8-N}{RGB}{69,117,180}
\definecolor{RdYlBu-9-1}{RGB}{215,48,39}
\definecolor{RdYlBu-9-B}{RGB}{215,48,39}
\definecolor{RdYlBu-9-2}{RGB}{244,109,67}
\definecolor{RdYlBu-9-D}{RGB}{244,109,67}
\definecolor{RdYlBu-9-3}{RGB}{253,174,97}
\definecolor{RdYlBu-9-F}{RGB}{253,174,97}
\definecolor{RdYlBu-9-4}{RGB}{254,224,144}
\definecolor{RdYlBu-9-G}{RGB}{254,224,144}
\definecolor{RdYlBu-9-5}{RGB}{255,255,191}
\definecolor{RdYlBu-9-H}{RGB}{255,255,191}
\definecolor{RdYlBu-9-6}{RGB}{224,243,248}
\definecolor{RdYlBu-9-I}{RGB}{224,243,248}
\definecolor{RdYlBu-9-7}{RGB}{171,217,233}
\definecolor{RdYlBu-9-J}{RGB}{171,217,233}
\definecolor{RdYlBu-9-8}{RGB}{116,173,209}
\definecolor{RdYlBu-9-L}{RGB}{116,173,209}
\definecolor{RdYlBu-9-9}{RGB}{69,117,180}
\definecolor{RdYlBu-9-N}{RGB}{69,117,180}
\definecolor{RdYlBu-10-1}{RGB}{165,0,38}
\definecolor{RdYlBu-10-A}{RGB}{165,0,38}
\definecolor{RdYlBu-10-2}{RGB}{215,48,39}
\definecolor{RdYlBu-10-B}{RGB}{215,48,39}
\definecolor{RdYlBu-10-3}{RGB}{244,109,67}
\definecolor{RdYlBu-10-D}{RGB}{244,109,67}
\definecolor{RdYlBu-10-4}{RGB}{253,174,97}
\definecolor{RdYlBu-10-F}{RGB}{253,174,97}
\definecolor{RdYlBu-10-5}{RGB}{254,224,144}
\definecolor{RdYlBu-10-G}{RGB}{254,224,144}
\definecolor{RdYlBu-10-6}{RGB}{224,243,248}
\definecolor{RdYlBu-10-I}{RGB}{224,243,248}
\definecolor{RdYlBu-10-7}{RGB}{171,217,233}
\definecolor{RdYlBu-10-J}{RGB}{171,217,233}
\definecolor{RdYlBu-10-8}{RGB}{116,173,209}
\definecolor{RdYlBu-10-L}{RGB}{116,173,209}
\definecolor{RdYlBu-10-9}{RGB}{69,117,180}
\definecolor{RdYlBu-10-N}{RGB}{69,117,180}
\definecolor{RdYlBu-10-10}{RGB}{49,54,149}
\definecolor{RdYlBu-10-O}{RGB}{49,54,149}
\definecolor{RdYlBu-11-1}{RGB}{165,0,38}
\definecolor{RdYlBu-11-A}{RGB}{165,0,38}
\definecolor{RdYlBu-11-2}{RGB}{215,48,39}
\definecolor{RdYlBu-11-B}{RGB}{215,48,39}
\definecolor{RdYlBu-11-3}{RGB}{244,109,67}
\definecolor{RdYlBu-11-D}{RGB}{244,109,67}
\definecolor{RdYlBu-11-4}{RGB}{253,174,97}
\definecolor{RdYlBu-11-F}{RGB}{253,174,97}
\definecolor{RdYlBu-11-5}{RGB}{254,224,144}
\definecolor{RdYlBu-11-G}{RGB}{254,224,144}
\definecolor{RdYlBu-11-6}{RGB}{255,255,191}
\definecolor{RdYlBu-11-H}{RGB}{255,255,191}
\definecolor{RdYlBu-11-7}{RGB}{224,243,248}
\definecolor{RdYlBu-11-I}{RGB}{224,243,248}
\definecolor{RdYlBu-11-8}{RGB}{171,217,233}
\definecolor{RdYlBu-11-J}{RGB}{171,217,233}
\definecolor{RdYlBu-11-9}{RGB}{116,173,209}
\definecolor{RdYlBu-11-L}{RGB}{116,173,209}
\definecolor{RdYlBu-11-10}{RGB}{69,117,180}
\definecolor{RdYlBu-11-N}{RGB}{69,117,180}
\definecolor{RdYlBu-11-11}{RGB}{49,54,149}
\definecolor{RdYlBu-11-O}{RGB}{49,54,149}
\definecolor{Spectral-3-1}{RGB}{252,141,89}
\definecolor{Spectral-3-E}{RGB}{252,141,89}
\definecolor{Spectral-3-2}{RGB}{255,255,191}
\definecolor{Spectral-3-H}{RGB}{255,255,191}
\definecolor{Spectral-3-3}{RGB}{153,213,148}
\definecolor{Spectral-3-K}{RGB}{153,213,148}
\definecolor{Spectral-4-1}{RGB}{215,25,28}
\definecolor{Spectral-4-C}{RGB}{215,25,28}
\definecolor{Spectral-4-2}{RGB}{253,174,97}
\definecolor{Spectral-4-F}{RGB}{253,174,97}
\definecolor{Spectral-4-3}{RGB}{171,221,164}
\definecolor{Spectral-4-J}{RGB}{171,221,164}
\definecolor{Spectral-4-4}{RGB}{43,131,186}
\definecolor{Spectral-4-M}{RGB}{43,131,186}
\definecolor{Spectral-5-1}{RGB}{215,25,28}
\definecolor{Spectral-5-C}{RGB}{215,25,28}
\definecolor{Spectral-5-2}{RGB}{253,174,97}
\definecolor{Spectral-5-F}{RGB}{253,174,97}
\definecolor{Spectral-5-3}{RGB}{255,255,191}
\definecolor{Spectral-5-H}{RGB}{255,255,191}
\definecolor{Spectral-5-4}{RGB}{171,221,164}
\definecolor{Spectral-5-J}{RGB}{171,221,164}
\definecolor{Spectral-5-5}{RGB}{43,131,186}
\definecolor{Spectral-5-M}{RGB}{43,131,186}
\definecolor{Spectral-6-1}{RGB}{213,62,79}
\definecolor{Spectral-6-B}{RGB}{213,62,79}
\definecolor{Spectral-6-2}{RGB}{252,141,89}
\definecolor{Spectral-6-E}{RGB}{252,141,89}
\definecolor{Spectral-6-3}{RGB}{254,224,139}
\definecolor{Spectral-6-G}{RGB}{254,224,139}
\definecolor{Spectral-6-4}{RGB}{230,245,152}
\definecolor{Spectral-6-I}{RGB}{230,245,152}
\definecolor{Spectral-6-5}{RGB}{153,213,148}
\definecolor{Spectral-6-K}{RGB}{153,213,148}
\definecolor{Spectral-6-6}{RGB}{50,136,189}
\definecolor{Spectral-6-N}{RGB}{50,136,189}
\definecolor{Spectral-7-1}{RGB}{213,62,79}
\definecolor{Spectral-7-B}{RGB}{213,62,79}
\definecolor{Spectral-7-2}{RGB}{252,141,89}
\definecolor{Spectral-7-E}{RGB}{252,141,89}
\definecolor{Spectral-7-3}{RGB}{254,224,139}
\definecolor{Spectral-7-G}{RGB}{254,224,139}
\definecolor{Spectral-7-4}{RGB}{255,255,191}
\definecolor{Spectral-7-H}{RGB}{255,255,191}
\definecolor{Spectral-7-5}{RGB}{230,245,152}
\definecolor{Spectral-7-I}{RGB}{230,245,152}
\definecolor{Spectral-7-6}{RGB}{153,213,148}
\definecolor{Spectral-7-K}{RGB}{153,213,148}
\definecolor{Spectral-7-7}{RGB}{50,136,189}
\definecolor{Spectral-7-N}{RGB}{50,136,189}
\definecolor{Spectral-8-1}{RGB}{213,62,79}
\definecolor{Spectral-8-B}{RGB}{213,62,79}
\definecolor{Spectral-8-2}{RGB}{244,109,67}
\definecolor{Spectral-8-D}{RGB}{244,109,67}
\definecolor{Spectral-8-3}{RGB}{253,174,97}
\definecolor{Spectral-8-F}{RGB}{253,174,97}
\definecolor{Spectral-8-4}{RGB}{254,224,139}
\definecolor{Spectral-8-G}{RGB}{254,224,139}
\definecolor{Spectral-8-5}{RGB}{230,245,152}
\definecolor{Spectral-8-I}{RGB}{230,245,152}
\definecolor{Spectral-8-6}{RGB}{171,221,164}
\definecolor{Spectral-8-J}{RGB}{171,221,164}
\definecolor{Spectral-8-7}{RGB}{102,194,165}
\definecolor{Spectral-8-L}{RGB}{102,194,165}
\definecolor{Spectral-8-8}{RGB}{50,136,189}
\definecolor{Spectral-8-N}{RGB}{50,136,189}
\definecolor{Spectral-9-1}{RGB}{213,62,79}
\definecolor{Spectral-9-B}{RGB}{213,62,79}
\definecolor{Spectral-9-2}{RGB}{244,109,67}
\definecolor{Spectral-9-D}{RGB}{244,109,67}
\definecolor{Spectral-9-3}{RGB}{253,174,97}
\definecolor{Spectral-9-F}{RGB}{253,174,97}
\definecolor{Spectral-9-4}{RGB}{254,224,139}
\definecolor{Spectral-9-G}{RGB}{254,224,139}
\definecolor{Spectral-9-5}{RGB}{255,255,191}
\definecolor{Spectral-9-H}{RGB}{255,255,191}
\definecolor{Spectral-9-6}{RGB}{230,245,152}
\definecolor{Spectral-9-I}{RGB}{230,245,152}
\definecolor{Spectral-9-7}{RGB}{171,221,164}
\definecolor{Spectral-9-J}{RGB}{171,221,164}
\definecolor{Spectral-9-8}{RGB}{102,194,165}
\definecolor{Spectral-9-L}{RGB}{102,194,165}
\definecolor{Spectral-9-9}{RGB}{50,136,189}
\definecolor{Spectral-9-N}{RGB}{50,136,189}
\definecolor{Spectral-10-1}{RGB}{158,1,66}
\definecolor{Spectral-10-A}{RGB}{158,1,66}
\definecolor{Spectral-10-2}{RGB}{213,62,79}
\definecolor{Spectral-10-B}{RGB}{213,62,79}
\definecolor{Spectral-10-3}{RGB}{244,109,67}
\definecolor{Spectral-10-D}{RGB}{244,109,67}
\definecolor{Spectral-10-4}{RGB}{253,174,97}
\definecolor{Spectral-10-F}{RGB}{253,174,97}
\definecolor{Spectral-10-5}{RGB}{254,224,139}
\definecolor{Spectral-10-G}{RGB}{254,224,139}
\definecolor{Spectral-10-6}{RGB}{230,245,152}
\definecolor{Spectral-10-I}{RGB}{230,245,152}
\definecolor{Spectral-10-7}{RGB}{171,221,164}
\definecolor{Spectral-10-J}{RGB}{171,221,164}
\definecolor{Spectral-10-8}{RGB}{102,194,165}
\definecolor{Spectral-10-L}{RGB}{102,194,165}
\definecolor{Spectral-10-9}{RGB}{50,136,189}
\definecolor{Spectral-10-N}{RGB}{50,136,189}
\definecolor{Spectral-10-10}{RGB}{94,79,162}
\definecolor{Spectral-10-O}{RGB}{94,79,162}
\definecolor{Spectral-11-1}{RGB}{158,1,66}
\definecolor{Spectral-11-A}{RGB}{158,1,66}
\definecolor{Spectral-11-2}{RGB}{213,62,79}
\definecolor{Spectral-11-B}{RGB}{213,62,79}
\definecolor{Spectral-11-3}{RGB}{244,109,67}
\definecolor{Spectral-11-D}{RGB}{244,109,67}
\definecolor{Spectral-11-4}{RGB}{253,174,97}
\definecolor{Spectral-11-F}{RGB}{253,174,97}
\definecolor{Spectral-11-5}{RGB}{254,224,139}
\definecolor{Spectral-11-G}{RGB}{254,224,139}
\definecolor{Spectral-11-6}{RGB}{255,255,191}
\definecolor{Spectral-11-H}{RGB}{255,255,191}
\definecolor{Spectral-11-7}{RGB}{230,245,152}
\definecolor{Spectral-11-I}{RGB}{230,245,152}
\definecolor{Spectral-11-8}{RGB}{171,221,164}
\definecolor{Spectral-11-J}{RGB}{171,221,164}
\definecolor{Spectral-11-9}{RGB}{102,194,165}
\definecolor{Spectral-11-L}{RGB}{102,194,165}
\definecolor{Spectral-11-10}{RGB}{50,136,189}
\definecolor{Spectral-11-N}{RGB}{50,136,189}
\definecolor{Spectral-11-11}{RGB}{94,79,162}
\definecolor{Spectral-11-O}{RGB}{94,79,162}
\definecolor{RdYlGn-3-1}{RGB}{252,141,89}
\definecolor{RdYlGn-3-E}{RGB}{252,141,89}
\definecolor{RdYlGn-3-2}{RGB}{255,255,191}
\definecolor{RdYlGn-3-H}{RGB}{255,255,191}
\definecolor{RdYlGn-3-3}{RGB}{145,207,96}
\definecolor{RdYlGn-3-K}{RGB}{145,207,96}
\definecolor{RdYlGn-4-1}{RGB}{215,25,28}
\definecolor{RdYlGn-4-C}{RGB}{215,25,28}
\definecolor{RdYlGn-4-2}{RGB}{253,174,97}
\definecolor{RdYlGn-4-F}{RGB}{253,174,97}
\definecolor{RdYlGn-4-3}{RGB}{166,217,106}
\definecolor{RdYlGn-4-J}{RGB}{166,217,106}
\definecolor{RdYlGn-4-4}{RGB}{26,150,65}
\definecolor{RdYlGn-4-M}{RGB}{26,150,65}
\definecolor{RdYlGn-5-1}{RGB}{215,25,28}
\definecolor{RdYlGn-5-C}{RGB}{215,25,28}
\definecolor{RdYlGn-5-2}{RGB}{253,174,97}
\definecolor{RdYlGn-5-F}{RGB}{253,174,97}
\definecolor{RdYlGn-5-3}{RGB}{255,255,191}
\definecolor{RdYlGn-5-H}{RGB}{255,255,191}
\definecolor{RdYlGn-5-4}{RGB}{166,217,106}
\definecolor{RdYlGn-5-J}{RGB}{166,217,106}
\definecolor{RdYlGn-5-5}{RGB}{26,150,65}
\definecolor{RdYlGn-5-M}{RGB}{26,150,65}
\definecolor{RdYlGn-6-1}{RGB}{215,48,39}
\definecolor{RdYlGn-6-B}{RGB}{215,48,39}
\definecolor{RdYlGn-6-2}{RGB}{252,141,89}
\definecolor{RdYlGn-6-E}{RGB}{252,141,89}
\definecolor{RdYlGn-6-3}{RGB}{254,224,139}
\definecolor{RdYlGn-6-G}{RGB}{254,224,139}
\definecolor{RdYlGn-6-4}{RGB}{217,239,139}
\definecolor{RdYlGn-6-I}{RGB}{217,239,139}
\definecolor{RdYlGn-6-5}{RGB}{145,207,96}
\definecolor{RdYlGn-6-K}{RGB}{145,207,96}
\definecolor{RdYlGn-6-6}{RGB}{26,152,80}
\definecolor{RdYlGn-6-N}{RGB}{26,152,80}
\definecolor{RdYlGn-7-1}{RGB}{215,48,39}
\definecolor{RdYlGn-7-B}{RGB}{215,48,39}
\definecolor{RdYlGn-7-2}{RGB}{252,141,89}
\definecolor{RdYlGn-7-E}{RGB}{252,141,89}
\definecolor{RdYlGn-7-3}{RGB}{254,224,139}
\definecolor{RdYlGn-7-G}{RGB}{254,224,139}
\definecolor{RdYlGn-7-4}{RGB}{255,255,191}
\definecolor{RdYlGn-7-H}{RGB}{255,255,191}
\definecolor{RdYlGn-7-5}{RGB}{217,239,139}
\definecolor{RdYlGn-7-I}{RGB}{217,239,139}
\definecolor{RdYlGn-7-6}{RGB}{145,207,96}
\definecolor{RdYlGn-7-K}{RGB}{145,207,96}
\definecolor{RdYlGn-7-7}{RGB}{26,152,80}
\definecolor{RdYlGn-7-N}{RGB}{26,152,80}
\definecolor{RdYlGn-8-1}{RGB}{215,48,39}
\definecolor{RdYlGn-8-B}{RGB}{215,48,39}
\definecolor{RdYlGn-8-2}{RGB}{244,109,67}
\definecolor{RdYlGn-8-D}{RGB}{244,109,67}
\definecolor{RdYlGn-8-3}{RGB}{253,174,97}
\definecolor{RdYlGn-8-F}{RGB}{253,174,97}
\definecolor{RdYlGn-8-4}{RGB}{254,224,139}
\definecolor{RdYlGn-8-G}{RGB}{254,224,139}
\definecolor{RdYlGn-8-5}{RGB}{217,239,139}
\definecolor{RdYlGn-8-I}{RGB}{217,239,139}
\definecolor{RdYlGn-8-6}{RGB}{166,217,106}
\definecolor{RdYlGn-8-J}{RGB}{166,217,106}
\definecolor{RdYlGn-8-7}{RGB}{102,189,99}
\definecolor{RdYlGn-8-L}{RGB}{102,189,99}
\definecolor{RdYlGn-8-8}{RGB}{26,152,80}
\definecolor{RdYlGn-8-N}{RGB}{26,152,80}
\definecolor{RdYlGn-9-1}{RGB}{215,48,39}
\definecolor{RdYlGn-9-B}{RGB}{215,48,39}
\definecolor{RdYlGn-9-2}{RGB}{244,109,67}
\definecolor{RdYlGn-9-D}{RGB}{244,109,67}
\definecolor{RdYlGn-9-3}{RGB}{253,174,97}
\definecolor{RdYlGn-9-F}{RGB}{253,174,97}
\definecolor{RdYlGn-9-4}{RGB}{254,224,139}
\definecolor{RdYlGn-9-G}{RGB}{254,224,139}
\definecolor{RdYlGn-9-5}{RGB}{255,255,191}
\definecolor{RdYlGn-9-H}{RGB}{255,255,191}
\definecolor{RdYlGn-9-6}{RGB}{217,239,139}
\definecolor{RdYlGn-9-I}{RGB}{217,239,139}
\definecolor{RdYlGn-9-7}{RGB}{166,217,106}
\definecolor{RdYlGn-9-J}{RGB}{166,217,106}
\definecolor{RdYlGn-9-8}{RGB}{102,189,99}
\definecolor{RdYlGn-9-L}{RGB}{102,189,99}
\definecolor{RdYlGn-9-9}{RGB}{26,152,80}
\definecolor{RdYlGn-9-N}{RGB}{26,152,80}
\definecolor{RdYlGn-10-1}{RGB}{165,0,38}
\definecolor{RdYlGn-10-A}{RGB}{165,0,38}
\definecolor{RdYlGn-10-2}{RGB}{215,48,39}
\definecolor{RdYlGn-10-B}{RGB}{215,48,39}
\definecolor{RdYlGn-10-3}{RGB}{244,109,67}
\definecolor{RdYlGn-10-D}{RGB}{244,109,67}
\definecolor{RdYlGn-10-4}{RGB}{253,174,97}
\definecolor{RdYlGn-10-F}{RGB}{253,174,97}
\definecolor{RdYlGn-10-5}{RGB}{254,224,139}
\definecolor{RdYlGn-10-G}{RGB}{254,224,139}
\definecolor{RdYlGn-10-6}{RGB}{217,239,139}
\definecolor{RdYlGn-10-I}{RGB}{217,239,139}
\definecolor{RdYlGn-10-7}{RGB}{166,217,106}
\definecolor{RdYlGn-10-J}{RGB}{166,217,106}
\definecolor{RdYlGn-10-8}{RGB}{102,189,99}
\definecolor{RdYlGn-10-L}{RGB}{102,189,99}
\definecolor{RdYlGn-10-9}{RGB}{26,152,80}
\definecolor{RdYlGn-10-N}{RGB}{26,152,80}
\definecolor{RdYlGn-10-10}{RGB}{0,104,55}
\definecolor{RdYlGn-10-O}{RGB}{0,104,55}
\definecolor{RdYlGn-11-1}{RGB}{165,0,38}
\definecolor{RdYlGn-11-A}{RGB}{165,0,38}
\definecolor{RdYlGn-11-2}{RGB}{215,48,39}
\definecolor{RdYlGn-11-B}{RGB}{215,48,39}
\definecolor{RdYlGn-11-3}{RGB}{244,109,67}
\definecolor{RdYlGn-11-D}{RGB}{244,109,67}
\definecolor{RdYlGn-11-4}{RGB}{253,174,97}
\definecolor{RdYlGn-11-F}{RGB}{253,174,97}
\definecolor{RdYlGn-11-5}{RGB}{254,224,139}
\definecolor{RdYlGn-11-G}{RGB}{254,224,139}
\definecolor{RdYlGn-11-6}{RGB}{255,255,191}
\definecolor{RdYlGn-11-H}{RGB}{255,255,191}
\definecolor{RdYlGn-11-7}{RGB}{217,239,139}
\definecolor{RdYlGn-11-I}{RGB}{217,239,139}
\definecolor{RdYlGn-11-8}{RGB}{166,217,106}
\definecolor{RdYlGn-11-J}{RGB}{166,217,106}
\definecolor{RdYlGn-11-9}{RGB}{102,189,99}
\definecolor{RdYlGn-11-L}{RGB}{102,189,99}
\definecolor{RdYlGn-11-10}{RGB}{26,152,80}
\definecolor{RdYlGn-11-N}{RGB}{26,152,80}
\definecolor{RdYlGn-11-11}{RGB}{0,104,55}
\definecolor{RdYlGn-11-O}{RGB}{0,104,55}
\definecolor{Set3-3-1}{RGB}{141,211,199}
\definecolor{Set3-3-A}{RGB}{141,211,199}
\definecolor{Set3-3-2}{RGB}{255,255,179}
\definecolor{Set3-3-B}{RGB}{255,255,179}
\definecolor{Set3-3-3}{RGB}{190,186,218}
\definecolor{Set3-3-C}{RGB}{190,186,218}
\definecolor{Set3-4-1}{RGB}{141,211,199}
\definecolor{Set3-4-A}{RGB}{141,211,199}
\definecolor{Set3-4-2}{RGB}{255,255,179}
\definecolor{Set3-4-B}{RGB}{255,255,179}
\definecolor{Set3-4-3}{RGB}{190,186,218}
\definecolor{Set3-4-C}{RGB}{190,186,218}
\definecolor{Set3-4-4}{RGB}{251,128,114}
\definecolor{Set3-4-D}{RGB}{251,128,114}
\definecolor{Set3-5-1}{RGB}{141,211,199}
\definecolor{Set3-5-A}{RGB}{141,211,199}
\definecolor{Set3-5-2}{RGB}{255,255,179}
\definecolor{Set3-5-B}{RGB}{255,255,179}
\definecolor{Set3-5-3}{RGB}{190,186,218}
\definecolor{Set3-5-C}{RGB}{190,186,218}
\definecolor{Set3-5-4}{RGB}{251,128,114}
\definecolor{Set3-5-D}{RGB}{251,128,114}
\definecolor{Set3-5-5}{RGB}{128,177,211}
\definecolor{Set3-5-E}{RGB}{128,177,211}
\definecolor{Set3-6-1}{RGB}{141,211,199}
\definecolor{Set3-6-A}{RGB}{141,211,199}
\definecolor{Set3-6-2}{RGB}{255,255,179}
\definecolor{Set3-6-B}{RGB}{255,255,179}
\definecolor{Set3-6-3}{RGB}{190,186,218}
\definecolor{Set3-6-C}{RGB}{190,186,218}
\definecolor{Set3-6-4}{RGB}{251,128,114}
\definecolor{Set3-6-D}{RGB}{251,128,114}
\definecolor{Set3-6-5}{RGB}{128,177,211}
\definecolor{Set3-6-E}{RGB}{128,177,211}
\definecolor{Set3-6-6}{RGB}{253,180,98}
\definecolor{Set3-6-F}{RGB}{253,180,98}
\definecolor{Set3-7-1}{RGB}{141,211,199}
\definecolor{Set3-7-A}{RGB}{141,211,199}
\definecolor{Set3-7-2}{RGB}{255,255,179}
\definecolor{Set3-7-B}{RGB}{255,255,179}
\definecolor{Set3-7-3}{RGB}{190,186,218}
\definecolor{Set3-7-C}{RGB}{190,186,218}
\definecolor{Set3-7-4}{RGB}{251,128,114}
\definecolor{Set3-7-D}{RGB}{251,128,114}
\definecolor{Set3-7-5}{RGB}{128,177,211}
\definecolor{Set3-7-E}{RGB}{128,177,211}
\definecolor{Set3-7-6}{RGB}{253,180,98}
\definecolor{Set3-7-F}{RGB}{253,180,98}
\definecolor{Set3-7-7}{RGB}{179,222,105}
\definecolor{Set3-7-G}{RGB}{179,222,105}
\definecolor{Set3-8-1}{RGB}{141,211,199}
\definecolor{Set3-8-A}{RGB}{141,211,199}
\definecolor{Set3-8-2}{RGB}{255,255,179}
\definecolor{Set3-8-B}{RGB}{255,255,179}
\definecolor{Set3-8-3}{RGB}{190,186,218}
\definecolor{Set3-8-C}{RGB}{190,186,218}
\definecolor{Set3-8-4}{RGB}{251,128,114}
\definecolor{Set3-8-D}{RGB}{251,128,114}
\definecolor{Set3-8-5}{RGB}{128,177,211}
\definecolor{Set3-8-E}{RGB}{128,177,211}
\definecolor{Set3-8-6}{RGB}{253,180,98}
\definecolor{Set3-8-F}{RGB}{253,180,98}
\definecolor{Set3-8-7}{RGB}{179,222,105}
\definecolor{Set3-8-G}{RGB}{179,222,105}
\definecolor{Set3-8-8}{RGB}{252,205,229}
\definecolor{Set3-8-H}{RGB}{252,205,229}
\definecolor{Set3-9-1}{RGB}{141,211,199}
\definecolor{Set3-9-A}{RGB}{141,211,199}
\definecolor{Set3-9-2}{RGB}{255,255,179}
\definecolor{Set3-9-B}{RGB}{255,255,179}
\definecolor{Set3-9-3}{RGB}{190,186,218}
\definecolor{Set3-9-C}{RGB}{190,186,218}
\definecolor{Set3-9-4}{RGB}{251,128,114}
\definecolor{Set3-9-D}{RGB}{251,128,114}
\definecolor{Set3-9-5}{RGB}{128,177,211}
\definecolor{Set3-9-E}{RGB}{128,177,211}
\definecolor{Set3-9-6}{RGB}{253,180,98}
\definecolor{Set3-9-F}{RGB}{253,180,98}
\definecolor{Set3-9-7}{RGB}{179,222,105}
\definecolor{Set3-9-G}{RGB}{179,222,105}
\definecolor{Set3-9-8}{RGB}{252,205,229}
\definecolor{Set3-9-H}{RGB}{252,205,229}
\definecolor{Set3-9-9}{RGB}{217,217,217}
\definecolor{Set3-9-I}{RGB}{217,217,217}
\definecolor{Set3-10-1}{RGB}{141,211,199}
\definecolor{Set3-10-A}{RGB}{141,211,199}
\definecolor{Set3-10-2}{RGB}{255,255,179}
\definecolor{Set3-10-B}{RGB}{255,255,179}
\definecolor{Set3-10-3}{RGB}{190,186,218}
\definecolor{Set3-10-C}{RGB}{190,186,218}
\definecolor{Set3-10-4}{RGB}{251,128,114}
\definecolor{Set3-10-D}{RGB}{251,128,114}
\definecolor{Set3-10-5}{RGB}{128,177,211}
\definecolor{Set3-10-E}{RGB}{128,177,211}
\definecolor{Set3-10-6}{RGB}{253,180,98}
\definecolor{Set3-10-F}{RGB}{253,180,98}
\definecolor{Set3-10-7}{RGB}{179,222,105}
\definecolor{Set3-10-G}{RGB}{179,222,105}
\definecolor{Set3-10-8}{RGB}{252,205,229}
\definecolor{Set3-10-H}{RGB}{252,205,229}
\definecolor{Set3-10-9}{RGB}{217,217,217}
\definecolor{Set3-10-I}{RGB}{217,217,217}
\definecolor{Set3-10-10}{RGB}{188,128,189}
\definecolor{Set3-10-J}{RGB}{188,128,189}
\definecolor{Set3-11-1}{RGB}{141,211,199}
\definecolor{Set3-11-A}{RGB}{141,211,199}
\definecolor{Set3-11-2}{RGB}{255,255,179}
\definecolor{Set3-11-B}{RGB}{255,255,179}
\definecolor{Set3-11-3}{RGB}{190,186,218}
\definecolor{Set3-11-C}{RGB}{190,186,218}
\definecolor{Set3-11-4}{RGB}{251,128,114}
\definecolor{Set3-11-D}{RGB}{251,128,114}
\definecolor{Set3-11-5}{RGB}{128,177,211}
\definecolor{Set3-11-E}{RGB}{128,177,211}
\definecolor{Set3-11-6}{RGB}{253,180,98}
\definecolor{Set3-11-F}{RGB}{253,180,98}
\definecolor{Set3-11-7}{RGB}{179,222,105}
\definecolor{Set3-11-G}{RGB}{179,222,105}
\definecolor{Set3-11-8}{RGB}{252,205,229}
\definecolor{Set3-11-H}{RGB}{252,205,229}
\definecolor{Set3-11-9}{RGB}{217,217,217}
\definecolor{Set3-11-I}{RGB}{217,217,217}
\definecolor{Set3-11-10}{RGB}{188,128,189}
\definecolor{Set3-11-J}{RGB}{188,128,189}
\definecolor{Set3-11-11}{RGB}{204,235,197}
\definecolor{Set3-11-K}{RGB}{204,235,197}
\definecolor{Set3-12-1}{RGB}{141,211,199}
\definecolor{Set3-12-A}{RGB}{141,211,199}
\definecolor{Set3-12-2}{RGB}{255,255,179}
\definecolor{Set3-12-B}{RGB}{255,255,179}
\definecolor{Set3-12-3}{RGB}{190,186,218}
\definecolor{Set3-12-C}{RGB}{190,186,218}
\definecolor{Set3-12-4}{RGB}{251,128,114}
\definecolor{Set3-12-D}{RGB}{251,128,114}
\definecolor{Set3-12-5}{RGB}{128,177,211}
\definecolor{Set3-12-E}{RGB}{128,177,211}
\definecolor{Set3-12-6}{RGB}{253,180,98}
\definecolor{Set3-12-F}{RGB}{253,180,98}
\definecolor{Set3-12-7}{RGB}{179,222,105}
\definecolor{Set3-12-G}{RGB}{179,222,105}
\definecolor{Set3-12-8}{RGB}{252,205,229}
\definecolor{Set3-12-H}{RGB}{252,205,229}
\definecolor{Set3-12-9}{RGB}{217,217,217}
\definecolor{Set3-12-I}{RGB}{217,217,217}
\definecolor{Set3-12-10}{RGB}{188,128,189}
\definecolor{Set3-12-J}{RGB}{188,128,189}
\definecolor{Set3-12-11}{RGB}{204,235,197}
\definecolor{Set3-12-K}{RGB}{204,235,197}
\definecolor{Set3-12-12}{RGB}{255,237,111}
\definecolor{Set3-12-L}{RGB}{255,237,111}
\definecolor{Pastel1-3-1}{RGB}{251,180,174}
\definecolor{Pastel1-3-A}{RGB}{251,180,174}
\definecolor{Pastel1-3-2}{RGB}{179,205,227}
\definecolor{Pastel1-3-B}{RGB}{179,205,227}
\definecolor{Pastel1-3-3}{RGB}{204,235,197}
\definecolor{Pastel1-3-C}{RGB}{204,235,197}
\definecolor{Pastel1-4-1}{RGB}{251,180,174}
\definecolor{Pastel1-4-A}{RGB}{251,180,174}
\definecolor{Pastel1-4-2}{RGB}{179,205,227}
\definecolor{Pastel1-4-B}{RGB}{179,205,227}
\definecolor{Pastel1-4-3}{RGB}{204,235,197}
\definecolor{Pastel1-4-C}{RGB}{204,235,197}
\definecolor{Pastel1-4-4}{RGB}{222,203,228}
\definecolor{Pastel1-4-D}{RGB}{222,203,228}
\definecolor{Pastel1-5-1}{RGB}{251,180,174}
\definecolor{Pastel1-5-A}{RGB}{251,180,174}
\definecolor{Pastel1-5-2}{RGB}{179,205,227}
\definecolor{Pastel1-5-B}{RGB}{179,205,227}
\definecolor{Pastel1-5-3}{RGB}{204,235,197}
\definecolor{Pastel1-5-C}{RGB}{204,235,197}
\definecolor{Pastel1-5-4}{RGB}{222,203,228}
\definecolor{Pastel1-5-D}{RGB}{222,203,228}
\definecolor{Pastel1-5-5}{RGB}{254,217,166}
\definecolor{Pastel1-5-E}{RGB}{254,217,166}
\definecolor{Pastel1-6-1}{RGB}{251,180,174}
\definecolor{Pastel1-6-A}{RGB}{251,180,174}
\definecolor{Pastel1-6-2}{RGB}{179,205,227}
\definecolor{Pastel1-6-B}{RGB}{179,205,227}
\definecolor{Pastel1-6-3}{RGB}{204,235,197}
\definecolor{Pastel1-6-C}{RGB}{204,235,197}
\definecolor{Pastel1-6-4}{RGB}{222,203,228}
\definecolor{Pastel1-6-D}{RGB}{222,203,228}
\definecolor{Pastel1-6-5}{RGB}{254,217,166}
\definecolor{Pastel1-6-E}{RGB}{254,217,166}
\definecolor{Pastel1-6-6}{RGB}{255,255,204}
\definecolor{Pastel1-6-F}{RGB}{255,255,204}
\definecolor{Pastel1-7-1}{RGB}{251,180,174}
\definecolor{Pastel1-7-A}{RGB}{251,180,174}
\definecolor{Pastel1-7-2}{RGB}{179,205,227}
\definecolor{Pastel1-7-B}{RGB}{179,205,227}
\definecolor{Pastel1-7-3}{RGB}{204,235,197}
\definecolor{Pastel1-7-C}{RGB}{204,235,197}
\definecolor{Pastel1-7-4}{RGB}{222,203,228}
\definecolor{Pastel1-7-D}{RGB}{222,203,228}
\definecolor{Pastel1-7-5}{RGB}{254,217,166}
\definecolor{Pastel1-7-E}{RGB}{254,217,166}
\definecolor{Pastel1-7-6}{RGB}{255,255,204}
\definecolor{Pastel1-7-F}{RGB}{255,255,204}
\definecolor{Pastel1-7-7}{RGB}{229,216,189}
\definecolor{Pastel1-7-G}{RGB}{229,216,189}
\definecolor{Pastel1-8-1}{RGB}{251,180,174}
\definecolor{Pastel1-8-A}{RGB}{251,180,174}
\definecolor{Pastel1-8-2}{RGB}{179,205,227}
\definecolor{Pastel1-8-B}{RGB}{179,205,227}
\definecolor{Pastel1-8-3}{RGB}{204,235,197}
\definecolor{Pastel1-8-C}{RGB}{204,235,197}
\definecolor{Pastel1-8-4}{RGB}{222,203,228}
\definecolor{Pastel1-8-D}{RGB}{222,203,228}
\definecolor{Pastel1-8-5}{RGB}{254,217,166}
\definecolor{Pastel1-8-E}{RGB}{254,217,166}
\definecolor{Pastel1-8-6}{RGB}{255,255,204}
\definecolor{Pastel1-8-F}{RGB}{255,255,204}
\definecolor{Pastel1-8-7}{RGB}{229,216,189}
\definecolor{Pastel1-8-G}{RGB}{229,216,189}
\definecolor{Pastel1-8-8}{RGB}{253,218,236}
\definecolor{Pastel1-8-H}{RGB}{253,218,236}
\definecolor{Pastel1-9-1}{RGB}{251,180,174}
\definecolor{Pastel1-9-A}{RGB}{251,180,174}
\definecolor{Pastel1-9-2}{RGB}{179,205,227}
\definecolor{Pastel1-9-B}{RGB}{179,205,227}
\definecolor{Pastel1-9-3}{RGB}{204,235,197}
\definecolor{Pastel1-9-C}{RGB}{204,235,197}
\definecolor{Pastel1-9-4}{RGB}{222,203,228}
\definecolor{Pastel1-9-D}{RGB}{222,203,228}
\definecolor{Pastel1-9-5}{RGB}{254,217,166}
\definecolor{Pastel1-9-E}{RGB}{254,217,166}
\definecolor{Pastel1-9-6}{RGB}{255,255,204}
\definecolor{Pastel1-9-F}{RGB}{255,255,204}
\definecolor{Pastel1-9-7}{RGB}{229,216,189}
\definecolor{Pastel1-9-G}{RGB}{229,216,189}
\definecolor{Pastel1-9-8}{RGB}{253,218,236}
\definecolor{Pastel1-9-H}{RGB}{253,218,236}
\definecolor{Pastel1-9-9}{RGB}{242,242,242}
\definecolor{Pastel1-9-I}{RGB}{242,242,242}
\definecolor{Set1-3-1}{RGB}{228,26,28}
\definecolor{Set1-3-A}{RGB}{228,26,28}
\definecolor{Set1-3-2}{RGB}{55,126,184}
\definecolor{Set1-3-B}{RGB}{55,126,184}
\definecolor{Set1-3-3}{RGB}{77,175,74}
\definecolor{Set1-3-C}{RGB}{77,175,74}
\definecolor{Set1-4-1}{RGB}{228,26,28}
\definecolor{Set1-4-A}{RGB}{228,26,28}
\definecolor{Set1-4-2}{RGB}{55,126,184}
\definecolor{Set1-4-B}{RGB}{55,126,184}
\definecolor{Set1-4-3}{RGB}{77,175,74}
\definecolor{Set1-4-C}{RGB}{77,175,74}
\definecolor{Set1-4-4}{RGB}{152,78,163}
\definecolor{Set1-4-D}{RGB}{152,78,163}
\definecolor{Set1-5-1}{RGB}{228,26,28}
\definecolor{Set1-5-A}{RGB}{228,26,28}
\definecolor{Set1-5-2}{RGB}{55,126,184}
\definecolor{Set1-5-B}{RGB}{55,126,184}
\definecolor{Set1-5-3}{RGB}{77,175,74}
\definecolor{Set1-5-C}{RGB}{77,175,74}
\definecolor{Set1-5-4}{RGB}{152,78,163}
\definecolor{Set1-5-D}{RGB}{152,78,163}
\definecolor{Set1-5-5}{RGB}{255,127,0}
\definecolor{Set1-5-E}{RGB}{255,127,0}
\definecolor{Set1-6-1}{RGB}{228,26,28}
\definecolor{Set1-6-A}{RGB}{228,26,28}
\definecolor{Set1-6-2}{RGB}{55,126,184}
\definecolor{Set1-6-B}{RGB}{55,126,184}
\definecolor{Set1-6-3}{RGB}{77,175,74}
\definecolor{Set1-6-C}{RGB}{77,175,74}
\definecolor{Set1-6-4}{RGB}{152,78,163}
\definecolor{Set1-6-D}{RGB}{152,78,163}
\definecolor{Set1-6-5}{RGB}{255,127,0}
\definecolor{Set1-6-E}{RGB}{255,127,0}
\definecolor{Set1-6-6}{RGB}{255,255,51}
\definecolor{Set1-6-F}{RGB}{255,255,51}
\definecolor{Set1-7-1}{RGB}{228,26,28}
\definecolor{Set1-7-A}{RGB}{228,26,28}
\definecolor{Set1-7-2}{RGB}{55,126,184}
\definecolor{Set1-7-B}{RGB}{55,126,184}
\definecolor{Set1-7-3}{RGB}{77,175,74}
\definecolor{Set1-7-C}{RGB}{77,175,74}
\definecolor{Set1-7-4}{RGB}{152,78,163}
\definecolor{Set1-7-D}{RGB}{152,78,163}
\definecolor{Set1-7-5}{RGB}{255,127,0}
\definecolor{Set1-7-E}{RGB}{255,127,0}
\definecolor{Set1-7-6}{RGB}{255,255,51}
\definecolor{Set1-7-F}{RGB}{255,255,51}
\definecolor{Set1-7-7}{RGB}{166,86,40}
\definecolor{Set1-7-G}{RGB}{166,86,40}
\definecolor{Set1-8-1}{RGB}{228,26,28}
\definecolor{Set1-8-A}{RGB}{228,26,28}
\definecolor{Set1-8-2}{RGB}{55,126,184}
\definecolor{Set1-8-B}{RGB}{55,126,184}
\definecolor{Set1-8-3}{RGB}{77,175,74}
\definecolor{Set1-8-C}{RGB}{77,175,74}
\definecolor{Set1-8-4}{RGB}{152,78,163}
\definecolor{Set1-8-D}{RGB}{152,78,163}
\definecolor{Set1-8-5}{RGB}{255,127,0}
\definecolor{Set1-8-E}{RGB}{255,127,0}
\definecolor{Set1-8-6}{RGB}{255,255,51}
\definecolor{Set1-8-F}{RGB}{255,255,51}
\definecolor{Set1-8-7}{RGB}{166,86,40}
\definecolor{Set1-8-G}{RGB}{166,86,40}
\definecolor{Set1-8-8}{RGB}{247,129,191}
\definecolor{Set1-8-H}{RGB}{247,129,191}
\definecolor{Set1-9-1}{RGB}{228,26,28}
\definecolor{Set1-9-A}{RGB}{228,26,28}
\definecolor{Set1-9-2}{RGB}{55,126,184}
\definecolor{Set1-9-B}{RGB}{55,126,184}
\definecolor{Set1-9-3}{RGB}{77,175,74}
\definecolor{Set1-9-C}{RGB}{77,175,74}
\definecolor{Set1-9-4}{RGB}{152,78,163}
\definecolor{Set1-9-D}{RGB}{152,78,163}
\definecolor{Set1-9-5}{RGB}{255,127,0}
\definecolor{Set1-9-E}{RGB}{255,127,0}
\definecolor{Set1-9-6}{RGB}{255,255,51}
\definecolor{Set1-9-F}{RGB}{255,255,51}
\definecolor{Set1-9-7}{RGB}{166,86,40}
\definecolor{Set1-9-G}{RGB}{166,86,40}
\definecolor{Set1-9-8}{RGB}{247,129,191}
\definecolor{Set1-9-H}{RGB}{247,129,191}
\definecolor{Set1-9-9}{RGB}{153,153,153}
\definecolor{Set1-9-I}{RGB}{153,153,153}
\definecolor{Pastel2-3-1}{RGB}{179,226,205}
\definecolor{Pastel2-3-A}{RGB}{179,226,205}
\definecolor{Pastel2-3-2}{RGB}{253,205,172}
\definecolor{Pastel2-3-B}{RGB}{253,205,172}
\definecolor{Pastel2-3-3}{RGB}{203,213,232}
\definecolor{Pastel2-3-C}{RGB}{203,213,232}
\definecolor{Pastel2-4-1}{RGB}{179,226,205}
\definecolor{Pastel2-4-A}{RGB}{179,226,205}
\definecolor{Pastel2-4-2}{RGB}{253,205,172}
\definecolor{Pastel2-4-B}{RGB}{253,205,172}
\definecolor{Pastel2-4-3}{RGB}{203,213,232}
\definecolor{Pastel2-4-C}{RGB}{203,213,232}
\definecolor{Pastel2-4-4}{RGB}{244,202,228}
\definecolor{Pastel2-4-D}{RGB}{244,202,228}
\definecolor{Pastel2-5-1}{RGB}{179,226,205}
\definecolor{Pastel2-5-A}{RGB}{179,226,205}
\definecolor{Pastel2-5-2}{RGB}{253,205,172}
\definecolor{Pastel2-5-B}{RGB}{253,205,172}
\definecolor{Pastel2-5-3}{RGB}{203,213,232}
\definecolor{Pastel2-5-C}{RGB}{203,213,232}
\definecolor{Pastel2-5-4}{RGB}{244,202,228}
\definecolor{Pastel2-5-D}{RGB}{244,202,228}
\definecolor{Pastel2-5-5}{RGB}{230,245,201}
\definecolor{Pastel2-5-E}{RGB}{230,245,201}
\definecolor{Pastel2-6-1}{RGB}{179,226,205}
\definecolor{Pastel2-6-A}{RGB}{179,226,205}
\definecolor{Pastel2-6-2}{RGB}{253,205,172}
\definecolor{Pastel2-6-B}{RGB}{253,205,172}
\definecolor{Pastel2-6-3}{RGB}{203,213,232}
\definecolor{Pastel2-6-C}{RGB}{203,213,232}
\definecolor{Pastel2-6-4}{RGB}{244,202,228}
\definecolor{Pastel2-6-D}{RGB}{244,202,228}
\definecolor{Pastel2-6-5}{RGB}{230,245,201}
\definecolor{Pastel2-6-E}{RGB}{230,245,201}
\definecolor{Pastel2-6-6}{RGB}{255,242,174}
\definecolor{Pastel2-6-F}{RGB}{255,242,174}
\definecolor{Pastel2-7-1}{RGB}{179,226,205}
\definecolor{Pastel2-7-A}{RGB}{179,226,205}
\definecolor{Pastel2-7-2}{RGB}{253,205,172}
\definecolor{Pastel2-7-B}{RGB}{253,205,172}
\definecolor{Pastel2-7-3}{RGB}{203,213,232}
\definecolor{Pastel2-7-C}{RGB}{203,213,232}
\definecolor{Pastel2-7-4}{RGB}{244,202,228}
\definecolor{Pastel2-7-D}{RGB}{244,202,228}
\definecolor{Pastel2-7-5}{RGB}{230,245,201}
\definecolor{Pastel2-7-E}{RGB}{230,245,201}
\definecolor{Pastel2-7-6}{RGB}{255,242,174}
\definecolor{Pastel2-7-F}{RGB}{255,242,174}
\definecolor{Pastel2-7-7}{RGB}{241,226,204}
\definecolor{Pastel2-7-G}{RGB}{241,226,204}
\definecolor{Pastel2-8-1}{RGB}{179,226,205}
\definecolor{Pastel2-8-A}{RGB}{179,226,205}
\definecolor{Pastel2-8-2}{RGB}{253,205,172}
\definecolor{Pastel2-8-B}{RGB}{253,205,172}
\definecolor{Pastel2-8-3}{RGB}{203,213,232}
\definecolor{Pastel2-8-C}{RGB}{203,213,232}
\definecolor{Pastel2-8-4}{RGB}{244,202,228}
\definecolor{Pastel2-8-D}{RGB}{244,202,228}
\definecolor{Pastel2-8-5}{RGB}{230,245,201}
\definecolor{Pastel2-8-E}{RGB}{230,245,201}
\definecolor{Pastel2-8-6}{RGB}{255,242,174}
\definecolor{Pastel2-8-F}{RGB}{255,242,174}
\definecolor{Pastel2-8-7}{RGB}{241,226,204}
\definecolor{Pastel2-8-G}{RGB}{241,226,204}
\definecolor{Pastel2-8-8}{RGB}{204,204,204}
\definecolor{Pastel2-8-H}{RGB}{204,204,204}
\definecolor{Set2-3-1}{RGB}{102,194,165}
\definecolor{Set2-3-A}{RGB}{102,194,165}
\definecolor{Set2-3-2}{RGB}{252,141,98}
\definecolor{Set2-3-B}{RGB}{252,141,98}
\definecolor{Set2-3-3}{RGB}{141,160,203}
\definecolor{Set2-3-C}{RGB}{141,160,203}
\definecolor{Set2-4-1}{RGB}{102,194,165}
\definecolor{Set2-4-A}{RGB}{102,194,165}
\definecolor{Set2-4-2}{RGB}{252,141,98}
\definecolor{Set2-4-B}{RGB}{252,141,98}
\definecolor{Set2-4-3}{RGB}{141,160,203}
\definecolor{Set2-4-C}{RGB}{141,160,203}
\definecolor{Set2-4-4}{RGB}{231,138,195}
\definecolor{Set2-4-D}{RGB}{231,138,195}
\definecolor{Set2-5-1}{RGB}{102,194,165}
\definecolor{Set2-5-A}{RGB}{102,194,165}
\definecolor{Set2-5-2}{RGB}{252,141,98}
\definecolor{Set2-5-B}{RGB}{252,141,98}
\definecolor{Set2-5-3}{RGB}{141,160,203}
\definecolor{Set2-5-C}{RGB}{141,160,203}
\definecolor{Set2-5-4}{RGB}{231,138,195}
\definecolor{Set2-5-D}{RGB}{231,138,195}
\definecolor{Set2-5-5}{RGB}{166,216,84}
\definecolor{Set2-5-E}{RGB}{166,216,84}
\definecolor{Set2-6-1}{RGB}{102,194,165}
\definecolor{Set2-6-A}{RGB}{102,194,165}
\definecolor{Set2-6-2}{RGB}{252,141,98}
\definecolor{Set2-6-B}{RGB}{252,141,98}
\definecolor{Set2-6-3}{RGB}{141,160,203}
\definecolor{Set2-6-C}{RGB}{141,160,203}
\definecolor{Set2-6-4}{RGB}{231,138,195}
\definecolor{Set2-6-D}{RGB}{231,138,195}
\definecolor{Set2-6-5}{RGB}{166,216,84}
\definecolor{Set2-6-E}{RGB}{166,216,84}
\definecolor{Set2-6-6}{RGB}{255,217,47}
\definecolor{Set2-6-F}{RGB}{255,217,47}
\definecolor{Set2-7-1}{RGB}{102,194,165}
\definecolor{Set2-7-A}{RGB}{102,194,165}
\definecolor{Set2-7-2}{RGB}{252,141,98}
\definecolor{Set2-7-B}{RGB}{252,141,98}
\definecolor{Set2-7-3}{RGB}{141,160,203}
\definecolor{Set2-7-C}{RGB}{141,160,203}
\definecolor{Set2-7-4}{RGB}{231,138,195}
\definecolor{Set2-7-D}{RGB}{231,138,195}
\definecolor{Set2-7-5}{RGB}{166,216,84}
\definecolor{Set2-7-E}{RGB}{166,216,84}
\definecolor{Set2-7-6}{RGB}{255,217,47}
\definecolor{Set2-7-F}{RGB}{255,217,47}
\definecolor{Set2-7-7}{RGB}{229,196,148}
\definecolor{Set2-7-G}{RGB}{229,196,148}
\definecolor{Set2-8-1}{RGB}{102,194,165}
\definecolor{Set2-8-A}{RGB}{102,194,165}
\definecolor{Set2-8-2}{RGB}{252,141,98}
\definecolor{Set2-8-B}{RGB}{252,141,98}
\definecolor{Set2-8-3}{RGB}{141,160,203}
\definecolor{Set2-8-C}{RGB}{141,160,203}
\definecolor{Set2-8-4}{RGB}{231,138,195}
\definecolor{Set2-8-D}{RGB}{231,138,195}
\definecolor{Set2-8-5}{RGB}{166,216,84}
\definecolor{Set2-8-E}{RGB}{166,216,84}
\definecolor{Set2-8-6}{RGB}{255,217,47}
\definecolor{Set2-8-F}{RGB}{255,217,47}
\definecolor{Set2-8-7}{RGB}{229,196,148}
\definecolor{Set2-8-G}{RGB}{229,196,148}
\definecolor{Set2-8-8}{RGB}{179,179,179}
\definecolor{Set2-8-H}{RGB}{179,179,179}
\definecolor{Dark2-3-1}{RGB}{27,158,119}
\definecolor{Dark2-3-A}{RGB}{27,158,119}
\definecolor{Dark2-3-2}{RGB}{217,95,2}
\definecolor{Dark2-3-B}{RGB}{217,95,2}
\definecolor{Dark2-3-3}{RGB}{117,112,179}
\definecolor{Dark2-3-C}{RGB}{117,112,179}
\definecolor{Dark2-4-1}{RGB}{27,158,119}
\definecolor{Dark2-4-A}{RGB}{27,158,119}
\definecolor{Dark2-4-2}{RGB}{217,95,2}
\definecolor{Dark2-4-B}{RGB}{217,95,2}
\definecolor{Dark2-4-3}{RGB}{117,112,179}
\definecolor{Dark2-4-C}{RGB}{117,112,179}
\definecolor{Dark2-4-4}{RGB}{231,41,138}
\definecolor{Dark2-4-D}{RGB}{231,41,138}
\definecolor{Dark2-5-1}{RGB}{27,158,119}
\definecolor{Dark2-5-A}{RGB}{27,158,119}
\definecolor{Dark2-5-2}{RGB}{217,95,2}
\definecolor{Dark2-5-B}{RGB}{217,95,2}
\definecolor{Dark2-5-3}{RGB}{117,112,179}
\definecolor{Dark2-5-C}{RGB}{117,112,179}
\definecolor{Dark2-5-4}{RGB}{231,41,138}
\definecolor{Dark2-5-D}{RGB}{231,41,138}
\definecolor{Dark2-5-5}{RGB}{102,166,30}
\definecolor{Dark2-5-E}{RGB}{102,166,30}
\definecolor{Dark2-6-1}{RGB}{27,158,119}
\definecolor{Dark2-6-A}{RGB}{27,158,119}
\definecolor{Dark2-6-2}{RGB}{217,95,2}
\definecolor{Dark2-6-B}{RGB}{217,95,2}
\definecolor{Dark2-6-3}{RGB}{117,112,179}
\definecolor{Dark2-6-C}{RGB}{117,112,179}
\definecolor{Dark2-6-4}{RGB}{231,41,138}
\definecolor{Dark2-6-D}{RGB}{231,41,138}
\definecolor{Dark2-6-5}{RGB}{102,166,30}
\definecolor{Dark2-6-E}{RGB}{102,166,30}
\definecolor{Dark2-6-6}{RGB}{230,171,2}
\definecolor{Dark2-6-F}{RGB}{230,171,2}
\definecolor{Dark2-7-1}{RGB}{27,158,119}
\definecolor{Dark2-7-A}{RGB}{27,158,119}
\definecolor{Dark2-7-2}{RGB}{217,95,2}
\definecolor{Dark2-7-B}{RGB}{217,95,2}
\definecolor{Dark2-7-3}{RGB}{117,112,179}
\definecolor{Dark2-7-C}{RGB}{117,112,179}
\definecolor{Dark2-7-4}{RGB}{231,41,138}
\definecolor{Dark2-7-D}{RGB}{231,41,138}
\definecolor{Dark2-7-5}{RGB}{102,166,30}
\definecolor{Dark2-7-E}{RGB}{102,166,30}
\definecolor{Dark2-7-6}{RGB}{230,171,2}
\definecolor{Dark2-7-F}{RGB}{230,171,2}
\definecolor{Dark2-7-7}{RGB}{166,118,29}
\definecolor{Dark2-7-G}{RGB}{166,118,29}
\definecolor{Dark2-8-1}{RGB}{27,158,119}
\definecolor{Dark2-8-A}{RGB}{27,158,119}
\definecolor{Dark2-8-2}{RGB}{217,95,2}
\definecolor{Dark2-8-B}{RGB}{217,95,2}
\definecolor{Dark2-8-3}{RGB}{117,112,179}
\definecolor{Dark2-8-C}{RGB}{117,112,179}
\definecolor{Dark2-8-4}{RGB}{231,41,138}
\definecolor{Dark2-8-D}{RGB}{231,41,138}
\definecolor{Dark2-8-5}{RGB}{102,166,30}
\definecolor{Dark2-8-E}{RGB}{102,166,30}
\definecolor{Dark2-8-6}{RGB}{230,171,2}
\definecolor{Dark2-8-F}{RGB}{230,171,2}
\definecolor{Dark2-8-7}{RGB}{166,118,29}
\definecolor{Dark2-8-G}{RGB}{166,118,29}
\definecolor{Dark2-8-8}{RGB}{102,102,102}
\definecolor{Dark2-8-H}{RGB}{102,102,102}
\definecolor{Paired-3-1}{RGB}{166,206,227}
\definecolor{Paired-3-A}{RGB}{166,206,227}
\definecolor{Paired-3-2}{RGB}{31,120,180}
\definecolor{Paired-3-B}{RGB}{31,120,180}
\definecolor{Paired-3-3}{RGB}{178,223,138}
\definecolor{Paired-3-C}{RGB}{178,223,138}
\definecolor{Paired-4-1}{RGB}{166,206,227}
\definecolor{Paired-4-A}{RGB}{166,206,227}
\definecolor{Paired-4-2}{RGB}{31,120,180}
\definecolor{Paired-4-B}{RGB}{31,120,180}
\definecolor{Paired-4-3}{RGB}{178,223,138}
\definecolor{Paired-4-C}{RGB}{178,223,138}
\definecolor{Paired-4-4}{RGB}{51,160,44}
\definecolor{Paired-4-D}{RGB}{51,160,44}
\definecolor{Paired-5-1}{RGB}{166,206,227}
\definecolor{Paired-5-A}{RGB}{166,206,227}
\definecolor{Paired-5-2}{RGB}{31,120,180}
\definecolor{Paired-5-B}{RGB}{31,120,180}
\definecolor{Paired-5-3}{RGB}{178,223,138}
\definecolor{Paired-5-C}{RGB}{178,223,138}
\definecolor{Paired-5-4}{RGB}{51,160,44}
\definecolor{Paired-5-D}{RGB}{51,160,44}
\definecolor{Paired-5-5}{RGB}{251,154,153}
\definecolor{Paired-5-E}{RGB}{251,154,153}
\definecolor{Paired-6-1}{RGB}{166,206,227}
\definecolor{Paired-6-A}{RGB}{166,206,227}
\definecolor{Paired-6-2}{RGB}{31,120,180}
\definecolor{Paired-6-B}{RGB}{31,120,180}
\definecolor{Paired-6-3}{RGB}{178,223,138}
\definecolor{Paired-6-C}{RGB}{178,223,138}
\definecolor{Paired-6-4}{RGB}{51,160,44}
\definecolor{Paired-6-D}{RGB}{51,160,44}
\definecolor{Paired-6-5}{RGB}{251,154,153}
\definecolor{Paired-6-E}{RGB}{251,154,153}
\definecolor{Paired-6-6}{RGB}{227,26,28}
\definecolor{Paired-6-F}{RGB}{227,26,28}
\definecolor{Paired-7-1}{RGB}{166,206,227}
\definecolor{Paired-7-A}{RGB}{166,206,227}
\definecolor{Paired-7-2}{RGB}{31,120,180}
\definecolor{Paired-7-B}{RGB}{31,120,180}
\definecolor{Paired-7-3}{RGB}{178,223,138}
\definecolor{Paired-7-C}{RGB}{178,223,138}
\definecolor{Paired-7-4}{RGB}{51,160,44}
\definecolor{Paired-7-D}{RGB}{51,160,44}
\definecolor{Paired-7-5}{RGB}{251,154,153}
\definecolor{Paired-7-E}{RGB}{251,154,153}
\definecolor{Paired-7-6}{RGB}{227,26,28}
\definecolor{Paired-7-F}{RGB}{227,26,28}
\definecolor{Paired-7-7}{RGB}{253,191,111}
\definecolor{Paired-7-G}{RGB}{253,191,111}
\definecolor{Paired-8-1}{RGB}{166,206,227}
\definecolor{Paired-8-A}{RGB}{166,206,227}
\definecolor{Paired-8-2}{RGB}{31,120,180}
\definecolor{Paired-8-B}{RGB}{31,120,180}
\definecolor{Paired-8-3}{RGB}{178,223,138}
\definecolor{Paired-8-C}{RGB}{178,223,138}
\definecolor{Paired-8-4}{RGB}{51,160,44}
\definecolor{Paired-8-D}{RGB}{51,160,44}
\definecolor{Paired-8-5}{RGB}{251,154,153}
\definecolor{Paired-8-E}{RGB}{251,154,153}
\definecolor{Paired-8-6}{RGB}{227,26,28}
\definecolor{Paired-8-F}{RGB}{227,26,28}
\definecolor{Paired-8-7}{RGB}{253,191,111}
\definecolor{Paired-8-G}{RGB}{253,191,111}
\definecolor{Paired-8-8}{RGB}{255,127,0}
\definecolor{Paired-8-H}{RGB}{255,127,0}
\definecolor{Paired-9-1}{RGB}{166,206,227}
\definecolor{Paired-9-A}{RGB}{166,206,227}
\definecolor{Paired-9-2}{RGB}{31,120,180}
\definecolor{Paired-9-B}{RGB}{31,120,180}
\definecolor{Paired-9-3}{RGB}{178,223,138}
\definecolor{Paired-9-C}{RGB}{178,223,138}
\definecolor{Paired-9-4}{RGB}{51,160,44}
\definecolor{Paired-9-D}{RGB}{51,160,44}
\definecolor{Paired-9-5}{RGB}{251,154,153}
\definecolor{Paired-9-E}{RGB}{251,154,153}
\definecolor{Paired-9-6}{RGB}{227,26,28}
\definecolor{Paired-9-F}{RGB}{227,26,28}
\definecolor{Paired-9-7}{RGB}{253,191,111}
\definecolor{Paired-9-G}{RGB}{253,191,111}
\definecolor{Paired-9-8}{RGB}{255,127,0}
\definecolor{Paired-9-H}{RGB}{255,127,0}
\definecolor{Paired-9-9}{RGB}{202,178,214}
\definecolor{Paired-9-I}{RGB}{202,178,214}
\definecolor{Paired-10-1}{RGB}{166,206,227}
\definecolor{Paired-10-A}{RGB}{166,206,227}
\definecolor{Paired-10-2}{RGB}{31,120,180}
\definecolor{Paired-10-B}{RGB}{31,120,180}
\definecolor{Paired-10-3}{RGB}{178,223,138}
\definecolor{Paired-10-C}{RGB}{178,223,138}
\definecolor{Paired-10-4}{RGB}{51,160,44}
\definecolor{Paired-10-D}{RGB}{51,160,44}
\definecolor{Paired-10-5}{RGB}{251,154,153}
\definecolor{Paired-10-E}{RGB}{251,154,153}
\definecolor{Paired-10-6}{RGB}{227,26,28}
\definecolor{Paired-10-F}{RGB}{227,26,28}
\definecolor{Paired-10-7}{RGB}{253,191,111}
\definecolor{Paired-10-G}{RGB}{253,191,111}
\definecolor{Paired-10-8}{RGB}{255,127,0}
\definecolor{Paired-10-H}{RGB}{255,127,0}
\definecolor{Paired-10-9}{RGB}{202,178,214}
\definecolor{Paired-10-I}{RGB}{202,178,214}
\definecolor{Paired-10-10}{RGB}{106,61,154}
\definecolor{Paired-10-J}{RGB}{106,61,154}
\definecolor{Paired-11-1}{RGB}{166,206,227}
\definecolor{Paired-11-A}{RGB}{166,206,227}
\definecolor{Paired-11-2}{RGB}{31,120,180}
\definecolor{Paired-11-B}{RGB}{31,120,180}
\definecolor{Paired-11-3}{RGB}{178,223,138}
\definecolor{Paired-11-C}{RGB}{178,223,138}
\definecolor{Paired-11-4}{RGB}{51,160,44}
\definecolor{Paired-11-D}{RGB}{51,160,44}
\definecolor{Paired-11-5}{RGB}{251,154,153}
\definecolor{Paired-11-E}{RGB}{251,154,153}
\definecolor{Paired-11-6}{RGB}{227,26,28}
\definecolor{Paired-11-F}{RGB}{227,26,28}
\definecolor{Paired-11-7}{RGB}{253,191,111}
\definecolor{Paired-11-G}{RGB}{253,191,111}
\definecolor{Paired-11-8}{RGB}{255,127,0}
\definecolor{Paired-11-H}{RGB}{255,127,0}
\definecolor{Paired-11-9}{RGB}{202,178,214}
\definecolor{Paired-11-I}{RGB}{202,178,214}
\definecolor{Paired-11-10}{RGB}{106,61,154}
\definecolor{Paired-11-J}{RGB}{106,61,154}
\definecolor{Paired-11-11}{RGB}{255,255,153}
\definecolor{Paired-11-K}{RGB}{255,255,153}
\definecolor{Paired-12-1}{RGB}{166,206,227}
\definecolor{Paired-12-A}{RGB}{166,206,227}
\definecolor{Paired-12-2}{RGB}{31,120,180}
\definecolor{Paired-12-B}{RGB}{31,120,180}
\definecolor{Paired-12-3}{RGB}{178,223,138}
\definecolor{Paired-12-C}{RGB}{178,223,138}
\definecolor{Paired-12-4}{RGB}{51,160,44}
\definecolor{Paired-12-D}{RGB}{51,160,44}
\definecolor{Paired-12-5}{RGB}{251,154,153}
\definecolor{Paired-12-E}{RGB}{251,154,153}
\definecolor{Paired-12-6}{RGB}{227,26,28}
\definecolor{Paired-12-F}{RGB}{227,26,28}
\definecolor{Paired-12-7}{RGB}{253,191,111}
\definecolor{Paired-12-G}{RGB}{253,191,111}
\definecolor{Paired-12-8}{RGB}{255,127,0}
\definecolor{Paired-12-H}{RGB}{255,127,0}
\definecolor{Paired-12-9}{RGB}{202,178,214}
\definecolor{Paired-12-I}{RGB}{202,178,214}
\definecolor{Paired-12-10}{RGB}{106,61,154}
\definecolor{Paired-12-J}{RGB}{106,61,154}
\definecolor{Paired-12-11}{RGB}{255,255,153}
\definecolor{Paired-12-K}{RGB}{255,255,153}
\definecolor{Paired-12-12}{RGB}{177,89,40}
\definecolor{Paired-12-L}{RGB}{177,89,40}
\definecolor{Accent-3-1}{RGB}{127,201,127}
\definecolor{Accent-3-A}{RGB}{127,201,127}
\definecolor{Accent-3-2}{RGB}{190,174,212}
\definecolor{Accent-3-B}{RGB}{190,174,212}
\definecolor{Accent-3-3}{RGB}{253,192,134}
\definecolor{Accent-3-C}{RGB}{253,192,134}
\definecolor{Accent-4-1}{RGB}{127,201,127}
\definecolor{Accent-4-A}{RGB}{127,201,127}
\definecolor{Accent-4-2}{RGB}{190,174,212}
\definecolor{Accent-4-B}{RGB}{190,174,212}
\definecolor{Accent-4-3}{RGB}{253,192,134}
\definecolor{Accent-4-C}{RGB}{253,192,134}
\definecolor{Accent-4-4}{RGB}{255,255,153}
\definecolor{Accent-4-D}{RGB}{255,255,153}
\definecolor{Accent-5-1}{RGB}{127,201,127}
\definecolor{Accent-5-A}{RGB}{127,201,127}
\definecolor{Accent-5-2}{RGB}{190,174,212}
\definecolor{Accent-5-B}{RGB}{190,174,212}
\definecolor{Accent-5-3}{RGB}{253,192,134}
\definecolor{Accent-5-C}{RGB}{253,192,134}
\definecolor{Accent-5-4}{RGB}{255,255,153}
\definecolor{Accent-5-D}{RGB}{255,255,153}
\definecolor{Accent-5-5}{RGB}{56,108,176}
\definecolor{Accent-5-E}{RGB}{56,108,176}
\definecolor{Accent-6-1}{RGB}{127,201,127}
\definecolor{Accent-6-A}{RGB}{127,201,127}
\definecolor{Accent-6-2}{RGB}{190,174,212}
\definecolor{Accent-6-B}{RGB}{190,174,212}
\definecolor{Accent-6-3}{RGB}{253,192,134}
\definecolor{Accent-6-C}{RGB}{253,192,134}
\definecolor{Accent-6-4}{RGB}{255,255,153}
\definecolor{Accent-6-D}{RGB}{255,255,153}
\definecolor{Accent-6-5}{RGB}{56,108,176}
\definecolor{Accent-6-E}{RGB}{56,108,176}
\definecolor{Accent-6-6}{RGB}{240,2,127}
\definecolor{Accent-6-F}{RGB}{240,2,127}
\definecolor{Accent-7-1}{RGB}{127,201,127}
\definecolor{Accent-7-A}{RGB}{127,201,127}
\definecolor{Accent-7-2}{RGB}{190,174,212}
\definecolor{Accent-7-B}{RGB}{190,174,212}
\definecolor{Accent-7-3}{RGB}{253,192,134}
\definecolor{Accent-7-C}{RGB}{253,192,134}
\definecolor{Accent-7-4}{RGB}{255,255,153}
\definecolor{Accent-7-D}{RGB}{255,255,153}
\definecolor{Accent-7-5}{RGB}{56,108,176}
\definecolor{Accent-7-E}{RGB}{56,108,176}
\definecolor{Accent-7-6}{RGB}{240,2,127}
\definecolor{Accent-7-F}{RGB}{240,2,127}
\definecolor{Accent-7-7}{RGB}{191,91,23}
\definecolor{Accent-7-G}{RGB}{191,91,23}
\definecolor{Accent-8-1}{RGB}{127,201,127}
\definecolor{Accent-8-A}{RGB}{127,201,127}
\definecolor{Accent-8-2}{RGB}{190,174,212}
\definecolor{Accent-8-B}{RGB}{190,174,212}
\definecolor{Accent-8-3}{RGB}{253,192,134}
\definecolor{Accent-8-C}{RGB}{253,192,134}
\definecolor{Accent-8-4}{RGB}{255,255,153}
\definecolor{Accent-8-D}{RGB}{255,255,153}
\definecolor{Accent-8-5}{RGB}{56,108,176}
\definecolor{Accent-8-E}{RGB}{56,108,176}
\definecolor{Accent-8-6}{RGB}{240,2,127}
\definecolor{Accent-8-F}{RGB}{240,2,127}
\definecolor{Accent-8-7}{RGB}{191,91,23}
\definecolor{Accent-8-G}{RGB}{191,91,23}
\definecolor{Accent-8-8}{RGB}{102,102,102}
\definecolor{Accent-8-H}{RGB}{102,102,102}

\definecolor{white}{rgb}{1.000,1.000,1.000}
\definecolor{snow}{rgb}{1.000,0.979,0.979}
\definecolor{honeydew}{rgb}{0.938,1.000,0.938}
\definecolor{mintcream}{rgb}{0.958,1.000,0.979}
\definecolor{azure}{rgb}{0.938,1.000,1.000}
\definecolor{aliceblue}{rgb}{0.938,0.971,1.000}
\definecolor{ghostwhite}{rgb}{0.971,0.971,1.000}
\definecolor{whitesmoke}{rgb}{0.958,0.958,0.958}
\definecolor{seashell}{rgb}{1.000,0.958,0.930}
\definecolor{beige}{rgb}{0.958,0.958,0.859}
\definecolor{oldlace}{rgb}{0.992,0.958,0.898}
\definecolor{floralwhite}{rgb}{1.000,0.979,0.938}
\definecolor{ivory}{rgb}{1.000,1.000,0.938}
\definecolor{antiquewhite}{rgb}{0.979,0.918,0.840}
\definecolor{linen}{rgb}{0.979,0.938,0.898}
\definecolor{lavenderblush}{rgb}{1.000,0.938,0.958}
\definecolor{mistyrose}{rgb}{1.000,0.891,0.879}
\definecolor{gray}{rgb}{0.500,0.500,0.500}
\definecolor{sgray}{rgb}{0.211,0.211,0.211}
\definecolor{gainsboro}{rgb}{0.859,0.859,0.859}
\definecolor{lightgray}{rgb}{0.824,0.824,0.824}
\definecolor{silver}{rgb}{0.750,0.750,0.750}
\definecolor{darkgray}{rgb}{0.660,0.660,0.660}
\definecolor{dimgray}{rgb}{0.410,0.410,0.410}
\definecolor{lightslategray}{rgb}{0.465,0.531,0.598}
\definecolor{slategray}{rgb}{0.438,0.500,0.562}
\definecolor{darkslategray}{rgb}{0.184,0.309,0.309}
\definecolor{black}{rgb}{0.000,0.000,0.000}
\definecolor{red}{rgb}{1.000,0.000,0.000}
\definecolor{lightsalmon}{rgb}{1.000,0.625,0.477}
\definecolor{salmon}{rgb}{0.979,0.500,0.445}
\definecolor{darksalmon}{rgb}{0.910,0.586,0.477}
\definecolor{lightcoral}{rgb}{0.938,0.500,0.500}
\definecolor{indianred}{rgb}{0.801,0.359,0.359}
\definecolor{crimson}{rgb}{0.859,0.078,0.234}
\definecolor{firebrick}{rgb}{0.695,0.133,0.133}
\definecolor{darkred}{rgb}{0.543,0.000,0.000}
\definecolor{pink}{rgb}{1.000,0.750,0.793}
\definecolor{lightpink}{rgb}{1.000,0.711,0.754}
\definecolor{hotpink}{rgb}{1.000,0.410,0.703}
\definecolor{deeppink}{rgb}{1.000,0.078,0.574}
\definecolor{palevioletred}{rgb}{0.855,0.438,0.574}
\definecolor{mediumvioletred}{rgb}{0.777,0.082,0.520}
\definecolor{orange}{rgb}{1.000,0.645,0.000}
\definecolor{darkorange}{rgb}{1.000,0.547,0.000}
\definecolor{coral}{rgb}{1.000,0.496,0.312}
\definecolor{tomato}{rgb}{1.000,0.387,0.277}
\definecolor{orangered}{rgb}{1.000,0.270,0.000}
\definecolor{yellow}{rgb}{1.000,1.000,0.000}
\definecolor{lightyellow}{rgb}{1.000,1.000,0.875}
\definecolor{lemonchiffon}{rgb}{1.000,0.979,0.801}
\definecolor{lightgoldenrodyellow}{rgb}{0.979,0.979,0.820}
\definecolor{papayawhip}{rgb}{1.000,0.934,0.832}
\definecolor{moccasin}{rgb}{1.000,0.891,0.707}
\definecolor{peachpuff}{rgb}{1.000,0.852,0.723}
\definecolor{palegoldenrod}{rgb}{0.930,0.906,0.664}
\definecolor{khaki}{rgb}{0.938,0.898,0.547}
\definecolor{darkkhaki}{rgb}{0.738,0.715,0.418}
\definecolor{gold}{rgb}{1.000,0.840,0.000}
\definecolor{brown}{rgb}{0.645,0.164,0.164}
\definecolor{cornsilk}{rgb}{1.000,0.971,0.859}
\definecolor{blanchedalmond}{rgb}{1.000,0.918,0.801}
\definecolor{bisque}{rgb}{1.000,0.891,0.766}
\definecolor{navajowhite}{rgb}{1.000,0.867,0.676}
\definecolor{wheat}{rgb}{0.958,0.867,0.699}
\definecolor{burlywood}{rgb}{0.867,0.719,0.527}
\definecolor{tan}{rgb}{0.820,0.703,0.547}
\definecolor{rosybrown}{rgb}{0.734,0.559,0.559}
\definecolor{sandybrown}{rgb}{0.954,0.641,0.375}
\definecolor{goldenrod}{rgb}{0.852,0.645,0.125}
\definecolor{darkgoldenrod}{rgb}{0.719,0.523,0.043}
\definecolor{peru}{rgb}{0.801,0.520,0.246}
\definecolor{chocolate}{rgb}{0.820,0.410,0.117}
\definecolor{saddlebrown}{rgb}{0.543,0.270,0.074}
\definecolor{sienna}{rgb}{0.625,0.320,0.176}
\definecolor{maroon}{rgb}{0.500,0.000,0.000}
\definecolor{green}{rgb}{0.000,0.500,0.000}
\definecolor{palegreen}{rgb}{0.594,0.983,0.594}
\definecolor{lightgreen}{rgb}{0.562,0.930,0.562}
\definecolor{yellowgreen}{rgb}{0.602,0.801,0.195}
\definecolor{greenyellow}{rgb}{0.676,1.000,0.184}
\definecolor{chartreuse}{rgb}{0.496,1.000,0.000}
\definecolor{lawngreen}{rgb}{0.484,0.988,0.000}
\definecolor{lime}{rgb}{0.000,1.000,0.000}
\definecolor{limegreen}{rgb}{0.195,0.801,0.195}
\definecolor{mediumspringgreen}{rgb}{0.000,0.979,0.602}
\definecolor{springgreen}{rgb}{0.000,1.000,0.496}
\definecolor{mediumaquamarine}{rgb}{0.398,0.801,0.664}
\definecolor{aquamarine}{rgb}{0.496,1.000,0.828}
\definecolor{lightseagreen}{rgb}{0.125,0.695,0.664}
\definecolor{mediumseagreen}{rgb}{0.234,0.699,0.441}
\definecolor{seagreen}{rgb}{0.180,0.543,0.340}
\definecolor{darkseagreen}{rgb}{0.559,0.734,0.559}
\definecolor{forestgreen}{rgb}{0.133,0.543,0.133}
\definecolor{darkgreen}{rgb}{0.000,0.391,0.000}
\definecolor{olivedrab}{rgb}{0.418,0.555,0.137}
\definecolor{olive}{rgb}{0.500,0.500,0.000}
\definecolor{darkolivegreen}{rgb}{0.332,0.418,0.184}
\definecolor{teal}{rgb}{0.000,0.500,0.500}
\definecolor{blue}{rgb}{0.000,0.000,1.000}
\definecolor{lightblue}{rgb}{0.676,0.844,0.898}
\definecolor{powderblue}{rgb}{0.688,0.875,0.898}
\definecolor{paleturquoise}{rgb}{0.684,0.930,0.930}
\definecolor{turquoise}{rgb}{0.250,0.875,0.812}
\definecolor{mediumturquoise}{rgb}{0.281,0.816,0.797}
\definecolor{darkturquoise}{rgb}{0.000,0.805,0.816}
\definecolor{lightcyan}{rgb}{0.875,1.000,1.000}
\definecolor{cyan}{rgb}{0.000,1.000,1.000}
\definecolor{aqua}{rgb}{0.000,1.000,1.000}
\definecolor{darkcyan}{rgb}{0.000,0.543,0.543}
\definecolor{cadetblue}{rgb}{0.371,0.617,0.625}
\definecolor{lightsteelblue}{rgb}{0.688,0.766,0.867}
\definecolor{steelblue}{rgb}{0.273,0.508,0.703}
\definecolor{lightskyblue}{rgb}{0.527,0.805,0.979}
\definecolor{skyblue}{rgb}{0.527,0.805,0.918}
\definecolor{deepskyblue}{rgb}{0.000,0.746,1.000}
\definecolor{dodgerblue}{rgb}{0.117,0.562,1.000}
\definecolor{cornflowerblue}{rgb}{0.391,0.582,0.926}
\definecolor{royalblue}{rgb}{0.254,0.410,0.879}
\definecolor{mediumblue}{rgb}{0.000,0.000,0.801}
\definecolor{darkblue}{rgb}{0.000,0.000,0.543}
\definecolor{navy}{rgb}{0.000,0.000,0.500}
\definecolor{midnightblue}{rgb}{0.098,0.098,0.438}
\definecolor{purple}{rgb}{0.500,0.000,0.500}
\definecolor{lavender}{rgb}{0.898,0.898,0.979}
\definecolor{thistle}{rgb}{0.844,0.746,0.844}
\definecolor{plum}{rgb}{0.863,0.625,0.863}
\definecolor{violet}{rgb}{0.930,0.508,0.930}
\definecolor{orchid}{rgb}{0.852,0.438,0.836}
\definecolor{fuchsia}{rgb}{1.000,0.000,1.000}
\definecolor{magenta}{rgb}{1.000,0.000,1.000}
\definecolor{mediumorchid}{rgb}{0.727,0.332,0.824}
\definecolor{mediumpurple}{rgb}{0.574,0.438,0.855}
\definecolor{amethyst}{rgb}{0.598,0.398,0.797}
\definecolor{blueviolet}{rgb}{0.539,0.168,0.883}
\definecolor{darkviolet}{rgb}{0.578,0.000,0.824}
\definecolor{darkorchid}{rgb}{0.598,0.195,0.797}
\definecolor{darkmagenta}{rgb}{0.543,0.000,0.543}
\definecolor{slateblue}{rgb}{0.414,0.352,0.801}
\definecolor{darkslateblue}{rgb}{0.281,0.238,0.543}
\definecolor{mediumslateblue}{rgb}{0.480,0.406,0.930}
\definecolor{indigo}{rgb}{0.293,0.000,0.508}
\definecolor{grey}{rgb}{0.500,0.500,0.500}
\definecolor{lightgrey}{rgb}{0.824,0.824,0.824}
\definecolor{darkgrey}{rgb}{0.660,0.660,0.660}
\definecolor{dimgrey}{rgb}{0.410,0.410,0.410}
\definecolor{lightslategrey}{rgb}{0.465,0.531,0.598}
\definecolor{slategrey}{rgb}{0.438,0.500,0.562}
\definecolor{darkslategrey}{rgb}{0.184,0.309,0.309}
\definecolor{rosa}{rgb}{1,0.5,0.5} 

\definecolor{fadv}{rgb}{1.000,0.547,0.000}
\definecolor{sym}{rgb}{0.500,0.500,0.500}
\definecolor{fdel}{rgb}{0.254,0.410,0.879}

\definecolor{sym2}{rgb}{1.000,1.000,1.000}
\definecolor{fadv}{RGB}{140,81,10}
\definecolor{padv}{RGB}{216,179,101}
\definecolor{ladv}{RGB}{246,232,195}
\definecolor{sym}{RGB}{245,245,245}
\definecolor{ldel}{RGB}{199,234,229}
\definecolor{pdel}{RGB}{90,180,172}
\definecolor{fdel}{RGB}{1,102,94}

\definecolor{gro}{RGB}{27,158,119}
\definecolor{em}{RGB}{217,95,2}
\definecolor{lo}{rgb}{117,112,179}

\definecolor{f1p5}{RGB}{229,245,224}
\definecolor{f1p20}{RGB}{199,233,192}
\definecolor{f1p35}{RGB}{161,217,155}
\definecolor{f1p45}{RGB}{116,196,118}
\definecolor{f1p65}{RGB}{65,171,93}
\definecolor{f1p89}{RGB}{35,139,69}
\definecolor{f1psine}{RGB}{0,109,44}

\definecolor{f2p5}{RGB}{254,230,206}
\definecolor{f2p20}{RGB}{253,208,162}
\definecolor{f2p35}{RGB}{253,174,107}
\definecolor{f2p45}{RGB}{253,141,60}
\definecolor{f2p65}{RGB}{241,105,19}
\definecolor{f2p89}{RGB}{217,72,1}
\definecolor{f2psine}{RGB}{166,54,3}

\definecolor{f3p5}{RGB}{222,235,247}
\definecolor{f3p20}{RGB}{198,219,239}
\definecolor{f3p35}{RGB}{158,202,225}
\definecolor{f3p45}{RGB}{107,174,214}
\definecolor{f3p65}{RGB}{66,146,198}
\definecolor{f3p89}{RGB}{33,113,181}
\definecolor{f3psine}{RGB}{8,81,156}

\newlength{\heightofx}

\definecolor{FURU}{named}{Reds-L}
\definecolor{FURD}{named}{Reds-H}
\definecolor{FDRD}{named}{Blues-H}
\definecolor{FDRU}{named}{Blues-L}
\def\lc{lift \texttt{+} cruise }

\title{Propeller-wing interactions of a lift + cruise eVTOL configuration during transition}

\author{Oliver Westcott$^{1*}$}
\contact{$^*$O.C.S.Westcott@soton.ac.uk}

\author{Robert Entwistle$^1$}

\author{Swathi Krishna$^1$}

\affil{$^1$Engineering and Physical Sciences, University of Southampton, Southampton, SO17 1BJ, UK}

\date{\today}

\begin{document}

\newcommand{\markerone}{\raisebox{0.5pt}{\tikz{\node[draw=black,scale=0.4,circle,fill=black](){};}}}
\newcommand{\markertwo}{\raisebox{0.5pt}{\tikz{\node[draw=black,scale=0.4,regular polygon, regular polygon sides=3,fill=none,rotate=0](){};}}}
\newcommand{\greenline}{\raisebox{2pt}{\tikz{\draw[-,Greens-F,solid,line width = 2pt](0,0) -- (2mm,0);}}}

\tikzset{liftmotorthrust/.style={color=black, mark=triangle*, mark options={solid,fill=black}}}

\tikzset{frontmotorthrust/.style={color=mts1, mark=diamond*,mark options={solid,fill=mts1}}}
\tikzset{rearmotorthrust/.style={color=mts7,solid, mark=diamond*, mark options={solid,fill=mts7}}}

\tikzset{dpactive/.style={color=mts5, mark=*,mark options={solid,fill=mts5}}}
\tikzset{dpinactive/.style={color=mts4, mark=*, solid, mark options={solid,fill=mts4}}}

\tikzset{lpactive/.style={color=mts3, mark=square*, solid, mark options={solid,fill=mts3}}}
\tikzset{lpinactive/.style={color=mts2, mark=square*, solid, mark options={solid,fill=mts2}}}

\tikzset{motorthrustsum/.style={color=mts6,mark=diamond, mark options={solid,fill=mts6}}}

\newcommand{\liftmotorthrust}{\raisebox{0pt}{\tikz[baseline=-0.5ex]
\draw[PuOr-M, dashed] (0,0) -- (0.5,0)
node[pos=0.5, inner sep=0pt, outer sep=0pt] 
{\tikz \draw[PuOr-M, fill=PuOr-M,solid] (0,0.09) -- (-0.09,-0.09) -- (0.09,-0.09) -- cycle;};}}

\newcommand{\allmotorthrust}{\raisebox{0pt}{\tikz[baseline=-0.5ex] 
\draw[blue] (0,0) -- (0.4,0) 
node[pos=0.5, draw=none, inner sep=0pt, outer sep=0pt, fill=blue, diamond, mark=*, minimum size=7pt]{};}}

\newcommand{\lpact}{\raisebox{0pt}{\tikz[baseline=-0.5ex] 
\draw[black] (0,0) -- (0.4,0) 
node[pos=0.5, draw=none, inner sep=0pt, outer sep=0pt, fill=black, mark=square*, minimum size=5pt]{};}}

\newcommand{\lpinact}{\raisebox{0pt}{\tikz[baseline=-0.5ex] 
\draw[magenta] (0,0) -- (0.4,0) 
node[pos=0.5, draw=none, inner sep=0pt, outer sep=0pt, fill=magenta, mark=square*, minimum size=5pt]{};}}

\newcommand{\frontthrust}{\raisebox{0pt}{\tikz[baseline=-0.5ex] 
\draw[maroon] (0,0) -- (0.4,0) 
node[pos=0.5, draw=none, inner sep=0pt, outer sep=0pt, fill=maroon, diamond, mark=*, minimum size=7pt]{};}}

\newcommand{\rearthrust}{\raisebox{0pt}{\tikz[baseline=-0.5ex] 
\draw[teal] (0,0) -- (0.4,0) 
node[pos=0.5, draw=none, inner sep=0pt, outer sep=0pt, fill=teal, diamond, mark=*, minimum size=7pt]{};}}

\newcommand{\dpact}{\raisebox{0pt}{\tikz[baseline=-0.5ex] 
\draw[Greens-M] (0,0) -- (0.4,0) 
node[pos=0.5, draw=none, inner sep=0pt, outer sep=0pt, fill=Greens-M,circle, mark=*, minimum size=5pt]{};}}

\newcommand{\dpinact}{\raisebox{0pt}{\tikz[baseline=-0.5ex] 
\draw[RdBu-M] (0,0) -- (0.4,0) 
node[pos=0.5, draw=none, inner sep=0pt, outer sep=0pt, fill=RdBu-M, circle, mark=*, minimum size=5pt]{};}}

\newcommand{\fdrushort}{\raisebox{0pt}{\tikz[baseline=-0.5ex,inner sep=0pt,outer sep=0pt]{\draw[FDRU,dashed] (-0.1,0) -- (0.4,0); \node[shape=regular polygon,regular polygon sides=5,minimum size=7pt,fill=none,draw=none] (p) at (0.15,0) {}; \draw[line width=0.5pt,color=FDRU] (p.corner 1) -- (p.corner 2) -- (p.corner 3) -- (p.corner 4) -- (p.corner 5) -- cycle;}}}

\newcommand{\furdshort}{\raisebox{0pt}{\tikz[baseline=-0.5ex,inner sep=0pt,outer sep=0pt]{\draw[FURD,dashed] (-0.1,0) -- (0.4,0); \node[shape=regular polygon,regular polygon sides=5,minimum size=7pt,fill=none,draw=none] (p) at (0.15,0) {}; \draw[line width=0.5pt,color=FURD] (p.corner 1) -- (p.corner 2) -- (p.corner 3) -- (p.corner 4) -- (p.corner 5) -- cycle;}}}

\newcommand{\furushort}{\raisebox{0pt}{\tikz[baseline=-0.5ex,inner sep=0pt,outer sep=0pt]{\draw[FURU,dashed] (-0.1,0) -- (0.4,0); \node[shape=regular polygon,regular polygon sides=5,minimum size=7pt,fill=none,draw=none] (p) at (0.15,0) {}; \draw[line width=0.5pt,color=FURU] (p.corner 1) -- (p.corner 2) -- (p.corner 3) -- (p.corner 4) -- (p.corner 5) -- cycle;}}}

\newcommand{\fdrulong}{\raisebox{0pt}{\tikz[baseline=-0.5ex,inner sep=0pt,outer sep=0pt]{\draw[FDRU] (-0.1,0) -- (0.4,0); \node[shape=regular polygon,regular polygon sides=5,minimum size=7pt,fill=FDRU,draw=none] (p) at (0.15,0) {}; \draw[line width=0.5pt,color=FDRU] (p.corner 1) -- (p.corner 2) -- (p.corner 3) -- (p.corner 4) -- (p.corner 5) -- cycle;}}}

\newcommand{\furulong}{\raisebox{0pt}{\tikz[baseline=-0.5ex,inner sep=0pt,outer sep=0pt]{\draw[FURU] (-0.1,0) -- (0.4,0); \node[shape=regular polygon,regular polygon sides=5,minimum size=7pt,fill=FURU,draw=none] (p) at (0.15,0) {}; \draw[line width=0.5pt,color=FURU] (p.corner 1) -- (p.corner 2) -- (p.corner 3) -- (p.corner 4) -- (p.corner 5) -- cycle;}}}

\newcommand{\fdrdshort}{\raisebox{0pt}{ \tikz[baseline=-0.5ex,inner sep=0pt,outer sep=0pt]{\draw[FDRD,dashed] (-0.1,0) -- (0.4,0); \node[shape=regular polygon,regular polygon sides=5,minimum size=7pt,fill=none,draw=none] (p) at (0.15,0) {}; \draw[line width=0.5pt,color=FDRD] (p.corner 1) -- (p.corner 2) -- (p.corner 3) -- (p.corner 4) -- (p.corner 5) -- cycle;}}}

\newcommand{\fdrdlong}{\raisebox{0pt}{\tikz[baseline=-0.5ex,inner sep=0pt,outer sep=0pt]{\draw[FDRD] (-0.1,0) -- (0.4,0); \node[shape=regular polygon,regular polygon sides=5,minimum size=7pt,fill=FDRD,draw=none] (p) at (0.15,0) {}; \draw[line width=0.5pt,color=FDRD] (p.corner 1) -- (p.corner 2) -- (p.corner 3) -- (p.corner 4) -- (p.corner 5) -- cycle;}}}

\newcommand{\furdlong}{\raisebox{0pt}{\tikz[baseline=-0.5ex,inner sep=0pt,outer sep=0pt]{\draw[FURD] (-0.1,0) -- (0.4,0); \node[shape=regular polygon,regular polygon sides=5,minimum size=7pt,fill=FURD,draw=none] (p) at (0.15,0) {}; \draw[line width=0.5pt,color=FURD] (p.corner 1) -- (p.corner 2) -- (p.corner 3) -- (p.corner 4) -- (p.corner 5) -- cycle;}}}

	\maketitle
	
	\begin{abstract}
		\footnotesize
The wakes of a \lc aircraft’s vertical lift propellers interact with the wing during the transition from hover to forward flight. This study investigates how variations in axial and longitudinal separations between the vertical lift propellers and the wing affect aerodynamic performance during this critical phase. Axial variations were introduced by orienting the motors either upright or inverted at two longitudinal positions relative to the wing. Wind tunnel tests were conducted on the resulting \lc wing configurations over a Reynolds number range of 52,000 to 377,000, using tufts, pressure taps, and load cells to capture aerodynamic behavior. Results show that the early transition phase (front propeller advance ratios between 0 and 0.25) is particularly sensitive to configuration changes. Rear motor orientation had a pronounced effect: when oriented upright at the shortest longitudinal separation, the active lift propellers increased the lift-drag ratio of the design by approximately 4 units, whereas in the rear motor inverted case the design suffered a 4-point reduction in the lift-drag ratio. Decreasing the longitudinal separation amplified both the beneficial lift contribution of the rear propeller and the generally adverse effects of the front propeller. However, placing the front propeller closer to the wing in the upright orientation enhanced performance at advance ratios between 0.11 and 0.25, due to favorable interactions between the wing and the front propeller's trailing vortices.

\end{abstract}	

\nomenclature[01]{AR}{Aspect ratio [-]}
\nomenclature[02]{b}{Wingspan, [\si{\meter}]}
\nomenclature[03]{$C_{\mathrm{D}}$}{Drag coefficient [-]}
\nomenclature[04]{$C_{\mathrm{L}}$}{Lift coefficient [-]}
\nomenclature[05]{$C_{\mathrm{p}}$}{Pressure coefficient [-]}
\nomenclature[06]{c}{Wing chord length, [\si{\meter}]}
\nomenclature[07]{$D_{P}$}{Propeller diameter, [\si{\meter}]}
\nomenclature[09]{n}{Propeller rotation rate [revolutions per minute]}
\nomenclature[10]{N}{Number of propellers [-]}
\nomenclature[11]{$R_{P}$}{Propeller radius [\si{\meter}]}
\nomenclature[12]{S}{Wing area, [\si{\meter\squared}]}
\nomenclature[13]{$S_{P}$}{Single propeller area, [\si{\meter\squared}]}
\nomenclature[14]{t}{time, [\si{\second}]}
\nomenclature[15]{$v_{h}$}{Ideal induced hover velocity, [\si{\meter\per\second}]}
\nomenclature[15]{$v_{i}$}{Induced propeller velocity, [\si{\meter\per\second}]}
\nomenclature[16]{$V_{\infty}$}{Freestream velocity, [\si{\meter\per\second}]}
\nomenclature[17]{$x_{B}$}{Boom length, [\si{\meter}]}
\nomenclature[18]{$\rho$}{Air density [\si{\kilogram\per\meter\cubed}]}
\nomenclature[19]{$\mu$}{Propeller advance ratio [-]}

\printnomenclature

\section{Introduction} \label{section:Introduction}
In alignment with the wider goals of improving the sustainability of aviation, electric Vertical Take-Off and Landing (eVTOL) developers have been leveraging advances in battery and electric power-train technology to target socio-economic issues such as traffic congestion and the efficient transportation of goods \cite{LONG2023102436}. Of the emerging configurations, the \lc design has gained popularity for its relative simplicity and the potential to decouple vertical and forward flight propulsion devices to allow each system to be tailored to its specific regime. Although the designers of these new vehicles have a rich history of conventional aircraft and rotorcraft design to guide decision-making, there are a number of new challenges that require further investigation - of these key research areas, understanding interactions of the vertical lifting propellers with each other and with the wing during the transition has been identified as important for improving aerodynamic performance \cite{johnson2018observations}.

In addition to the potential aerodynamic impacts, the axial and longitudinal positioning of the vertical lift system is expected to influence the noise generated by the design as blade vortices interact with each other \cite{silva2021practical} and other parts of the airframe \cite{rizzi2020urban}. A logical assumption based upon the ideas presented above is that minimizing the interaction of the propeller wakes with downstream components will be advantageous for the acoustic performance of the design. Other competing objectives in the design of VTOL aircraft include practical considerations such as the footprint of the aircraft and payload access - where decisions such as positioning the front VTOL motors below the wing would be expected to offer aerodynamic benefits but may also hinder passenger ingress and egress \cite{silva2018vtol}. For eVTOL aircraft pursuing certification, the proximity of lifting propellers may also be constrained by crash safety requirements \cite{EASA_crash_req}.

The following sections will describe the transition process in more detail for the \lc configuration and explore some of the existing literature for propeller-wing interactions.

\subsection{Transition process for lift + cruise configurations}
\begin{figure}[ht]
        \centering
            \begin{subfigure}[t]{0.495\textwidth}
            \centering
            \includegraphics[width=\linewidth]{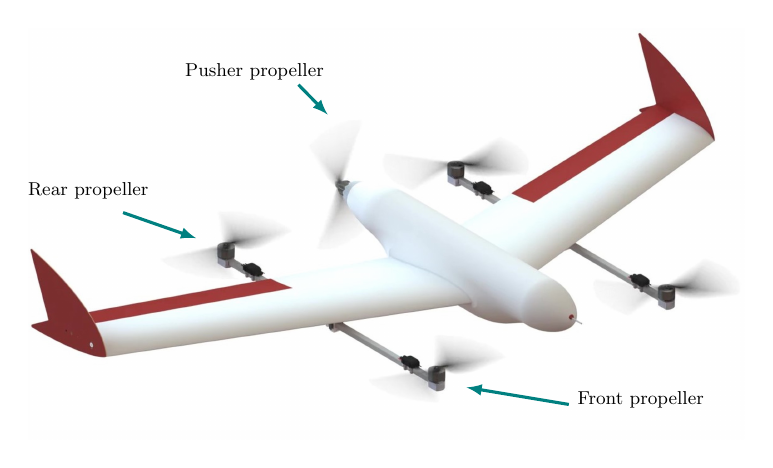}
            \caption{The Valerie \lc design.}
            \label{fig:V3_transition}
        \end{subfigure}
        \hfill
        \begin{subfigure}[t]{0.495\textwidth}
        \centering
        \includegraphics[width=\linewidth]{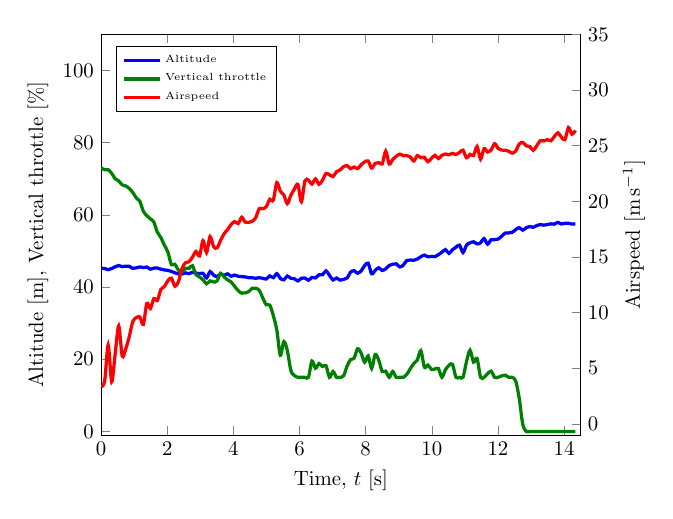}
            \caption[Airspeed, altitude and vertical throttle during the transition phase to forward flight from flight logs of the Valerie aircraft.]{Airspeed, altitude and vertical throttle during the transition phase to forward flight from flight logs of the Valerie aircraft. Note: the flight controller maintains a low throttle setting on the vertical lifting motors until the transition has been assured.}
            \label{fig:val_transition_log}
        \end{subfigure}\\
        \caption{The Valerie \lc design presented with a transition flight log.}
        \label{fig:V3_transition_and_log}
\end{figure}

The University of Southampton's custom-made Uncrewed Aerial Vehicle (UAV) named Valerie — a representative \lc design (Figure~\ref{fig:V3_transition})— uses thrust from vertical lifting propellers to supplement a deficit in wing lift when the aircraft is operating below its stall velocity \cite{boyd2021cascade}. The proportion of thrust required from the vertical lifting propellers is inversely proportional to the aircraft’s forward velocity and is modulated by a flight controller to maintain a constant altitude transition to or from forward flight (Figure~\ref{fig:val_transition_log}). For \lc designs that utilize segregated vertical and forward propulsion systems, the forward thrust component required to accelerate the aircraft from a hover into wing-borne forward flight is provided by a dedicated forward propulsion motor (Figure~\ref{fig:V3_transition}).

\subsection{Prior work} \label{section:prior_work_and_motivation}
Propeller-wing interactions in conventional fixed-wing aircraft have been extensively studied, primarily focusing on arrangements with the propeller mounted in front of a wing (tractor configuration) and those mounted aft of the wing (pusher configuration). In the tractor configuration, the interaction of the propeller slipstream with the wing has been shown to increase the wing's lift generation capability by increasing the dynamic pressure over the wing. The inclination angle of a tractor propeller was also investigated, showing nose down inclinations to be beneficial to lifting performance by increasing the angle of attack, the inverse was also shown for nose up inclinations \cite{Delft_prop}. For configurations with the aft mounted pusher configuration, the entrainment of flow into the propeller's inlet has been shown to help prevent the formation of adverse pressure gradients by accelerating the flow over the suction surface of the wing \cite{catalano2004effects}.

Research on interactions between lifting surfaces and rotors has historically been dominated by studies of rotorcraft, examining phenomena such as rotor wake effects on helicopter empennages \cite{rotor_tail} and on the small lifting surfaces of compound helicopters \cite{rotor_compound}. In these studies, the downwash generated by the rotor reduced the effectiveness of the lifting surface. The aerodynamic performance of the surface was also dependent on the rotor advance ratio (the ratio of the freestream velocity to the the angular rate of the rotor) and the placement of the surface relative to the rotor as these factors determine the amount of rotor-wing interaction experienced as the wake is swept downstream. Research has also been conducted into prop-rotor wake interactions with the wings of vectoring thrust designs such as the wing download testing of the V-22 \cite{macveigh1985v} and more recently in the transition process of newer thrust-vectoring aircraft designs \cite{DROANDI2016116}. Although pertinent, it is difficult to make predictions relating to the transition performance of \lc designs using rotorcraft or large diameter prop-rotor studies because of the relative size differences between the rotor and the lifting surface, and the difference in orientation of the prop-rotor and its wake relative to the wing during the transition. 

Previous research on tandem lift fan models offers more directly applicable insights into the performance of \lc configurations \cite{hickey1966aerodynamic}. This work evaluated the aerodynamic impact of lift fans mounted fore and aft of an aircraft wing at increasing airspeeds and constant fan pressure ratio. The results suggested that the wake exiting the front fan causes a reduction in the angle of attack and subsequently the lift produced by the wing in the region behind the front fan position. Three front fan positions of varying vertical height were tested, with the fan mounted significantly below the wing producing the least lift reduction. In contrast to the front fan, the proximity of the rear fan to the wing is suggested to increase the wing's lift production in a manner similar to a jet flap - a mixing of the jet efflux with the oncoming flow to generate circulation around the wing with a similar effect to deploying a physical trailing edge device \cite{williams1961aerodynamics}. With both front and rear vertical lift fans operating, a net positive impact on wing lift production may be achievable with increasing ratios of the jet to freestream velocity ratio \cite{hickey1966aerodynamic}. 

Exploring the effects of jet exhaust locations on Vertical/Short Take-Off and Landing (V/STOL) aircraft performance, a wide range of jet placements relative to a wing surface were investigated at constant thrust and increasing airspeed \cite{carter1969effects}. Jet positions in front of the wing were found to always be detrimental and most severe above the plane of the wing chord. Aft jet positions were found to produce favorable increases in wing lift, with minimal sensitivity to vertical location. However, thrust output was not varied during the transition in either the tandem lift-fan or jet effect studies to reflect a realistic transition profile \cite{hickey1966aerodynamic, carter1969effects}. Without this information it is unclear how the aerodynamic performance of the wing would be affected as the lift production shifts between the vertical lift propellers and the wing as the airspeed increases. This gap in current understanding highlights the need for a systematic investigation into the transition process, which this research aims to address.

\subsubsection{Rigid propeller wakes}

\begin{figure}[h!]
    \centering
    \begin{tikzpicture}
    \node[anchor=south west, inner sep=0] (image) at (0,0) {\includegraphics[width=0.4\textwidth]{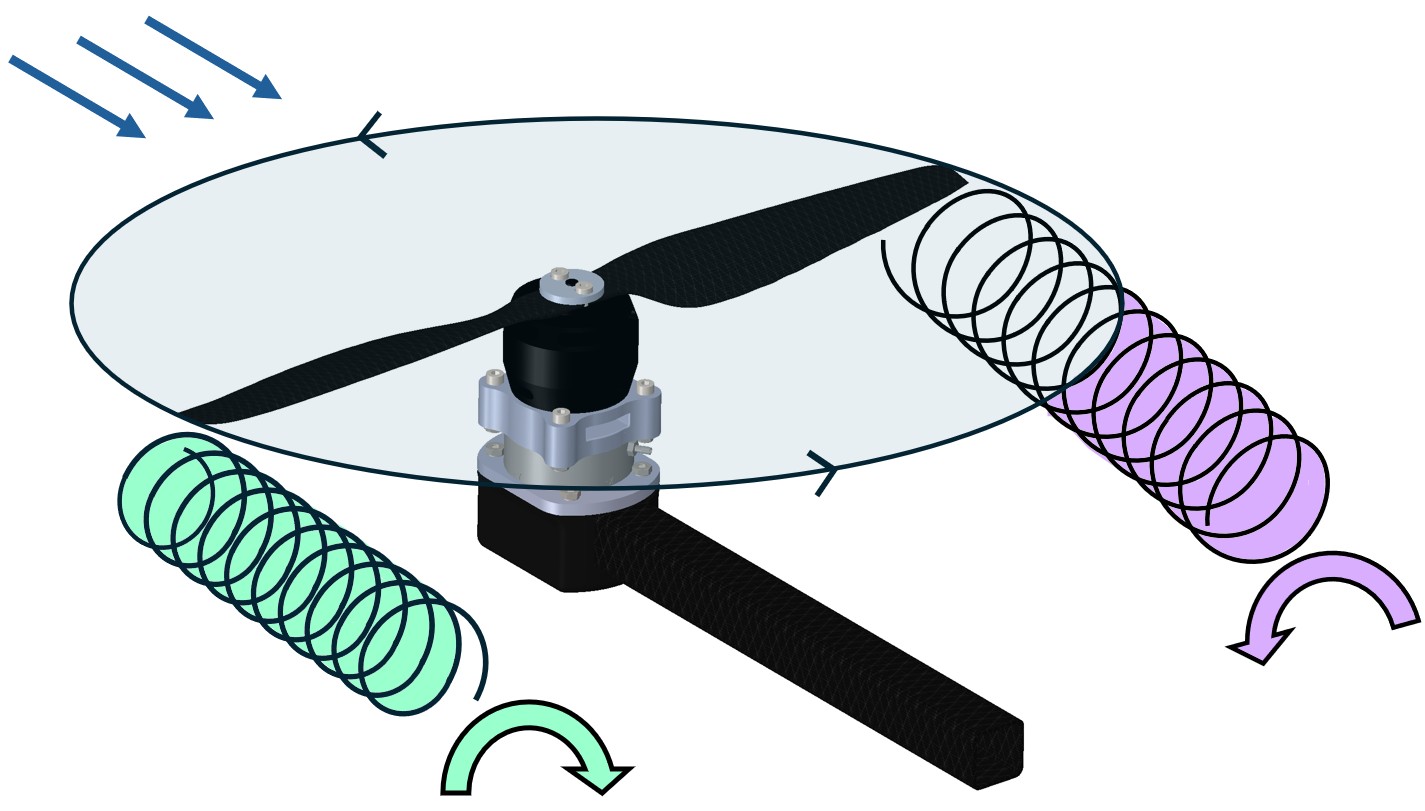}};
    \begin{scope}[x={(image.south east)}, y={(image.north west)}]
        \node[fill=white, opacity=0.7, text opacity=1] at (0.05,1.05) {$V_{\infty}$};
    \end{scope}
    \end{tikzpicture} 
    \begin{tikzpicture}
    \node[anchor=south west, inner sep=0] (image) at (0,0) {\includegraphics[width=0.54\textwidth]{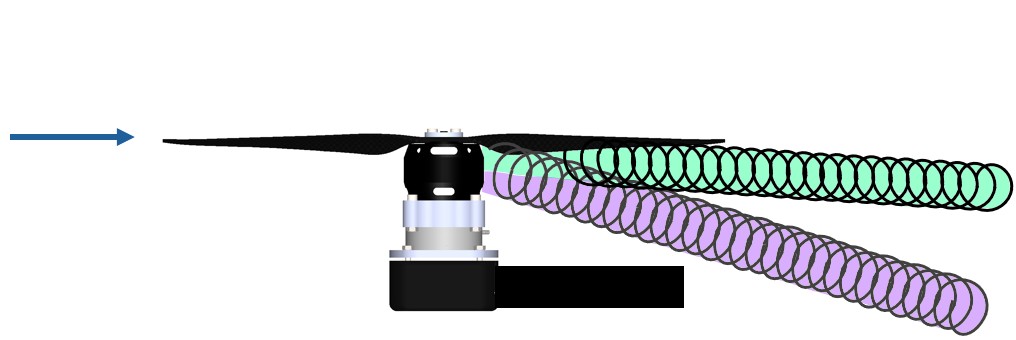}};
    \begin{scope}[x={(image.south east)}, y={(image.north west)}]
        \node[fill=white, opacity=0.7, text opacity=1] at (0.07,0.7) {$V_{\infty}$};
    \end{scope}
    \end{tikzpicture} 
    \caption{A diagrammatic depiction of trailing super-vortices from a rigid propeller in edgewise flight.}
    \label{fig:tip_vortices_rigid}
\end{figure}

A conventional rotor in edgewise flight at moderate advance ratios ($\mu \approx 0.05$–$0.20$) typically generates two trailing super-vortices as the rotor wake rolls up and is deflected aft by the freestream \cite{Johnson_2013_wakes}. The rigid vertical lift propellers commonly used on UAVs, modify this behavior due to their lack of conventional degrees of freedom, such as cyclic pitch, blade flapping, and lead-lag motion. At low advance ratios, loading is relatively symmetric, resulting in practically symmetrical trailing vortices. Without a rotor's degrees of freedom, lift in edgewise flight becomes increasingly asymmetric with increasing advance ratio - concentrating on the advancing side of the propeller disc, causing the vortex from the advancing side to experience stronger downwash and to convect downwards at a steeper angle. The trailing pair of super-vortices generate downwash between them and up wash outside the wake, producing a lateral shift of the wake in the direction of the advancing blade. These effects, as described in the work presented in \cite{rotor_tip_vortices}, are illustrated in Figure~\ref{fig:tip_vortices_rigid} in relation to the current experimental setup.

As these super-vortices convect downstream, they behave like a pair of counter-rotating streamwise vortices. The amount of downwards deflection that each vortex experiences is determined by the direction of the propeller’s rotation, advance ratio, and thrust output. Stream-wise orientated vortices have been widely studied, particularly in investigations to reduce induced drag in formation flight \cite{formation_flying, detailed_close_formation}, and for their potential to increase cyclic loading on downstream structures like tail fins and following wings through vortex breakdown \cite{fin_buffet, follower_wing}.

Analytical models based on lifting line theory \cite{hancock1971aerodynamic} and potential flow theory \cite{bodstein1996three} have been used to predict how streamwise vortices affect nearby lifting surfaces.
Experiments with a close-coupled wing-canard setup showed good agreement with theory when the vortex was well-separated from the surface, but reduced accuracy when stronger, closer interactions caused vortex breakdown. 
The impingement of a vortex on a wing can affect aerodynamic performance, increasing lift-to-drag ratio when impinging near the wing tip \cite{garmann2015interactions}. As vortex impingement moves inboard, downwash can reduce the effective angle of attack and create separation bubbles.

Increases in lift under the trailing vortex core, particularly over the aft 75\% of the chord, have been shown with a vertical offset between a 2D wing and a streamwise vortex \cite{patel1974some}. Suction intensifies on the up wash side of the trailing vortex, while pressure rises on the downwash side.

Most prior work focuses on a single streamwise vortex aligned with the freestream, typically generated by the tip of a canard or leading aircraft wing. In contrast, the current study deals with two contra-rotating super-vortices shed by the front vertical lift propeller as the aircraft transitions to forward flight. These vortices will not necessarily be aligned with the freestream — especially at low advance ratios.

The findings of Gordnier and Visbal \cite{gordnier1999numerical}, who studied delta-wing generated vortices interacting with a downstream plate, are particularly relevant to the understanding of the current work. In their study, they found that the trailing pair of vortices changed the span-wise lift distribution on the wing, with downwash generated between the vortices and up-wash at their exteriors. They also showed that the low-pressure region of the vortex core was detectable on the flat plate as it passed over its surface.

The insights from the literature described above will be used to support the interpretation of how the interactions between the vertical lift propeller wakes and the wing influence aerodynamic performance during the transition.

\FloatBarrier

\subsubsection{Study aims and motivation}

Following the surge in interest in the eVTOL domain, there are a multitude of research avenues to explore. In addition to conducting their own research into vehicle configurations like the \lc design \cite{liu2023high, liu2024high}, partners in NASA's Revolutionary Vertical Lift Technology (RVLT) project have outlined a set of impactful research topics to advance the design and operation of new VTOL aircraft \cite{johnson2018observations}. Currently there is a scarcity in the literature studying areas of interactional aerodynamics, particularly in experiments investigating physical models to generate validation data for computational tools \cite{silva2018vtol}.

The current study aims to address the lack of experimental data generated by systematic investigations of the interactions between active vertical lift propellers and the wing of a generic \lc design during transition. The vertical lift propeller thrust will be modulated to supplement the wing lift deficit and to maintain zero net pitching moment about the wing's quarter-chord - simulating a constant altitude, constant attitude transition from hover to forward flight. This approach contrasts with previous studies that maintained constant thrust at increasing airspeeds \cite{hickey1966aerodynamic, carter1969effects}. A combination of force, torque, pressure, and tuft measurement techniques will be employed to:

\begin{enumerate}
    \item Quantify the changes in aerodynamic performance induced by the action of a pair of vertical lift propellers in proximity to a wing during a transition; further exploring the impact of varying the axial and longitudinal separation of the vertical lift propellers to the wing.
    \item Identify the underlying physical mechanisms through which the vertical lift propellers influence the aerodynamic behavior of the configuration.
    \item Determine the relative impact of fore and aft motor placements on aerodynamic performance during the transition.

\end{enumerate}

\section{Methodology}
The following section describes the wind tunnel testing of a \lc wing and motor model designed around a use case on a UAV. The model had electric motors which were individually controllable to facilitate replication of a realistic transition maneuver. The model was also instrumented to record force and torque data using load cells, with a complimentary set of suction surface pressure and tuft measurements. All tests were carried out in the University of Southampton's  wind tunnel with a cross-section of 2.1 x 1.5 \si{\meter}. The tunnel is operated with a closed-return through a working section 4.1 \si{\meter} long and is capable of flow speeds up to 45 \si{\meter\per\second}.

\subsection{Wind tunnel model}

\begin{figure}[h!]
    \centering
    \includegraphics[width=0.95\textwidth]{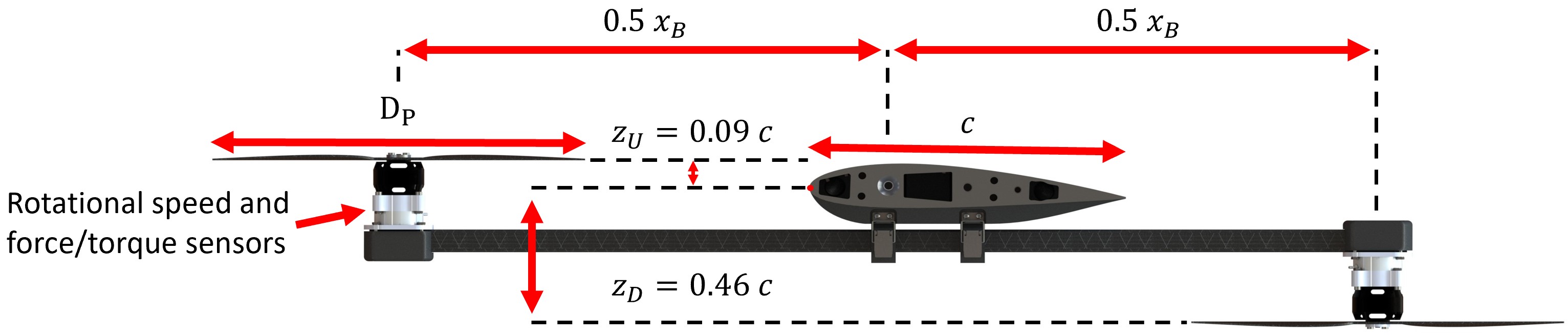}
    \caption[Propeller separation distances with motors upright and inverted.]{Possible arrangements of upright ($z_{U}$) and downwards ($z_{D}$) propeller separations from the wing leading edge as a proportion of the wing's chord length $c$. Also shown is the vertical lift propeller diameter $D_{P}$, and the separation between the propeller rotation axis and the wing's quarter-chord location (0.5~$x_{B}$), being symmetric fore and aft.}
    \label{fig:motor_sep_setting}
\end{figure}

A generic \lc wind tunnel model was designed based upon previous work scaling and simplifying the Valerie UAV wing and motor assembly \cite{WESTCOTT2023108552}. The baseline geometry of the wing and propeller remained unaltered from the previous study (Table~\ref{tab:pressure_model_characteristics}), but the motors were now powered and instrumented with their own ATI Mini 40 force and torque (F/T) sensors and the shaft of each motor was fitted with a magnetic encoder using a TLE4946-2K hall effect sensor to estimate the rotational speed of the motors. The wing was set at a constant setting angle of two degrees relative to the horizontal boom (Figure~\ref{fig:motor_sep_setting}).

\begin{table}[ht]
	\caption{Characteristics of the wind tunnel model.}
	\label{tab:pressure_model_characteristics}
	\centering
    \begin{tabular}{l l}
        \textbf{Parameter} & \textbf{Value} \\
		Wing span ($b$): & 0.9 \si{\meter} \\
        Aerofoil profile: & NACA 2415 \\
		Aspect ratio ($AR$): & 6.94\\
		Propeller diameter ($D_{P}$): & 0.31 \si{\meter} (12")\\
		Number of propellers ($N$): & 2\\
		Wing chord ($c$): & 0.26 \si{\meter} \\
		Propeller model: &  T-motor P12x4\\
		Motor model: &  T-motor U3 700 kV\\
        Electronic speed controller: & T-motor AT 55 A \\
    \end{tabular}
\end{table}

\begin{figure}[t!]
    \centering
    \begin{subfigure}[b]{0.49\textwidth}
        \centering
        \includegraphics[width=1\textwidth]{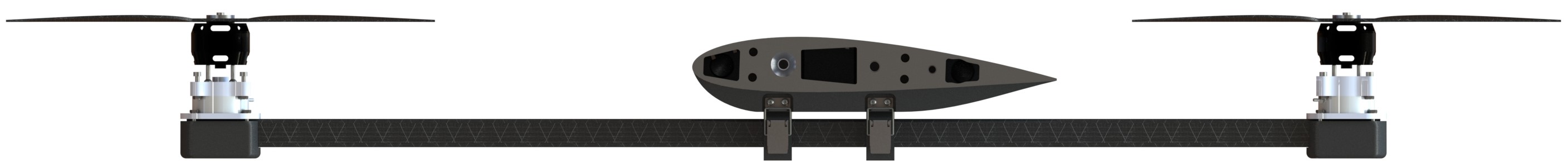}
        \caption{Front Up Rear Up (\textcolor{FURU}{FURU}) motor configuration.}
        \label{fig:both_upright_motor}   
    \end{subfigure}
    \begin{subfigure}[b]{0.49\textwidth}  
        \centering
        \includegraphics[width=1\textwidth]{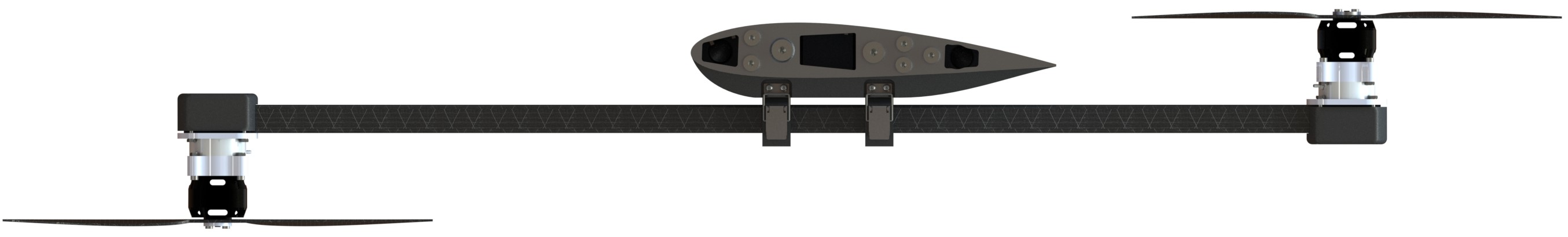}
        \caption{Front Down Rear Up (\textcolor{FDRU}{FDRU}) motor configuration.}
        \label{fig:front_inverted_motor}  
    \end{subfigure}

    \begin{subfigure}[b]{0.49\textwidth}   
        \centering
        \includegraphics[width=1\textwidth]{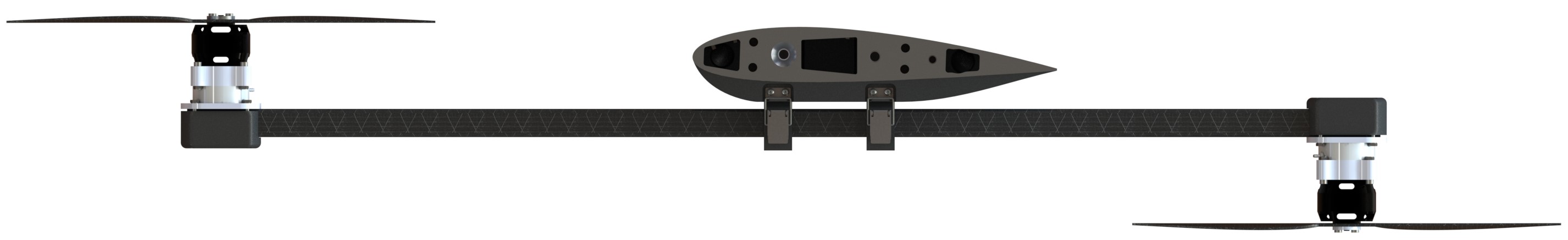}
        \caption{Front Up Rear Down (\textcolor{FURD}{FURD}) motor configuration.}
        \label{fig:rear_inverted_motor}
    \end{subfigure}
    \begin{subfigure}[b]{0.49\textwidth}   
        \centering
        \includegraphics[width=1\textwidth]{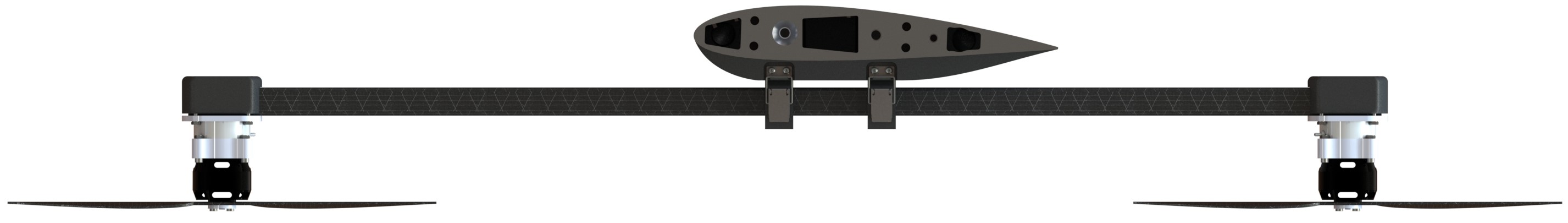}
        \caption{Front Down Rear Down (\textcolor{FDRD}{FDRD}) motor configuration.}
        \label{fig:both_inverted_motor}  
    \end{subfigure}
    \caption{Alternative vertical placements for the motors and propellers.}
    \label{fig:inverted_motor_choices}
\end{figure}

The model was designed to be modular, allowing the motor position to be easily changed between upright and inverted (Figure~\ref{fig:inverted_motor_choices}). The four resulting combinations were defined as: Front Up Rear Up (\textcolor{FURU}{FURU}), Front Up Rear Down (\textcolor{FURD}{FURD}), Front Down Rear Up (\textcolor{FDRU}{FDRU}), and Front Down Rear Down (\textcolor{FDRD}{FDRD}). The upright and inverted motor positions resulted in a separation between the plane of the propeller disc and the leading edge of the wing of $z_U=0.09~c$ and $z_D=0.46~c$ respectively,  where $c$ is the wing's chord length (Figure~\ref{fig:motor_sep_setting}).

\begin{figure}[ht]
        \centering
        \begin{subfigure}{\linewidth}
                \centering
                \includegraphics[width=0.49\textwidth]{Figures/newFURU.JPG}
                \caption{B1 boom length ($x_{B} = 0.8 \ \si{\meter}$)}
                \label{fig:original_boom_length}
        \end{subfigure}
        \begin{subfigure}{\linewidth}
                \centering
                \includegraphics[width=0.40\textwidth]{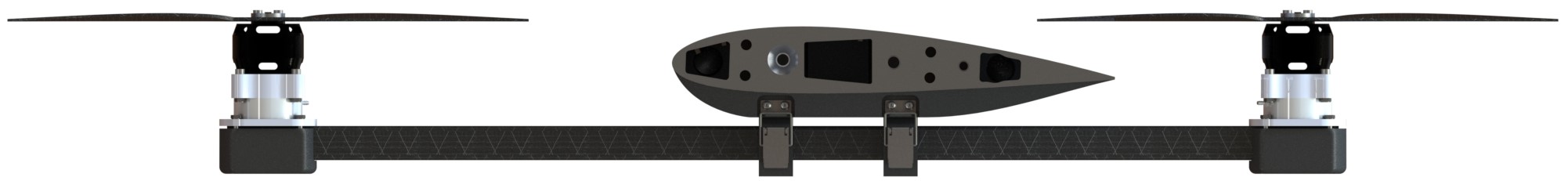}
                \caption{B2 boom length (($x_{B} = 0.6 \ \si{\meter}$)}
                \label{fig:short_boom_length}
        \end{subfigure}
        \caption{Longitudinal motor and propeller positions investigated in this study.}
        \label{fig:boom_choices}
\end{figure}

The longitudinal separation between the rotational axes of each propeller configuration ($x_B$) could also be varied between two positions by changing the length of the boom (Figure~\ref{fig:boom_choices}). The two boom lengths of $x_{B} = 0.8$ \si{\meter} (B1) and $x_{B} = 0.6$ \si{\meter} (B2) provide symmetric longitudinal motor placements relative to the wing's quarter-chord location, being the assumed centre of gravity location.

In the shortest case (B2), it is worth noting that a propeller separation of 0.60 \si{\meter} is chosen to produce an overlap of $30\%$ of the rear blade's radius with the trailing edge of the wing. The purpose of this is to explore how an overlapping rear propeller impacts wing performance. This aspect of the design was not covered in previous work \cite{hickey1966aerodynamic}, but does appear to be present in some commercial eVTOL vehicles,

\begin{table}[h!]
	\caption[Propeller center separations relative to the wing's leading edge as a proportion of wing chord length.]{Propeller center separations relative to the wing's leading edge as a proportion of wing chord length $c$. Here FU is Front Up, FD is Front Down, RU is Rear Up and RD is Rear Down.}
	\label{tab:orientations_xc_zc}
	\centering
        \resizebox{0.8\textwidth}{!}{
	\begin{tabular}{c c c c c c c c c}
	   & \textbf{FU B1} & \textbf{FD B1} & \textbf{FU B2} & \textbf{FD B2} & \textbf{RU B1} & \textbf{RD B1} & \textbf{RU B2} & \textbf{RD B2} \\
        $x/c$ & -1.29 & -1.29  & -0.91 & -0.91 & 1.79 & 1.79 & 1.41 & 1.41 \\
        $z/c$ & 0.09 & -0.46 & 0.09 & -0.46 & 0.09 & -0.46 & 0.09 & -0.46  \\
	\end{tabular}}
\end{table}

\begin{figure}[ht]
        \centering
            \begin{subfigure}[t]{0.45\textwidth}
            \begin{tikzpicture}
            \node[anchor=south west,inner sep=0] (image) at (0,0) {\includegraphics[width=\textwidth]{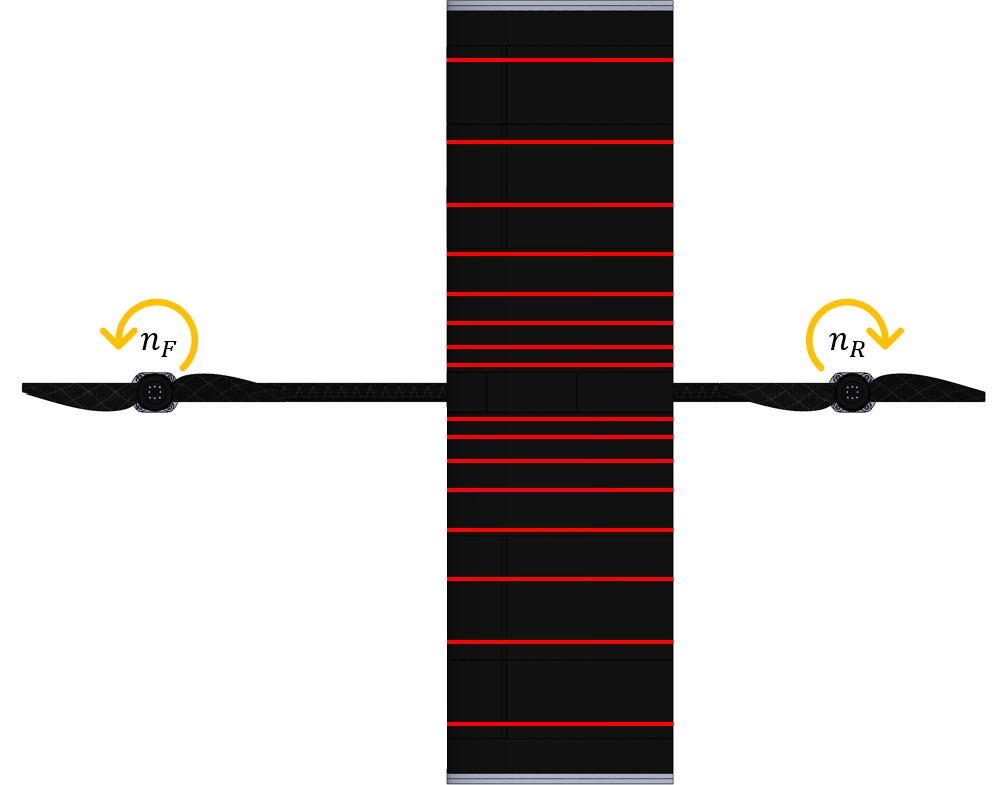}};
            \begin{scope}[x={(image.south east)},y={(image.north west)}]
            \node [anchor=south west,coordinate] (start) at (0.22,0.6) {};
            \draw [-latex, line width=1.5pt, black] (start) -- ++(0.0,0.2);
            \draw [-latex, line width=1.5pt, black] (start) -- ++(0.15,0.0);
            \node [anchor=west] (yco) at (0.195,0.83) {\small $y$};
            \node [anchor=west] (xco) at (0.36,0.6) {\small $x$};
            
            \draw[Oranges-G, dashed, line width=1.5pt] (0.22,0.6) -- (0.45,0.5);
            \filldraw[Oranges-G] (0.22,0.6) circle (2pt);
            \filldraw[Oranges-G] (0.45,0.5) circle (2pt);
            
            \end{scope}
            \end{tikzpicture}%
            \caption[Span-wise pressure tap distribution.]{Planes representing the span-wise distribution of pressure taps and direction of rotation for the front and rear propeller at rotation rates $n_{F}$ and $n_{R}$ respectively.}
            \label{fig:span_wise_taps}
        \end{subfigure}
        \hspace{1cm}
        \begin{subfigure}[t]{0.35\textwidth}
        \centering
            \begin{tikzpicture}
            \node[anchor=south west,inner sep=0] (image) at (0,0) {\includegraphics[width=\textwidth]{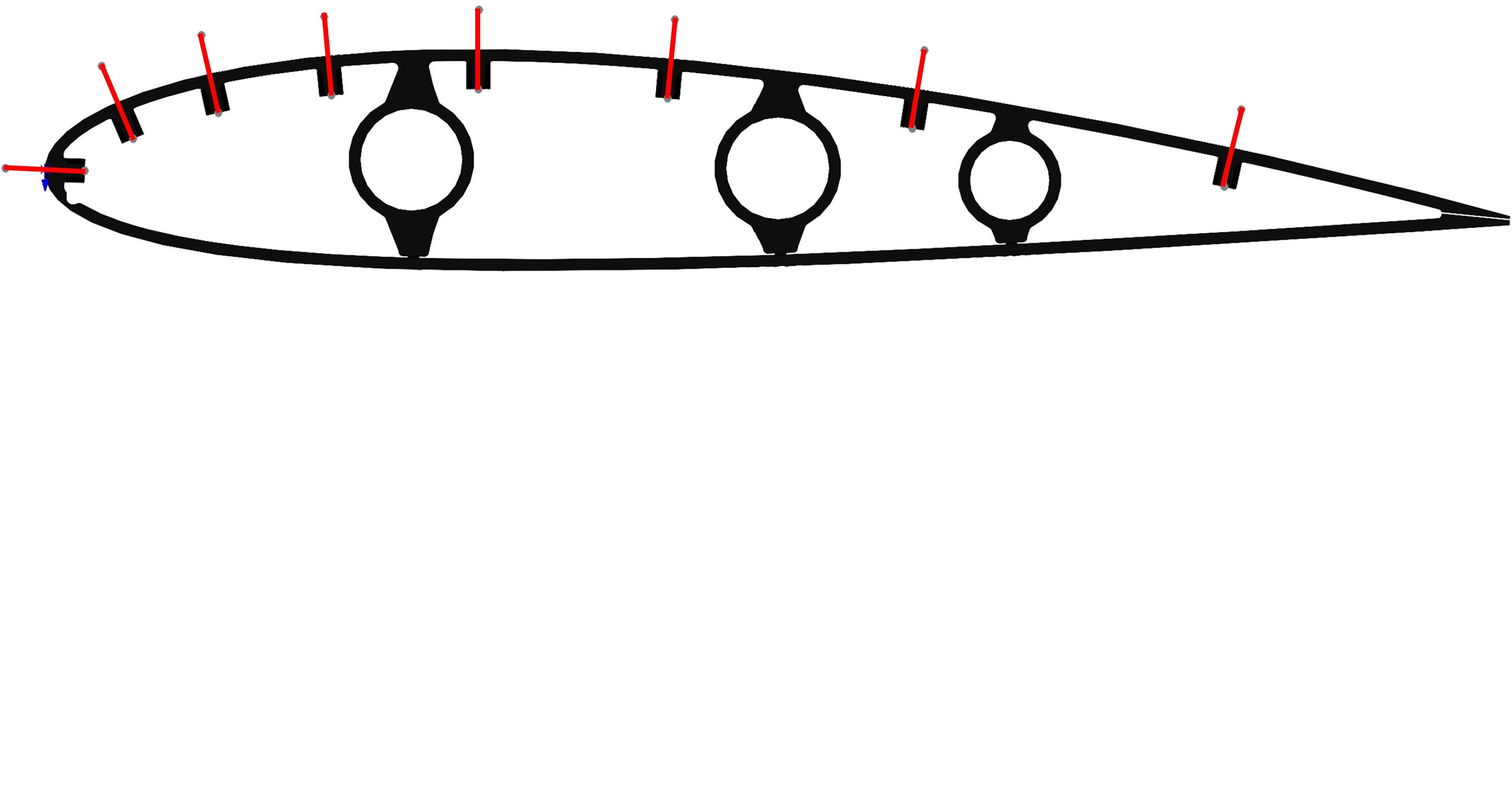}};
            \begin{scope}[x={(image.south east)},y={(image.north west)}]
            \node [anchor=south west,coordinate] (start) at (0.0,1.12) {};
            \draw [-latex, line width=1.5pt, black] (start) -- ++(0.0,0.38);
            \draw [-latex, line width=1.5pt, black] (start) -- ++(0.2,0.0);
            \node [anchor=west] (zco) at (-0.04,1.55) {\small $z$};
            \node [anchor=west] (xco) at (0.18,1.12) {\small $x$};

            \draw[Oranges-G, dashed, line width=1.5pt] (0.0,1.12) -- (0.04,0.78);
            \filldraw[Oranges-G] (0.0,1.12) circle (2pt);
            \filldraw[Oranges-G] (0.04,0.78) circle (2pt);
            
            \end{scope}
            \end{tikzpicture}%
            \caption{Chord-wise distribution of the pressure taps.}
            \label{fig:chord_wise_taps}
        \end{subfigure}
        \caption{Wind tunnel model with pressure tapping locations.}
        \label{fig:pressure_tap_loc_cad}
\end{figure}

Initial prototyping of a wing model showed it would be infeasible to install a high density of pressure taps in both the pressure and suction surfaces of the wing model due to the minimum radius of curvature requirements of the pressure tubing and the access required for installation and maintenance. Therefore, pressure taps were installed only on the suction surface of the wing as this is where the majority of the propeller wake interaction with the wing was expected to occur.
Most interaction between the front propeller's wake and the wing was expected in the region directly behind the front propeller, as the freestream velocity cants the wake rearwards. As a result, the pressure taps were arranged with a higher density in the center of the span that decreased exponentially towards the wing tips - noting that it was not possible to place taps directly behind the propeller's center due to structure in the wing (Figure \ref{fig:span_wise_taps}). The chord-wise location of the taps were exponentially distributed between the leading edge of the wing profile and a maximum chord-wise location of 250 mm (0.81 $c$). This rear limit was chosen to allow the depth of the aft most tap to be located within the profile's thickness. The chord-wise pressure tap locations were chosen to not coincide with the location of the spars that pass through the design and to provide an increased density of taps around the leading of the profile as this is where the largest pressure gradients would likely occur (Figure \ref{fig:chord_wise_taps}).
The wing surface was constructed from 3D printed Poly-lactic Acid (PLA) plastic, with holes for 128 pressure taps pre-located in the design. The surface of the model was lightly sanded after printing to remove layer lines and then filled with an automotive spray filler to improve the surface finish. The tubes used for the taps were machined steel of 0.46 mm internal diameter and 1.59 mm external diameter. The ports on the wing surface are divided into two symmetrical banks of 64 pressure ports about the mid-span location, one bank belonging to the outboard section (positive span coordinates) of the span and the other 64 to the inboard section (negative span coordinates). The Scanivalve ZOC 33 pressure scanner used to collect the pressure data only contains 64 measurement ports from which it can record data in one acquisition. To collect data from the entire wing, a quick disconnect block is fitted to each bank of ports and swapped between measurements. A summary of the pressure tap locations is given in Table~\ref{tab:Pressure_tap_dist}.

\begin{table}[h]
	\caption[Pressure tap distribution.]{Pressure tap distribution as proportions of the wing's chord and span lengths $c$ \& $b$, measured from the leading edge and wingspan center-line respectively.}
	\label{tab:Pressure_tap_dist}
	\centering
        \resizebox{0.65\textwidth}{!}{
	\begin{tabular}{l c c c c c c c c}
		\hline
        \hline
		Chord-wise ($x/c$) [\%] & 0 & 5 & 11 & 19 & 30 & 43 & 60 & 81 \\
        Span-wise $\pm$($y/b$) [\%] & 3 & 6 & 9 & 13 & 17 & 24 & 32 & 42 \\
		\hline
        \hline
	\end{tabular}}
\end{table}

\begin{figure}[ht]
        \centering
            \begin{subfigure}[t]{0.441\textwidth}
            \centering
            \includegraphics[width=\textwidth]{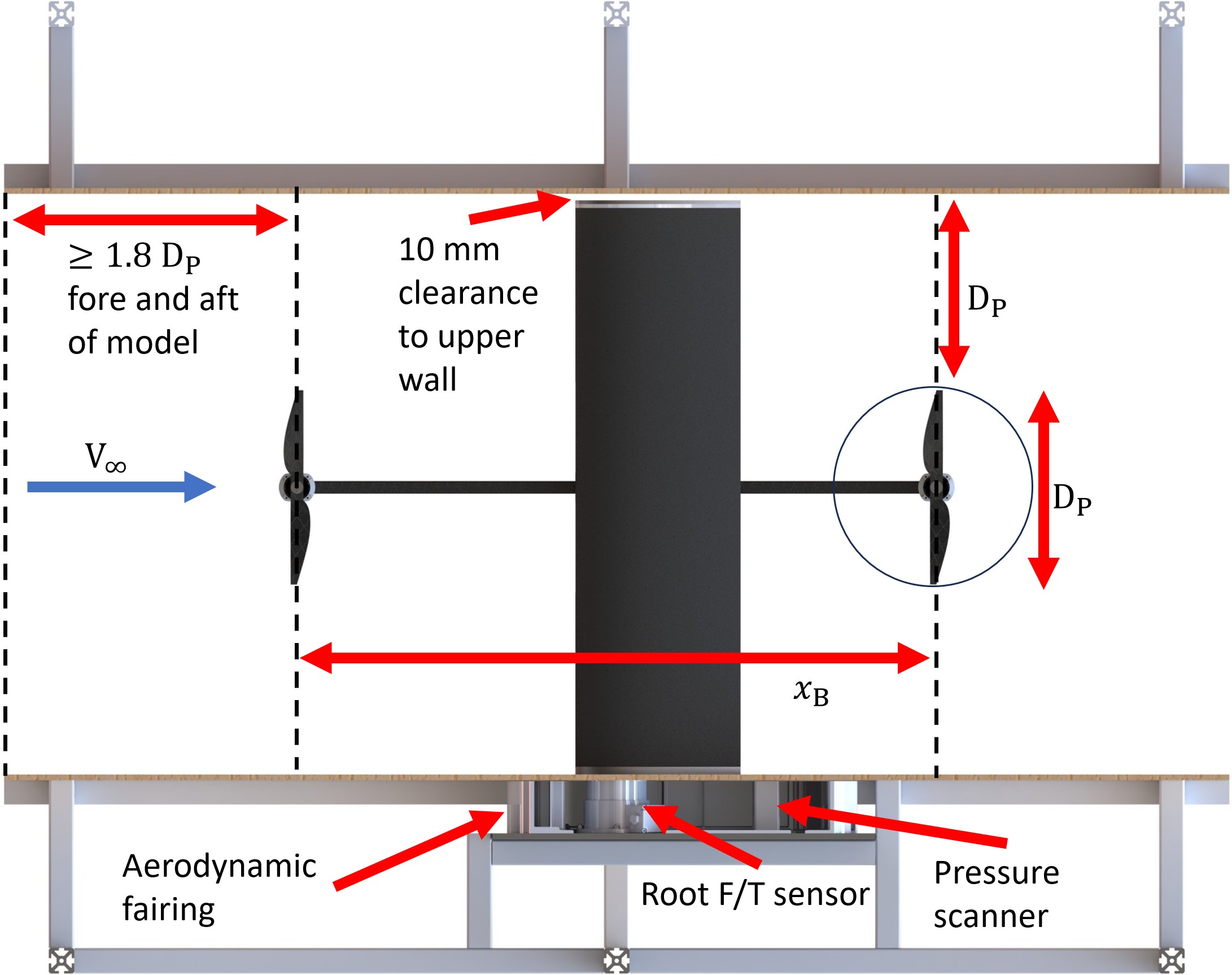}
            \caption[Side view of the model installed within the reduced working section of the tunnel.]{Side view of the model installed within the reduced working section of the tunnel with reference dimensions in relation to the propeller diameter $D_{P}$ and propeller separation distance $x_{B}$.}
            \label{fig:annotated_side_view_render}
        \end{subfigure}
        \hfill
        \begin{subfigure}[t]{0.49\textwidth}
            \centering
            \includegraphics[width=\textwidth]{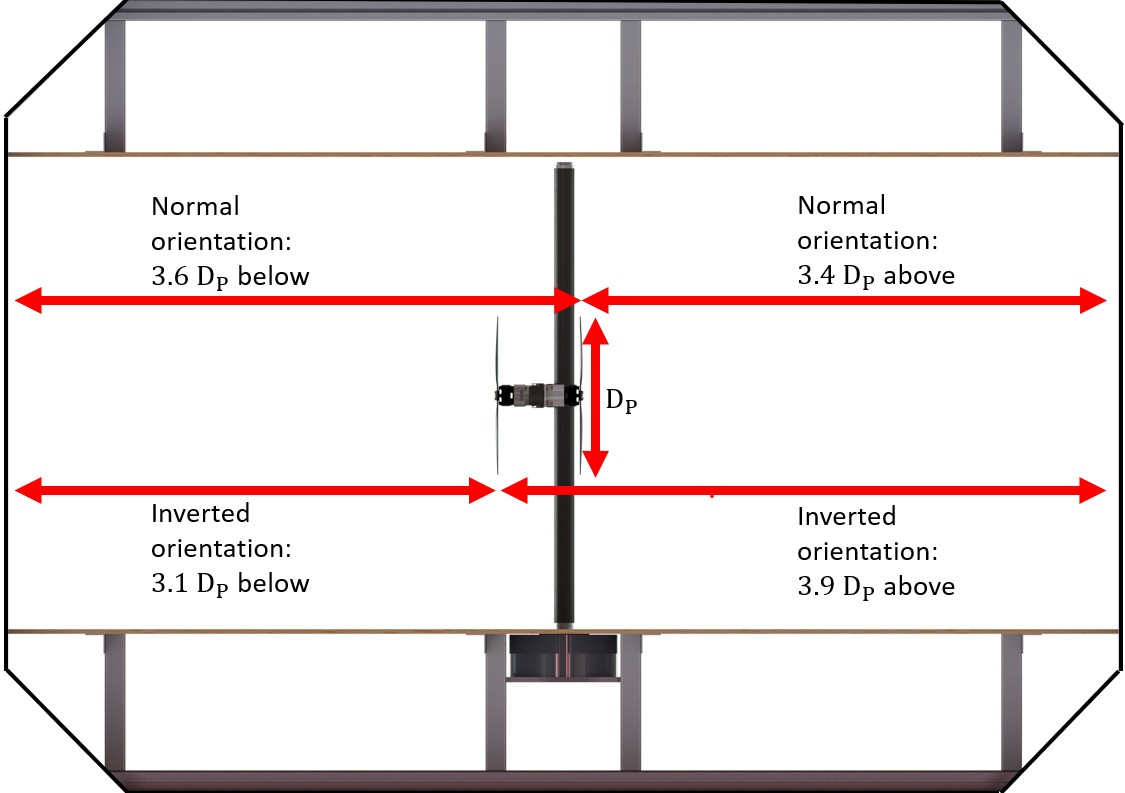}
            \caption[Front view of the model installed within the reduced working section of the tunnel]{Front view of the model installed within the reduced working section of the tunnel. The separations of the upright and inverted propeller discs are shown in relation to the propeller diameter $D_{P}$.}
            \label{fig:annotated_front_view_render}
        \end{subfigure}
        \caption{Wind tunnel assembly diagrams.}
        \label{fig:wind_tunnel_assembly_diagrams}
\end{figure}

The normal working section of the wind tunnel was reduced by adding a separate floor and ceiling panel that coincided with the root of the model whilst providing 10 mm (0.04 $c$) of clearance at the wing tip (Figure ~\ref{fig:annotated_side_view_render}). The panels were added to limit the effect of the wing's tip vortex whilst providing clearance to prevent the wing tip grounding against the ceiling panel. A small clearance cut was also made in the lower panel to allow the model to pass through without interference. In addition to the F/T sensors installed under each motor, a larger ATI Delta IP65 SI-165-15 F/T sensor was installed at the root of the model at a fixed angle of zero degrees, with the attached assembly adopting the same angle. Using the root sensor along with the sensors beneath each motor made it possible to infer the wing’s performance in the presence of the active vertical lift propellers.

The operation of propellers in close proximity to surfaces can result in an increase in the thrust output of the propeller in a phenomenon known as `ground effect'. The strength of the ground effect decreases with the increasing separation of the propeller from the ground plane, with experimentation and theory suggesting that the ground effect is mostly absent at separations of $1.5~D_{P}$ or greater \cite{knight1941analysis,betz1937ground,cheeseman1955effect,johnson_2013_vertical}. Additionally, the forward velocity encountered by a rotor in forward flight is found to reduce the ground effect even further, as the wake is swept downstream \cite{johnson_2013_forward}. Figure~\ref{fig:annotated_front_view_render} shows a front view of the tunnel model in relation to the side walls. Separations between the plane of the propeller disc in both the upright and inverted orientations were significantly ($3.1~D_{P}$, at minimum) above the $1.5~D_{P}$ threshold, and therefore the influence of any ground effect was suspected to be negligible. In addition to the spacing above and below the propellers, the floor panels also extended 1.8 propeller diameters in front of and behind the propeller axes of rotation with the longest boom length.

\begin{figure}[ht]
        \centering
            \begin{subfigure}[t]{0.4445\textwidth}
            \centering
            \includegraphics[width=\textwidth]{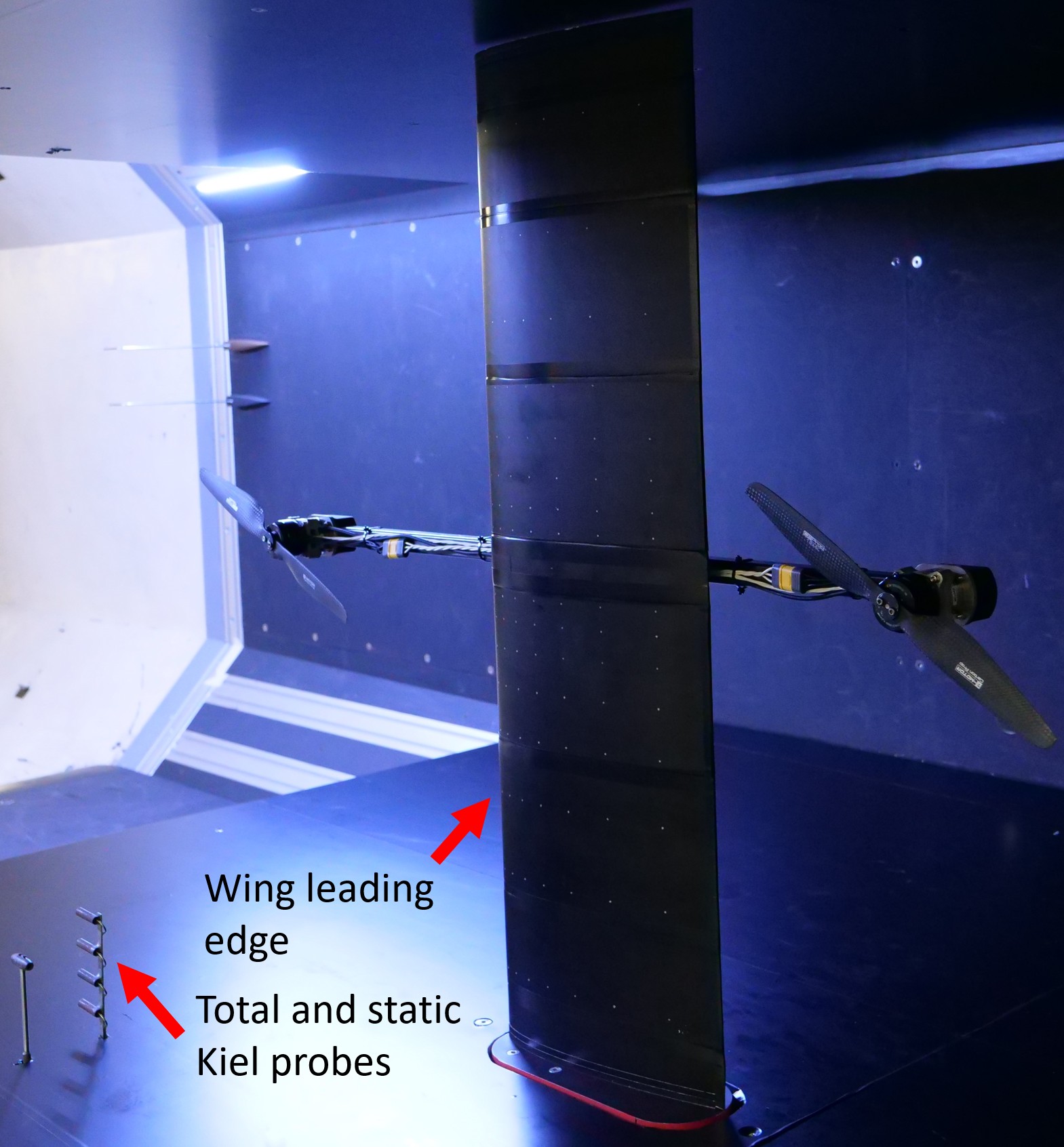}
            \caption[Surface pressure measurement model. Note the presence of the total and static pressure Kiel tubes within the reduced working section.]{Surface pressure measurement model. Note the presence of the total and static pressure Kiel probes within the reduced working section.}
            \label{fig:annotated_pressure_tubes}
        \end{subfigure}
        \hfill
        \begin{subfigure}[t]{0.46\textwidth}
            \centering
                \begin{tikzpicture}
                \node[anchor=south west, inner sep=0] (img) at (0,0) {
                    \includegraphics[width=\linewidth]{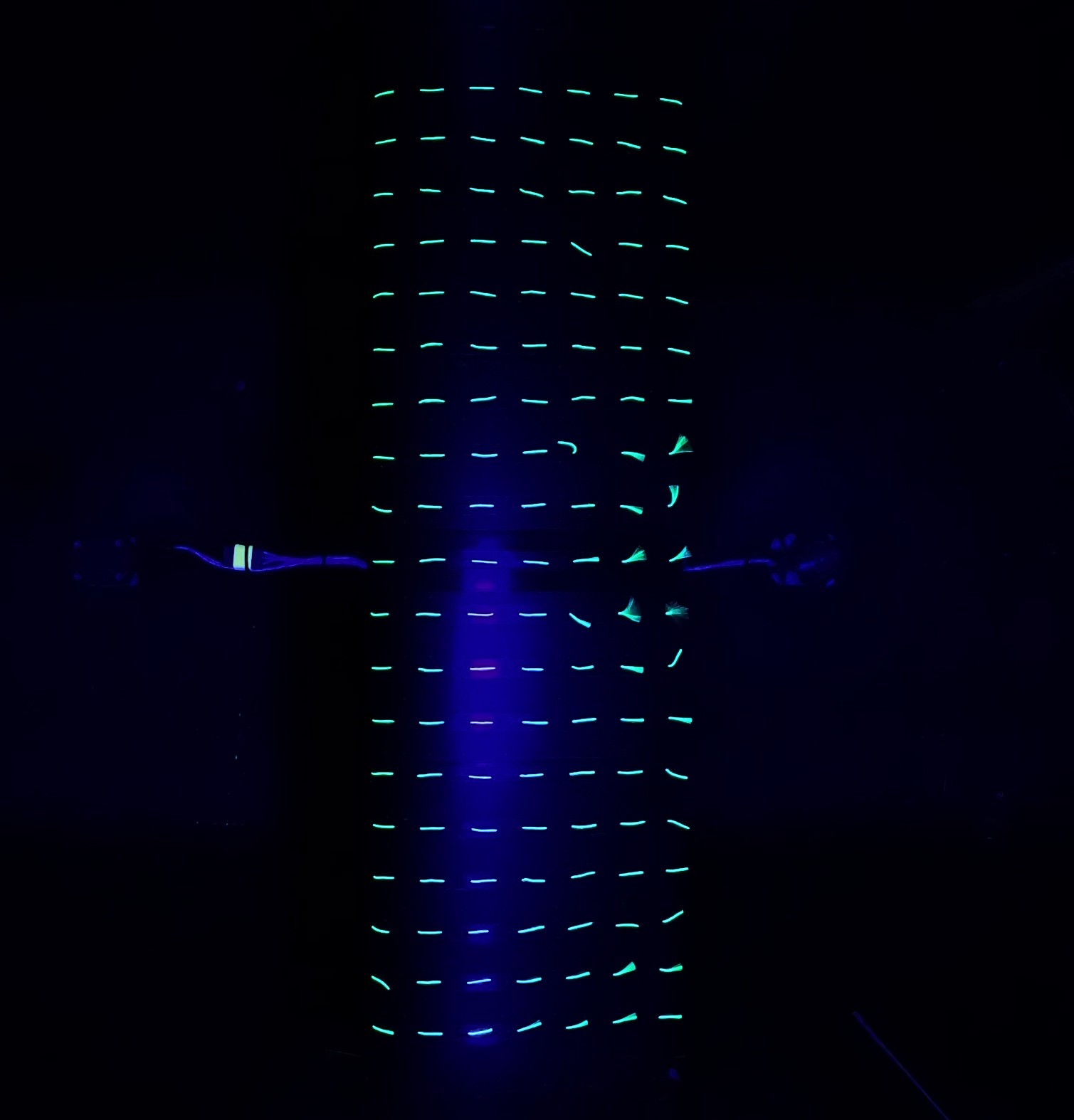}
                  };
                \draw[white, very thick, dotted]
                    (4.05,0.54) rectangle (5.05,1.25);
                \end{tikzpicture}
            \caption[Tufts mounted to the model and illuminated with UV light]{Tufts mounted to the model and illuminated with UV light. A small amount of flow leakage is present in the indicated region for most tuft images and is a result of the clearance cut in the floor panel.}
            \label{fig:tufts_setup}
        \end{subfigure}
        \caption{Transition model mounted in the wind tunnel for force, torque, surface pressure and tuft measurements.}
\end{figure}

The introduction of the new floor and ceiling panels, in addition to the structure needed to support them and the measurement apparatus, resulted in a blockage by frontal area of approximately 10\% of the working section. To account for the influence of the blockage on the freestream velocity measurements, total and static pressure Kiel probes were installed on the new floor panel one wing chord length in front and one wing chord length above the suction surface of the wing model (Figure~\ref{fig:annotated_pressure_tubes}). The Kiel probes utilize cylindrical enclosures around and in front of the measurement ports to align the flow being measured and to prevent the propeller wakes influencing the velocity readings. The readings from the Kiel probes within the new working section were then used to calculate the freestream velocity ($V_{\infty}$) at the model location and as a reference static pressure for the pressure scanner readings on the wing. Two measurement ports on the pressure scanner were allocated for the Kiel tube measurements, meaning the pressure taps at 19\% \& 30\% chord locations of the 42\% span position were left unrecorded (Table~\ref{tab:Pressure_tap_dist}) - the data for these unrecorded pressure taps would later be approximated by linear interpolation of the neighboring points.

\subsection{Testing procedure}
A representative \lc UAV was sized for the wing geometry of the model, ensuring the experimental results were based upon the requirements of a realistic transition profile. The result of the sizing process was an estimate of the take-off mass of the UAV (Table~\ref{tab:full_aircraft_model_characteristics}), which was then used as a target weight to balance during each phase of the transition to forward flight.

The transition from hover to fully established wing-borne flight was simulated by running the wind tunnel at increasing velocities of 0, 3, 5, 7.5, 10, 12.5, 15 \& 17.5 \si{\meter\per\second} as measured in the wind tunnel's main working section. The readings from the Kiel probes were used to calculate the velocity at the model location for subsequent analysis and were consistently slightly larger than the main working section velocities due to the introduced contraction in the section. The velocity readings obtained from the Kiel probes could not be used to set the wind tunnel velocity as the result was only obtained after post processing. Instead the main working section readings for velocity were consistently set at the above values for each configuration.

Prior to each measurement, the readings from the F/T sensors and pressure scanner were zeroed with no flow. The wind tunnel was then brought up to the measurement velocity and the throttle of the front and rear motors adjusted using the readings from the F/T sensors at the root of the wing and on each motor to target zero excess lift and zero pitching moment about the model's quarter-chord location. A threshold for beginning the acquisition was set around a moving mean for values of excess lift and pitching moment being at or within 3\% of their target (zero-excess lift and zero-net pitching moment). Once the acquisition criteria were met, the setup was observed for a further 30 seconds to ensure the process had settled, then load cell and pressure data were recorded for 120 seconds — capturing around 1600 convection cycles at the lowest freestream velocity. Forces and torques were acquired through three NI PXI-4472 acquisition cards mounted in a NI PXIE-1062Q chassis at a sampling rate of 40 \si{\kilo\hertz}. Pressure data for each port was acquired at 2 \si{\kilo\hertz} and then averaged and stored every 0.5 seconds. The rotational speeds of the front and rear motor were calculated using quadrature encoding on a micro-controller, with encoder signals sampled at  20 \si{\hertz}. Atmospheric conditions in the wind tunnel were also recorded for each measurement to standardize results between samples.

To determine the effects induced by the action of the vertical lift propellers in proximity to the wing during the transition, a series of tests were also conducted with the propellers inactive and locked parallel to the oncoming flow. The velocities tested with propellers inactive were the same as the active transition case but with an additional point at 22.5 \si{\meter\per\second} to represent fully established wing-borne flight.

Tuft measurements were also carried out to visualize the flow over the wing. 133 fluorescent tufts of 120S sewing thread, each 2 cm in length and approximately 400 microns thick were attached to the suction surface of the wing as it underwent the same test conditions used for the active propeller transition cases (Figure~\ref{fig:tufts_setup}). The tests were conducted in a separate campaign because the adhesive pads used to secure the tufts would obscure some pressure measurement ports and the tufts themselves may have had small influences on the pressure readings. The tufts were arranged in a uniformly spaced grid of 19 span-wise rows, with each row containing 7 equally spaced tufts in the chord-wise direction starting from the wing leading edge.

\subsection{Data processing techniques}
\subsubsection{Propeller data}
The advance ratio of a rotor $\mu$, is a key parameter for understanding the shape and propagation of the rotor's wake in edgewise flight \cite{Johnson_2013_wakes}. To capitalize on existing literature for explaining observed phenomena in the following discussions, the advance ratio of the front vertical lift propeller is selected as a reference value. The advance ratio of the front propeller is defined as in (Eq.~\ref{eq:mu_f}), where $n_{F}$ is the rotation rate of the front vertical lifting propeller in radians per second and $R_{P}$ is the radius of the propeller.

\begin{equation}
    \mathit{\mu}= \frac{V_{\infty}}{n_{F}R_{P}} \label{eq:mu_f}
\end{equation}

\subsubsection{Pressure data}
Pressure data is conventionally converted to a pressure coefficient $C_{p}$ (Eq.~\ref{eq:CP_normal}), by normalizing the difference between the pressure reading $p$ and the static pressure reference $p_{\infty}$, with the dynamic pressure of the freestream $q = 0.5 \rho V_{\infty}^2$.

\begin{equation}
    \mathit{C_{p}}= \frac{p-p_{\infty}}{0.5 \rho V_{\infty}^2} \label{eq:CP_normal}
\end{equation}

In the current study, Eq.~\ref{eq:CP_normal} would neglect the influence of the active lifting propellers on the pressure measurements. To address this, the following method develops an approach to normalize the recorded pressure difference by a quantity that reflects the contribution of both the freestream velocity and the propeller wakes. Momentum theory for a rotor in forward flight has an analytical solution for the induced velocity $v_{i}$, through the rotor when the freestream velocity is parallel to the disc \cite{johnson_2013_forward} (Eq.~\ref{eq:vi}). For this wind tunnel model, the boom and propeller plane are mounted parallel to the oncoming flow making the application of Eq.~\ref{eq:vi} reasonable:

\begin{equation}
    \mathit{v_{i}^2}= \frac{-V_{\infty}}{2} + \sqrt{(\frac{V_{\infty}}{2})^{2} + v_{h}^4} \label{eq:vi}
\end{equation}

Where the ideal hover induced propeller velocity $v_{h}$, is given by Eq.~\ref{eq:vh}:

\begin{equation}
    \mathit{v_{h}}= \sqrt{\frac{T}{2 \rho S_{P}}} \label{eq:vh}
\end{equation}

With $T$ being the thrust output and $S_{P}$ the area of the propeller disc. For each of the test velocities, thrust values are read from the load cells under the front and rear motor. From these readings, values of $v_{h}$ and subsequently $v_{i}$ are calculated for the front and rear propellers. The induced velocity values for the front ($v_{iF}$) and rear ($v_{iR}$) propeller are then averaged to provide an average induced velocity $v_{iA}$ (Eq.~\ref{eq:viA}):

\begin{equation}
    \mathit{v_{iA}} = \frac{v_{iF} + v_{iR}}{2} \label{eq:viA}
\end{equation}

A sum of the squares of the freestream velocity and the average induced velocity through the propellers is then used to normalize the pressure differential as in Eq.~\ref{eq:CP_new}.

\begin{equation}
    \mathit{C_{p}}= \frac{p-p_{\infty}}{0.5 \rho (V_{\infty}^2+v_{iA}^2)} \label{eq:CP_new}
\end{equation}

\subsubsection{Aerodynamic data} \label{aerodynamic_data_processing}
The aerodynamic performance of the model with active vertical lift propellers is separated from the aerodynamic performance of the wing with stationary vertical lifting propellers to quantify the change in aerodynamic performance during the transition. The change in the lift and drag coefficients resulting from the action of the vertical lift propellers, denoted $\Delta C_{L}$ \& $\Delta C_{D}$ respectively, are calculated as in Eqs.~\ref{eq:delta_CL} \& \ref{eq:delta_CD}:

\begin{equation}
    \mathit{\Delta C_{L}} = \frac{F_{Y}-T_{F}-T_{R}-L_{s}}{qS}  \label{eq:delta_CL}
\end{equation}

\begin{equation}
    \mathit{\Delta C_{D}} = \frac{F_{X}-D_{s}}{qS}  \label{eq:delta_CD}
\end{equation}

Where $F_{X}$ and $F_{Y}$ are the total forces measured at the root of the model in the X and Y axes and $T_{F}$ \& $T_{R}$ are the thrust forces recorded for the front and rear motors respectively. $L_{S}$ \& $D_{S}$ are the lift and drag forces recorded on the model with the 
vertical lifting propellers locked parallel to the freestream and $qS$ is the product of the dynamic pressure and the wing reference area.

\subsubsection{Tuft data}
The tufts were illuminated with a UV light source and their movement recorded with a GoPro Hero12 camera for a duration of 30 seconds at 4K resolution and 60 frames per second. The camera to record the tufts was mounted to the tunnel wall perpendicular to the center span of the wing. Computer vision techniques from Python's OpenCV library were applied to extract the tuft position data from each frame of the recorded video \cite{opencv_library}. Images were first adjusted for glare and color channel saturation to increase contrast between the tufts and the wing. The angle of a tuft relative to the freestream direction was calculated in each frame using the line made by connecting the end points of the shape detected as a tuft by computer vision techniques - this is a similar method to that described in \cite{steinfurth2020tuft}. Angular data for each tuft was then compiled across all the frames of the video, removing outliers based upon a threshold of two standard deviations away from the mean. The angular deviation, being the difference between the extremity angles detected for each tuft throughout the recording, could then be plotted to study areas of the wing surface disturbed by interactions with the propeller wakes.

\section{Results and discussion}
The results presented below summarize the findings from the four propeller configurations tested across two boom lengths. This section begins with an analysis of the \textcolor{FURU}{FURU} B2 configuration, which exhibits several key features representative of the broader set of configurations tested. This analysis is then followed by a summary of the changes in aerodynamic performance induced by the presence of the active vertical lift propellers across the remaining longitudinal and axial configurations.

\subsection{Detailed analysis of the FURU B2 configuration}

\subsubsection{Transition profile} \label{subsection:exemplar transition profile}

\begin{figure}[b!]
  \centering
  
    \begin{subfigure}[b]{\textwidth}
        \centering
            \includegraphics[width=\linewidth]{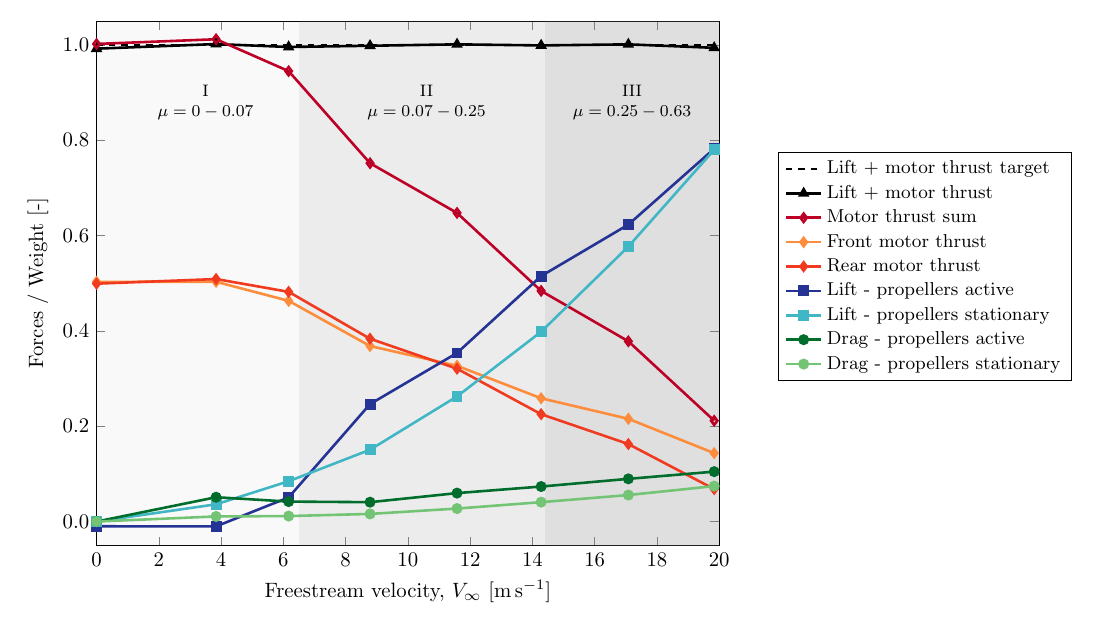}
        \caption{Forces non-dimensionalized by aircraft weight versus freestream velocity during the transition of the \textcolor{FURU}{FURU} B2 configuration. Here $\mu = V_{\infty}/n_{F}R_{P}$ is the advance ratio of the front propeller.}
        \label{fig:FURU transition summary}

        \end{subfigure}
 
  \vspace{1em}

  \begin{subfigure}[b]{\textwidth}
    \centering
        	\begin{tabular}{l c c c c c c c c}
        		\hline
        		\hline
        		Freestream velocity, $V_{\infty}$ [\si{\meter\per\second}] & 0.0 & 3.8 & 6.2 & 8.8 & 11.6 & 14.3 & 17.1 & 19.8  \\
                \hline
        		Front propeller RPM, $n_{F}$ [RPM] & 6575 & 6375 & 6010 & 5050 & 4420 & 3645 & 3070 & 2135 \\
                Rear propeller RPM, $n_{R}$ [RPM] & 6575 & 6530 & 6315 & 5730 & 5125 & 4145 & 3270 & 1670 \\
        		\hline
        		\hline
        	\end{tabular}
    \caption{\textcolor{FURU}{FURU} B2 front and rear propeller rotation rates during the transition.}
    \label{tab:FURU_B2_RPMS}
  \end{subfigure}

  \caption{\textcolor{FURU}{FURU} B2 transition summary.}
  \label{fig:FURU_B2_transition_summary_with_RPM_table}
\end{figure}

The \textcolor{FURU}{FURU} B2 configuration is presented as an exemplar of the experimental transition profile for the wind tunnel model produced. The transition profile has been divided in to three stages of front propeller advance ratio: I ($\mu =0-0.07$), II ($\mu =0.07-0.25$), and III ($\mu =0.25-0.63$), which broadly capture the transients in aerodynamic performance induced by the action of the propeller wakes on the wing. Figure~\ref{fig:FURU transition summary} will be used to provide a general overview of the transition profile and to detail trends in aerodynamic forces during the transition. Further insights into the underlying flow mechanisms will be explored in Section~\ref{subsubsection:exemplar_surface_measurements} through surface pressure and tuft flow visualization on the suction surface.

At the start of the transition, the vertical lift motors support nearly 100\% of the aircraft’s weight, as the wing - with or without active vertical lift propellers, is unable to produce sufficient lift below its stall velocity (Figure~\ref{fig:FURU transition summary}). As the airspeed increases, the wing generates increasing amounts of lift allowing the thrust from the vertical lift motors to gradually decrease whilst ensuring the aircraft weight is consistently supported between the wing lift and motor thrusts. By the end of the test sequence the wing carries approximately 80\% of the aircraft's weight whilst the vertical lift motors make up the remaining 20\%.

In stage I when the advance ratio ($\mu$) is low, the strength of the front propeller wake is high in comparison to the freestream velocity, deflecting the wing's incoming flow downwards to produce a reduction in the effective angle of attack experienced by the central section of wing behind the front propeller. This results in an average reduction of 84\% in wing lift when the propellers are active compared to when they are stationary. During this stage, drag in the active propeller case is, on average, three times higher than in the stationary case, likely due to the increased induced power required at higher thrust levels \cite{johnson_2013_forward}. As the transition approaches the end of Stage I ($V_{\infty} = 6.2$\si{\meter\per\second}), increasing wing lift reduces the vertical lift propeller thrust requirement, thereby lowering the induced power requirement and its associated drag.

In Stage II, the model with active vertical lift propellers produces, on average, 43\% more lift than the model with stationary propellers. As discussed in Section~\ref{section:prior_work_and_motivation}, this lift increase is primarily attributed to the rear propeller wake acting similarly to a jet flap, enhancing wing lift. Additionally, as suggested in Section~\ref{section:prior_work_and_motivation}, 
interactions between the super-vortices from the front propeller wake and the wing surface may be further contributing to lift enhancement — a hypothesis examined in more detail in Section~\ref{subsubsection:exemplar_surface_measurements}. Despite the increased lift, the drag in the active propeller case does not rise proportionally and maintains a relatively steady 91\% offset compared to the stationary propeller case throughout Stages II and III.

In Stage III, as the thrust output decreases, the additional lift generated in the active propeller case gradually tends back towards the stationary propeller case.

The rotational speed (Table~\ref{tab:FURU_B2_RPMS}) and thrust outputs of the front and rear propellers are similar throughout stage I and most of II, but start to diverge at freestream velocities above 11.6 \si{\meter\per\second}. As the freestream velocity increases, the wing's nose-down pitching moment increases requiring the front motor thrust to increase relative to the rear motor to maintain zero net pitching moment about the model's quarter-chord. The front propeller generally achieves this requirement with a lower RPM due to its cleaner inflow and the increasing amounts of translational lift it experiences with increasing freestream velocity. In the case of a complete \lc aircraft during transition, the control surfaces on the wing and or tail would become effective as the airspeed increases, likely requiring less input from the motors in the latter stage of transition than is depicted here.

The following section correlates suction surface pressure and tuft flow measurements with the changes in aerodynamic forces observed during transition.

\subsubsection{Correlating surface pressure and tuft data} \label{subsubsection:exemplar_surface_measurements}

\begin{figure}[t!]
        \centering
        \begin{subfigure}[t]{\textwidth}
            \centering
            \includegraphics[width=\linewidth]{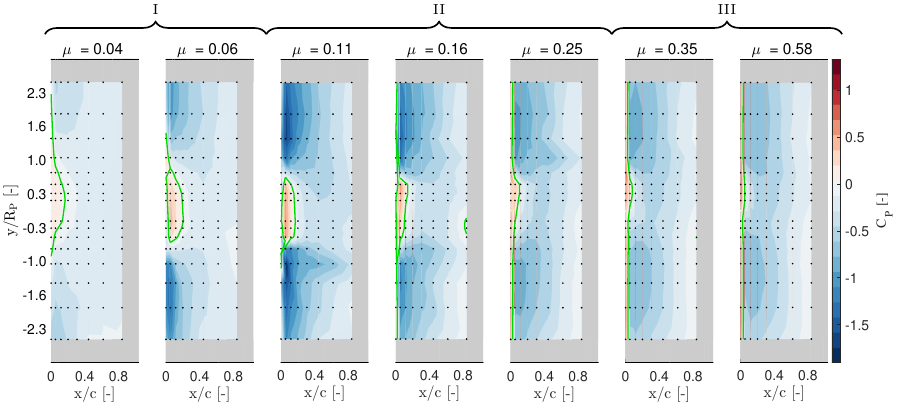}
            \caption[Suction wing surface pressure coefficient plots during the transition of the \textcolor{FURU}{FURU} B2 configuration.]{Wing suction surface pressure coefficient plots during the transition of the \textcolor{FURU}{FURU} B2 configuration, where \protect\markerone \ markings denote the location of a pressure tap and contours marked with \protect\greenline \ denote the $C_{p} = 0$ crossing location.}
            \label{fig:FURU_short_cp}
            \end{subfigure}
        \begin{subfigure}[t]{\textwidth}
            \centering
            \includegraphics[width=\linewidth]{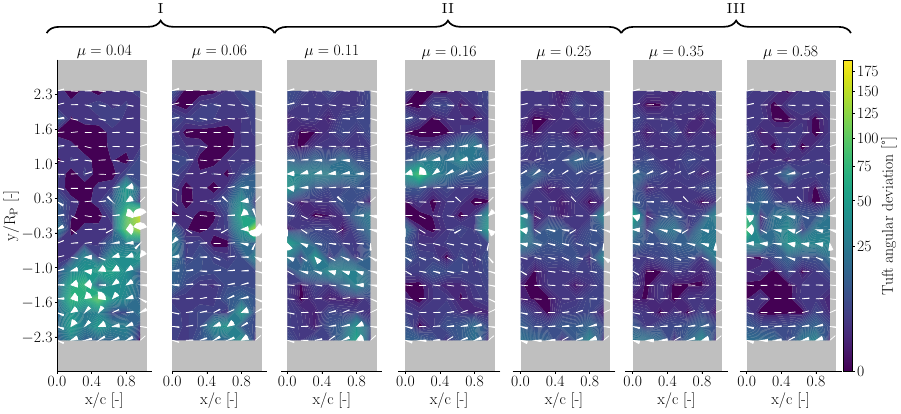}
            \caption{Wing suction surface tuft disturbances during the transition of the \textcolor{FURU}{FURU} B2 configuration.}
            \label{fig:FURU_short_tuft}
        \end{subfigure}
        \caption{Pressure coefficient and tuft disturbance during the transition of the \textcolor{FURU}{FURU} B2 configuration.}
        \label{fig:FURU_s_comp}
\end{figure}

The suction surface pressure coefficient plots for each transition stage of the \textcolor{FURU}{FURU} B2 configuration are presented in Figure~\ref{fig:FURU_short_cp}. A key feature across these plots is a region of positive pressure near the leading edge, bounded in the spanwise direction by regions of negative pressure. The boundary between these regions is indicated by a line (\protect\greenline) representing the linearly interpolated $C_p = 0$ contour from measured data.

Behind the area of positive pressure at the leading edge of the wing exists a channel like section of weakly negative pressure coefficient between $\pm1.0 \ \mathrm{y/R_{P}}$. As the area of positive pressure shrinks and intensifies during the transition, the areas of strongly negative pressure coefficient on each side of this channel redistribute to fill the area more evenly. The areas of most negative pressure coefficient on the suction surface are concentrated towards the outer portions of span ($\pm1.0 - 2.3 \ \mathrm{y/R_{P}}$) - with most suction generated across the first 25\% of the wing chord as would be expected of a conventional wing generating lift.

An analysis of the tuft plots indicates that flow disturbances on the wing’s suction surface are predominantly confined within a spanwise range of $\pm 1.6 \ \mathrm{y/R_{P}}$ (Figure~\ref{fig:FURU_short_tuft}). These disturbances generally extend over a significant portion of the chord and tend to become more concentrated near the centerline as the transition progresses. Minor deviations from this pattern are observed, such as localized flow leakage through the lower wind tunnel wall at $-2.3 \ \mathrm{y/R_{P}}$. The spanwise extent of these tuft disturbances closely corresponds to the boundaries of key features in the pressure coefficient distributions (Figure~\ref{fig:FURU_short_cp}), suggesting that variations in suction surface pressure are largely attributable to the influence of the vertical lift propeller wakes.

In stage I, the two trailing super-vortices in the front propeller wake are expected to generate a region of downwash along the symmetry plane between them. Combined with the downward trajectory of the mean flow in the front propeller’s wake, this reduces the effective angle of attack experienced by the central section of the wing located downstream. As discussed in Section~\ref{section:prior_work_and_motivation}, this interaction can suppress lift development in the affected region. Without pressure-side measurements or flow visualization, it is not possible to confirm whether this reduction in angle of attack leads to pressure-side separation, as observed in comparable numerical studies \cite{gordnier1999numerical}.

Regions of negative pressure coefficient develop between spanwise positions of $\pm 1.0 - 2.3 \ \mathrm{y/R_{P}}$ and within the first 30\% of the wing chord, strengthening progressively throughout stage I. Outboard of the vortex core location, upwash generated by vortex rotation may increase the local effective angle of attack, reducing pressure near the leading edge \cite{patel1974some}. This mechanism appears active in the present study, where a negative-positive-negative spanwise pressure distribution forms. The net aerodynamic effect during stage I, is a significant lift penalty for the active propeller case relative to the stationary configuration, as shown in Figure~\ref{fig:FURU transition summary}.

Tuft measurements in stage I (Figure~\ref{fig:FURU_short_tuft}) show substantial disturbance at negative span loactions for $\mu = 0.04$, corresponding to the influence of the retreating front propeller blade. However, the absence of a corresponding pressure coefficient variation suggests that this disturbance has limited aerodynamic impact. As the advance ratio increases, the affected region disappears, indicating that the freestream velocity quickly dominates the flow in the outer wing sections.

Additional tuft disturbance is observed along the trailing edge between $\pm 0.3 \ \mathrm{y/R_{P}}$, spanning the aft 30\% of the chord. This is likely caused by reversed flow from the impinging rear propeller wake. Tufts in this region orient upstream and outward, indicating radial spreading of the wake along the trailing edge. Despite the visible disturbance, the impact on the pressure distribution at the trailing edge remains limited during stage I.

As the transition progresses into stage II, two streamwise disturbances appear in the tuft visualization at $\mu = 0.11$, straddling a central region of undisturbed tufts (Figure~\ref{fig:FURU_short_tuft}). Based upon their spanwise locations at $\pm 1.0 \ \mathrm{y/R_{P}}$, these disturbances are believed to correspond to the trailing super-vortices from the front propeller wake convecting over the suction surface of the wing. Disturbances to the tufts caused by the interaction with the vortices originate near the leading edge at $\pm 1.0 \ \mathrm{y/R_{P}}$ and spread laterally to $\pm 1.6 \ \mathrm{y/R_{P}}$ by the time they reach the trailing edge.

The suction effect of the streamwise vortex core on the negative span section at $\mu = 0.11$ is discernible in the pressure coefficient distribution at $-1.0 \ \mathrm{y/R_{P}}$ (Figure~\ref{fig:FURU_short_cp}). This observation aligns with previously reported findings where pressure reductions across the leading edge were attributed to the up wash generated by the rotational nature of a vortex \cite{patel1974some}. It is also consistent with Gordnier and Visbal’s numerical results, which showed that a pair of streamwise vortices convecting over a surface generate a local pressure deficit beneath their cores and a widening footprint as they propagate downstream \cite{gordnier1999numerical}.

As the super-vortices align more closely with the freestream direction at increasing advance ratios, the associated up wash and downwash components exert greater influence on the wing's effective angle of attack. The increasing influence of the vortices is balanced with a reduction in their strength at increasing airspeeds as the thrust from the vertical lift propellers is gradually reduced. These two contrasting effects act to stabilize the lift augmentation experienced by the wing across stage II to a relatively consistent level (Figure~\ref{fig:FURU transition summary}).

At higher advance ratios within stage II, the rigid nature of the vertical lift propeller blades increasingly affects the vertical position and symmetry of the super-vortices shed by the advancing and retreating blades. At $\mu = 0.11$, both vortices register comparably in the tuft plots (Figure~\ref{fig:FURU_short_tuft}). However, the suction associated with the vortex core shed by the advancing blade at $1.0 \ \mathrm{y/R_{P}}$ is less apparent in the corresponding pressure distribution. This weaker response suggests that the advancing blade's vortex may not be traveling cleanly over the wing's suction surface, perhaps impinging on the wings leading edge instead. By $\mu = 0.16$, the vortex shed by the retreating blade becomes less prominent, while the stronger downwash from the advancing blade’s vortex keeps it in close proximity to the wing's suction surface - generating a region of low pressure beneath its path (Figure~\ref{fig:FURU_short_cp}).

During stage II, the wing with active vertical lift propellers produces on average 43\% more lift than the wing with stationary propellers, with lift increasing nearly linearly throughout this region (Figure~\ref{fig:FURU transition summary}). At $\mu = 0.16$, a transition between positive and negative $C_p$ values is observed near $-0.3 \ \mathrm{y/R_{P}}$ and 80\% chord—immediately upstream of the advancing side of the rear propeller (Figure~\ref{fig:FURU_short_cp}). The location of this $C_p$ crossing site suggests the development of an adverse pressure gradient that may be promoting flow separation near the wing’s trailing edge. Indications of separated flow at this location and airspeed ($V_{\infty}=11.6$ \si{\meter\per\second}) are further supported by a slight reduction in the lift trend observed in the active-propeller case during stage II (Figure~\ref{fig:FURU transition summary}).

As the transition progresses into stage III, tuft measurements indicate that the front propeller wake begins to break down and convect downstream, impinging on the wing surface near $-0.3 \ \mathrm{y/R_{P}}$ (Figure~\ref{fig:FURU_short_tuft}). Regions of positive pressure and tuft disturbance are observed at the leading edge of the central span, corresponding with the now disordered wake structures scattering across the suction surface (Figures~\ref{fig:FURU_short_cp} and \ref{fig:FURU_short_tuft}). This behavior aligns with the trend showing the lift generated by the wing with active vertical lift propellers converging towards that of the stationary configuration as forward airspeed increases and propeller rotation rates decrease (Figure~\ref{fig:FURU transition summary}).

The \textcolor{FURU}{FURU} B2 configuration shows a reduction in aerodynamic performance during stage I, followed by a region of improved performance during stage II, that ultimately tends towards that of the stationary propeller configuration at the end of the transition. Suction surface pressure and tuft data indicate that super-vortices from the front vertical lift propeller’s wake interact with the wing, while adverse pressure gradients generated by the rear propeller form near the trailing edge at moderate advance ratios. The following section investigates how axial and longitudinal variations in motor positioning influences these flow features and their impact on aerodynamic performance during the transition.

\subsection{Effect of wing-motor axial and longitudinal separation}
By comparing each configuration with active propellers to its stationary-propeller counterpart, the direct impact of each propeller's operation and its sensitivity to positional changes can be evaluated. The analysis discussed here is supported by the figures in Appendix~\ref{section:additional_configs}.

\subsubsection{Effect of active vertical lift propellers on aerodynamic performance metrics} \label{subsection:performance summary}
\begin{figure}[h!]
        \centering
        \begin{subfigure}[t]{0.5\textwidth}
            \centering
            \includegraphics[width=\linewidth]{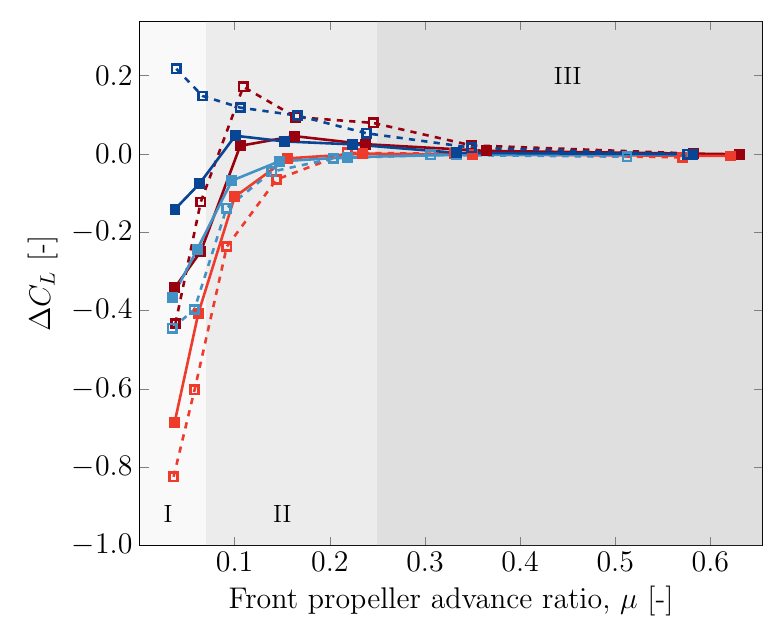}
        \caption{Deltas in the lift coefficient caused by the active lifting propellers in comparison to the same configuration with stationary vertical lift propellers.}
        \label{fig:delta CL summary}
            \end{subfigure}
        \hfill
        \begin{subfigure}[t]{0.483\textwidth}
            \centering
            \includegraphics[width=\linewidth]{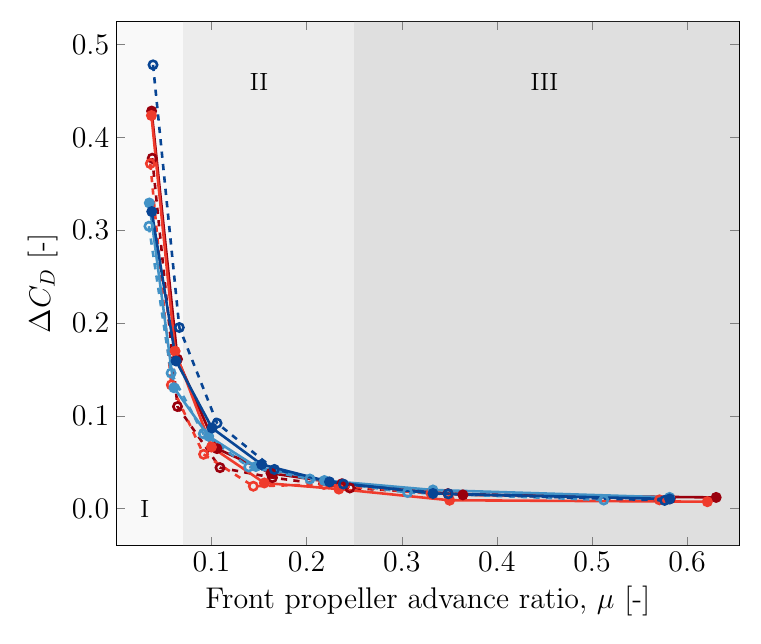}
        \caption{Deltas in the drag coefficient caused by the active lifting propellers in comparison to the same configuration with stationary vertical lift propellers.}
        \label{fig:delta CD summary}
        \end{subfigure}
        
        \vspace{1em}
        
        \begin{subfigure}[t]{0.495\textwidth}
        \centering
            \includegraphics[width=\linewidth]{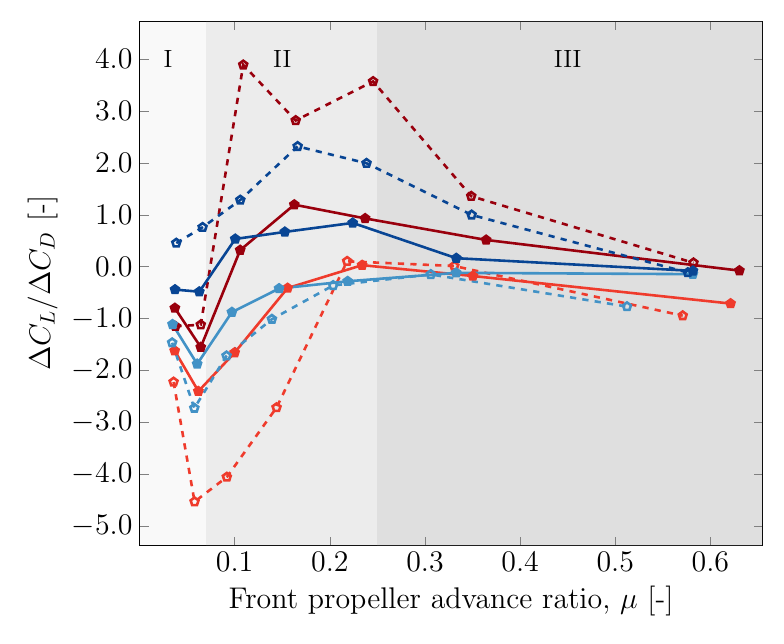}
        \caption{Deltas in lift-drag ratio caused by the active lifting propellers in comparison to the same configuration with stationary vertical lift propellers.}
        \label{fig:delta CL_CD summary}
        \end{subfigure}
        \hfill
        \raisebox{1cm}{
        \begin{subfigure}[t]{0.4\textwidth}
        \centering
        \includegraphics[width=\linewidth]{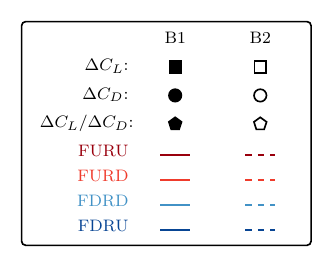}
        \end{subfigure}}
        \caption{Comparison of the deltas in aerodynamic performance for each configuration during the transition.}
        \label{fig:All_configs_aero_comp}
        
\end{figure}

To evaluate the overall aerodynamic impact of the active vertical lift propellers on each configuration, a delta lift-drag ratio (${\Delta C_{L}}/{\Delta C_{D}}$) is introduced. This metric captures the relative change in aerodynamic efficiency, with $\Delta C_{L}$ and $\Delta C_{D}$ calculated as described in Section~\ref{aerodynamic_data_processing} by subtracting the stationary propeller case from the active propeller case across the same airspeed range. The transition profiles presented in the following figures are divided into the same three advance ratio stages introduced in Section~\ref{subsection:exemplar transition profile}. The data presented for the $\Delta C_{L}$ and $\Delta C_{D}$ values shown in Figures~\ref{fig:delta CL summary} and~\ref{fig:delta CD summary} are used to provide insight into the ${\Delta C_{L}}/{\Delta C_{D}}$ trends shown in Figure~\ref{fig:delta CL_CD summary} - which will serve as the primary metric for assessing changes in aerodynamic performance.

Generally, large changes in $\Delta C_{L}$ and $\Delta C_{D}$ are observed at very low advance ratios, which then tend back towards zero as the advance ratio increases. During stage I, the corresponding ${\Delta C_{L}}/{\Delta C_{D}}$ values are negative for all configurations except the \textcolor{FDRU}{FDRU} B2 (\fdrushort) case, which produces small positive values. Most configurations show an increasingly negative trend in ${\Delta C_{L}}/{\Delta C_{D}}$ during this stage, with the \textcolor{FURD}{FURD} B2 (\furdshort) configuration demonstrating the worst performance by the end of stage I. In contrast, both the \textcolor{FDRU}{FDRU} B2 and \textcolor{FURU}{FURU} B2 (\furushort) configurations exhibit positive-trending ${\Delta C_{L}}/{\Delta C_{D}}$ values during stage I; although the \textcolor{FURU}{FURU} B2 trend is only weakly positive and still within the negative range.

Configurations with the largest or least negative ${\Delta C_{L}}/{\Delta C_{D}}$ values all feature rear motors mounted in the upwards orientation. This trend is also evident in the $\Delta C_{L}$ data presented in Figure~\ref{fig:delta CL summary}.

The $\Delta C_{D}$ trends reveal that configurations generating the largest absolute ${\Delta C_{L}}$ values also tend to experience the highest levels of drag at very low advance ratios. Across all configurations, $\Delta C_{D}$ reduces sharply over stage I, with most values decreasing by at least half of their initial values. The $\Delta C_{D}$ values then tend towards zero with increasing advance ratio, showing little variation between configurations.

The onset of stage II marks an inflection point in the ${\Delta C_{L}}/{\Delta C_{D}}$ trends for all configurations, with the exception of \textcolor{FDRU}{FDRU} B2 and \textcolor{FURU}{FURU} B2, which continue their previous upward trend with increasing advance ratio. Unlike in stage I, the \textcolor{FURU}{FURU} B2 configuration now outperforms the \textcolor{FDRU}{FDRU} B2 configuration in terms of ${\Delta C_{L}}/{\Delta C_{D}}$, with a large increase between the end of stage I and stage II. This sharp increase corresponds to the \textcolor{FURU}{FURU} B2 configuration generating larger ${\Delta C_{L}}$ values than the \textcolor{FDRU}{FDRU} B2 case, while also maintaining the lowest $\Delta C_{D}$ values among all configurations. In contrast, the \textcolor{FDRU}{FDRU} B2 configuration continues to exhibit the highest drag values but remains the only configuration to sustain a positive ${\Delta C_{L}}/{\Delta C_{D}}$ throughout the entire transition.

The rapid increase in ${\Delta C_{L}}/{\Delta C_{D}}$ observed in the \textcolor{FURU}{FURU} B2 configuration is believed to be the result of lift augmentation caused by the front propeller's trailing super-vortices - which are detected passing over the suction surface  as the performance increase occurs (Figure~\ref{fig:FURU_short_tuft}). An investigation of these vortex interactions with the wing's surface is carried out in the following sections.

At $\mu = 0.16$, a decrease in the ${\Delta C_{L}}/{\Delta C_{D}}$ of the \textcolor{FURU}{FURU} B2 configuration occurs. This reduction in lift is reflected in the ${\Delta C_{L}}$ data and is attributed to flow separation at the trailing edge of the wing, likely caused by the impinging wake from the rear vertical lift propeller, as discussed in Section~\ref{subsubsection:exemplar_surface_measurements}.

The divergence in ${\Delta C_{L}}/{\Delta C_{D}}$ trends between configurations with upwards and downwards mounted rear motors identified in stage I becomes more pronounced during stage II. Configurations with upwards-oriented rear motors—\textcolor{FDRU}{FDRU} (B1, B2) (\fdrulong, \fdrushort) and \textcolor{FURU}{FURU} (B1, B2) (\furulong, \furushort) — achieve positive ${\Delta C_{L}}/{\Delta C_{D}}$ values, while those with downwards-oriented rear motors remain largely negative. Within the upward-mounted group, the B2 configurations demonstrate stronger increases in ${\Delta C_{L}}/{\Delta C_{D}}$ compared to their B1 counterparts, indicating that reduced longitudinal separation enhances the aerodynamic performance.

The \textcolor{FURU}{FURU} B1 configuration does not exhibit the reduction in ${\Delta C_{L}}/{\Delta C_{D}}$ observed in the B2 variant at $\mu = 0.16$, implying that greater separation between the rear vertical lift propeller and the wing may mitigate the onset of flow separation due to an adverse pressure gradient. Configurations with downward orientated rear motors - \textcolor{FDRD}{FDRD} (B1, B2) (\fdrdlong, \fdrdshort) and \textcolor{FURD}{FURD} B1 (\furdlong) initially show asymptotic trends in ${\Delta C_{L}}/{\Delta C_{D}}$ that tend towards zero as they approach the performance of the model with stationary propellers in wing-borne flight. Beyond $\mu = 0.3$, only the \textcolor{FDRD}{FDRD} B1 configuration maintains this trend, while \textcolor{FDRD}{FDRD} B2 and \textcolor{FURD}{FURD} B1 display a secondary decline in ${\Delta C_{L}}/{\Delta C_{D}}$.

The continued asymptotic behavior of the \textcolor{FDRD}{FDRD} B1 configuration suggests that the front propeller wake is sufficiently separated from the wing at higher advance ratios, resulting in minimal interaction with the wing. In contrast, the \textcolor{FURD}{FURD} B2 configuration is consistently the worst performing configuration, with a relatively weak increase in ${\Delta C_{L}}/{\Delta C_{D}}$ followed by a secondary decline by the end of stage II.

Previous studies on tandem lifting fan systems showed a linear increase in net lift with increasing jet-to-freestream velocity ratios \cite{hickey1966aerodynamic}. However, those analyses were based upon fixed fan pressure ratios at increasing airspeeds. In the present study, by modeling realistic thrust variation between motors over the transition, a more complex and nonlinear relationship emerges. It has also been shown here that the aerodynamic performance of the configuration appears highly sensitive to the axial and longitudinal spacing between the motors.

In summary, configurations with upwards orientated rear motors tend to reach peak ${\Delta C_{L}}/{\Delta C_{D}}$ values earlier and more centrally within stage II. In contrast, configurations with downwards-oriented rear motors achieve their best performance at higher advance ratios, when the thrust contribution from the vertical lift propellers is reduced. This distinction suggests that configurations with upright rear motors benefit from aerodynamic interactions between the propeller wakes and the wing, whereas those with downward-facing rear motors primarily endure the transition without leveraging any aerodynamic performance. The observed non-linear relationship between ${\Delta C_{L}}/{\Delta C_{D}}$ and advance ratio ($\mu$) indicates a complex series of interactions between the wing and the propeller wakes. These interactions are further investigated in the following sections through wing suction surface pressure and tuft measurements, correlating observations to changes in aerodynamic performance.

\subsubsection{Summary of suction surface pressure and tuft analyses} \label{subsection:cp and tuft comp all configs}
This section correlates suction surface pressure and tuft measurements to changes in aerodynamic performance, examining the role of propeller wake interactions with the wing during the transition.

\begin{figure}[b!]
  \centering
  
    \begin{subfigure}[b]{\textwidth}
    \centering
    \includegraphics[width=\linewidth]{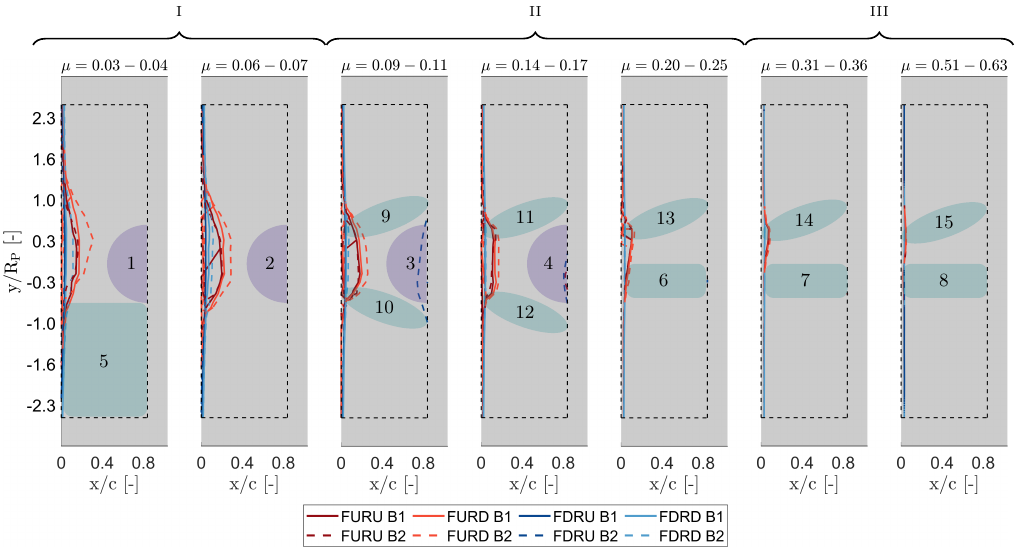}
    \caption{$C_{p}=0$ crossing lines for each configuration plotted within the sampled area of the wing's suction surface (\hbox to 0.35cm{\tikz[baseline=-0.5ex] \draw[black, dashed, line width=1pt] (0,0) -- (0.35,0);}), with the location of prominent flow features detected from their tuft plots also highlighted. Plots are labeled with an advance ratio ($\mu$) range, capturing the range of front propeller advance ratios present for each of the individual configurations.}
    \label{fig:Crossing_data_all}
  \end{subfigure}
  
  \vspace{1em}

  \begin{subfigure}[b]{\textwidth}
    \centering
                        
        \begin{tikzpicture}
          \useasboundingbox (0,0) rectangle (14,1); 
        
          \draw[decorate, decoration={brace, amplitude=6pt}, thick]
            (3.42,0) -- (5.8,0) node[midway, yshift=16pt, font=\scriptsize, text width=4cm, align=center] {Rear propeller \\ wake impingement};
        
          \draw[decorate, decoration={brace, amplitude=6pt}, thick]
            (5.82,0) -- (8.35,0) node[midway, yshift=16pt, font=\scriptsize, text width=4cm, align=center] {Front propeller \\ wake interaction};
        
          \draw[decorate, decoration={brace, amplitude=6pt}, thick]
            (8.37,0) -- (13.85,0) node[midway, yshift=16pt, font=\scriptsize, text width=6cm, align=center] {Front propeller \\ trailing tip vortices};
        \end{tikzpicture}
            
            \vspace{0.25em}
            
        	\begin{tabular}{l c c c c c c c c c c c c c c c}
        		\hline
        		\hline
        		Feature number & 1 & 2 & 3 & 4 & 5 & 6 & 7 & 8 & 9 & 10 & 11 & 12 & 13 & 14 & 15\\
                \hline
                
        		\textcolor{FURU}{FURU} B1 & \tikz\draw[draw=black, fill=none] (0,0) circle (3pt); & \tikz\draw[draw=black, fill=none] (0,0) circle (3pt); & - & - &  - &  \tikz\draw[draw=black, fill=none] (0,0) circle (3pt); & \tikz\fill[Greens-J] (0,0) circle (3pt); & \tikz\fill[Greens-J] (0,0) circle (3pt); & - & \tikz\fill[Greens-J] (0,0) circle (3pt); & \tikz\draw[draw=black, fill=none] (0,0) circle (3pt); & \tikz\draw[draw=black, fill=none] (0,0) circle (3pt); & \tikz\fill[Greens-J] (0,0) circle (3pt); & \tikz\fill[Greens-J] (0,0) circle (3pt); & \tikz\draw[draw=black, fill=none] (0,0) circle (3pt); \\
                
                \textcolor{FURU}{FURU} B2 & \tikz\fill[Greens-J] (0,0) circle (3pt); & \tikz\fill[Greens-J] (0,0) circle (3pt); & \tikz\fill[Greens-J] (0,0) circle (3pt); & \tikz\fill[Greens-J] (0,0) circle (3pt); & \tikz\fill[Greens-J] (0,0) circle (3pt); & \tikz\fill[Greens-J] (0,0) circle (3pt); & \tikz\fill[Greens-J] (0,0) circle (3pt); & \tikz\fill[Greens-J] (0,0) circle (3pt); & \tikz\fill[Greens-J] (0,0) circle (3pt); & \tikz\fill[Greens-J] (0,0) circle (3pt); & \tikz\fill[Greens-J] (0,0) circle (3pt); & \tikz\draw[draw=black, fill=none] (0,0) circle (3pt); & \tikz\fill[Greens-J] (0,0) circle (3pt); & \tikz\draw[draw=black, fill=none] (0,0) circle (3pt); & \tikz\draw[draw=black, fill=none] (0,0) circle (3pt); \\
                
                \textcolor{FURD}{FURD} B1 & - & - & - & - & - & \tikz\draw[draw=black, fill=none] (0,0) circle (3pt); & \tikz\fill[Greens-J] (0,0) circle (3pt); & \tikz\fill[Greens-J] (0,0) circle (3pt); & - & \tikz\fill[Greens-J] (0,0) circle (3pt); & \tikz\draw[draw=black, fill=none] (0,0) circle (3pt); & \tikz\fill[Greens-J] (0,0) circle (3pt); & \tikz\fill[Greens-J] (0,0) circle (3pt); & \tikz\fill[Greens-J] (0,0) circle (3pt); & \tikz\draw[draw=black, fill=none] (0,0) circle (3pt); \\
                
                \textcolor{FURD}{FURD} B2 & - & - & - & - & \tikz\fill[Greens-J] (0,0) circle (3pt); & \tikz\draw[draw=black, fill=none] (0,0) circle (3pt); & \tikz\fill[Greens-J] (0,0) circle (3pt); & \tikz\fill[Greens-J] (0,0) circle (3pt); & - & \tikz\fill[Greens-J] (0,0) circle (3pt); & \tikz\fill[Greens-J] (0,0) circle (3pt); & \tikz\fill[Greens-J] (0,0) circle (3pt); & \tikz\fill[Greens-J] (0,0) circle (3pt); & \tikz\draw[draw=black, fill=none] (0,0) circle (3pt); & \tikz\draw[draw=black, fill=none] (0,0) circle (3pt); \\
                
                \textcolor{FDRU}{FDRU} B1 & - & - & - & - & - & - & - & - & - & - & - & - & - & - & - \\
                
                \textcolor{FDRU}{FDRU} B2 & \tikz\fill[Greens-J] (0,0) circle (3pt); & \tikz\fill[Greens-J] (0,0) circle (3pt); & \tikz\fill[Greens-J] (0,0) circle (3pt); & \tikz\fill[Greens-J] (0,0) circle (3pt); & - & - & - & - & - & - & - & - & - & - & - \\
                
                \textcolor{FDRD}{FDRD} B1 & - & - & - & - & - & - & - & - & - & - & - & - & - & - & - \\
                
                \textcolor{FDRD}{FDRD} B2 & - & - & - & - & - & - & - & - & - & - & - & - & - & - & - \\

        		\hline
        		\hline
        	\end{tabular}
    \caption{Transition feature summary, where \tikz\fill[Greens-J] (0,0) circle (3pt); represents a strong indication of a feature, \tikz\draw[draw=black, fill=none] (0,0) circle (3pt); represents a weak indication of a feature and (--) shows no indication of the feature.}
    \label{tab:Features_present}
  \end{subfigure}

  \caption{Transition feature summary.}
  \label{fig:Features_present_table_and_plot}
\end{figure}

Figure~\ref{fig:Features_present_table_and_plot} presents a composite plot that overlays the $C_{p}=0$ crossing locations on the suction surface for each configuration. Prominent flow features identified through tuft analysis are marked as regions 1–15 and their presence is recorded in Table~\ref{tab:Features_present} as either strongly present (clearly identifiable tuft disturbance), weakly present (noticeable but less distinct), or not present (no identifiable disturbance).

A clear distinction emerges between configurations with the front motor mounted upwards (\textcolor{FURU}{FURU}, \textcolor{FURD}{FURD}) and those with the front motor mounted downwards (\textcolor{FDRU}{FDRU}, \textcolor{FDRD}{FDRD}) as seen in Figure~\ref{fig:Crossing_data_all}. Configurations with an upwards orientated front motor consistently develop a region of positive pressure coefficient at the wing's leading edge, spanning approximately $-1.0$ to $1.6 \ \mathrm{y/R_{P}}$. This region typically extends up to around 17\% of the chord, except in the \textcolor{FURD}{FURD} B2 configuration (\hbox to 0.35cm{\tikz[baseline=-0.5ex] \draw[FURD, dashed, line width=1pt] (0,0) -- (0.35,0);}), where it reaches around 29\%. As the transition progresses, this positive pressure region contracts toward $\pm 0.3 \ \mathrm{y/R_{P}}$, consistent with the evolution described for the \textcolor{FURU}{FURU} B2 configuration in Section~\ref{subsubsection:exemplar_surface_measurements}.

In contrast, configurations with downwards orientated front motors (\textcolor{FDRU}{FDRU}, \textcolor{FDRD}{FDRD}) exhibit very little evidence of positive pressure regions forming at the wing's leading edge, and only when the longitudinal motor separation is short (B2). For example, the \textcolor{FDRD}{FDRD} B2 configuration (\hbox to 0.35cm{\tikz[baseline=-0.5ex] \draw[FDRD, dashed, line width=1pt] (0,0) -- (0.35,0);}) develops a small region of positive $C_p$ at the leading edge between $-1.0$ and $1.6 \ \mathrm{y/R_{P}}$ during stage I, with a chord wise extent of approximately 11\%. This region practically disappears by the start of stage II, suggesting that as the freestream velocity increases, it dominates the flow redirecting tendency of the front propeller’s downward wake. The \textcolor{FDRU}{FDRU} B2 configuration does not exhibit this leading edge pressure feature, despite identical front motor placement. This difference is attributed to the rear propeller - which being mounted upright likely draws airflow over the wing’s upper surface in a manner similar to a conventional pusher propeller, mitigating the velocity deficit over the suction surface of the wing induced by the front propeller’s wake.

Only the \textcolor{FURU}{FURU} and \textcolor{FDRU}{FDRU} B2 configurations show evidence of an adverse pressure gradient at the trailing edge during stage II. For the \textcolor{FURU}{FURU} B2 configuration, this was previously observed at $\mu = 0.16$. In the \textcolor{FDRU}{FDRU} B2 configuration it first appears at $\mu = 0.11$, spanning laterally across $\pm 1.0 \ \mathrm{y/R_{P}}$ and the latter part of the chord. While the full extent of the adverse pressure gradient is unclear due to the sparsely distributed pressure taps around the trailing edge, the adverse pressure regions of each configuration begin to coincide by the middle of stage II. Notably, the \textcolor{FDRU}{FDRU} B2 configuration does not show an associated lift penalty during this period, as seen in Figure~\ref{fig:delta CL summary}, suggesting that the rear propeller’s lift augmenting effect remains dominant.

As expected, Table~\ref{tab:Features_present} confirms that only the B2 configurations with overlapping rear motors (\textcolor{FURU}{FURU} and \textcolor{FDRU}{FDRU}) exhibit any tuft disturbances indicative of rear wake impingement. Additionally, all configurations with downwards orientated front motors, with the exception of \textcolor{FDRU}{FDRU} B2, show no signs of tuft disturbance on the suction surface throughout the transition. In contrast, all upwards orientated front motor configurations display signs of wake interaction.

Based on the prior work discussed in Section~\ref{section:prior_work_and_motivation}, the following expectations guided this experimental campaign:

\begin{enumerate}
    \item Configurations with the front motor orientated downwards and the rear motor orientated upwards (\textcolor{FDRU}{FDRU}) would generate the largest increases in lift due the combined jet flap effect and the axial displacement of the front motor below the wing's surface - minimizing the lift reducing tendency of the mean flow in the front propeller's wake.
    \item  Configurations with the front motor orientated upwards and the rear downwards (\textcolor{FURD}{FURD}) were expected to produce the worst lifting performance for the inverse of the reasoning presented in 1.
    \item The lifting performance of the remaining configurations were expected to lie somewhere in between the \textcolor{FURD}{FURD} and \textcolor{FDRU}{FDRU} configurations.
    \item Reducing the longitudinal separation between the motors was expected to preserve the above ranking, but to intensify the lift inducing properties of the rear propeller and the lift reducing tendencies of the front propeller.
\end{enumerate}

These expectations were generally confirmed by the experimental data, apart from the \textcolor{FURU}{FURU} B2 configuration which outperformed the \textcolor{FDRU}{FDRU} B2 configuration in terms of peak ${\Delta C_{L}}/{\Delta C_{D}}$ values. A key difference from earlier studies on tandem lifting fans, which formed much of the initial understanding for this investigation, is the identification of super-vortices generated by the front vertical lift propeller that travel across the suction surface of the wing. In these previous studies, the lifting fans were located in the same vertical plane as the wing or below it and were enclosed within shrouds - conditions likely to suppress the propagation of tip vortices toward the downstream wing \cite{hickey1966aerodynamic}. The formation of a vortex pair due to the rolling up of jet exhaust downstream of the nozzle exit has also been described in studies of jet-based vertical lift systems \cite{carter1969effects}. 
Although, in this study, all jet positions in front of the wing had a negative impact on wing performance, suggesting a different set of interactions occurred.

It is evident that the wake of the front vertical lift propeller interacts with the wing's suction surface in the \textcolor{FURU}{FURU} B2 configuration, but does not significantly in the \textcolor{FDRU}{FDRU} B2 configuration (Table~\ref{tab:Features_present}).

\begin{figure}[h!]
    \centering
    \includegraphics[width=\linewidth]{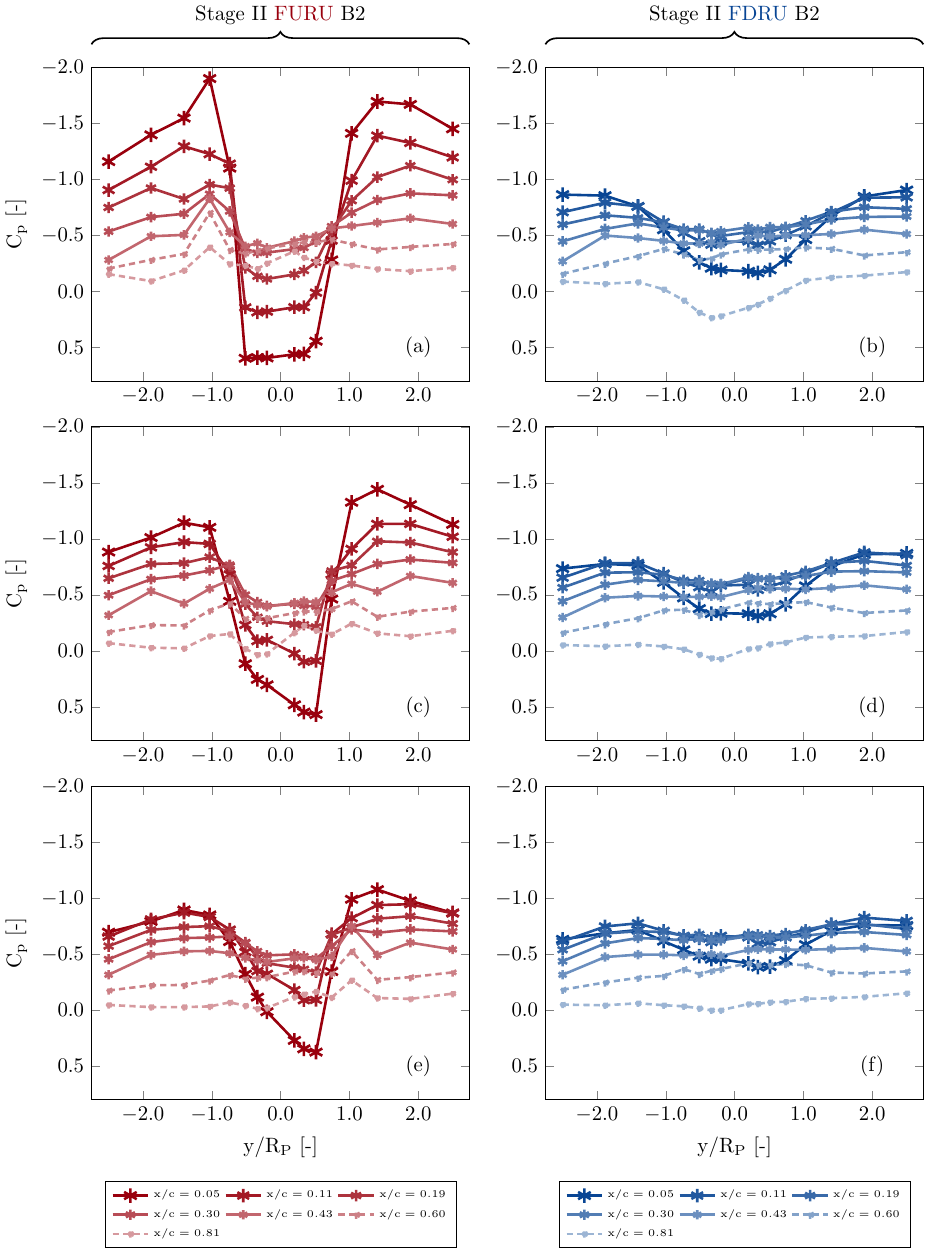}
    \vskip 10pt
    \begin{tabular}{@{} l l l l l l@{}}
        (a) $\mu = 0.11$  & (b) $\mu = 0.11$  & (c) $\mu = 0.16$ & (d) $\mu = 0.17$ &(e) $\mu = 0.25$ & (f) $\mu = 0.24$ \\
    \end{tabular}
    \caption{Wing suction surface $C_{p}$ distributions plotted over the wing span and chord during stage II of the transition for the \textcolor{FURU}{FURU} \& \textcolor{FDRU}{FDRU} B2 configurations. Sub-figures are labeled with their respective front propeller advance ratio ($\mu$).}
    \label{fig:cp_vs_span_FURU_FDRU}
\end{figure}

The spanwise variation in suction surface pressure coefficient (Figure~\ref{fig:cp_vs_span_FURU_FDRU}) reveals that the \textcolor{FURU}{FURU} B2 configuration exhibits patterns similar to the streamwise vortices over a two-dimensional wing discussed in \cite{patel1974some}. In the current study, the spanwise $C_{p}$ variation appears as a mirrored and slightly asymmetric version of the result presented in \cite{patel1974some}. This asymmetry is attributed to the presence of two contra-rotating super-vortices in the front propeller's wake, with the vortex on the advancing blade side generating stronger downwash due to the additional translational lift that the advancing blade experiences.

The \textcolor{FURU}{FURU} B2 configuration displays increased suction between $\pm1$–$2.5 \ \mathrm{y/R_{P}}$ and regions of positive pressure between $\pm1.0\ \mathrm{y/R_{P}}$, consistent with the up wash and downwash induced by the rotating vortices. These effects are most prominent near the leading edge of the wing, decaying at larger chord-wise positions. A bias toward the positive span (advancing blade) positions is generally evident, resulting from the increased strength of the vortex generated on that side.

The \textcolor{FDRU}{FDRU} B2 configuration shows far weaker interactions with the front propeller wake, as the vortices in this configuration are vertically displaced below the wing. Nonetheless, a slight reduction in $C_{p}$ between $\pm1.0 \ \mathrm{y/R_{P}}$ is still observed, likely caused by the downwash between the contra-rotating vortices, as seen in Figure~\ref{fig:cp_vs_span_FURU_FDRU}. Additionally, an area of positive pressure is observed at the wing's trailing edge, also spanning $\pm1.0 \ \mathrm{y/R_{P}}$, which results from rear propeller's wake impingement on the suction surface.

At $\mu = 0.11$, a distinct suction peak is visible at $\mathrm{y/R_{P}} = -1.0$ (Figure~\ref{fig:cp_vs_span_FURU_FDRU}a) for the \textcolor{FURU}{FURU} B2 configuration. This peak is generated by the retreating blade’s vortex core as it travels tangentially, in the streamwise direction over the wing's chord and closely resembles observations in prior work \cite{patel1974some}. As the transition progresses through stage II, the suction peak from the advancing blade's vortex core emerges at $\mathrm{y/R_{P}} = 1.0$, while the influence of the retreating blade’s vortex core diminishes. This sequence of vortex interactions aligns with the tuft disturbance evolution described earlier in Figure~\ref{fig:FURU_short_tuft}.

The axial displacement of the front propeller above the wing in the \textcolor{FURU}{FURU} B2 configuration appears to enhance aerodynamic performance by facilitating beneficial interaction between the front propeller's super-vortices and the wing's suction surface. This effect is not replicated in the \textcolor{FDRU}{FDRU} B2 configuration, where the front propeller's wake passes beneath the wing. While the pressure and tuft measurements do not include data from the pressure side of the wing, and thus cannot confirm the state of the flow on that surface, it is clear that the suction side interactions in the \textcolor{FURU}{FURU} B2 configuration are sufficient to yield higher ${\Delta C_{L}}/{\Delta C_{D}}$ values than the \textcolor{FDRU}{FDRU} B2 configuration during stages II and III.

\FloatBarrier

\section{Conclusion}
This study has successfully developed, tested, and analyzed a representative \lc wing design with a range of common motor configurations. The transition profile was designed to reflect the weight and thrust requirements of a realistic UAV, approximating a constant-altitude, constant-attitude trajectory from hover to forward flight. A novel aspect of this work is the active modulation of front and rear motor throttles to achieve zero excess lift and zero net pitching moment at increasing airspeeds. This feature of the experimental campaign, not evident in prior studies, has enabled the identification of a unique and configuration-sensitive relationship between aerodynamic performance and the front propeller’s advance ratio.

Analysis of suction surface pressure and tuft data revealed how the mean flow and trailing super-vortices from the front propeller influence the wing’s lift production. Between $\pm 1.0 \ \mathrm{y/R_{P}}$, downwash reduces lift, while outboard up wash enhances it by strengthening negative surface pressures. In configurations where the front motor is mounted above the wing plane, the low-pressure vortex cores additionally reinforce suction directly beneath their path - most notably in the shorter boom case. Configurations with upwards-oriented rear motors consistently outperformed those with downwards-oriented ones in terms of ${\Delta C_{L}}/{\Delta C_{D}}$, with only the former achieving positive values across the transition. However, in the shorter boom configurations, flow impingement effects from upwards orientated rear propellers appear to induce adverse pressure gradients, potentially leading to flow separation near the trailing edge. In future designs, placing a trailing-edge flap on the wing below the rear propeller could prevent the formation of an adverse pressure gradient, whilst further increasing wing lift during the transition through jet-flap effects.

Decreasing the separation distance between the motors by switching between the long and short boom lengths intensified existing aerodynamic trends by moving the propeller wakes into closer proximity with the wing. The \textcolor{FURU}{FURU} configuration at the shortest boom length achieved the highest ${\Delta C_{L}}/{\Delta C_{D}}$ during transition by leveraging vortex-induced lift enhancement, that was not present in the \textcolor{FDRU}{FDRU} configuration at the same boom length - the latter also suffered a drag penalty due to its split motor orientations. While improved transition performance is a priority, holistic VTOL aircraft design will also require consideration of factors such as noise, mechanical complexity, and forward flight efficiency when the propellers are stationary. In addition, the performance advantages of the \textcolor{FURU}{FURU} configuration with the short boom length appears tied to interactions between the front propeller’s trailing super-vortices and the wing. Prior studies suggest that such vortex systems could increase cyclic loads on downstream structures, possibly impacting downstream high-lift devices or control surfaces. These additional cyclic loads might necessitate a heavier structure, stiffer actuators, or more frequent inspections - factors that may offset the aerodynamic gains observed in transition in favor of the consistently positive performance of the \textcolor{FDRU}{FDRU} B2 configuration.

The results were segmented into three stages of advance ratio to capture distinct aerodynamic behavior during the transition. Of these, stage II emerged as the most critical, containing the majority of the transient aerodynamic behavior. The relationship between the maximum lift-drag ratio induced by the active vertical lift propellers and the advance ratio at which it occurs may be a quantity of interest for VTOL aircraft designers and operators in the future. At present, when the airspace is relatively clear and the advent of large urban air mobility networks is still imminent, transition profiles for VTOL vehicles between hover and forward flight may be purely dictated by the operator, regulator and airspace controller as required. However, in the future, it may be that noise restrictions and congestion around airports, urban air mobility hubs and delivery/transport UAV operations requires the imposition of more prescriptive flight profiles for the beginning and end of VTOL operations. These new profiles may be similar in nature to the patterns flown by conventional aircraft approaching or departing airports and may require the aircraft to operate at airspeeds that necessitate the use of the vertical lifting motors for lift and or control. In these scenarios the advance ratios contained within stage II could be considered in an analogous way to the best range and minimum power speeds referenced by operators of conventional aircraft to improve performance.

In the motor configurations tested here, varying the rear motor orientation between its upright and inverted positions, a vertical displacement of $0.47~D_{P}$ or 143 \si{\milli\meter}, resulted in a change in the lift to drag ratio of
approximately 8 during early stage II transitions. This substantial variation underscores that even modest changes in motor placement may result in relatively large impacts on aerodynamic performance. Accordingly, careful consideration of vertical lift motor placement relative to the wing should be made in the early stages of transitioning VTOL aircraft development.

\section*{ACKNOWLEDGEMENTS}
This research was supported by the Soton UAV Triple Helix Unit (THU) at the University of Southampton. The authors would also like to thank the 7x5 wind tunnel team, experimental officers and members of the ISVR for their expertise and advice.

\bibliographystyle{apalike}
\bibliography{citations}


\appendix

\setcounter{table}{0}
\renewcommand{\thetable}{A\arabic{table}}
\section{UAV Sizing}
\begin{table}[ht]
	\caption{Mass budget for an example \lc UAV.}
	\label{tab:full_aircraft_model_characteristics}
	\centering
        \resizebox{0.8\textwidth}{!}{
	\begin{tabular}{c l c}
		\hline
		\hline
		Quantity & Item &  Total mass [\si{\kilogram}] \\
		\hline
        4 & T-motor U3 700 kV & 0.51 \\
        4 & T-motor P12x4 & 0.06 \\
        5 & T-motor AT 55 A  & 0.20 \\
        1 & T-motor 3520 850 kV & 0.22 \\
        1 & APC 13x6.5 propeller  & 0.03 \\
        1 & 6250 mAh VTOL 4S battery  & 0.67 \\
        1 & 5800 mAh forward flight 4S battery  & 0.52 \\
        1 & Airspeed sensor  & 0.01 \\
        1 & Cube orange flight controller  & 0.07 \\
        1 & Here 3+ GPS  & 0.05 \\
        1 & RC receiver  & 0.02 \\
         & Payload  & 0.4 (10\% aircraft mass) \\
         & Ancillary mass, wiring  & 0.4 (10\% aircraft mass) \\
         & Structural mass  & 1.1 (24\% aircraft mass) \\
        \hline
        Total aircraft mass: & & 4.0 \\
		\hline
        \hline
	\end{tabular}}
\end{table}
\section{Supplementary configuration data} \label{section:additional_configs}

\subsection{B1 configurations}

\begin{figure}[h!]
  \centering

  \begin{subfigure}[b]{\textwidth}
    \centering
    \includegraphics[width=\linewidth]{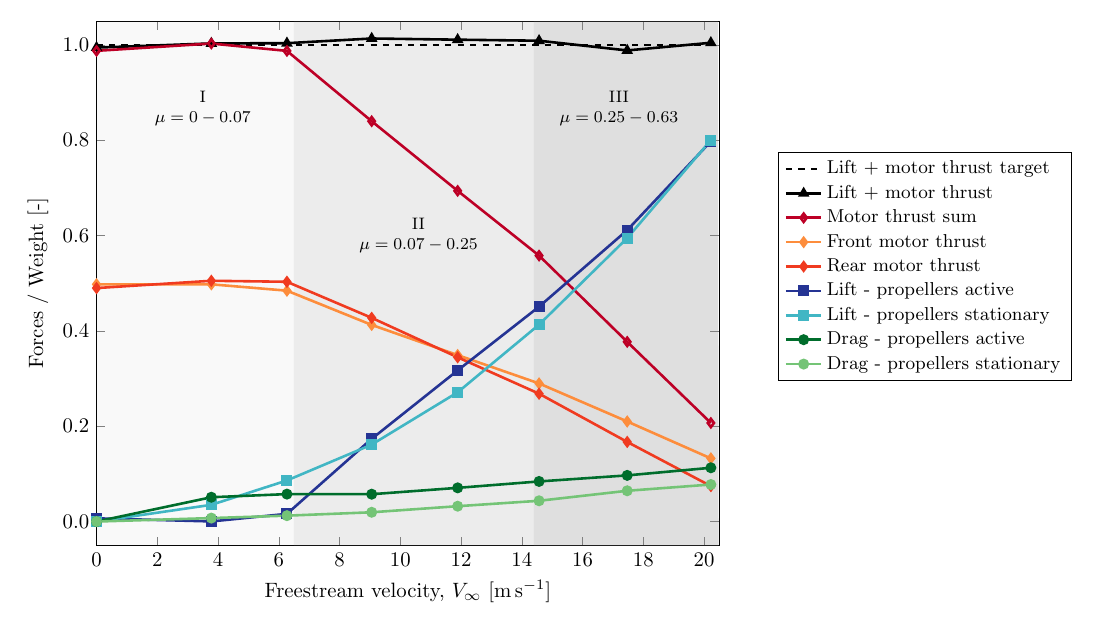}
        \caption{Forces non-dimensionalized by aircraft weight versus freestream velocity during the transition of the \textcolor{Reds-L}{FURU} B1. Here $\mu = V_{\infty}/n_{F}R_{P}$ is the advance ratio of the front propeller.}
        \label{fig:FURU_n transition summary}
  \end{subfigure}

  \vspace{1em}

  \begin{subfigure}[b]{\textwidth}
    \centering
        	\begin{tabular}{l c c c c c c c c}
        		\hline
        		\hline
        		Freestream velocity, $V_{\infty}$ [\si{\meter\per\second}] & 0.0 & 3.8 & 6.3 & 9.1 & 11.9 & 14.6 & 17.5 & 20.2  \\
                \hline
        		Front propeller RPM, $n_{F}$ [RPM] & 6440 & 6380 & 6105 & 5365 & 4580 & 3850 & 3005 & 2010 \\
                Rear propeller RPM, $n_{R}$ [RPM]  & 6440 & 6515 & 6320 & 5770 & 5130 & 4235 & 3165 & 1685 \\
        		\hline
        		\hline
        	\end{tabular}
    \caption{\textcolor{Reds-L}{FURU} B1 front and rear propeller rotation rates during the transition.}
    \label{tab:FURU_B1_RPMS}
  \end{subfigure}

  \caption{\textcolor{Reds-L}{FURU} B1 transition summary.}
  \label{fig:FURU_B1_transition_summary_with_RPM_table}
\end{figure}

\begin{figure}[h!]
        \centering
        \begin{subfigure}[t]{\textwidth}
            \centering
                \includegraphics[width=\linewidth]{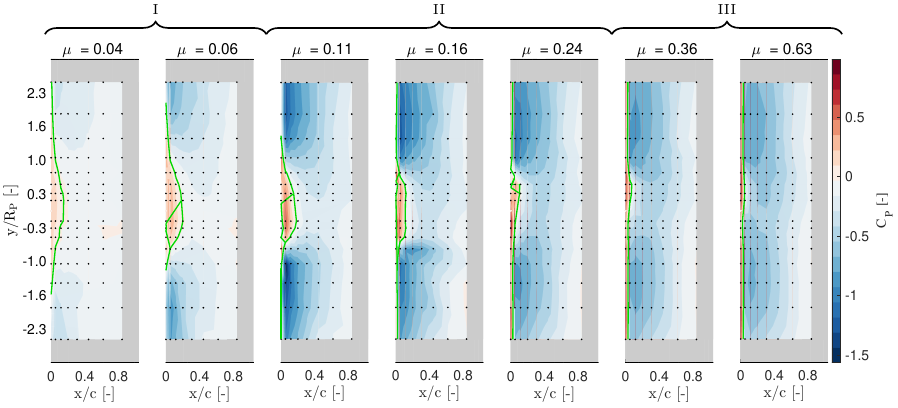}
            \caption[Suction wing surface pressure coefficient plots during the transition of the \textcolor{Reds-L}{FURU} B1 configuration.]{Suction wing surface pressure coefficient plots during the transition of the \textcolor{Reds-L}{FURU} B1 configuration, where \protect\markerone \ markings denote the location of a pressure tap and contours marked with \protect\greenline \ denote the $C_{P} = 0$ crossing location.}
            \label{fig:FURU_normal_cp}
            \end{subfigure}
        \begin{subfigure}[t]{\textwidth}
            \centering
            \includegraphics[width=\linewidth]{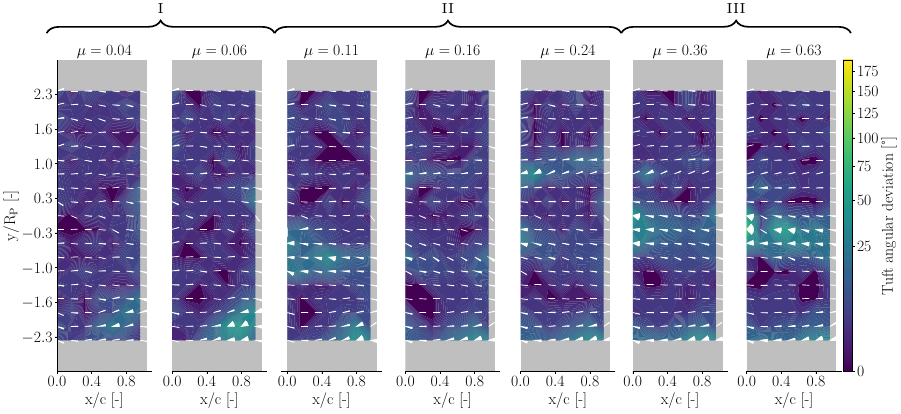}
            \caption{Suction wing surface tuft disturbances during the transition of the \textcolor{Reds-L}{FURU} B1 configuration.}
            \label{fig:FURU_normal_tuft}
        \end{subfigure}
        \caption{Pressure coefficient and tuft disturbance plots during the transition of the \textcolor{Reds-L}{FURU} B1 configuration.}
        \label{fig:FURU_n_comp}
\end{figure}

\begin{figure}[h!]
  \centering

  \begin{subfigure}[b]{\textwidth}
    \centering
    \includegraphics[width=\linewidth]{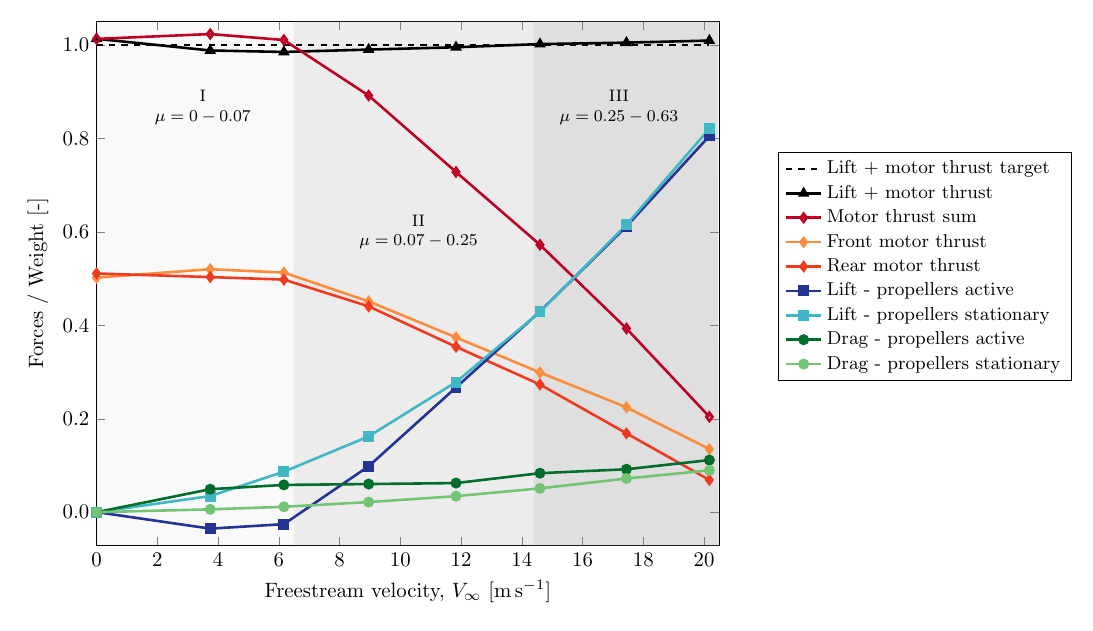}
        \caption{Forces non-dimensionalized by aircraft weight versus freestream velocity during the transition of the \textcolor{Reds-H}{FURD} B1 configuration. Here $\mu = V_{\infty}/n_{F}R_{P}$ is the advance ratio of the front propeller.}
        \label{fig:FURD_n transition summary}
  \end{subfigure}

  \vspace{1em}

  \begin{subfigure}[b]{\textwidth}
    \centering
        	\begin{tabular}{l c c c c c c c c}
        		\hline
        		\hline
        		Freestream velocity, $V_{\infty}$ [\si{\meter\per\second}] & 0.0 & 3.7 & 6.2 & 9.0 & 11.8 & 14.6 & 17.4 & 20.2  \\
                \hline
        		Front propeller RPM, $n_{F}$ [RPM] & 6430 & 6350 & 6250 & 5615 & 4765 & 3910 & 3120 & 2035 \\
                Rear propeller RPM, $n_{R}$ [RPM]  & 6430 & 6585 & 6460 & 6195 & 5310 & 4015 & 3045 & 1520 \\
        		\hline
        		\hline
        	\end{tabular}
    \caption{\textcolor{Reds-H}{FURD} B1 front and rear propeller rotation rates during the transition.}
    \label{tab:FURD_B1_RPMS}
  \end{subfigure}

  \caption{\textcolor{Reds-H}{FURD} B1 transition summary.}
  \label{fig:FURD_B1_transition_summary_with_RPM_table}
\end{figure}

\begin{figure}[h!]
        \centering
        \begin{subfigure}[t]{\textwidth}
            \centering
                \includegraphics[width=\linewidth]{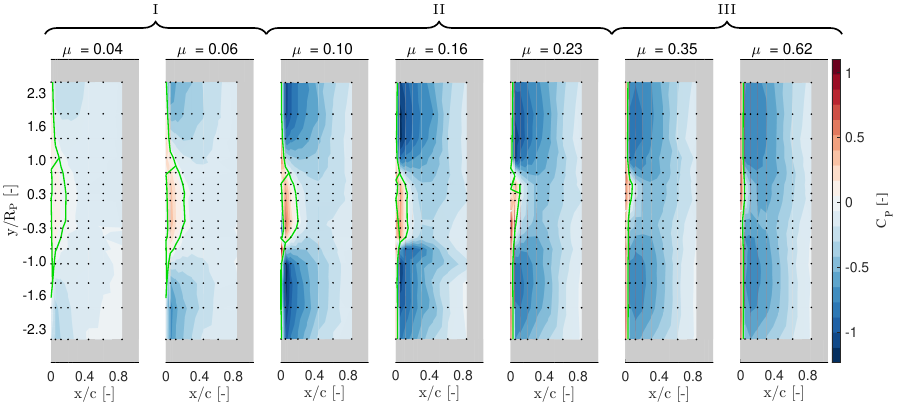}
            \caption[Suction wing surface pressure coefficient plots during the transition of the \textcolor{Reds-H}{FURD} B1 configuration.]{Suction wing surface pressure coefficient plots during the transition of the \textcolor{Reds-H}{FURD} B1 configuration, where \protect\markerone \ markings denote the location of a pressure tap and contours marked with \protect\greenline \ denote the $C_{P} = 0$ crossing location.}
            \label{fig:FURD_normal_cp}
            \end{subfigure}
        \begin{subfigure}[t]{\textwidth}
            \centering
            \includegraphics[width=\linewidth]{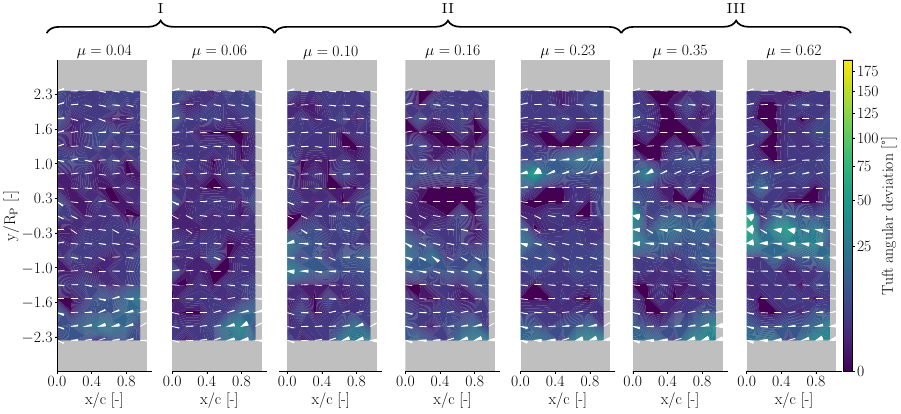}
            \caption{Suction wing surface tuft disturbances during the transition of the \textcolor{Reds-H}{FURD} B1 configuration.}
            \label{fig:FURD_normal_tuft}
        \end{subfigure}
        \caption{Pressure coefficient and tuft disturbance plots during the transition of the \textcolor{Reds-H}{FURD} B1 configuration.}
        \label{fig:FURD_n_comp}
\end{figure}

\begin{figure}[h!]
  \centering

  \begin{subfigure}[b]{\textwidth}
    \centering
    \includegraphics[width=\linewidth]{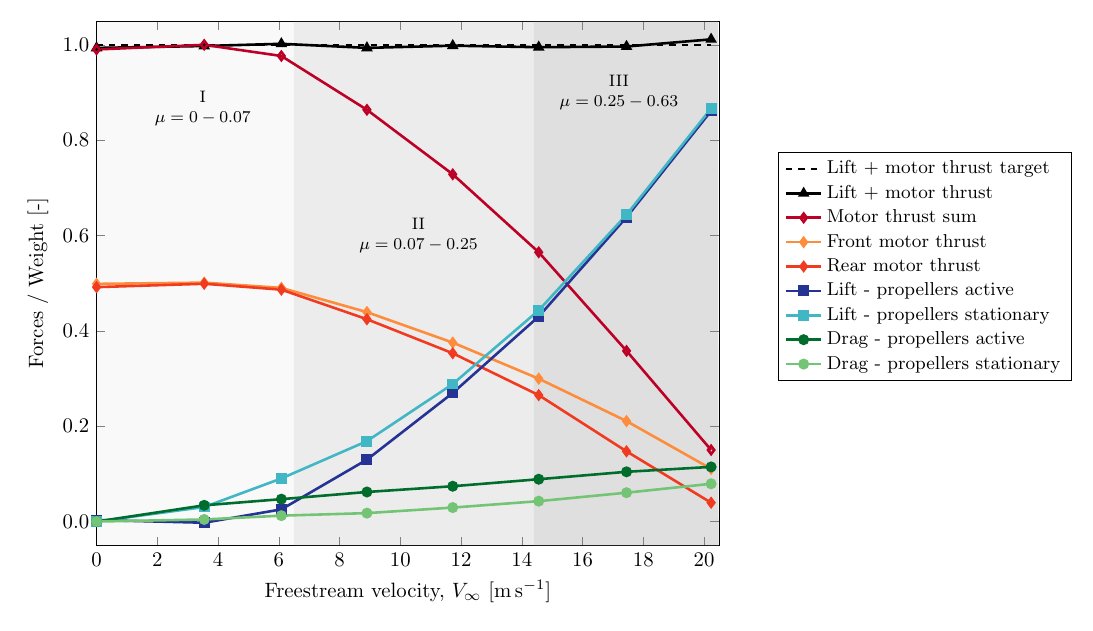}
        \caption{Forces non-dimensionalized by aircraft weight versus freestream velocity during the transition of the \textcolor{Blues-H}{FDRD} B1 configuration. Here $\mu = V_{\infty}/n_{F}R_{P}$ is the advance ratio of the front propeller.}
        \label{fig:FDRD_n transition summary}
  \end{subfigure}

  \vspace{1em}

  \begin{subfigure}[b]{\textwidth}
    \centering
        	\begin{tabular}{l c c c c c c c c}
        		\hline
        		\hline
        		Freestream velocity, $V_{\infty}$ [\si{\meter\per\second}] & 0.0 & 3.5 & 6.1 & 8.9 & 11.7 & 14.6 & 17.4 & 20.2  \\
                \hline
        		Front propeller RPM, $n_{F}$ [RPM] & 6465 & 6405 & 6295 & 5745 & 5010 & 4170 & 3285 & 2180 \\
                Rear propeller RPM, $n_{R}$ [RPM]  & 6465 & 6495 & 6395 & 5920 & 5350 & 4420 & 3100 & 1065 \\
        		\hline
        		\hline
        	\end{tabular}
    \caption{\textcolor{Blues-H}{FDRD} B1 front and rear propeller rotation rates during the transition.}
    \label{tab:FDRD_B1_RPMS}
  \end{subfigure}

  \caption{\textcolor{Blues-H}{FDRD} B1 transition summary.}
  \label{fig:FDRD_B1_transition_summary_with_RPM_table}
\end{figure}

\begin{figure}[h!]
        \centering
        \begin{subfigure}[t]{\textwidth}
            \centering
                \includegraphics[width=\linewidth]{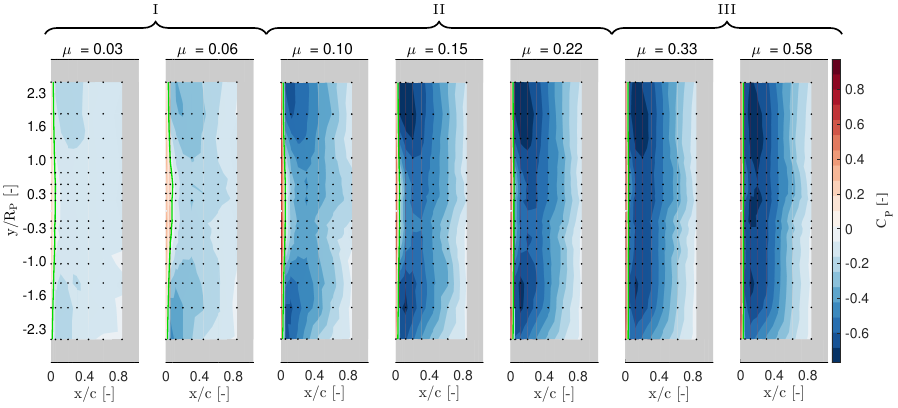}
            \caption[Suction wing surface pressure coefficient plots during the transition of the \textcolor{Blues-H}{FDRD} B1 configuration.]{Suction wing surface pressure coefficient plots during the transition of the \textcolor{Blues-H}{FDRD} B1 configuration, where \protect\markerone \ markings denote the location of a pressure tap and contours marked with \protect\greenline \ denote the $C_{P} = 0$ crossing location.}
            \label{fig:FDRD_normal_cp}
            \end{subfigure}
        \begin{subfigure}[t]{\textwidth}
            \centering
            \includegraphics[width=\linewidth]{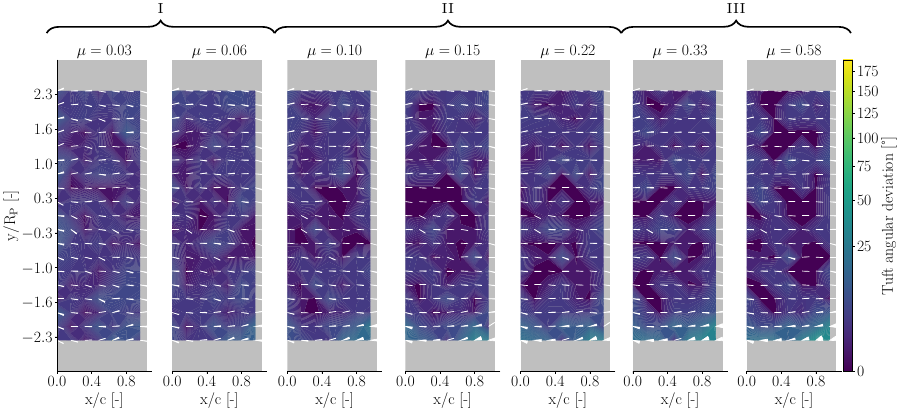}
            \caption{Suction wing surface tuft disturbances during the transition of the \textcolor{Blues-H}{FDRD} B1 B1 configuration.}
            \label{fig:FDRD_normal_tuft}
        \end{subfigure}
        \caption{Pressure coefficient and tuft disturbance plots during the transition of the \textcolor{Blues-H}{FDRD} B1 configuration.}
        \label{fig:FDRD_n_comp}
\end{figure}

\begin{figure}[h!]
  \centering

  \begin{subfigure}[b]{\textwidth}
    \centering
    \includegraphics[width=\linewidth]{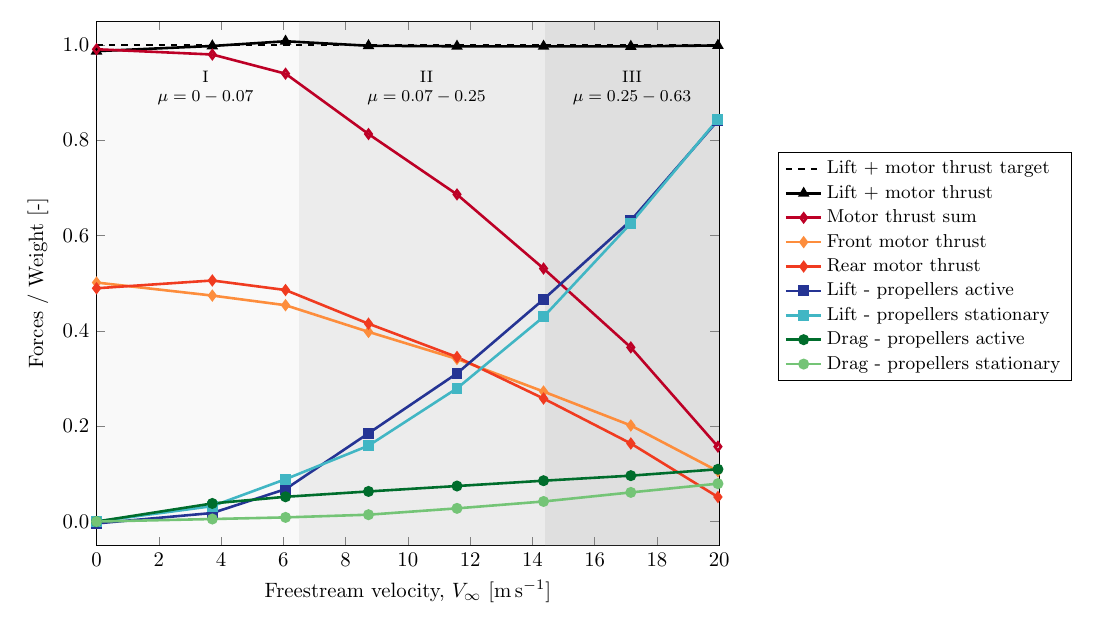}
        \caption{Forces non-dimensionalized by aircraft weight versus freestream velocity during the transition of the \textcolor{Blues-L}{FDRU} B1 configuration. Here $\mu = V_{\infty}/n_{F}R_{P}$ is the advance ratio of the front propeller.}
        \label{fig:FDRU_n transition summary}
  \end{subfigure}

  \vspace{1em}

  \begin{subfigure}[b]{\textwidth}
    \centering
        	\begin{tabular}{l c c c c c c c c}
        		\hline
        		\hline
        		Freestream velocity, $V_{\infty}$ [\si{\meter\per\second}] & 0.0 & 3.7 & 6.1 & 8.7 & 11.6 & 14.4 & 17.2 & 20.0  \\
                \hline
        		Front propeller RPM, $n_{F}$ [RPM] & 6465 & 6250 & 6055 & 5440 & 4750 & 4015 & 3230 & 2150 \\
                Rear propeller RPM, $n_{R}$ [RPM]  & 6435 & 6355 & 6195 & 5640 & 4910 & 4000 & 2840 & 985 \\
        		\hline
        		\hline
        	\end{tabular}
    \caption{\textcolor{Blues-L}{FDRU} B1 front and rear propeller rotation rates during the transition.}
    \label{tab:FDRU_B1_RPMS}
  \end{subfigure}

  \caption{\textcolor{Blues-L}{FDRU} B1 transition summary.}
  \label{fig:FDRU_B1_transition_summary_with_RPM_table}
\end{figure}

\begin{figure}[h!]
        \centering
        \begin{subfigure}[t]{\textwidth}
            \centering
                \includegraphics[width=\linewidth]{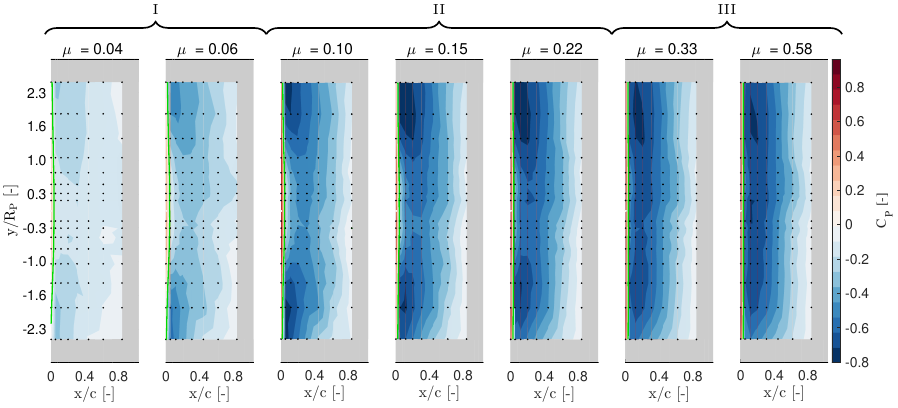}
            \caption[Suction wing surface pressure coefficient plots during the transition of the \textcolor{Blues-L}{FDRU} B1 configuration.]{Suction wing surface pressure coefficient plots during the transition of the \textcolor{Blues-L}{FDRU} B1 configuration, where \protect\markerone \ markings denote the location of a pressure tap and contours marked with \protect\greenline \ denote the $C_{P} = 0$ crossing location.}
            \label{fig:FDRU_normal_cp}
            \end{subfigure}
        \begin{subfigure}[t]{\textwidth}
            \centering
            \includegraphics[width=\linewidth]{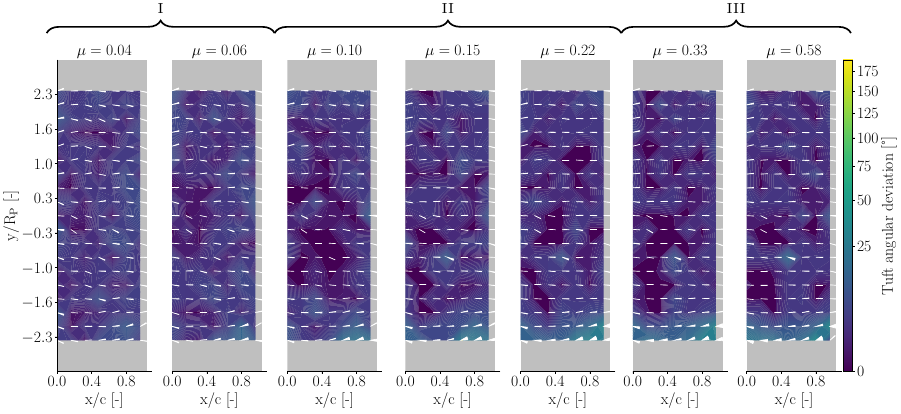}
            \caption{Suction wing surface tuft disturbances during the transition of the \textcolor{Blues-L}{FDRU} B1 configuration.}
            \label{fig:FDRU_normal_tuft}
        \end{subfigure}
        \caption{Pressure coefficient and tuft disturbance plots during the transition of the \textcolor{Blues-L}{FDRU} B1 configuration.}
        \label{fig:FDRU_n_comp}
\end{figure}

\clearpage

\subsection{B2 configurations}

\begin{figure}[h!]
  \centering

  \begin{subfigure}[b]{\textwidth}
    \centering
    \includegraphics[width=\linewidth]{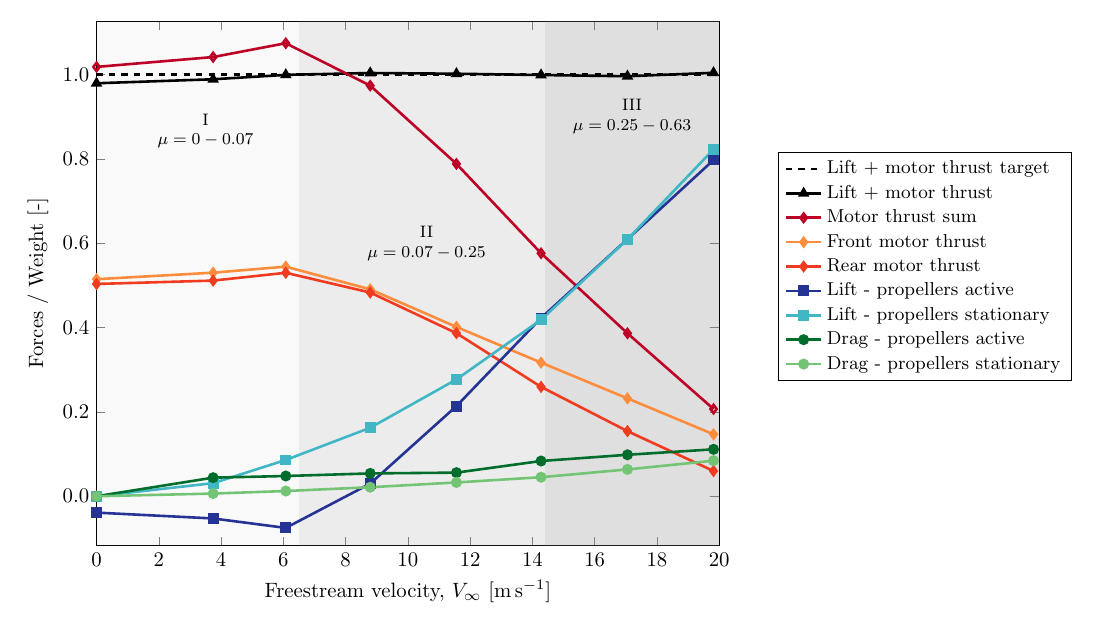}
        \caption{Forces non-dimensionalized by aircraft weight versus freestream velocity during the transition of the \textcolor{Reds-H}{FURD} B2 configuration. Here $\mu = V_{\infty}/n_{F}R_{P}$ is the advance ratio of the front propeller.}
        \label{fig:FURD_s transition summary}
  \end{subfigure}

  \vspace{1em}

  \begin{subfigure}[b]{\textwidth}
    \centering
        	\begin{tabular}{l c c c c c c c c}
        		\hline
        		\hline
        		Freestream velocity, $V_{\infty}$ [\si{\meter\per\second}] & 0.0 & 3.7 & 6.1 & 8.8 & 11.6 & 14.3 & 17.1 & 19.8  \\
                \hline
        		Front propeller RPM, $n_{F}$ [RPM] & 6635 & 6560 & 6585 & 6005 & 5030 & 4100 & 3230 & 2175 \\
                Rear propeller RPM, $n_{R}$ [RPM]  & 6610 & 6820 & 6875 & 6760 & 5575 & 4230 & 3060 & 1385 \\
        		\hline
        		\hline
        	\end{tabular}
    \caption{\textcolor{Reds-H}{FURD} B2 front and rear propeller rotation rates during the transition.}
    \label{tab:FURD_B2_RPMS}
  \end{subfigure}

  \caption{\textcolor{Reds-H}{FURD} B2 transition summary.}
  \label{fig:FURD_B2_transition_summary_with_RPM_table}
\end{figure}

\begin{figure}[h!]
        \centering
        \begin{subfigure}[t]{\textwidth}
            \centering
                \includegraphics[width=\linewidth]{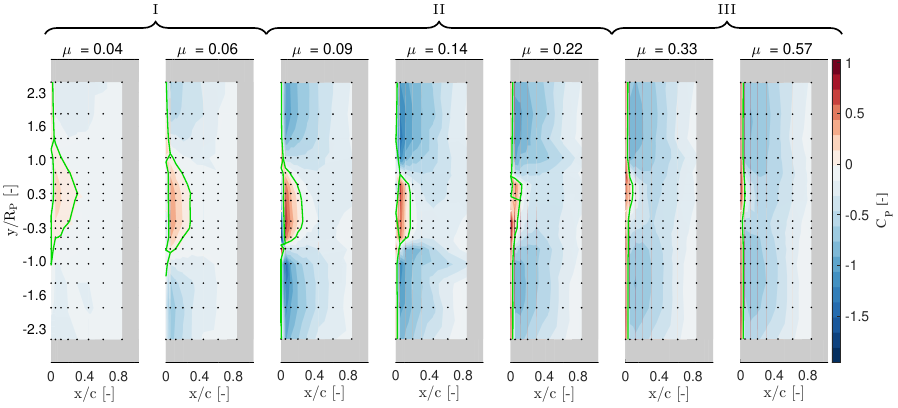}
            \caption[Suction wing surface pressure coefficient plots during the transition of the \textcolor{Reds-H}{FURD} B2 configuration.]{Suction wing surface pressure coefficient plots during the transition of the \textcolor{Reds-H}{FURD} B2 configuration, where \protect\markerone \ markings denote the location of a pressure tap and contours marked with \protect\greenline \ denote the $C_{P} = 0$ crossing location.}
            \label{fig:FURD_short_cp}
            \end{subfigure}
        \begin{subfigure}[t]{\textwidth}
            \centering
            \includegraphics[width=\linewidth]{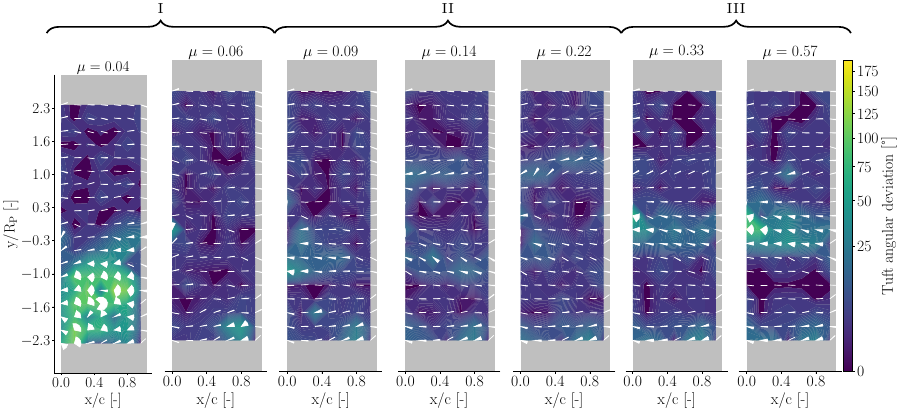}
            \caption{Suction wing surface tuft disturbances during the transition of the \textcolor{Reds-H}{FURD} B2 configuration.}
            \label{fig:FURD_short_tuft}
        \end{subfigure}
        \caption{Pressure coefficient and tuft disturbance plots during the transition of the \textcolor{Reds-H}{FURD} B2 configuration.}
        \label{fig:FURD_s_comp}
\end{figure}

\begin{figure}[h!]
  \centering

  \begin{subfigure}[b]{\textwidth}
    \centering
        \includegraphics[width=\linewidth]{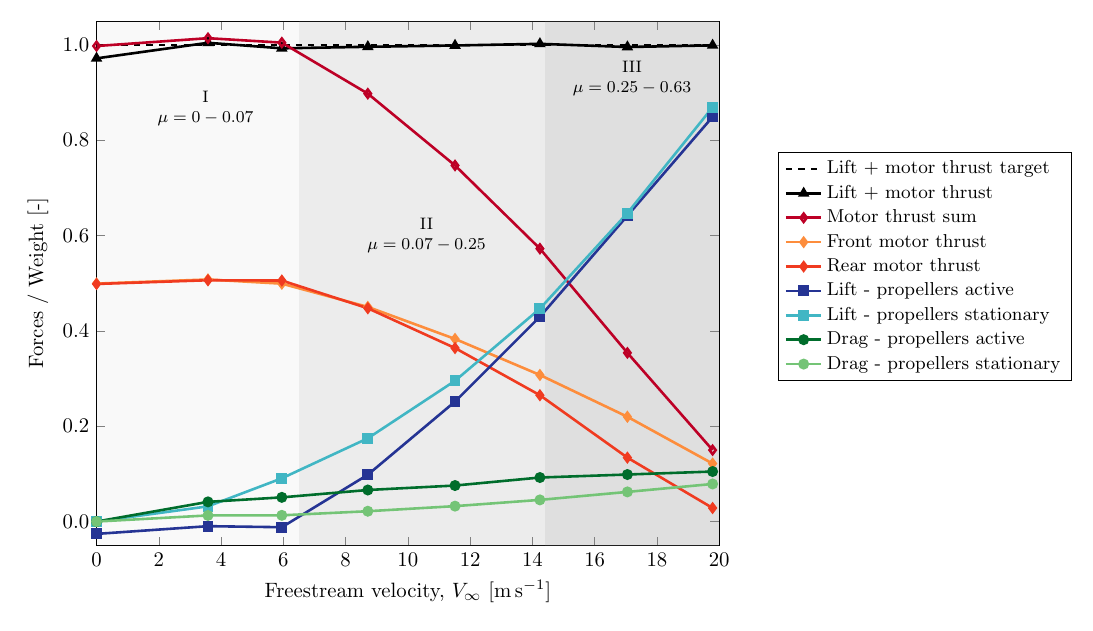}
        \caption{Forces non-dimensionalized by aircraft weight versus freestream velocity during the transition of the \textcolor{Blues-H}{FDRD} B2 configuration. Here $\mu = V_{\infty}/n_{F}R_{P}$ is the advance ratio of the front propeller.}
        \label{fig:FDRD_s transition summary}
  \end{subfigure}

  \vspace{1em}

  \begin{subfigure}[b]{\textwidth}
    \centering
        	\begin{tabular}{l c c c c c c c c}
        		\hline
        		\hline
        		Freestream velocity, $V_{\infty}$ [\si{\meter\per\second}] & 0.0 & 3.6 & 6.0 & 8.7 & 11.5 & 14.2 & 17.1 & 19.8  \\
                \hline
        		Front propeller RPM, $n_{F}$ [RPM] & 6745 & 6545 & 6475 & 5970 & 5185 & 4385 & 3490 & 2420 \\
                Rear propeller RPM, $n_{R}$ [RPM]  & 6665 & 6670 & 6575 & 6180 & 5590 & 4655 & 3180 & 860 \\
        		\hline
        		\hline
        	\end{tabular}
    \caption{\textcolor{Blues-H}{FDRD} B2 front and rear propeller rotation rates during the transition.}
    \label{tab:FDRD_B2_RPMS}
  \end{subfigure}

  \caption{\textcolor{Blues-H}{FDRD} B2 transition summary.}
  \label{fig:FDRD_B2_transition_summary_with_RPM_table}
\end{figure}

\begin{figure}[h!]
        \centering
        \begin{subfigure}[t]{\textwidth}
            \centering
            \includegraphics[width=\linewidth]{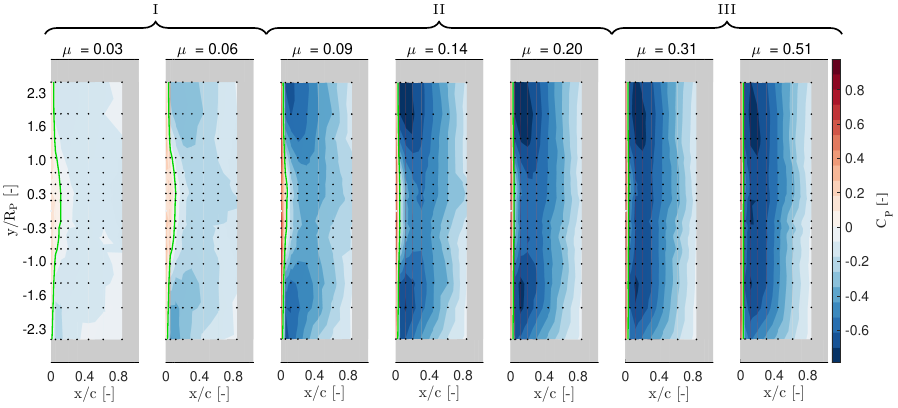}
            \caption[Suction wing surface pressure coefficient plots during the transition of the \textcolor{Blues-H}{FDRD} B2 configuration.]{Suction wing surface pressure coefficient plots during the transition of the \textcolor{Blues-H}{FDRD} B2 configuration, where \protect\markerone \ markings denote the location of a pressure tap and contours marked with \protect\greenline \ denote the $C_{P} = 0$ crossing location.}
            \label{fig:FDRD_short_cp}
            \end{subfigure}
        \begin{subfigure}[t]{\textwidth}
            \centering
            \includegraphics[width=\linewidth]{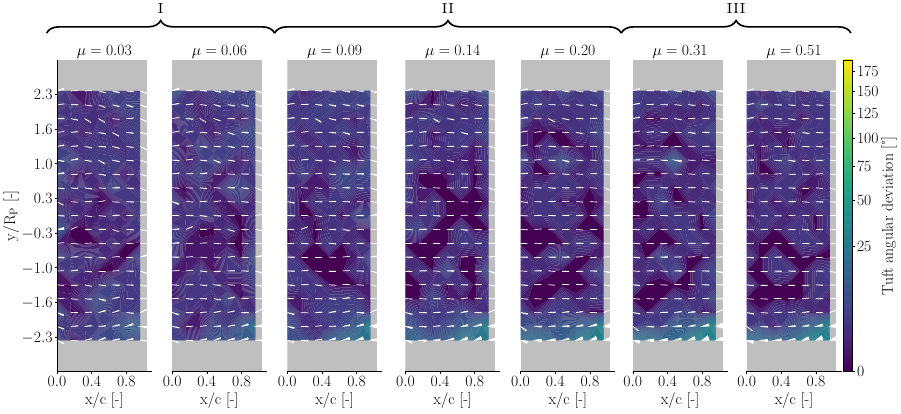}
            \caption{Suction wing surface tuft disturbances during the transition of the \textcolor{Blues-H}{FDRD} B2 configuration.}
            \label{fig:FDRD_short_tuft}
        \end{subfigure}
        \caption{Pressure coefficient and tuft disturbance plots during the transition of the \textcolor{Blues-H}{FDRD} B2 configuration.}
        \label{fig:FDRD_s_comp}
\end{figure}

\begin{figure}[h!]
  \centering

  \begin{subfigure}[b]{\textwidth}
    \centering
        \includegraphics[width=\linewidth]{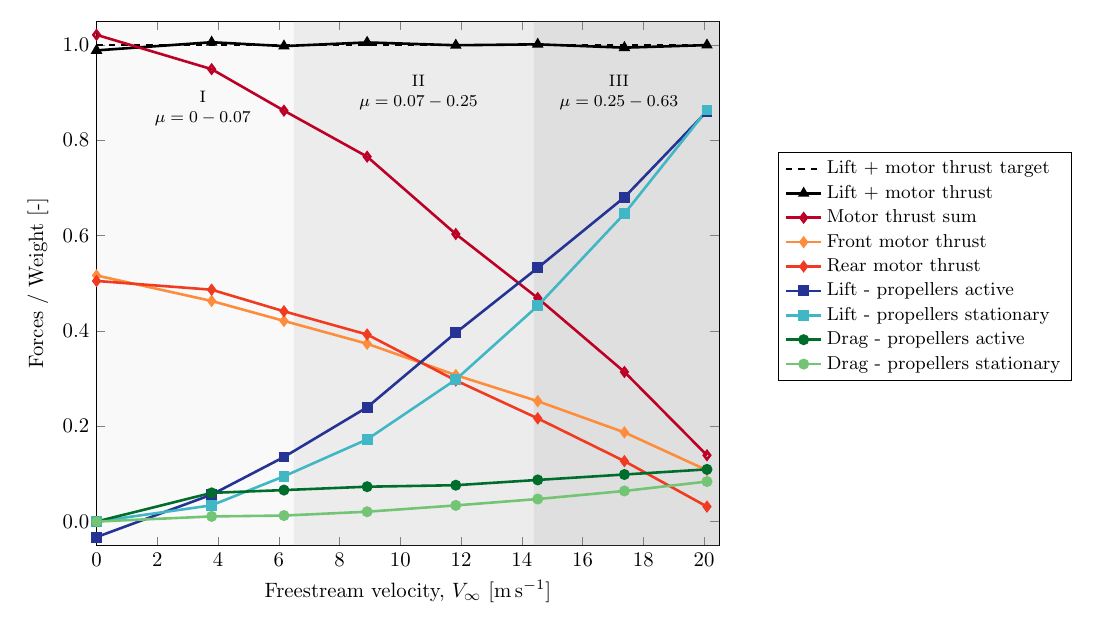}
        \caption{Forces non-dimensionalized by aircraft weight versus freestream velocity during the transition of the \textcolor{Blues-L}{FDRU} B2 configuration. Here $\mu = V_{\infty}/n_{F}R_{P}$ is the advance ratio of the front propeller.}
        \label{fig:FDRU_s transition summary}
  \end{subfigure}

  \vspace{1em}

  \begin{subfigure}[b]{\textwidth}
    \centering
        	\begin{tabular}{l c c c c c c c c}
        		\hline
        		\hline
        		Freestream velocity, $V_{\infty}$ [\si{\meter\per\second}] & 0.0 & 3.8 & 6.2 & 8.9 & 11.8 & 14.5 & 17.4 & 20.1  \\
                \hline
        		Front propeller RPM, $n_{F}$ [RPM] & 6650 & 6150 & 5835 & 5270 & 4460 & 3815 & 3120 & 2185 \\
                Rear propeller RPM, $n_{R}$ [RPM]  & 6520 & 6245 & 5955 & 5500 & 4585 & 3685 & 2545 & 675 \\
        		\hline
        		\hline
        	\end{tabular}
    \caption{\textcolor{Blues-L}{FDRU} B2 front and rear propeller rotation rates during the transition.}
    \label{tab:FDRU_B2_RPMS}
  \end{subfigure}

  \caption{\textcolor{Blues-L}{FDRU} B2 transition summary.}
  \label{fig:FDRU_B2_transition_summary_with_RPM_table}
\end{figure}

\begin{figure}[h!]
        \centering
        \begin{subfigure}[t]{\textwidth}
            \centering
                \includegraphics[width=\linewidth]{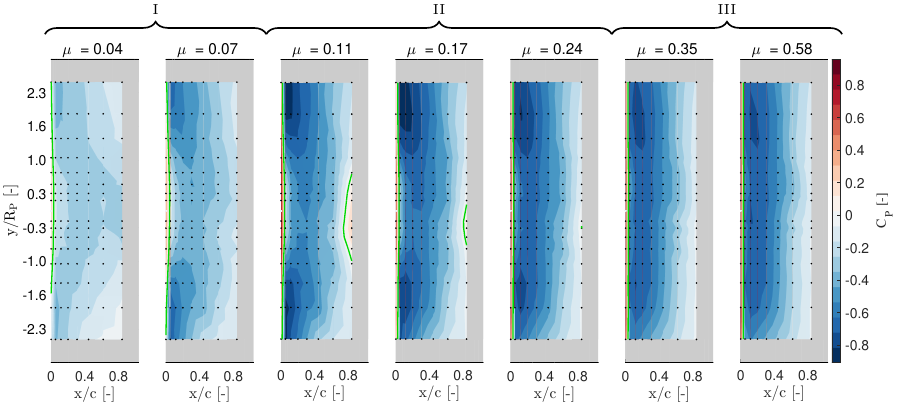}
            \caption[Suction wing surface pressure coefficient plots during the transition of the \textcolor{Blues-L}{FDRU} B2 configuration.]{Suction wing surface pressure coefficient plots during the transition of the \textcolor{Blues-L}{FDRU} B2 configuration, where \protect\markerone \ markings denote the location of a pressure tap and contours marked with \protect\greenline \ denote the $C_{P} = 0$ crossing location.}
            \label{fig:FDRU_short_cp}
            \end{subfigure}
        \begin{subfigure}[t]{\textwidth}
            \centering
            \includegraphics[width=\linewidth]{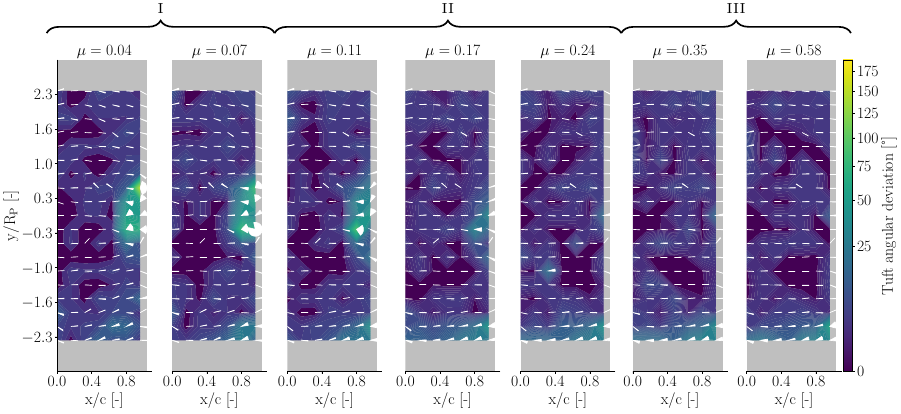}
            \caption{Suction wing surface tuft disturbances during the transition of the \textcolor{Blues-L}{FDRU} B2 configuration.}
            \label{fig:FDRU_short_tuft}
        \end{subfigure}
        \caption{Pressure coefficient and tuft disturbance plots during the transition of the \textcolor{Blues-L}{FDRU} B2 configuration.}
        \label{fig:FDRU_s_comp}
\end{figure}

\end{document}